\documentclass[aps,pra,twocolumn,groupedaddress]{revtex4-2}

\usepackage{epsfig}
\usepackage{amssymb}
\usepackage{color,pstcol}

\newcommand{\ymodifartitionehundredthirtyninevonestepone}[1]
{#1}
\newcommand{\ymodifartitionehundredthirtyninevonesteptwo}[1]
{#1}
\begin{document}

%
%
%



\newcommand{\ysssrcnb}
          {$\Ysssrcnb$}
\newcommand{\Ysssrcnb}
           {M}

\newcommand{\ysssensnb}
          {$\Ysssensnb$}
\newcommand{\Ysssensnb}
           {P}

\newcommand{\yssoutsepsystsigassocconttimecentnb}
{$\Yssoutsepsystsigassocconttimecentnb$}
\newcommand{\Yssoutsepsystsigassocconttimecentnb}
{L}

\newcommand{\yssbothtimevalnb}
          {$\Yssbothtimevalnb$}
\newcommand{\Yssbothtimevalnb}
           {N}

\newcommand{\yssbothfreqvalnb}
          {$\Yssbothfreqvalnb$}
\newcommand{\Yssbothfreqvalnb}
           {\Yssbothtimevalnb^{\prime}}

\newcommand{\yssbothscalevalnb}
          {$\Yssbothscalevalnb$}
\newcommand{\Yssbothscalevalnb}
           {\Yssbothtimevalnb^{\prime \prime}}



\newcommand{\yssconttimeval}
          {$\Yssconttimeval$}
\newcommand{\Yssconttimeval}
           {t}

\newcommand{\yssconttimevalsubone}
          {$\Yssconttimevalsubone$}
\newcommand{\Yssconttimevalsubone}
           {\Yssconttimeval _1}

\newcommand{\yssconttimevalsubtwo}
          {$\Yssconttimevalsubtwo$}
\newcommand{\Yssconttimevalsubtwo}
           {\Yssconttimeval _2}

\newcommand{\ysssubconttimeval}
          {$\Ysssubconttimeval$}
\newcommand{\Ysssubconttimeval}
           {p}

\newcommand{\yssconttimevalsubstd}
          {$\Yssconttimevalsubstd$}
\newcommand{\Yssconttimevalsubstd}
           {\Yssconttimeval _{\Ysssubconttimeval}}

\newcommand{\ysssubotherconttimeval}
          {$\Ysssubotherconttimeval$}
\newcommand{\Ysssubotherconttimeval}
           {q}

\newcommand{\yssconttimevalsubstdother}
          {$\Yssconttimevalsubstdother$}
\newcommand{\Yssconttimevalsubstdother}
           {\Yssconttimeval _{\Ysssubotherconttimeval}}

\newcommand{\yssconttimevalother}
          {$\Yssconttimevalother$}
\newcommand{\Yssconttimevalother}
           {t^{\prime}}

\newcommand{\yssconttimevalothersubone}
          {$\Yssconttimevalothersubone$}
\newcommand{\Yssconttimevalothersubone}
           {\Yssconttimevalother _1}

\newcommand{\yssconttimevalothersubtwo}
          {$\Yssconttimevalothersubtwo$}
\newcommand{\Yssconttimevalothersubtwo}
           {\Yssconttimevalother _2}





\newcommand{\yssdisctimeval}
          {$\Yssdisctimeval$}
\newcommand{\Yssdisctimeval}
           {n}

\newcommand{\yssdisctimevalsubone}
          {$\Yssdisctimevalsubone$}
\newcommand{\Yssdisctimevalsubone}
           {\Yssdisctimeval _1}

\newcommand{\yssdisctimevalsubtwo}
          {$\Yssdisctimevalsubtwo$}
\newcommand{\Yssdisctimevalsubtwo}
           {\Yssdisctimeval _2}

\newcommand{\ysssubdisctimeval}
          {$\Ysssubdisctimeval$}
\newcommand{\Ysssubdisctimeval}
           {i}

\newcommand{\yssdisctimevalsubstd}
          {$\Yssdisctimevalsubstd$}
\newcommand{\Yssdisctimevalsubstd}
           {\Yssdisctimeval _{\Ysssubdisctimeval}}

\newcommand{\ysssubotherdisctimeval}
          {$\Ysssubotherdisctimeval$}
\newcommand{\Ysssubotherdisctimeval}
           {j}

\newcommand{\yssdisctimevalsubstdother}
          {$\Yssdisctimevalsubstdother$}
\newcommand{\Yssdisctimevalsubstdother}
           {\Yssdisctimeval _{\Ysssubotherdisctimeval}}

\newcommand{\yssdisctimevalother}
          {$\Yssdisctimevalother$}
\newcommand{\Yssdisctimevalother}
           {m}

\newcommand{\yssdisctimevalothersubone}
          {$\Yssdisctimevalothersubone$}
\newcommand{\Yssdisctimevalothersubone}
           {\Yssdisctimevalother _1}

\newcommand{\yssdisctimevalothersubtwo}
          {$\Yssdisctimevalothersubtwo$}
\newcommand{\Yssdisctimevalothersubtwo}
           {\Yssdisctimevalother _2}


\newcommand{\yssdisctimelag}
          {$\Yssdisctimelag$}
\newcommand{\Yssdisctimelag}
           {l}

\newcommand{\yssdisctimelagsubone}
          {$\Yssdisctimelagsubone$}
\newcommand{\Yssdisctimelagsubone}
           {\Yssdisctimelag _1}

\newcommand{\yssdisctimelagsubtwo}
          {$\Yssdisctimelagsubtwo$}
\newcommand{\Yssdisctimelagsubtwo}
           {\Yssdisctimelag _2}


\newcommand{\ysscontfreqval}
          {$\Ysscontfreqval$}
\newcommand{\Ysscontfreqval}
           {\omega}

\newcommand{\ysssubcontfreqval}
          {$\Ysssubcontfreqval$}
\newcommand{\Ysssubcontfreqval}
           {\Ysssubconttimeval^{\prime}}

\newcommand{\ysscontfreqvalsubstd}
          {$\Ysscontfreqvalsubstd$}
\newcommand{\Ysscontfreqvalsubstd}
           {\Ysscontfreqval _{\Ysssubcontfreqval}}



\newcommand{\ysstimescalecontshiftval}
          {$\Ysstimescalecontshiftval$}
\newcommand{\Ysstimescalecontshiftval}
           {%
       \tau
       }

\newcommand{\ysstimescalecontscaleval}
          {$\Ysstimescalecontscaleval$}
\newcommand{\Ysstimescalecontscaleval}
           {%
       d
       }

\newcommand{\ysstimescalecontscaleexp}
          {$\Ysstimescalecontscaleexp$}
\newcommand{\Ysstimescalecontscaleexp}
           {%
       j
       }

\newcommand{\ysssubtimescalecontscaleval}
          {$\Ysssubtimescalecontscaleval$}
\newcommand{\Ysssubtimescalecontscaleval}
           {\Ysssubconttimeval^{\prime \prime}}

\newcommand{\ysstimescalecontscalevalsubstd}
          {$\Ysstimescalecontscalevalsubstd$}
\newcommand{\Ysstimescalecontscalevalsubstd}
           {\Ysstimescalecontscaleval _{\Ysssubtimescalecontscaleval}}

\newcommand{\ysstimescalecontmotherwav}
          {$\Ysstimescalecontmotherwav$}
\newcommand{\Ysstimescalecontmotherwav}
           {\psi}

\newcommand{\ysstimescalecontmotherwavval}
          {$\Ysstimescalecontmotherwavval$}
\newcommand{\Ysstimescalecontmotherwavval}
           {\Ysstimescalecontmotherwav ( \Yssconttimeval )}

\newcommand{\ysstimescalecontscaleshiftwav}
          {$\Ysstimescalecontscaleshiftwav$}
\newcommand{\Ysstimescalecontscaleshiftwav}
           {\Ysstimescalecontmotherwav
        _{\Ysstimescalecontshiftval , \Ysstimescalecontscaleval}
       }

\newcommand{\ysstimescalecontscaleshiftwavval}
          {$\Ysstimescalecontscaleshiftwavval$}
\newcommand{\Ysstimescalecontscaleshiftwavval}
           {\Ysstimescalecontscaleshiftwav ( \Yssconttimeval )}

\newcommand{\ysstimescalecontcoefnot}
          {$\Ysstimescalecontcoefnot$}
\newcommand{\Ysstimescalecontcoefnot}
           {W}


\newcommand{\ysssepsystlag}
          {$\Ysssepsystlag$}
\newcommand{\Ysssepsystlag}
           {k}

\newcommand{\ysscomplfiltlag}
          {$\Ysscomplfiltlag$}
\newcommand{\Ysscomplfiltlag}
           {m}

\newcommand{\ysscomplfiltlagother}
          {$\Ysscomplfiltlagother$}
\newcommand{\Ysscomplfiltlagother}
           {l}


\newcommand{\ysswayindex}
          {$\Ysswayindex$}
\newcommand{\Ysswayindex}
           {i}

\newcommand{\ysswayindexother}
          {$\Ysswayindexother$}
\newcommand{\Ysswayindexother}
           {j}

%
\newcommand{\ysswayindexothervertwo}
          {$\Ysswayindexothervertwo$}
\newcommand{\Ysswayindexothervertwo}
           {\Ysswayindexother^{\prime}}
%
\newcommand{\ysswayindexotherverthree}
          {$\Ysswayindexotherverthree$}
\newcommand{\Ysswayindexotherverthree}
           {\Ysswayindexother^{\prime\prime}}
%

\newcommand{\ysswayindexthird}
          {$\Ysswayindexthird$}
\newcommand{\Ysswayindexthird}
           {k}

%
\newcommand{\ysswayindexthirdvertwo}
          {$\Ysswayindexthirdvertwo$}
\newcommand{\Ysswayindexthirdvertwo}
           {\Ysswayindexthird^{\prime}}
%
\newcommand{\ysswayindexthirdverthree}
          {$\Ysswayindexthirdverthree$}
\newcommand{\Ysswayindexthirdverthree}
           {\Ysswayindexthird^{\prime\prime}}
%

\newcommand{\ysswayindexfourth}
          {$\Ysswayindexfourth$}
\newcommand{\Ysswayindexfourth}
           {\ell}

%
\newcommand{\ysswayindexfifth}
          {$\Ysswayindexfifth$}
\newcommand{\Ysswayindexfifth}
           {m}
%




\newcommand{\ysssrcsiginnovdisctimecentvec}
          {$\Ysssrcsiginnovdisctimecentvec$}
\newcommand{\Ysssrcsiginnovdisctimecentvec}
           {p}

\newcommand{\ysssrcsiginnovdisctimecentvecval}
          {$\Ysssrcsiginnovdisctimecentvecval$}
\newcommand{\Ysssrcsiginnovdisctimecentvecval}
           {\Ysssrcsiginnovdisctimecentvec (\Yssdisctimeval)}

\newcommand{\ysssrcsiginnnovdisctimecentone}
          {$\Ysssrcsiginnovdisctimecentone$}
\newcommand{\Ysssrcsiginnovdisctimecentone}
           {\Ysssrcsiginnovdisctimecentvec _{1}}

\newcommand{\ysssrcsiginnovdisctimecentoneval}
          {$\Ysssrcsiginnovdisctimecentoneval$}
\newcommand{\Ysssrcsiginnovdisctimecentoneval}
           {\Ysssrcsiginnovdisctimecentone (\Yssdisctimeval)}

\newcommand{\ysssrcsiginnnovdisctimecenttwo}
          {$\Ysssrcsiginnovdisctimecenttwo$}
\newcommand{\Ysssrcsiginnovdisctimecenttwo}
           {\Ysssrcsiginnovdisctimecentvec _{2}}

\newcommand{\ysssrcsiginnovdisctimecenttwoval}
          {$\Ysssrcsiginnovdisctimecenttwoval$}
\newcommand{\Ysssrcsiginnovdisctimecenttwoval}
           {\Ysssrcsiginnovdisctimecenttwo (\Yssdisctimeval)}

\newcommand{\ysssrcsiginnovdisctimecentindex}
          {$\Ysssrcsiginnovdisctimecentindex$}
\newcommand{\Ysssrcsiginnovdisctimecentindex}
           {\Ysssrcsiginnovdisctimecentvec _{\Ysswayindex}}

\newcommand{\ysssrcsiginnovdisctimecentindexval}
          {$\Ysssrcsiginnovdisctimecentindexval$}
\newcommand{\Ysssrcsiginnovdisctimecentindexval}
           {\Ysssrcsiginnovdisctimecentindex (\Yssdisctimeval)}

\newcommand{\ysssrcsiginnovdisctimecentindexother}
          {$\Ysssrcsiginnovdisctimecentindexother$}
\newcommand{\Ysssrcsiginnovdisctimecentindexother}
           {\Ysssrcsiginnovdisctimecentvec _{\Ysswayindexother}}

\newcommand{\ysssrcsiginnovdisctimecentindexotherval}
          {$\Ysssrcsiginnovdisctimecentindexotherval$}
\newcommand{\Ysssrcsiginnovdisctimecentindexotherval}
           {\Ysssrcsiginnovdisctimecentindexother (\Yssdisctimeval)}


\newcommand{\ysssrcsiginnovdisctimecentveczt}
          {$\Ysssrcsiginnovdisctimecentveczt$}
\newcommand{\Ysssrcsiginnovdisctimecentveczt}
           {P}

\newcommand{\ysssrcsiginnovdisctimecentvecztval}
          {$\Ysssrcsiginnovdisctimecentvecztval$}
\newcommand{\Ysssrcsiginnovdisctimecentvecztval}
           {\Ysssrcsiginnovdisctimecentveczt (z)}

\newcommand{\ysssrcsiginnovdisctimecentindexzt}
          {$\Ysssrcsiginnovdisctimecentindexzt$}
\newcommand{\Ysssrcsiginnovdisctimecentindexzt}
           {\Ysssrcsiginnovdisctimecentveczt _{\Ysswayindex}}

\newcommand{\ysssrcsiginnovdisctimecentindexztval}
          {$\Ysssrcsiginnovdisctimecentindexztval$}
\newcommand{\Ysssrcsiginnovdisctimecentindexztval}
           {\Ysssrcsiginnovdisctimecentindexzt (z)}



\newcommand{\ysssrcsigconttimecentvec}
          {$\Ysssrcsigconttimecentvec$}
\newcommand{\Ysssrcsigconttimecentvec}
           {s}

\newcommand{\ysssrcsigconttimecentvecval}
          {$\Ysssrcsigconttimecentvecval$}
\newcommand{\Ysssrcsigconttimecentvecval}
           {\Ysssrcsigconttimecentvec (\Yssconttimeval)}

\newcommand{\ysssrcsigconttimecentone}
          {$\Ysssrcsigconttimecentone$}
\newcommand{\Ysssrcsigconttimecentone}
           {\Ysssrcsigconttimecentvec _{1}}

\newcommand{\ysssrcsigconttimecentoneval}
          {$\Ysssrcsigconttimecentoneval$}
\newcommand{\Ysssrcsigconttimecentoneval}
           {\Ysssrcsigconttimecentone (\Yssconttimeval)}

\newcommand{\ysssrcsigconttimecenttwo}
          {$\Ysssrcsigconttimecenttwo$}
\newcommand{\Ysssrcsigconttimecenttwo}
           {\Ysssrcsigconttimecentvec _{2}}

\newcommand{\ysssrcsigconttimecenttwoval}
          {$\Ysssrcsigconttimecenttwoval$}
\newcommand{\Ysssrcsigconttimecenttwoval}
           {\Ysssrcsigconttimecenttwo (\Yssconttimeval)}

\newcommand{\ysssrcsigconttimecentthree}
          {$\Ysssrcsigconttimecentthree$}
\newcommand{\Ysssrcsigconttimecentthree}
           {\Ysssrcsigconttimecentvec _{3}}

\newcommand{\ysssrcsigconttimecentthreeval}
          {$\Ysssrcsigconttimecentthreeval$}
\newcommand{\Ysssrcsigconttimecentthreeval}
           {\Ysssrcsigconttimecentthree (\Yssconttimeval)}

\newcommand{\ysssrcsigconttimecentindex}
          {$\Ysssrcsigconttimecentindex$}
\newcommand{\Ysssrcsigconttimecentindex}
           {\Ysssrcsigconttimecentvec _{\Ysswayindex}}

\newcommand{\ysssrcsigconttimecentindexval}
          {$\Ysssrcsigconttimecentindexval$}
\newcommand{\Ysssrcsigconttimecentindexval}
           {\Ysssrcsigconttimecentindex (\Yssconttimeval)}

\newcommand{\ysssrcsigconttimecentindexother}
          {$\Ysssrcsigconttimecentindexother$}
\newcommand{\Ysssrcsigconttimecentindexother}
           {\Ysssrcsigconttimecentvec _{\Ysswayindexother}}

\newcommand{\ysssrcsigconttimecentindexotherval}
          {$\Ysssrcsigconttimecentindexotherval$}
\newcommand{\Ysssrcsigconttimecentindexotherval}
           {\Ysssrcsigconttimecentindexother (\Yssconttimeval)}

\newcommand{\ysssrcsigconttimecentindexthird}
          {$\Ysssrcsigconttimecentindexthird$}
\newcommand{\Ysssrcsigconttimecentindexthird}
           {\Ysssrcsigconttimecentvec _{\Ysswayindexthird}}

\newcommand{\ysssrcsigconttimecentindexthirdval}
          {$\Ysssrcsigconttimecentindexthirdval$}
\newcommand{\Ysssrcsigconttimecentindexthirdval}
           {\Ysssrcsigconttimecentindexthird (\Yssconttimeval)}

\newcommand{\ysssrcsigconttimecentindexfourth}
          {$\Ysssrcsigconttimecentindexfourth$}
\newcommand{\Ysssrcsigconttimecentindexfourth}
           {\Ysssrcsigconttimecentvec _{\Ysswayindexfourth}}

\newcommand{\ysssrcsigconttimecentindexfourthval}
          {$\Ysssrcsigconttimecentindexfourthval$}
\newcommand{\Ysssrcsigconttimecentindexfourthval}
           {\Ysssrcsigconttimecentindexfourth (\Yssconttimeval)}

\newcommand{\ysssrcsigconttimecentlast}
          {$\Ysssrcsigconttimecentlast$}
\newcommand{\Ysssrcsigconttimecentlast}
           {\Ysssrcsigconttimecentvec _{\Ysssrcnb}}

\newcommand{\ysssrcsigconttimecentlastval}
          {$\Ysssrcsigconttimecentlastval$}
\newcommand{\Ysssrcsigconttimecentlastval}
           {\Ysssrcsigconttimecentlast (\Yssconttimeval)}



\newcommand{\ysssrcsigconttimecentvecrand}
          {$\Ysssrcsigconttimecentvecrand$}
\newcommand{\Ysssrcsigconttimecentvecrand}
           {S}

\newcommand{\ysssrcsigconttimecentonerand}
          {$\Ysssrcsigconttimecentonerand$}
\newcommand{\Ysssrcsigconttimecentonerand}
           {\Ysssrcsigconttimecentvecrand _{1}}




\newcommand{\ysssrcsigassocconttimecentvec}
          {$\Ysssrcsigassocconttimecentvec$}
\newcommand{\Ysssrcsigassocconttimecentvec}
           {\Ysssrcsigconttimecentvec ^{\prime}}

\newcommand{\ysssrcsigassocconttimecentvecval}
          {$\Ysssrcsigassocconttimecentvecval$}
\newcommand{\Ysssrcsigassocconttimecentvecval}
           {\Ysssrcsigassocconttimecentvec (\Yssconttimeval)}

\newcommand{\ysssrcsigassocconttimecentone}
          {$\Ysssrcsigassocconttimecentone$}
\newcommand{\Ysssrcsigassocconttimecentone}
           {\Ysssrcsigassocconttimecentvec _{1}}

\newcommand{\ysssrcsigassocconttimecentoneval}
          {$\Ysssrcsigassocconttimecentoneval$}
\newcommand{\Ysssrcsigassocconttimecentoneval}
           {\Ysssrcsigassocconttimecentone (\Yssconttimeval)}

\newcommand{\ysssrcsigassocconttimecenttwo}
          {$\Ysssrcsigassocconttimecenttwo$}
\newcommand{\Ysssrcsigassocconttimecenttwo}
           {\Ysssrcsigassocconttimecentvec _{2}}

\newcommand{\ysssrcsigassocconttimecenttwoval}
          {$\Ysssrcsigassocconttimecenttwoval$}
\newcommand{\Ysssrcsigassocconttimecenttwoval}
           {\Ysssrcsigassocconttimecenttwo (\Yssconttimeval)}

\newcommand{\ysssrcsigassocconttimecentthree}
          {$\Ysssrcsigassocconttimecentthree$}
\newcommand{\Ysssrcsigassocconttimecentthree}
           {\Ysssrcsigassocconttimecentvec _{3}}

\newcommand{\ysssrcsigassocconttimecentthreeval}
          {$\Ysssrcsigassocconttimecentthreeval$}
\newcommand{\Ysssrcsigassocconttimecentthreeval}
           {\Ysssrcsigassocconttimecentthree (\Yssconttimeval)}

\newcommand{\ysssrcsigassocconttimecentindex}
          {$\Ysssrcsigassocconttimecentindex$}
\newcommand{\Ysssrcsigassocconttimecentindex}
           {\Ysssrcsigassocconttimecentvec _{\Ysswayindex}}

\newcommand{\ysssrcsigassocconttimecentindexval}
          {$\Ysssrcsigassocconttimecentindexval$}
\newcommand{\Ysssrcsigassocconttimecentindexval}
           {\Ysssrcsigassocconttimecentindex (\Yssconttimeval)}

\newcommand{\ysssrcsigassocconttimecentindexother}
          {$\Ysssrcsigassocconttimecentindexother$}
\newcommand{\Ysssrcsigassocconttimecentindexother}
           {\Ysssrcsigassocconttimecentvec _{\Ysswayindexother}}

\newcommand{\ysssrcsigassocconttimecentindexotherval}
          {$\Ysssrcsigassocconttimecentindexotherval$}
\newcommand{\Ysssrcsigassocconttimecentindexotherval}
           {\Ysssrcsigassocconttimecentindexother (\Yssconttimeval)}

\newcommand{\ysssrcsigassocconttimecentindexthird}
          {$\Ysssrcsigassocconttimecentindexthird$}
\newcommand{\Ysssrcsigassocconttimecentindexthird}
           {\Ysssrcsigassocconttimecentvec _{\Ysswayindexthird}}

\newcommand{\ysssrcsigassocconttimecentindexthirdval}
          {$\Ysssrcsigassocconttimecentindexthirdval$}
\newcommand{\Ysssrcsigassocconttimecentindexthirdval}
           {\Ysssrcsigassocconttimecentindexthird (\Yssconttimeval)}

\newcommand{\ysssrcsigassocconttimecentindexfourth}
{$\Ysssrcsigassocconttimecentindexfourth$}
\newcommand{\Ysssrcsigassocconttimecentindexfourth}
{\Ysssrcsigassocconttimecentvec _{\Ysswayindexfourth}}

\newcommand{\ysssrcsigassocconttimecentindexfourthval}
{$\Ysssrcsigassocconttimecentindexfourthval$}
\newcommand{\Ysssrcsigassocconttimecentindexfourthval}
{\Ysssrcsigassocconttimecentindexfourth (\Yssconttimeval)}

\newcommand{\ysssrcsigassocconttimecentlast}
          {$\Ysssrcsigassocconttimecentlast$}
\newcommand{\Ysssrcsigassocconttimecentlast}
           {\Ysssrcsigassocconttimecentvec _{\Ysssrcnb}}

\newcommand{\ysssrcsigassocconttimecentlastval}
          {$\Ysssrcsigassocconttimecentlastval$}
\newcommand{\Ysssrcsigassocconttimecentlastval}
           {\Ysssrcsigassocconttimecentlast (\Yssconttimeval)}



\newcommand{\ysssrcsigassocconttimecentvecrand}
          {$\Ysssrcsigassocconttimecentvecrand$}
\newcommand{\Ysssrcsigassocconttimecentvecrand}
           {S ^{\prime}}

\newcommand{\ysssrcsigassocconttimecentonerand}
          {$\Ysssrcsigassocconttimecentonerand$}
\newcommand{\Ysssrcsigassocconttimecentonerand}
           {\Ysssrcsigassocconttimecentvecrand _{1}}

\newcommand{\ysssrcsigassocconttimecenttworand}
          {$\Ysssrcsigassocconttimecenttworand$}
\newcommand{\Ysssrcsigassocconttimecenttworand}
           {\Ysssrcsigassocconttimecentvecrand _{2}}


\newcommand{\ysssrcsigassocconttimecentindexrand}
          {$\Ysssrcsigassocconttimecentindexrand$}
\newcommand{\Ysssrcsigassocconttimecentindexrand}
           {\Ysssrcsigassocconttimecentvecrand _{\Ysswayindex}}


\newcommand{\ysssrcsigassocconttimecentlastrand}
          {$\Ysssrcsigassocconttimecentlastrand$}
\newcommand{\Ysssrcsigassocconttimecentlastrand}
           {\Ysssrcsigassocconttimecentvecrand _{\Ysssrcnb}}



\newcommand{\ysssrcsigconttimecontfreqnoncentvec}
          {$\Ysssrcsigconttimecontfreqnoncentvec$}
\newcommand{\Ysssrcsigconttimecontfreqnoncentvec}
           {S}

\newcommand{\ysssrcsigconttimecontfreqnoncentvecval}
          {$\Ysssrcsigconttimecontfreqnoncentvecval$}
\newcommand{\Ysssrcsigconttimecontfreqnoncentvecval}
           {\Ysssrcsigconttimecontfreqnoncentvec ( \Yssconttimeval , \Ysscontfreqval ) }

\newcommand{\ysssrcsigconttimecontfreqnoncentone}
          {$\Ysssrcsigconttimecontfreqnoncentone$}
\newcommand{\Ysssrcsigconttimecontfreqnoncentone}
           {\Ysssrcsigconttimecontfreqnoncentvec _{1}}

\newcommand{\ysssrcsigconttimecontfreqnoncentoneval}
          {$\Ysssrcsigconttimecontfreqnoncentoneval$}
\newcommand{\Ysssrcsigconttimecontfreqnoncentoneval}
           {\Ysssrcsigconttimecontfreqnoncentone ( \Yssconttimeval , \Ysscontfreqval ) }

\newcommand{\ysssrcsigconttimecontfreqnoncenttwo}
          {$\Ysssrcsigconttimecontfreqnoncenttwo$}
\newcommand{\Ysssrcsigconttimecontfreqnoncenttwo}
           {\Ysssrcsigconttimecontfreqnoncentvec _{2}}

\newcommand{\ysssrcsigconttimecontfreqnoncenttwoval}
          {$\Ysssrcsigconttimecontfreqnoncenttwoval$}
\newcommand{\Ysssrcsigconttimecontfreqnoncenttwoval}
           {\Ysssrcsigconttimecontfreqnoncenttwo ( \Yssconttimeval , \Ysscontfreqval ) }

\newcommand{\ysssrcsigconttimecontfreqnoncentindex}
          {$\Ysssrcsigconttimecontfreqnoncentindex$}
\newcommand{\Ysssrcsigconttimecontfreqnoncentindex}
           {\Ysssrcsigconttimecontfreqnoncentvec _{\Ysswayindex}}

\newcommand{\ysssrcsigconttimecontfreqnoncentindexval}
          {$\Ysssrcsigconttimecontfreqnoncentindexval$}
\newcommand{\Ysssrcsigconttimecontfreqnoncentindexval}
           {\Ysssrcsigconttimecontfreqnoncentindex ( \Yssconttimeval , \Ysscontfreqval ) }

\newcommand{\ysssrcsigconttimecontfreqnoncentindexother}
          {$\Ysssrcsigconttimecontfreqnoncentindexother$}
\newcommand{\Ysssrcsigconttimecontfreqnoncentindexother}
           {\Ysssrcsigconttimecontfreqnoncentvec _{\Ysswayindexother}}

\newcommand{\ysssrcsigconttimecontfreqnoncentindexotherval}
          {$\Ysssrcsigconttimecontfreqnoncentindexotherval$}
\newcommand{\Ysssrcsigconttimecontfreqnoncentindexotherval}
           {\Ysssrcsigconttimecontfreqnoncentindexother ( \Yssconttimeval , \Ysscontfreqval ) }

\newcommand{\ysssrcsigconttimecontfreqnoncentindexthird}
          {$\Ysssrcsigconttimecontfreqnoncentindexthird$}
\newcommand{\Ysssrcsigconttimecontfreqnoncentindexthird}
           {\Ysssrcsigconttimecontfreqnoncentvec _{\Ysswayindexthird}}

\newcommand{\ysssrcsigconttimecontfreqnoncentindexthirdval}
          {$\Ysssrcsigconttimecontfreqnoncentindexthirdval$}
\newcommand{\Ysssrcsigconttimecontfreqnoncentindexthirdval}
           {\Ysssrcsigconttimecontfreqnoncentindexthird ( \Yssconttimeval , \Ysscontfreqval ) }

\newcommand{\ysssrcsigconttimecontfreqnoncentindexfourth}
          {$\Ysssrcsigconttimecontfreqnoncentindexfourth$}
\newcommand{\Ysssrcsigconttimecontfreqnoncentindexfourth}
           {\Ysssrcsigconttimecontfreqnoncentvec _{\Ysswayindexfourth}}

\newcommand{\ysssrcsigconttimecontfreqnoncentindexfourthval}
          {$\Ysssrcsigconttimecontfreqnoncentindexfourthval$}
\newcommand{\Ysssrcsigconttimecontfreqnoncentindexfourthval}
           {\Ysssrcsigconttimecontfreqnoncentindexfourth ( \Yssconttimeval , \Ysscontfreqval ) }


\newcommand{\ysssrcsigconttimecontscalecentindexother}
          {$\Ysssrcsigconttimecontscalecentindexother$}
\newcommand{\Ysssrcsigconttimecontscalecentindexother}
           {\Ysstimescalecontcoefnot _{\Ysssrcsigconttimecentvec _{\Ysswayindexother}}}

\newcommand{\ysssrcsigconttimecontscalecentindexotherval}
          {$\Ysssrcsigconttimecontscalecentindexotherval$}
\newcommand{\Ysssrcsigconttimecontscalecentindexotherval}
           {\Ysssrcsigconttimecontscalecentindexother
        ( \Ysstimescalecontshiftval , \Ysstimescalecontscaleval )
           }

\newcommand{\ysssrcsigconttimecontscalecentindexothervalsubstd}
          {$\Ysssrcsigconttimecontscalecentindexothervalsubstd$}
\newcommand{\Ysssrcsigconttimecontscalecentindexothervalsubstd}
           {\Ysssrcsigconttimecontscalecentindexother
        ( \Ysstimescalecontshiftval , \Ysstimescalecontscalevalsubstd )
           }



\newcommand{\ysssrcsigdisctimecentmatrix}
          {$\Ysssrcsigdisctimecentmatrix$}
\newcommand{\Ysssrcsigdisctimecentmatrix}
%
%
%
           {S}


%
\newcommand{\ysssrcsigdisctimecentmatrixadapt}
          {$\Ysssrcsigdisctimecentmatrixadapt$}
\newcommand{\Ysssrcsigdisctimecentmatrixadapt}
           {\check{S}_
L
}
%


\newcommand{\ysssrcsigdisctimecentmatrixestim}
          {$\Ysssrcsigdisctimecentmatrixestim$}
\newcommand{\Ysssrcsigdisctimecentmatrixestim}
           {\hat{\Ysssrcsigdisctimecentmatrix}}


\newcommand{\ysssrcsigdisctimecentvec}
          {$\Ysssrcsigdisctimecentvec$}
\newcommand{\Ysssrcsigdisctimecentvec}
           {s}

%
\newcommand{\ysssrcsigdisctimecentvecestim}
          {$\Ysssrcsigdisctimecentvecestim$}
\newcommand{\Ysssrcsigdisctimecentvecestim}
           {\hat{\Ysssrcsigdisctimecentvec}}

\newcommand{\ysssrcsigdisctimecentvecval}
          {$\Ysssrcsigdisctimecentvecval$}
\newcommand{\Ysssrcsigdisctimecentvecval}
           {\Ysssrcsigdisctimecentvec (\Yssdisctimeval)}

\newcommand{\ysssrcsigdisctimecentone}
          {$\Ysssrcsigdisctimecentone$}
\newcommand{\Ysssrcsigdisctimecentone}
           {\Ysssrcsigdisctimecentvec _{1}}

\newcommand{\ysssrcsigdisctimecentoneval}
          {$\Ysssrcsigdisctimecentoneval$}
\newcommand{\Ysssrcsigdisctimecentoneval}
           {\Ysssrcsigdisctimecentone (\Yssdisctimeval)}

\newcommand{\ysssrcsigdisctimecenttwo}
          {$\Ysssrcsigdisctimecenttwo$}
\newcommand{\Ysssrcsigdisctimecenttwo}
           {\Ysssrcsigdisctimecentvec _{2}}

\newcommand{\ysssrcsigdisctimecenttwoval}
          {$\Ysssrcsigdisctimecenttwoval$}
\newcommand{\Ysssrcsigdisctimecenttwoval}
           {\Ysssrcsigdisctimecenttwo (\Yssdisctimeval)}

\newcommand{\ysssrcsigdisctimecentthree}
          {$\Ysssrcsigdisctimecentthree$}
\newcommand{\Ysssrcsigdisctimecentthree}
           {\Ysssrcsigdisctimecentvec _{3}}

\newcommand{\ysssrcsigdisctimecentthreeval}
          {$\Ysssrcsigdisctimecentthreeval$}
\newcommand{\Ysssrcsigdisctimecentthreeval}
           {\Ysssrcsigdisctimecentthree (\Yssdisctimeval)}

\newcommand{\ysssrcsigdisctimecentindex}
          {$\Ysssrcsigdisctimecentindex$}
\newcommand{\Ysssrcsigdisctimecentindex}
           {\Ysssrcsigdisctimecentvec _{\Ysswayindex}}

\newcommand{\ysssrcsigdisctimecentindexval}
          {$\Ysssrcsigdisctimecentindexval$}
\newcommand{\Ysssrcsigdisctimecentindexval}
           {\Ysssrcsigdisctimecentindex (\Yssdisctimeval)}

\newcommand{\ysssrcsigdisctimecentindexother}
          {$\Ysssrcsigdisctimecentindexother$}
\newcommand{\Ysssrcsigdisctimecentindexother}
           {\Ysssrcsigdisctimecentvec _{\Ysswayindexother}}

%
\newcommand{\ysssrcsigdisctimecentindexotherestim}
          {$\Ysssrcsigdisctimecentindexotherestim$}
\newcommand{\Ysssrcsigdisctimecentindexotherestim}
           {\Ysssrcsigdisctimecentvecestim _{\Ysswayindexother}}

\newcommand{\ysssrcsigdisctimecentindexotherval}
          {$\Ysssrcsigdisctimecentindexotherval$}
\newcommand{\Ysssrcsigdisctimecentindexotherval}
           {\Ysssrcsigdisctimecentindexother (\Yssdisctimeval)}

%
\newcommand{\ysssrcsigdisctimecentindexothervalestim}
          {$\Ysssrcsigdisctimecentindexothervalestim$}
\newcommand{\Ysssrcsigdisctimecentindexothervalestim}
           {\Ysssrcsigdisctimecentindexotherestim (\Yssdisctimeval)}


\newcommand{\ysssrcsigdisctimecentindexothervertwo}
          {$\Ysssrcsigdisctimecentindexothervertwo$}
\newcommand{\Ysssrcsigdisctimecentindexothervertwo}
           {\Ysssrcsigdisctimecentvec _{\Ysswayindexothervertwo}}


\newcommand{\ysssrcsigdisctimecentindexthird}
          {$\Ysssrcsigdisctimecentindexthird$}
\newcommand{\Ysssrcsigdisctimecentindexthird}
           {\Ysssrcsigdisctimecentvec _{\Ysswayindexthird}}

\newcommand{\ysssrcsigdisctimecentindexthirdval}
          {$\Ysssrcsigdisctimecentindexthirdval$}
\newcommand{\Ysssrcsigdisctimecentindexthirdval}
           {\Ysssrcsigdisctimecentindexthird (\Yssdisctimeval)}


\newcommand{\ysssrcsigdisctimecentindexthirdvertwo}
          {$\Ysssrcsigdisctimecentindexthirdvertwo$}
\newcommand{\Ysssrcsigdisctimecentindexthirdvertwo}
           {\Ysssrcsigdisctimecentvec _{\Ysswayindexthirdvertwo}}

\newcommand{\ysssrcsigdisctimecentindexfourth}
          {$\Ysssrcsigdisctimecentindexfourth$}
\newcommand{\Ysssrcsigdisctimecentindexfourth}
           {\Ysssrcsigdisctimecentvec _{\Ysswayindexfourth}}


\newcommand{\ysssrcsigdisctimecentlast}
          {$\Ysssrcsigdisctimecentlast$}
\newcommand{\Ysssrcsigdisctimecentlast}
           {\Ysssrcsigdisctimecentvec _{\Ysssrcnb}}

\newcommand{\ysssrcsigdisctimecentlastval}
          {$\Ysssrcsigdisctimecentlastval$}
\newcommand{\Ysssrcsigdisctimecentlastval}
           {\Ysssrcsigdisctimecentlast (\Yssdisctimeval)}



\newcommand{\ysssrcsigassocdisctimecentvec}
          {$\Ysssrcsigassocdisctimecentvec$}
\newcommand{\Ysssrcsigassocdisctimecentvec}
           {\Ysssrcsigdisctimecentvec ^{\prime}}

\newcommand{\ysssrcsigassocdisctimecentindexother}
          {$\Ysssrcsigassocdisctimecentindexother$}
\newcommand{\Ysssrcsigassocdisctimecentindexother}
           {\Ysssrcsigassocdisctimecentvec _{\Ysswayindexother}}

\newcommand{\ysssrcsigassocdisctimecentindexotherval}
          {$\Ysssrcsigassocdisctimecentindexotherval$}
\newcommand{\Ysssrcsigassocdisctimecentindexotherval}
           {\Ysssrcsigassocdisctimecentindexother (\Yssdisctimeval)}



\newcommand{\ysssrcsigassoctwodisctimecentvec}
          {$\Ysssrcsigassoctwodisctimecentvec$}
\newcommand{\Ysssrcsigassoctwodisctimecentvec}
           {
\widetilde
{\Ysssrcsigdisctimecentvec}}

\newcommand{\ysssrcsigassoctwodisctimecentvecval}
          {$\Ysssrcsigassoctwodisctimecentvecval$}
\newcommand{\Ysssrcsigassoctwodisctimecentvecval}
           {\Ysssrcsigassoctwodisctimecentvec (\Yssdisctimeval)}

\newcommand{\ysssrcsigassoctwodisctimecentone}
{$\Ysssrcsigassoctwodisctimecentone$}
\newcommand{\Ysssrcsigassoctwodisctimecentone}
{\Ysssrcsigassoctwodisctimecentvec _{1}}

\newcommand{\ysssrcsigassoctwodisctimecentoneval}
{$\Ysssrcsigassoctwodisctimecentoneval$}
\newcommand{\Ysssrcsigassoctwodisctimecentoneval}
{\Ysssrcsigassoctwodisctimecentone (\Yssdisctimeval)}

\newcommand{\ysssrcsigassoctwodisctimecenttwo}
{$\Ysssrcsigassoctwodisctimecenttwo$}
\newcommand{\Ysssrcsigassoctwodisctimecenttwo}
{\Ysssrcsigassoctwodisctimecentvec _{2}}

\newcommand{\ysssrcsigassoctwodisctimecenttwoval}
{$\Ysssrcsigassoctwodisctimecenttwoval$}
\newcommand{\Ysssrcsigassoctwodisctimecenttwoval}
{\Ysssrcsigassoctwodisctimecenttwo (\Yssdisctimeval)}

\newcommand{\ysssrcsigassoctwodisctimecentindex}
{$\Ysssrcsigassoctwodisctimecentindex$}
\newcommand{\Ysssrcsigassoctwodisctimecentindex}
{\Ysssrcsigassoctwodisctimecentvec _{\Ysswayindex}}

\newcommand{\ysssrcsigassoctwodisctimecentindexval}
{$\Ysssrcsigassoctwodisctimecentindexval$}
\newcommand{\Ysssrcsigassoctwodisctimecentindexval}
{\Ysssrcsigassoctwodisctimecentindex (\Yssdisctimeval)}

\newcommand{\ysssrcsigassoctwodisctimecentindexother}
          {$\Ysssrcsigassoctwodisctimecentindexother$}
\newcommand{\Ysssrcsigassoctwodisctimecentindexother}
           {\Ysssrcsigassoctwodisctimecentvec _{\Ysswayindexother}}

\newcommand{\ysssrcsigassoctwodisctimecentindexotherval}
          {$\Ysssrcsigassoctwodisctimecentindexotherval$}
\newcommand{\Ysssrcsigassoctwodisctimecentindexotherval}
           {\Ysssrcsigassoctwodisctimecentindexother (\Yssdisctimeval)}


\newcommand{\ysssrcsigdisctimecontfreqnotonenoncentvec}
          {$\Ysssrcsigdisctimecontfreqnotonenoncentvec$}
\newcommand{\Ysssrcsigdisctimecontfreqnotonenoncentvec}
           {S}

\newcommand{\ysssrcsigdisctimecontfreqnotonenoncentvecval}
          {$\Ysssrcsigdisctimecontfreqnotonenoncentvecval$}
\newcommand{\Ysssrcsigdisctimecontfreqnotonenoncentvecval}
           {\Ysssrcsigdisctimecontfreqnotonenoncentvec ( \Ysscontfreqval ) }

\newcommand{\ysssrcsigdisctimecontfreqnotonenoncentone}
          {$\Ysssrcsigdisctimecontfreqnotonenoncentone$}
\newcommand{\Ysssrcsigdisctimecontfreqnotonenoncentone}
           {\Ysssrcsigdisctimecontfreqnotonenoncentvec _{1}}

\newcommand{\ysssrcsigdisctimecontfreqnotonenoncentoneval}
          {$\Ysssrcsigdisctimecontfreqnotonenoncentoneval$}
\newcommand{\Ysssrcsigdisctimecontfreqnotonenoncentoneval}
           {\Ysssrcsigdisctimecontfreqnotonenoncentone (\Ysscontfreqval ) }

\newcommand{\ysssrcsigdisctimecontfreqnotonenoncentindexother}
          {$\Ysssrcsigdisctimecontfreqnotonenoncentindexother$}
\newcommand{\Ysssrcsigdisctimecontfreqnotonenoncentindexother}
           {\Ysssrcsigdisctimecontfreqnotonenoncentvec _{\Ysswayindexother}}

\newcommand{\ysssrcsigdisctimecontfreqnotonenoncentindexotherval}
          {$\Ysssrcsigdisctimecontfreqnotonenoncentindexotherval$}
\newcommand{\Ysssrcsigdisctimecontfreqnotonenoncentindexotherval}
           {\Ysssrcsigdisctimecontfreqnotonenoncentindexother (\Ysscontfreqval ) }

\newcommand{\ysssrcsigdisctimecontfreqnotonenoncentlast}
          {$\Ysssrcsigdisctimecontfreqnotonenoncentlast$}
\newcommand{\Ysssrcsigdisctimecontfreqnotonenoncentlast}
           {\Ysssrcsigdisctimecontfreqnotonenoncentvec _{\Ysssrcnb}}

\newcommand{\ysssrcsigdisctimecontfreqnotonenoncentlastval}
          {$\Ysssrcsigdisctimecontfreqnotonenoncentlastval$}
\newcommand{\Ysssrcsigdisctimecontfreqnotonenoncentlastval}
           {\Ysssrcsigdisctimecontfreqnotonenoncentlast (\Ysscontfreqval ) }


\newcommand{\ysssrcsigassocdisctimecontfreqnotonenoncentvec}
          {$\Ysssrcsigassocdisctimecontfreqnotonenoncentvec$}
\newcommand{\Ysssrcsigassocdisctimecontfreqnotonenoncentvec}
           {\Ysssrcsigdisctimecontfreqnotonenoncentvec ^{\prime}}

\newcommand{\ysssrcsigassocdisctimecontfreqnotonenoncentvecval}
          {$\Ysssrcsigassocdisctimecontfreqnotonenoncentvecval$}
\newcommand{\Ysssrcsigassocdisctimecontfreqnotonenoncentvecval}
           {\Ysssrcsigassocdisctimecontfreqnotonenoncentvec ( \Ysscontfreqval ) }

\newcommand{\ysssrcsigassocdisctimecontfreqnotonenoncentone}
          {$\Ysssrcsigassocdisctimecontfreqnotonenoncentone$}
\newcommand{\Ysssrcsigassocdisctimecontfreqnotonenoncentone}
           {\Ysssrcsigassocdisctimecontfreqnotonenoncentvec _{1}}

\newcommand{\ysssrcsigassocdisctimecontfreqnotonenoncentoneval}
          {$\Ysssrcsigassocdisctimecontfreqnotonenoncentoneval$}
\newcommand{\Ysssrcsigassocdisctimecontfreqnotonenoncentoneval}
           {\Ysssrcsigassocdisctimecontfreqnotonenoncentone (\Ysscontfreqval ) }

\newcommand{\ysssrcsigassocdisctimecontfreqnotonenoncentindexother}
          {$\Ysssrcsigassocdisctimecontfreqnotonenoncentindexother$}
\newcommand{\Ysssrcsigassocdisctimecontfreqnotonenoncentindexother}
           {\Ysssrcsigassocdisctimecontfreqnotonenoncentvec _{\Ysswayindexother}}

\newcommand{\ysssrcsigassocdisctimecontfreqnotonenoncentindexotherval}
          {$\Ysssrcsigassocdisctimecontfreqnotonenoncentindexotherval$}
\newcommand{\Ysssrcsigassocdisctimecontfreqnotonenoncentindexotherval}
           {\Ysssrcsigassocdisctimecontfreqnotonenoncentindexother (\Ysscontfreqval ) }

\newcommand{\ysssrcsigassocdisctimecontfreqnotonenoncentlast}
          {$\Ysssrcsigassocdisctimecontfreqnotonenoncentlast$}
\newcommand{\Ysssrcsigassocdisctimecontfreqnotonenoncentlast}
           {\Ysssrcsigassocdisctimecontfreqnotonenoncentvec _{\Ysssrcnb}}

\newcommand{\ysssrcsigassocdisctimecontfreqnotonenoncentlastval}
          {$\Ysssrcsigassocdisctimecontfreqnotonenoncentlastval$}
\newcommand{\Ysssrcsigassocdisctimecontfreqnotonenoncentlastval}
           {\Ysssrcsigassocdisctimecontfreqnotonenoncentlast (\Ysscontfreqval ) }


\newcommand{\ysssrcsigdisctimecontfreqnoncentvec}
          {$\Ysssrcsigdisctimecontfreqnoncentvec$}
\newcommand{\Ysssrcsigdisctimecontfreqnoncentvec}
           {S}

\newcommand{\ysssrcsigdisctimecontfreqnoncentvecval}
          {$\Ysssrcsigdisctimecontfreqnoncentvecval$}
\newcommand{\Ysssrcsigdisctimecontfreqnoncentvecval}
           {\Ysssrcsigdisctimecontfreqnoncentvec ( \Yssdisctimeval , \Ysscontfreqval ) }

\newcommand{\ysssrcsigdisctimecontfreqnoncentone}
          {$\Ysssrcsigdisctimecontfreqnoncentone$}
\newcommand{\Ysssrcsigdisctimecontfreqnoncentone}
           {\Ysssrcsigdisctimecontfreqnoncentvec _{1}}

\newcommand{\ysssrcsigdisctimecontfreqnoncentoneval}
          {$\Ysssrcsigdisctimecontfreqnoncentoneval$}
\newcommand{\Ysssrcsigdisctimecontfreqnoncentoneval}
           {\Ysssrcsigdisctimecontfreqnoncentone ( \Yssdisctimeval , \Ysscontfreqval ) }

\newcommand{\ysssrcsigdisctimecontfreqnoncentindex}
          {$\Ysssrcsigdisctimecontfreqnoncentindex$}
\newcommand{\Ysssrcsigdisctimecontfreqnoncentindex}
           {\Ysssrcsigdisctimecontfreqnoncentvec _{\Ysswayindex}}

\newcommand{\ysssrcsigdisctimecontfreqnoncentindexval}
          {$\Ysssrcsigdisctimecontfreqnoncentindexval$}
\newcommand{\Ysssrcsigdisctimecontfreqnoncentindexval}
           {\Ysssrcsigdisctimecontfreqnoncentindex ( \Yssdisctimeval , \Ysscontfreqval ) }

\newcommand{\ysssrcsigdisctimecontfreqnoncentindexother}
          {$\Ysssrcsigdisctimecontfreqnoncentindexother$}
\newcommand{\Ysssrcsigdisctimecontfreqnoncentindexother}
           {\Ysssrcsigdisctimecontfreqnoncentvec _{\Ysswayindexother}}

\newcommand{\ysssrcsigdisctimecontfreqnoncentindexotherval}
          {$\Ysssrcsigdisctimecontfreqnoncentindexotherval$}
\newcommand{\Ysssrcsigdisctimecontfreqnoncentindexotherval}
           {\Ysssrcsigdisctimecontfreqnoncentindexother ( \Yssdisctimeval , \Ysscontfreqval ) }

\newcommand{\ysssrcsigdisctimecontfreqnoncentindexthird}
          {$\Ysssrcsigdisctimecontfreqnoncentindexthird$}
\newcommand{\Ysssrcsigdisctimecontfreqnoncentindexthird}
           {\Ysssrcsigdisctimecontfreqnoncentvec _{\Ysswayindexthird}}

\newcommand{\ysssrcsigdisctimecontfreqnoncentindexthirdval}
          {$\Ysssrcsigdisctimecontfreqnoncentindexthirdval$}
\newcommand{\Ysssrcsigdisctimecontfreqnoncentindexthirdval}
           {\Ysssrcsigdisctimecontfreqnoncentindexthird ( \Yssdisctimeval , \Ysscontfreqval ) }


\newcommand{\ysssrcsigassoctwodisctimecontfreqnoncentvec}
          {$\Ysssrcsigassoctwodisctimecontfreqnoncentvec$}
\newcommand{\Ysssrcsigassoctwodisctimecontfreqnoncentvec}
           {
\widetilde
{\Ysssrcsigdisctimecontfreqnoncentvec}}

\newcommand{\ysssrcsigassoctwodisctimecontfreqnoncentvecval}
          {$\Ysssrcsigassoctwodisctimecontfreqnoncentvecval$}
\newcommand{\Ysssrcsigassoctwodisctimecontfreqnoncentvecval}
           {\Ysssrcsigassoctwodisctimecontfreqnoncentvec ( \Yssdisctimeval , \Ysscontfreqval ) }

\newcommand{\ysssrcsigassoctwodisctimecontfreqnoncentindexother}
          {$\Ysssrcsigassoctwodisctimecontfreqnoncentindexother$}
\newcommand{\Ysssrcsigassoctwodisctimecontfreqnoncentindexother}
           {\Ysssrcsigassoctwodisctimecontfreqnoncentvec _{\Ysswayindexother}}

\newcommand{\ysssrcsigassoctwodisctimecontfreqnoncentindexotherval}
          {$\Ysssrcsigassoctwodisctimecontfreqnoncentindexotherval$}
\newcommand{\Ysssrcsigassoctwodisctimecontfreqnoncentindexotherval}
           {\Ysssrcsigassoctwodisctimecontfreqnoncentindexother ( \Yssdisctimeval , \Ysscontfreqval ) }


\newcommand{\ysssrcsigdisctimecentveczt}
          {$\Ysssrcsigdisctimecentveczt$}
\newcommand{\Ysssrcsigdisctimecentveczt}
           {S}

\newcommand{\ysssrcsigdisctimecentvecztval}
          {$\Ysssrcsigdisctimecentvecztval$}
\newcommand{\Ysssrcsigdisctimecentvecztval}
           {\Ysssrcsigdisctimecentveczt (z)}

\newcommand{\ysssrcsigdisctimecentonezt}
          {$\Ysssrcsigdisctimecentonezt$}
\newcommand{\Ysssrcsigdisctimecentonezt}
           {\Ysssrcsigdisctimecentveczt _{1}}

\newcommand{\ysssrcsigdisctimecentoneztval}
          {$\Ysssrcsigdisctimecentoneztval$}
\newcommand{\Ysssrcsigdisctimecentoneztval}
           {\Ysssrcsigdisctimecentonezt (z)}

\newcommand{\ysssrcsigdisctimecenttwozt}
          {$\Ysssrcsigdisctimecenttwozt$}
\newcommand{\Ysssrcsigdisctimecenttwozt}
           {\Ysssrcsigdisctimecentveczt _{2}}

\newcommand{\ysssrcsigdisctimecenttwoztval}
          {$\Ysssrcsigdisctimecenttwoztval$}
\newcommand{\Ysssrcsigdisctimecenttwoztval}
           {\Ysssrcsigdisctimecenttwozt (z)}

\newcommand{\ysssrcsigdisctimecentthreezt}
          {$\Ysssrcsigdisctimecentthreezt$}
\newcommand{\Ysssrcsigdisctimecentthreezt}
           {\Ysssrcsigdisctimecentveczt _{3}}

\newcommand{\ysssrcsigdisctimecentthreeztval}
          {$\Ysssrcsigdisctimecentthreeztval$}
\newcommand{\Ysssrcsigdisctimecentthreeztval}
           {\Ysssrcsigdisctimecentthreezt (z)}

\newcommand{\ysssrcsigdisctimecentindexzt}
          {$\Ysssrcsigdisctimecentindexzt$}
\newcommand{\Ysssrcsigdisctimecentindexzt}
           {\Ysssrcsigdisctimecentveczt _{\Ysswayindex}}

\newcommand{\ysssrcsigdisctimecentindexztval}
          {$\Ysssrcsigdisctimecentindexztval$}
\newcommand{\Ysssrcsigdisctimecentindexztval}
           {\Ysssrcsigdisctimecentindexzt (z)}

\newcommand{\ysssrcsigdisctimecentindexotherzt}
          {$\Ysssrcsigdisctimecentindexotherzt$}
\newcommand{\Ysssrcsigdisctimecentindexotherzt}
           {\Ysssrcsigdisctimecentveczt _{\Ysswayindexother}}

\newcommand{\ysssrcsigdisctimecentindexotherztval}
          {$\Ysssrcsigdisctimecentindexotherztval$}
\newcommand{\Ysssrcsigdisctimecentindexotherztval}
           {\Ysssrcsigdisctimecentindexotherzt (z)}


\newcommand{\ysssrcsigdisctimecentindexthirdzt}
          {$\Ysssrcsigdisctimecentindexthirdzt$}
\newcommand{\Ysssrcsigdisctimecentindexthirdzt}
           {\Ysssrcsigdisctimecentveczt _{\Ysswayindexthird}}

\newcommand{\ysssrcsigdisctimecentindexthirdztval}
          {$\Ysssrcsigdisctimecentindexthirdztval$}
\newcommand{\Ysssrcsigdisctimecentindexthirdztval}
           {\Ysssrcsigdisctimecentindexthirdzt (z)}


\newcommand{\ysssrcsigdisctimecentlastzt}
          {$\Ysssrcsigdisctimecentlastzt$}
\newcommand{\Ysssrcsigdisctimecentlastzt}
           {\Ysssrcsigdisctimecentveczt _{\Ysssrcnb}}

\newcommand{\ysssrcsigdisctimecentlastztval}
          {$\Ysssrcsigdisctimecentlastztval$}
\newcommand{\Ysssrcsigdisctimecentlastztval}
           {\Ysssrcsigdisctimecentlastzt (z)}

%
%
%
%
%
\newcommand{\yssquadsrcsigdisctimecentvec}
          {$\Yssquadsrcsigdisctimecentvec$}
\newcommand{\Yssquadsrcsigdisctimecentvec}
           {
p
}
%
%
%
\newcommand{\yssquadsrcsigdisctimecentvecval}
          {$\Yssquadsrcsigdisctimecentvecval$}
\newcommand{\Yssquadsrcsigdisctimecentvecval}
           {\Yssquadsrcsigdisctimecentvec (\Yssdisctimeval)}
%
%
%
\newcommand{\yssquadsrcsigdisctimecentindexotherindexthird}
          {$\Yssquadsrcsigdisctimecentindexotherindexthird$}
\newcommand{\Yssquadsrcsigdisctimecentindexotherindexthird}
{ 
\Yssquadsrcsigdisctimecentvec
_{\Ysswayindexother \Ysswayindexthird}}
%
\newcommand{\yssquadsrcsigdisctimecentindexotherindexthirdval}
          {$\Yssquadsrcsigdisctimecentindexotherindexthirdval$}
\newcommand{\Yssquadsrcsigdisctimecentindexotherindexthirdval}
           {\Yssquadsrcsigdisctimecentindexotherindexthird (\Yssdisctimeval)}
%
%
%
%
%
\newcommand{\yssextquadsrcsigdisctimecentmatrix}
          {$\Yssextquadsrcsigdisctimecentmatrix$}
\newcommand{\Yssextquadsrcsigdisctimecentmatrix}
           {
\widetilde
{\Ysssrcsigdisctimecentmatrix}}

%
%
\newcommand{\yssextquadsrcsigdisctimecentmatrixadapt}
          {$\Yssextquadsrcsigdisctimecentmatrixadapt$}
\newcommand{\Yssextquadsrcsigdisctimecentmatrixadapt}
           {\check{\Ysssrcsigdisctimecentmatrix}}
%
%

\newcommand{\yssextquadsrcsigdisctimecentmatrixestim}
          {$\Yssextquadsrcsigdisctimecentmatrixestim$}
\newcommand{\Yssextquadsrcsigdisctimecentmatrixestim}
%
           {\widehat{\Yssextquadsrcsigdisctimecentmatrix}}
%
%
%
\newcommand{\yssextquadsrcsigdisctimecentvec}
          {$\Yssextquadsrcsigdisctimecentvec$}
\newcommand{\Yssextquadsrcsigdisctimecentvec}
           {
\widetilde
{\Ysssrcsigdisctimecentvec}}
%
%
%
\newcommand{\yssextquadsrcsigdisctimecentvecval}
          {$\Yssextquadsrcsigdisctimecentvecval$}
\newcommand{\Yssextquadsrcsigdisctimecentvecval}
           {\Yssextquadsrcsigdisctimecentvec (\Yssdisctimeval)}
%

%
\newcommand{\yssextquadsrcsigdisctimecentmatrixestimnotvecallsampsrcnot}
          {$\Yssextquadsrcsigdisctimecentmatrixestimnotvecallsampsrcnot$}
\newcommand{\Yssextquadsrcsigdisctimecentmatrixestimnotvecallsampsrcnot}
%
{\check{s}}
%
%
%
%
\newcommand{\yssextquadsrcsigdisctimecentmatrixestimvecallsampsrcone}
          {$\Yssextquadsrcsigdisctimecentmatrixestimvecallsampsrcone$}
\newcommand{\Yssextquadsrcsigdisctimecentmatrixestimvecallsampsrcone}
{\Yssextquadsrcsigdisctimecentmatrixestimnotvecallsampsrcnot_{1}}
%
%
%
\newcommand{\yssextquadsrcsigdisctimecentmatrixestimvecallsampsrctwo}
          {$\Yssextquadsrcsigdisctimecentmatrixestimvecallsampsrctwo$}
\newcommand{\Yssextquadsrcsigdisctimecentmatrixestimvecallsampsrctwo}
{\Yssextquadsrcsigdisctimecentmatrixestimnotvecallsampsrcnot_{2}}
%
%
%
\newcommand{\yssextquadsrcsigdisctimecentmatrixestimvecallsampsrcindexother}
          {$\Yssextquadsrcsigdisctimecentmatrixestimvecallsampsrcindexother$}
\newcommand{\Yssextquadsrcsigdisctimecentmatrixestimvecallsampsrcindexother}
{\Yssextquadsrcsigdisctimecentmatrixestimnotvecallsampsrcnot_{\Ysswayindexother}}
%
%
%
\newcommand{\yssextquadsrcsigdisctimecentmatrixestimvecallsampsrcindexthird}
          {$\Yssextquadsrcsigdisctimecentmatrixestimvecallsampsrcindexthird$}
\newcommand{\Yssextquadsrcsigdisctimecentmatrixestimvecallsampsrcindexthird}
{\Yssextquadsrcsigdisctimecentmatrixestimnotvecallsampsrcnot_{\Ysswayindexthird}}
%
%
%
%
\newcommand{\yssextquadsrcsigdisctimecentmatrixestimvecallsampsrcindexfourth}
          {$\Yssextquadsrcsigdisctimecentmatrixestimvecallsampsrcindexfourth$}
\newcommand{\Yssextquadsrcsigdisctimecentmatrixestimvecallsampsrcindexfourth}
{\Yssextquadsrcsigdisctimecentmatrixestimnotvecallsampsrcnot_{\Ysswayindexfourth}}
%
%
%
\newcommand{\yssextquadsrcsigdisctimecentmatrixestimvecallsampsrcindexfifth}
          {$\Yssextquadsrcsigdisctimecentmatrixestimvecallsampsrcindexfifth$}
\newcommand{\Yssextquadsrcsigdisctimecentmatrixestimvecallsampsrcindexfifth}
{\Yssextquadsrcsigdisctimecentmatrixestimnotvecallsampsrcnot_{\Ysswayindexfifth}}
%
%
%
%
\newcommand{\yssextquadsrcsigdisctimecentmatrixestimvecallsampsrcoriglast}
          {$\Yssextquadsrcsigdisctimecentmatrixestimvecallsampsrcoriglast$}
\newcommand{\Yssextquadsrcsigdisctimecentmatrixestimvecallsampsrcoriglast}
{\Yssextquadsrcsigdisctimecentmatrixestimnotvecallsampsrcnot_{\Ysssrcnb}}
%



\newcommand{\yssmixsigconttimecentvec}
          {$\Yssmixsigconttimecentvec$}
\newcommand{\Yssmixsigconttimecentvec}
           {x}

\newcommand{\yssmixsigconttimecentvecval}
          {$\Yssmixsigconttimecentvecval$}
\newcommand{\Yssmixsigconttimecentvecval}
           {\Yssmixsigconttimecentvec (\Yssconttimeval)}

\newcommand{\yssmixsigconttimecentone}
          {$\Yssmixsigconttimecentone$}
\newcommand{\Yssmixsigconttimecentone}
           {\Yssmixsigconttimecentvec _{1}}

\newcommand{\yssmixsigconttimecentoneval}
          {$\Yssmixsigconttimecentoneval$}
\newcommand{\Yssmixsigconttimecentoneval}
           {\Yssmixsigconttimecentone (\Yssconttimeval)}

\newcommand{\yssmixsigconttimecenttwo}
          {$\Yssmixsigconttimecenttwo$}
\newcommand{\Yssmixsigconttimecenttwo}
           {\Yssmixsigconttimecentvec _{2}}

\newcommand{\yssmixsigconttimecenttwoval}
          {$\Yssmixsigconttimecenttwoval$}
\newcommand{\Yssmixsigconttimecenttwoval}
           {\Yssmixsigconttimecenttwo (\Yssconttimeval)}

\newcommand{\yssmixsigconttimecentindex}
          {$\Yssmixsigconttimecentindex$}
\newcommand{\Yssmixsigconttimecentindex}
           {\Yssmixsigconttimecentvec _{\Ysswayindex}}

\newcommand{\yssmixsigconttimecentindexval}
          {$\Yssmixsigconttimecentindexval$}
\newcommand{\Yssmixsigconttimecentindexval}
           {\Yssmixsigconttimecentindex (\Yssconttimeval)}

\newcommand{\yssmixsigconttimecentindexother}
          {$\Yssmixsigconttimecentindexother$}
\newcommand{\Yssmixsigconttimecentindexother}
           {\Yssmixsigconttimecentvec _{\Ysswayindexother}}

\newcommand{\yssmixsigconttimecentindexotherval}
          {$\Yssmixsigconttimecentindexotherval$}
\newcommand{\Yssmixsigconttimecentindexotherval}
           {\Yssmixsigconttimecentindexother (\Yssconttimeval)}

\newcommand{\yssmixsigconttimecentindexthird}
          {$\Yssmixsigconttimecentindexthird$}
\newcommand{\Yssmixsigconttimecentindexthird}
           {\Yssmixsigconttimecentvec _{\Ysswayindexthird}}

\newcommand{\yssmixsigconttimecentindexthirdval}
          {$\Yssmixsigconttimecentindexthirdval$}
\newcommand{\Yssmixsigconttimecentindexthirdval}
           {\Yssmixsigconttimecentindexthird (\Yssconttimeval)}

\newcommand{\yssmixsigconttimecentlast}
          {$\Yssmixsigconttimecentlast$}
\newcommand{\Yssmixsigconttimecentlast}
           {\Yssmixsigconttimecentvec _{\Ysssensnb}}

\newcommand{\yssmixsigconttimecentlastval}
          {$\Yssmixsigconttimecentlastval$}
\newcommand{\Yssmixsigconttimecentlastval}
           {\Yssmixsigconttimecentlast (\Yssconttimeval)}

\newcommand{\yssmixsigconttimecentsrcnb}
          {$\Yssmixsigconttimecentsrcnb$}
\newcommand{\Yssmixsigconttimecentsrcnb}
           {\Yssmixsigconttimecentvec _{\Ysssrcnb}}


\newcommand{\yssmixsigconttimecentsrcnbval}
          {$\Yssmixsigconttimecentsrcnbval$}
\newcommand{\Yssmixsigconttimecentsrcnbval}
           {\Yssmixsigconttimecentsrcnb (\Yssconttimeval)}


\newcommand{\yssmixsigassocconttimecentvec}
{$\Yssmixsigassocconttimecentvec$}
\newcommand{\Yssmixsigassocconttimecentvec}
{\Yssmixsigconttimecentvec ^{\prime}}

\newcommand{\yssmixsigassocconttimecentvecval}
{$\Yssmixsigassocconttimecentvecval$}
\newcommand{\Yssmixsigassocconttimecentvecval}
{\Yssmixsigassocconttimecentvec (\Yssconttimeval)}

\newcommand{\yssmixsigassocconttimecentone}
{$\Yssmixsigassocconttimecentone$}
\newcommand{\Yssmixsigassocconttimecentone}
{\Yssmixsigassocconttimecentvec _{1}}

\newcommand{\yssmixsigassocconttimecentoneval}
{$\Yssmixsigassocconttimecentoneval$}
\newcommand{\Yssmixsigassocconttimecentoneval}
{\Yssmixsigassocconttimecentone (\Yssconttimeval)}

\newcommand{\yssmixsigassocconttimecentindex}
{$\Yssmixsigassocconttimecentindex$}
\newcommand{\Yssmixsigassocconttimecentindex}
{\Yssmixsigassocconttimecentvec _{\Ysswayindex}}

\newcommand{\yssmixsigassocconttimecentindexval}
{$\Yssmixsigassocconttimecentindexval$}
\newcommand{\Yssmixsigassocconttimecentindexval}
{\Yssmixsigassocconttimecentindex (\Yssconttimeval)}

\newcommand{\yssmixsigassocconttimecentindexthird}
{$\Yssmixsigassocconttimecentindexthird$}
\newcommand{\Yssmixsigassocconttimecentindexthird}
{\Yssmixsigassocconttimecentvec _{\Ysswayindexthird}}

\newcommand{\yssmixsigassocconttimecentindexthirdval}
{$\Yssmixsigassocconttimecentindexthirdval$}
\newcommand{\Yssmixsigassocconttimecentindexthirdval}
{\Yssmixsigassocconttimecentindexthird (\Yssconttimeval)}


\newcommand{\yssmixsigconttimecontfreqnoncentvec}
          {$\Yssmixsigconttimecontfreqnoncentvec$}
\newcommand{\Yssmixsigconttimecontfreqnoncentvec}
           {X}

\newcommand{\yssmixsigconttimecontfreqnoncentvecval}
          {$\Yssmixsigconttimecontfreqnoncentvecval$}
\newcommand{\Yssmixsigconttimecontfreqnoncentvecval}
           {\Yssmixsigconttimecontfreqnoncentvec ( \Yssconttimeval , \Ysscontfreqval ) }

\newcommand{\yssmixsigconttimecontfreqnoncentone}
          {$\Yssmixsigconttimecontfreqnoncentone$}
\newcommand{\Yssmixsigconttimecontfreqnoncentone}
           {\Yssmixsigconttimecontfreqnoncentvec _{1}}

\newcommand{\yssmixsigconttimecontfreqnoncentoneval}
          {$\Yssmixsigconttimecontfreqnoncentoneval$}
\newcommand{\Yssmixsigconttimecontfreqnoncentoneval}
           {\Yssmixsigconttimecontfreqnoncentone ( \Yssconttimeval , \Ysscontfreqval ) }

\newcommand{\yssmixsigconttimecontfreqnoncenttwo}
          {$\Yssmixsigconttimecontfreqnoncenttwo$}
\newcommand{\Yssmixsigconttimecontfreqnoncenttwo}
           {\Yssmixsigconttimecontfreqnoncentvec _{2}}

\newcommand{\yssmixsigconttimecontfreqnoncenttwoval}
          {$\Yssmixsigconttimecontfreqnoncenttwoval$}
\newcommand{\Yssmixsigconttimecontfreqnoncenttwoval}
           {\Yssmixsigconttimecontfreqnoncenttwo ( \Yssconttimeval , \Ysscontfreqval ) }

\newcommand{\yssmixsigconttimecontfreqnoncentindex}
          {$\Yssmixsigconttimecontfreqnoncentindex$}
\newcommand{\Yssmixsigconttimecontfreqnoncentindex}
           {\Yssmixsigconttimecontfreqnoncentvec _{\Ysswayindex}}

\newcommand{\yssmixsigconttimecontfreqnoncentindexval}
          {$\Yssmixsigconttimecontfreqnoncentindexval$}
\newcommand{\Yssmixsigconttimecontfreqnoncentindexval}
           {\Yssmixsigconttimecontfreqnoncentindex ( \Yssconttimeval , \Ysscontfreqval ) }

\newcommand{\yssmixsigconttimecontfreqnoncentindexother}
          {$\Yssmixsigconttimecontfreqnoncentindexother$}
\newcommand{\Yssmixsigconttimecontfreqnoncentindexother}
           {\Yssmixsigconttimecontfreqnoncentvec _{\Ysswayindexother}}

\newcommand{\yssmixsigconttimecontfreqnoncentindexotherval}
          {$\Yssmixsigconttimecontfreqnoncentindexotherval$}
\newcommand{\Yssmixsigconttimecontfreqnoncentindexotherval}
           {\Yssmixsigconttimecontfreqnoncentindexother ( \Yssconttimeval , \Ysscontfreqval ) }

\newcommand{\yssmixsigconttimecontfreqnoncentindexthird}
          {$\Yssmixsigconttimecontfreqnoncentindexthird$}
\newcommand{\Yssmixsigconttimecontfreqnoncentindexthird}
           {\Yssmixsigconttimecontfreqnoncentvec _{\Ysswayindexthird}}

\newcommand{\yssmixsigconttimecontfreqnoncentindexthirdval}
          {$\Yssmixsigconttimecontfreqnoncentindexthirdval$}
\newcommand{\Yssmixsigconttimecontfreqnoncentindexthirdval}
           {\Yssmixsigconttimecontfreqnoncentindexthird ( \Yssconttimeval , \Ysscontfreqval ) }


\newcommand{\yssmixsigconttimecontscalecentindex}
          {$\Yssmixsigconttimecontscalecentindex$}
\newcommand{\Yssmixsigconttimecontscalecentindex}
           {\Ysstimescalecontcoefnot _{\Yssmixsigconttimecentvec _{\Ysswayindex}}}

\newcommand{\yssmixsigconttimecontscalecentindexval}
          {$\Yssmixsigconttimecontscalecentindexval$}
\newcommand{\Yssmixsigconttimecontscalecentindexval}
           {\Yssmixsigconttimecontscalecentindex
        ( \Ysstimescalecontshiftval , \Ysstimescalecontscaleval )
           }

\newcommand{\yssmixsigconttimecontscalecentindexvalsubstd}
          {$\Yssmixsigconttimecontscalecentindexvalsubstd$}
\newcommand{\Yssmixsigconttimecontscalecentindexvalsubstd}
           {\Yssmixsigconttimecontscalecentindex
        ( \Ysstimescalecontshiftval , \Ysstimescalecontscalevalsubstd )
           }



\newcommand{\yssmixsigdisctimecentmatrix}
          {$\Yssmixsigdisctimecentmatrix$}
\newcommand{\Yssmixsigdisctimecentmatrix}
           {X}


\newcommand{\yssmixsigdisctimecentvec}
          {$\Yssmixsigdisctimecentvec$}
\newcommand{\Yssmixsigdisctimecentvec}
           {x}

\newcommand{\yssmixsigdisctimecentvecval}
          {$\Yssmixsigdisctimecentvecval$}
\newcommand{\Yssmixsigdisctimecentvecval}
           {\Yssmixsigdisctimecentvec (\Yssdisctimeval)}

\newcommand{\yssmixsigdisctimecentone}
          {$\Yssmixsigdisctimecentone$}
\newcommand{\Yssmixsigdisctimecentone}
           {\Yssmixsigdisctimecentvec _{1}}

\newcommand{\yssmixsigdisctimecentoneval}
          {$\Yssmixsigdisctimecentoneval$}
\newcommand{\Yssmixsigdisctimecentoneval}
           {\Yssmixsigdisctimecentone (\Yssdisctimeval)}

\newcommand{\yssmixsigdisctimecenttwo}
          {$\Yssmixsigdisctimecenttwo$}
\newcommand{\Yssmixsigdisctimecenttwo}
           {\Yssmixsigdisctimecentvec _{2}}

\newcommand{\yssmixsigdisctimecenttwoval}
          {$\Yssmixsigdisctimecenttwoval$}
\newcommand{\Yssmixsigdisctimecenttwoval}
           {\Yssmixsigdisctimecenttwo (\Yssdisctimeval)}

\newcommand{\yssmixsigdisctimecentthree}
{$\Yssmixsigdisctimecentthree$}
\newcommand{\Yssmixsigdisctimecentthree}
{\Yssmixsigdisctimecentvec _{3}}

\newcommand{\yssmixsigdisctimecentthreeval}
{$\Yssmixsigdisctimecentthreeval$}
\newcommand{\Yssmixsigdisctimecentthreeval}
{\Yssmixsigdisctimecentthree (\Yssdisctimeval)}

\newcommand{\yssmixsigdisctimecentindex}
          {$\Yssmixsigdisctimecentindex$}
\newcommand{\Yssmixsigdisctimecentindex}
           {\Yssmixsigdisctimecentvec _{\Ysswayindex}}

\newcommand{\yssmixsigdisctimecentindexval}
          {$\Yssmixsigdisctimecentindexval$}
\newcommand{\Yssmixsigdisctimecentindexval}
           {\Yssmixsigdisctimecentindex (\Yssdisctimeval)}

\newcommand{\yssmixsigdisctimecentindexother}
{$\Yssmixsigdisctimecentindexother$}
\newcommand{\Yssmixsigdisctimecentindexother}
{\Yssmixsigdisctimecentvec _{\Ysswayindexother}}

\newcommand{\yssmixsigdisctimecentindexotherval}
{$\Yssmixsigdisctimecentindexotherval$}
\newcommand{\Yssmixsigdisctimecentindexotherval}
{\Yssmixsigdisctimecentindexother (\Yssdisctimeval)}

\newcommand{\yssmixsigdisctimecentlast}
          {$\Yssmixsigdisctimecentlast$}
\newcommand{\Yssmixsigdisctimecentlast}
           {\Yssmixsigdisctimecentvec _{\Ysssensnb}}

\newcommand{\yssmixsigdisctimecentlastval}
          {$\Yssmixsigdisctimecentlastval$}
\newcommand{\Yssmixsigdisctimecentlastval}
           {\Yssmixsigdisctimecentlast (\Yssdisctimeval)}

\newcommand{\yssmixsigdisctimecentsrcnb}
          {$\Yssmixsigdisctimecentsrcnb$}
\newcommand{\Yssmixsigdisctimecentsrcnb}
           {\Yssmixsigdisctimecentvec _{\Ysssrcnb}}


\newcommand{\yssmixsigdisctimecentsrcnbval}
          {$\Yssmixsigdisctimecentsrcnbval$}
\newcommand{\Yssmixsigdisctimecentsrcnbval}
           {\Yssmixsigdisctimecentsrcnb (\Yssdisctimeval)}


\newcommand{\yssmixsigdisctimecontfreqnotonenoncentvec}
          {$\Yssmixsigdisctimecontfreqnotonenoncentvec$}
\newcommand{\Yssmixsigdisctimecontfreqnotonenoncentvec}
           {X}

\newcommand{\yssmixsigdisctimecontfreqnotonenoncentvecval}
          {$\Yssmixsigdisctimecontfreqnotonenoncentvecval$}
\newcommand{\Yssmixsigdisctimecontfreqnotonenoncentvecval}
           {\Yssmixsigdisctimecontfreqnotonenoncentvec ( \Ysscontfreqval ) }

\newcommand{\yssmixsigdisctimecontfreqnotonenoncentone}
          {$\Yssmixsigdisctimecontfreqnotonenoncentone$}
\newcommand{\Yssmixsigdisctimecontfreqnotonenoncentone}
           {\Yssmixsigdisctimecontfreqnotonenoncentvec _{1}}

\newcommand{\yssmixsigdisctimecontfreqnotonenoncentoneval}
          {$\Yssmixsigdisctimecontfreqnotonenoncentoneval$}
\newcommand{\Yssmixsigdisctimecontfreqnotonenoncentoneval}
           {\Yssmixsigdisctimecontfreqnotonenoncentone (\Ysscontfreqval ) }

\newcommand{\yssmixsigdisctimecontfreqnotonenoncenttwo}
          {$\Yssmixsigdisctimecontfreqnotonenoncenttwo$}
\newcommand{\Yssmixsigdisctimecontfreqnotonenoncenttwo}
           {\Yssmixsigdisctimecontfreqnotonenoncentvec _{2}}

\newcommand{\yssmixsigdisctimecontfreqnotonenoncenttwoval}
          {$\Yssmixsigdisctimecontfreqnotonenoncenttwoval$}
\newcommand{\Yssmixsigdisctimecontfreqnotonenoncenttwoval}
           {\Yssmixsigdisctimecontfreqnotonenoncenttwo (\Ysscontfreqval ) }

\newcommand{\yssmixsigdisctimecontfreqnotonenoncentindex}
          {$\Yssmixsigdisctimecontfreqnotonenoncentindex$}
\newcommand{\Yssmixsigdisctimecontfreqnotonenoncentindex}
           {\Yssmixsigdisctimecontfreqnotonenoncentvec _{\Ysswayindex}}

\newcommand{\yssmixsigdisctimecontfreqnotonenoncentindexval}
          {$\Yssmixsigdisctimecontfreqnotonenoncentindexval$}
\newcommand{\Yssmixsigdisctimecontfreqnotonenoncentindexval}
           {\Yssmixsigdisctimecontfreqnotonenoncentindex (\Ysscontfreqval ) }

\newcommand{\yssmixsigdisctimecontfreqnotonenoncentsrcnb}
          {$\Yssmixsigdisctimecontfreqnotonenoncentsrcnb$}
\newcommand{\Yssmixsigdisctimecontfreqnotonenoncentsrcnb}
           {\Yssmixsigdisctimecontfreqnotonenoncentvec _{\Ysssrcnb}}

\newcommand{\yssmixsigdisctimecontfreqnotonenoncentsrcnbval}
          {$\Yssmixsigdisctimecontfreqnotonenoncentsrcnbval$}
\newcommand{\Yssmixsigdisctimecontfreqnotonenoncentsrcnbval}
           {\Yssmixsigdisctimecontfreqnotonenoncentsrcnb (\Ysscontfreqval ) }


\newcommand{\yssmixsigdisctimecontfreqnoncentvec}
          {$\Yssmixsigdisctimecontfreqnoncentvec$}
\newcommand{\Yssmixsigdisctimecontfreqnoncentvec}
           {X}

\newcommand{\yssmixsigdisctimecontfreqnoncentvecval}
          {$\Yssmixsigdisctimecontfreqnoncentvecval$}
\newcommand{\Yssmixsigdisctimecontfreqnoncentvecval}
           {\Yssmixsigdisctimecontfreqnoncentvec ( \Yssdisctimeval , \Ysscontfreqval ) }

\newcommand{\yssmixsigdisctimecontfreqnoncentone}
          {$\Yssmixsigdisctimecontfreqnoncentone$}
\newcommand{\Yssmixsigdisctimecontfreqnoncentone}
           {\Yssmixsigdisctimecontfreqnoncentvec _{1}}

\newcommand{\yssmixsigdisctimecontfreqnoncentoneval}
          {$\Yssmixsigdisctimecontfreqnoncentoneval$}
\newcommand{\Yssmixsigdisctimecontfreqnoncentoneval}
           {\Yssmixsigdisctimecontfreqnoncentone ( \Yssdisctimeval , \Ysscontfreqval ) }

\newcommand{\yssmixsigdisctimecontfreqnoncentindex}
          {$\Yssmixsigdisctimecontfreqnoncentindex$}
\newcommand{\Yssmixsigdisctimecontfreqnoncentindex}
           {\Yssmixsigdisctimecontfreqnoncentvec _{\Ysswayindex}}

\newcommand{\yssmixsigdisctimecontfreqnoncentindexval}
          {$\Yssmixsigdisctimecontfreqnoncentindexval$}
\newcommand{\Yssmixsigdisctimecontfreqnoncentindexval}
           {\Yssmixsigdisctimecontfreqnoncentindex ( \Yssdisctimeval , \Ysscontfreqval ) }

\newcommand{\yssmixsigdisctimecontfreqnoncentindexother}
          {$\Yssmixsigdisctimecontfreqnoncentindexother$}
\newcommand{\Yssmixsigdisctimecontfreqnoncentindexother}
           {\Yssmixsigdisctimecontfreqnoncentvec _{\Ysswayindexother}}

\newcommand{\yssmixsigdisctimecontfreqnoncentindexotherval}
          {$\Yssmixsigdisctimecontfreqnoncentindexotherval$}
\newcommand{\Yssmixsigdisctimecontfreqnoncentindexotherval}
           {\Yssmixsigdisctimecontfreqnoncentindexother ( \Yssdisctimeval , \Ysscontfreqval ) }


\newcommand{\yssmixsigdisctimecentveczt}
          {$\Yssmixsigdisctimecentveczt$}
\newcommand{\Yssmixsigdisctimecentveczt}
           {X}

\newcommand{\yssmixsigdisctimecentvecztval}
          {$\Yssmixsigdisctimecentvecztval$}
\newcommand{\Yssmixsigdisctimecentvecztval}
           {\Yssmixsigdisctimecentveczt (z)}

\newcommand{\yssmixsigdisctimecentonezt}
          {$\Yssmixsigdisctimecentonezt$}
\newcommand{\Yssmixsigdisctimecentonezt}
           {\Yssmixsigdisctimecentveczt _{1}}

\newcommand{\yssmixsigdisctimecentoneztval}
          {$\Yssmixsigdisctimecentoneztval$}
\newcommand{\Yssmixsigdisctimecentoneztval}
           {\Yssmixsigdisctimecentonezt (z)}

\newcommand{\yssmixsigdisctimecenttwozt}
          {$\Yssmixsigdisctimecenttwozt$}
\newcommand{\Yssmixsigdisctimecenttwozt}
           {\Yssmixsigdisctimecentveczt _{2}}

\newcommand{\yssmixsigdisctimecenttwoztval}
          {$\Yssmixsigdisctimecenttwoztval$}
\newcommand{\Yssmixsigdisctimecenttwoztval}
           {\Yssmixsigdisctimecenttwozt (z)}

\newcommand{\yssmixsigdisctimecentindexzt}
          {$\Yssmixsigdisctimecentindexzt$}
\newcommand{\Yssmixsigdisctimecentindexzt}
           {\Yssmixsigdisctimecentveczt _{\Ysswayindex}}

\newcommand{\yssmixsigdisctimecentindexztval}
          {$\Yssmixsigdisctimecentindexztval$}
\newcommand{\Yssmixsigdisctimecentindexztval}
           {\Yssmixsigdisctimecentindexzt (z)}

\newcommand{\yssmixsigdisctimecentlastzt}
          {$\Yssmixsigdisctimecentlastzt$}
\newcommand{\Yssmixsigdisctimecentlastzt}
           {\Yssmixsigdisctimecentveczt _{\Ysssensnb}}

\newcommand{\yssmixsigdisctimecentlastztval}
          {$\Yssmixsigdisctimecentlastztval$}
\newcommand{\Yssmixsigdisctimecentlastztval}
           {\Yssmixsigdisctimecentlastzt (z)}



\newcommand{\yssintermsepsystsigdisctimecentvec}
          {$\Yssintermsepsystsigdisctimecentvec$}
\newcommand{\Yssintermsepsystsigdisctimecentvec}
           {u}

\newcommand{\yssintermsepsystsigdisctimecentone}
          {$\Yssintermsepsystsigdisctimecentone$}
\newcommand{\Yssintermsepsystsigdisctimecentone}
           {\Yssintermsepsystsigdisctimecentvec _{1}}

\newcommand{\yssintermsepsystsigdisctimecentoneval}
          {$\Yssintermsepsystsigdisctimecentoneval$}
\newcommand{\Yssintermsepsystsigdisctimecentoneval}
           {\Yssintermsepsystsigdisctimecentone ( \Yssdisctimeval )}

\newcommand{\yssintermsepsystsigdisctimecenttwo}
          {$\Yssintermsepsystsigdisctimecenttwo$}
\newcommand{\Yssintermsepsystsigdisctimecenttwo}
           {\Yssintermsepsystsigdisctimecentvec _{2}}

\newcommand{\yssintermsepsystsigdisctimecenttwoval}
          {$\Yssintermsepsystsigdisctimecenttwoval$}
\newcommand{\Yssintermsepsystsigdisctimecenttwoval}
           {\Yssintermsepsystsigdisctimecenttwo ( \Yssdisctimeval )}


\newcommand{\yssintermsepsystsigdisctimecentthree}
{$\Yssintermsepsystsigdisctimecentthree$}
\newcommand{\Yssintermsepsystsigdisctimecentthree}
{\Yssintermsepsystsigdisctimecentvec _{3}}

\newcommand{\yssintermsepsystsigdisctimecentthreeval}
{$\Yssintermsepsystsigdisctimecentthreeval$}
\newcommand{\Yssintermsepsystsigdisctimecentthreeval}
{\Yssintermsepsystsigdisctimecentthree ( \Yssdisctimeval )}


\newcommand{\yssintermsepsystsigdisctimecentindex}
          {$\Yssintermsepsystsigdisctimecentindex$}
\newcommand{\Yssintermsepsystsigdisctimecentindex}
           {\Yssintermsepsystsigdisctimecentvec _{\Ysswayindex}}

\newcommand{\yssintermsepsystsigdisctimecentindexval}
          {$\Yssintermsepsystsigdisctimecentindexval$}
\newcommand{\Yssintermsepsystsigdisctimecentindexval}
           {\Yssintermsepsystsigdisctimecentindex ( \Yssdisctimeval )}

\newcommand{\yssintermsepsystsigdisctimecentindexother}
          {$\Yssintermsepsystsigdisctimecentindexother$}
\newcommand{\Yssintermsepsystsigdisctimecentindexother}
           {\Yssintermsepsystsigdisctimecentvec _{\Ysswayindexother}}


\newcommand{\yssoutsepsystsigconttimecentvec}
          {$\Yssoutsepsystsigconttimecentvec$}
\newcommand{\Yssoutsepsystsigconttimecentvec}
           {y}

\newcommand{\yssoutsepsystsigconttimecentvecval}
          {$\Yssoutsepsystsigconttimecentvecval$}
\newcommand{\Yssoutsepsystsigconttimecentvecval}
           {\Yssoutsepsystsigconttimecentvec (\Yssconttimeval)}

\newcommand{\yssoutsepsystsigconttimecentone}
          {$\Yssoutsepsystsigconttimecentone$}
\newcommand{\Yssoutsepsystsigconttimecentone}
           {\Yssoutsepsystsigconttimecentvec _{1}}

\newcommand{\yssoutsepsystsigconttimecentoneval}
          {$\Yssoutsepsystsigconttimecentoneval$}
\newcommand{\Yssoutsepsystsigconttimecentoneval}
           {\Yssoutsepsystsigconttimecentone (\Yssconttimeval)}


\newcommand{\yssoutsepsystsigconttimecenttwo}
          {$\Yssoutsepsystsigconttimecenttwo$}
\newcommand{\Yssoutsepsystsigconttimecenttwo}
           {\Yssoutsepsystsigconttimecentvec _{2}}

\newcommand{\yssoutsepsystsigconttimecenttwoval}
          {$\Yssoutsepsystsigconttimecenttwoval$}
\newcommand{\Yssoutsepsystsigconttimecenttwoval}
           {\Yssoutsepsystsigconttimecenttwo (\Yssconttimeval)}



\newcommand{\yssoutsepsystsigconttimecentthree}
{$\Yssoutsepsystsigconttimecentthree$}
\newcommand{\Yssoutsepsystsigconttimecentthree}
{\Yssoutsepsystsigconttimecentvec _{3}}

\newcommand{\yssoutsepsystsigconttimecentthreeval}
{$\Yssoutsepsystsigconttimecentthreeval$}
\newcommand{\Yssoutsepsystsigconttimecentthreeval}
{\Yssoutsepsystsigconttimecentthree (\Yssconttimeval)}


\newcommand{\yssoutsepsystsigconttimecentindex}
          {$\Yssoutsepsystsigconttimecentindex$}
\newcommand{\Yssoutsepsystsigconttimecentindex}
           {\Yssoutsepsystsigconttimecentvec _{\Ysswayindex}}

\newcommand{\yssoutsepsystsigconttimecentindexval}
          {$\Yssoutsepsystsigconttimecentindexval$}
\newcommand{\Yssoutsepsystsigconttimecentindexval}
           {\Yssoutsepsystsigconttimecentindex (\Yssconttimeval)}

\newcommand{\yssoutsepsystsigconttimecentindexother}
          {$\Yssoutsepsystsigconttimecentindexother$}
\newcommand{\Yssoutsepsystsigconttimecentindexother}
           {\Yssoutsepsystsigconttimecentvec _{\Ysswayindexother}}

\newcommand{\yssoutsepsystsigconttimecentindexotherval}
          {$\Yssoutsepsystsigconttimecentindexotherval$}
\newcommand{\Yssoutsepsystsigconttimecentindexotherval}
           {\Yssoutsepsystsigconttimecentindexother (\Yssconttimeval)}

\newcommand{\yssoutsepsystsigconttimecentindexfourth}
{$\Yssoutsepsystsigconttimecentindexfourth$}
\newcommand{\Yssoutsepsystsigconttimecentindexfourth}
{\Yssoutsepsystsigconttimecentvec _{\Ysswayindexfourth}}

\newcommand{\yssoutsepsystsigconttimecentindexfourthval}
{$\Yssoutsepsystsigconttimecentindexfourthval$}
\newcommand{\Yssoutsepsystsigconttimecentindexfourthval}
{\Yssoutsepsystsigconttimecentindexfourth (\Yssconttimeval)}


\newcommand{\yssoutsepsystsigconttimecentlast}
          {$\Yssoutsepsystsigconttimecentlast$}
\newcommand{\Yssoutsepsystsigconttimecentlast}
           {\Yssoutsepsystsigconttimecentvec _{\Ysssrcnb}}

\newcommand{\yssoutsepsystsigconttimecentlastval}
          {$\Yssoutsepsystsigconttimecentlastval$}
\newcommand{\Yssoutsepsystsigconttimecentlastval}
           {\Yssoutsepsystsigconttimecentlast (\Yssconttimeval)}




\newcommand{\yssoutsepsystsigconttimecentvecrand}
          {$\Yssoutsepsystsigconttimecentvecrand$}
\newcommand{\Yssoutsepsystsigconttimecentvecrand}
           {Y}

\newcommand{\yssoutsepsystsigconttimecentonerand}
          {$\Yssoutsepsystsigconttimecentonerand$}
\newcommand{\Yssoutsepsystsigconttimecentonerand}
           {\Yssoutsepsystsigconttimecentvecrand _{1}}

\newcommand{\yssoutsepsystsigconttimecenttworand}
          {$\Yssoutsepsystsigconttimecenttworand$}
\newcommand{\Yssoutsepsystsigconttimecenttworand}
           {\Yssoutsepsystsigconttimecentvecrand _{2}}

\newcommand{\yssoutsepsystsigconttimecentindexrand}
          {$\Yssoutsepsystsigconttimecentindexrand$}
\newcommand{\Yssoutsepsystsigconttimecentindexrand}
           {\Yssoutsepsystsigconttimecentvecrand _{\Ysswayindex}}


\newcommand{\yssoutsepsystsigconttimecentlastrand}
          {$\Yssoutsepsystsigconttimecentlastrand$}
\newcommand{\Yssoutsepsystsigconttimecentlastrand}
           {\Yssoutsepsystsigconttimecentvecrand _{\Ysssrcnb}}



\newcommand{\yssoutsepsystsigassocconttimecentvec}
          {$\Yssoutsepsystsigassocconttimecentvec$}
\newcommand{\Yssoutsepsystsigassocconttimecentvec}
           {\Yssoutsepsystsigconttimecentvec ^{\prime}}

\newcommand{\yssoutsepsystsigassocconttimecentvecval}
          {$\Yssoutsepsystsigassocconttimecentvecval$}
\newcommand{\Yssoutsepsystsigassocconttimecentvecval}
           {\Yssoutsepsystsigassocconttimecentvec (\Yssconttimeval)}

\newcommand{\yssoutsepsystsigassocconttimecentone}
          {$\Yssoutsepsystsigassocconttimecentone$}
\newcommand{\Yssoutsepsystsigassocconttimecentone}
           {\Yssoutsepsystsigassocconttimecentvec _{1}}

\newcommand{\yssoutsepsystsigassocconttimecentoneval}
          {$\Yssoutsepsystsigassocconttimecentoneval$}
\newcommand{\Yssoutsepsystsigassocconttimecentoneval}
           {\Yssoutsepsystsigassocconttimecentone (\Yssconttimeval)}

\newcommand{\yssoutsepsystsigassocconttimecentindex}
          {$\Yssoutsepsystsigassocconttimecentindex$}
\newcommand{\Yssoutsepsystsigassocconttimecentindex}
           {\Yssoutsepsystsigassocconttimecentvec _{\Ysswayindex}}

\newcommand{\yssoutsepsystsigassocconttimecentindexval}
          {$\Yssoutsepsystsigassocconttimecentindexval$}
\newcommand{\Yssoutsepsystsigassocconttimecentindexval}
           {\Yssoutsepsystsigassocconttimecentindex (\Yssconttimeval)}

\newcommand{\yssoutsepsystsigassocconttimecentindexother}
          {$\Yssoutsepsystsigassocconttimecentindexother$}
\newcommand{\Yssoutsepsystsigassocconttimecentindexother}
           {\Yssoutsepsystsigassocconttimecentvec _{\Ysswayindexother}}

\newcommand{\yssoutsepsystsigassocconttimecentindexotherval}
          {$\Yssoutsepsystsigassocconttimecentindexotherval$}
\newcommand{\Yssoutsepsystsigassocconttimecentindexotherval}
           {\Yssoutsepsystsigassocconttimecentindexother (\Yssconttimeval)}

\newcommand{\yssoutsepsystsigassocconttimecentindexfourth}
{$\Yssoutsepsystsigassocconttimecentindexfourth$}
\newcommand{\Yssoutsepsystsigassocconttimecentindexfourth}
{\Yssoutsepsystsigassocconttimecentvec _{\Ysswayindexfourth}}

\newcommand{\yssoutsepsystsigassocconttimecentindexfourthval}
{$\Yssoutsepsystsigassocconttimecentindexfourthval$}
\newcommand{\Yssoutsepsystsigassocconttimecentindexfourthval}
{\Yssoutsepsystsigassocconttimecentindexfourth (\Yssconttimeval)}


\newcommand{\yssoutsepsystsigdisctimecentvec}
          {$\Yssoutsepsystsigdisctimecentvec$}
\newcommand{\Yssoutsepsystsigdisctimecentvec}
           {y}

\newcommand{\yssoutsepsystsigdisctimecentvecval}
          {$\Yssoutsepsystsigdisctimecentvecval$}
\newcommand{\Yssoutsepsystsigdisctimecentvecval}
           {\Yssoutsepsystsigdisctimecentvec (\Yssdisctimeval)}

\newcommand{\yssoutsepsystsigdisctimecentvecvalother}
{$\Yssoutsepsystsigdisctimecentvecvalother$}
\newcommand{\Yssoutsepsystsigdisctimecentvecvalother}
{\Yssoutsepsystsigdisctimecentvec (\Yssdisctimevalother)}

\newcommand{\yssoutsepsystsigdisctimecentone}
          {$\Yssoutsepsystsigdisctimecentone$}
\newcommand{\Yssoutsepsystsigdisctimecentone}
           {\Yssoutsepsystsigdisctimecentvec _{1}}

\newcommand{\yssoutsepsystsigdisctimecentoneval}
          {$\Yssoutsepsystsigdisctimecentoneval$}
\newcommand{\Yssoutsepsystsigdisctimecentoneval}
           {\Yssoutsepsystsigdisctimecentone (\Yssdisctimeval)}

\newcommand{\yssoutsepsystsigdisctimecentonevalother}
{$\Yssoutsepsystsigdisctimecentonevalother$}
\newcommand{\Yssoutsepsystsigdisctimecentonevalother}
{\Yssoutsepsystsigdisctimecentone (\Yssdisctimevalother)}

\newcommand{\yssoutsepsystsigdisctimecenttwo}
{$\Yssoutsepsystsigdisctimecenttwo$}
\newcommand{\Yssoutsepsystsigdisctimecenttwo}
{\Yssoutsepsystsigdisctimecentvec _{2}}


\newcommand{\yssoutsepsystsigdisctimecenttwoval}
{$\Yssoutsepsystsigdisctimecenttwoval$}
\newcommand{\Yssoutsepsystsigdisctimecenttwoval}
{\Yssoutsepsystsigdisctimecenttwo (\Yssdisctimeval)}



\newcommand{\yssoutsepsystsigdisctimecenttwovalother}
{$\Yssoutsepsystsigdisctimecenttwovalother$}
\newcommand{\Yssoutsepsystsigdisctimecenttwovalother}
{\Yssoutsepsystsigdisctimecenttwo (\Yssdisctimevalother)}


\newcommand{\yssoutsepsystsigdisctimecentthree}
{$\Yssoutsepsystsigdisctimecentthree$}
\newcommand{\Yssoutsepsystsigdisctimecentthree}
{\Yssoutsepsystsigdisctimecentvec _{3}}

\newcommand{\yssoutsepsystsigdisctimecentindex}
          {$\Yssoutsepsystsigdisctimecentindex$}
\newcommand{\Yssoutsepsystsigdisctimecentindex}
           {\Yssoutsepsystsigdisctimecentvec _{\Ysswayindex}}

\newcommand{\yssoutsepsystsigdisctimecentindexval}
          {$\Yssoutsepsystsigdisctimecentindexval$}
\newcommand{\Yssoutsepsystsigdisctimecentindexval}
           {\Yssoutsepsystsigdisctimecentindex (\Yssdisctimeval)}

\newcommand{\yssoutsepsystsigdisctimecentindexvalother}
{$\Yssoutsepsystsigdisctimecentindexvalother$}
\newcommand{\Yssoutsepsystsigdisctimecentindexvalother}
{\Yssoutsepsystsigdisctimecentindex (\Yssdisctimevalother)}

\newcommand{\yssoutsepsystsigdisctimecentindexother}
          {$\Yssoutsepsystsigdisctimecentindexother$}
\newcommand{\Yssoutsepsystsigdisctimecentindexother}
           {\Yssoutsepsystsigdisctimecentvec _{\Ysswayindexother}}

\newcommand{\yssoutsepsystsigdisctimecentindexotherval}
          {$\Yssoutsepsystsigdisctimecentindexotherval$}
\newcommand{\Yssoutsepsystsigdisctimecentindexotherval}
           {\Yssoutsepsystsigdisctimecentindexother (\Yssdisctimeval)}

\newcommand{\yssoutsepsystsigdisctimecentindexothervalother}
{$\Yssoutsepsystsigdisctimecentindexothervalother$}
\newcommand{\Yssoutsepsystsigdisctimecentindexothervalother}
{\Yssoutsepsystsigdisctimecentindexother (\Yssdisctimevalother)}


\newcommand{\yssoutsepsystsigdisctimecentindexthird}
          {$\Yssoutsepsystsigdisctimecentindexthird$}
\newcommand{\Yssoutsepsystsigdisctimecentindexthird}
           {\Yssoutsepsystsigdisctimecentvec _{\Ysswayindexthird}}

\newcommand{\yssoutsepsystsigdisctimecentindexfourth}
          {$\Yssoutsepsystsigdisctimecentindexfourth$}
\newcommand{\Yssoutsepsystsigdisctimecentindexfourth}
           {\Yssoutsepsystsigdisctimecentvec _{\Ysswayindexfourth}}

\newcommand{\yssoutsepsystsigdisctimecentindexfourthval}
          {$\Yssoutsepsystsigdisctimecentindexfourthval$}
\newcommand{\Yssoutsepsystsigdisctimecentindexfourthval}
           {\Yssoutsepsystsigdisctimecentindexfourth (\Yssdisctimeval)}
           

\newcommand{\yssoutsepsystsigdisctimecentindexsensnb}
          {$\Yssoutsepsystsigdisctimecentindexsensnb$}
\newcommand{\Yssoutsepsystsigdisctimecentindexsensnb}
           {\Yssoutsepsystsigdisctimecentvec _{\Ysssensnb}}

\newcommand{\yssoutsepsystsigdisctimecentindexsensnbval}
          {$\Yssoutsepsystsigdisctimecentindexsensnbval$}
\newcommand{\Yssoutsepsystsigdisctimecentindexsensnbval}
           {\Yssoutsepsystsigdisctimecentindexsensnb (\Yssdisctimeval)}


\newcommand{\yssoutsepsystsigassocdisctimecontfreqnotonenoncentvec}
          {$\Yssoutsepsystsigassocdisctimecontfreqnotonenoncentvec$}
\newcommand{\Yssoutsepsystsigassocdisctimecontfreqnotonenoncentvec}
           {Y
       ^{\prime}}

\newcommand{\yssoutsepsystsigassocdisctimecontfreqnotonenoncentvecval}
          {$\Yssoutsepsystsigassocdisctimecontfreqnotonenoncentvecval$}
\newcommand{\Yssoutsepsystsigassocdisctimecontfreqnotonenoncentvecval}
           {\Yssoutsepsystsigassocdisctimecontfreqnotonenoncentvec ( \Ysscontfreqval ) }


\newcommand{\yssoutsepsystsigdisctimecentveczt}
          {$\Yssoutsepsystsigdisctimecentveczt$}
\newcommand{\Yssoutsepsystsigdisctimecentveczt}
           {Y}

\newcommand{\yssoutsepsystsigdisctimecentvecztval}
          {$\Yssoutsepsystsigdisctimecentvecztval$}
\newcommand{\Yssoutsepsystsigdisctimecentvecztval}
           {\Yssoutsepsystsigdisctimecentveczt (z)}

\newcommand{\yssoutsepsystsigdisctimecentonezt}
          {$\Yssoutsepsystsigdisctimecentonezt$}
\newcommand{\Yssoutsepsystsigdisctimecentonezt}
           {\Yssoutsepsystsigdisctimecentveczt _{1}}

\newcommand{\yssoutsepsystsigdisctimecentoneztval}
          {$\Yssoutsepsystsigdisctimecentoneztval$}
\newcommand{\Yssoutsepsystsigdisctimecentoneztval}
           {\Yssoutsepsystsigdisctimecentonezt (z)}

\newcommand{\yssoutsepsystsigdisctimecenttwozt}
          {$\Yssoutsepsystsigdisctimecenttwozt$}
\newcommand{\Yssoutsepsystsigdisctimecenttwozt}
           {\Yssoutsepsystsigdisctimecentveczt _{2}}

\newcommand{\yssoutsepsystsigdisctimecenttwoztval}
          {$\Yssoutsepsystsigdisctimecenttwoztval$}
\newcommand{\Yssoutsepsystsigdisctimecenttwoztval}
           {\Yssoutsepsystsigdisctimecenttwozt (z)}

\newcommand{\yssoutsepsystsigdisctimecentindexzt}
          {$\Yssoutsepsystsigdisctimecentindexzt$}
\newcommand{\Yssoutsepsystsigdisctimecentindexzt}
           {\Yssoutsepsystsigdisctimecentveczt _{\Ysswayindex}}

\newcommand{\yssoutsepsystsigdisctimecentindexztval}
          {$\Yssoutsepsystsigdisctimecentindexztval$}
\newcommand{\Yssoutsepsystsigdisctimecentindexztval}
           {\Yssoutsepsystsigdisctimecentindexzt (z)}

\newcommand{\yssoutsepsystsigdisctimecentindexsensnbzt}
          {$\Yssoutsepsystsigdisctimecentindexsensnbzt$}
\newcommand{\Yssoutsepsystsigdisctimecentindexsensnbzt}
           {\Yssoutsepsystsigdisctimecentveczt _{\Ysssensnb}}

\newcommand{\yssoutsepsystsigdisctimecentindexsensnbztval}
          {$\Yssoutsepsystsigdisctimecentindexsensnbztval$}
\newcommand{\Yssoutsepsystsigdisctimecentindexsensnbztval}
           {\Yssoutsepsystsigdisctimecentindexsensnbzt (z)}



\newcommand{\yssarbsigconttimenotone}
          {$\Yssarbsigconttimenotone$}
\newcommand{\Yssarbsigconttimenotone}
           {v}

\newcommand{\yssarbsigconttimenotoneval}
          {$\Yssarbsigconttimenotoneval$}
\newcommand{\Yssarbsigconttimenotoneval}
           {\Yssarbsigconttimenotone (\Yssconttimeval)}

\newcommand{\yssarbsigconttimenotoneindexone}
          {$\Yssarbsigconttimenotoneindexone$}
\newcommand{\Yssarbsigconttimenotoneindexone}
           {\Yssarbsigconttimenotone _{1}}

\newcommand{\yssarbsigconttimenotoneindexoneval}
          {$\Yssarbsigconttimenotoneindexoneval$}
\newcommand{\Yssarbsigconttimenotoneindexoneval}
           {\Yssarbsigconttimenotoneindexone (\Yssconttimeval)}

\newcommand{\yssarbsigconttimenotoneindextwo}
          {$\Yssarbsigconttimenotoneindextwo$}
\newcommand{\Yssarbsigconttimenotoneindextwo}
           {\Yssarbsigconttimenotone _{2}}

\newcommand{\yssarbsigconttimenotoneindextwoval}
          {$\Yssarbsigconttimenotoneindextwoval$}
\newcommand{\Yssarbsigconttimenotoneindextwoval}
           {\Yssarbsigconttimenotoneindextwo (\Yssconttimeval)}

\newcommand{\yssarbsigconttimenottwo}
          {$\Yssarbsigconttimenottwo$}
\newcommand{\Yssarbsigconttimenottwo}
           {w}


\newcommand{\yssarbsigconttimecontfreqnoncentnotone}
          {$\Yssarbsigconttimecontfreqnoncentnotone$}
\newcommand{\Yssarbsigconttimecontfreqnoncentnotone}
           {V}

\newcommand{\yssarbsigconttimecontfreqnoncentnotoneval}
          {$\Yssarbsigconttimecontfreqnoncentnotoneval$}
\newcommand{\Yssarbsigconttimecontfreqnoncentnotoneval}
           {\Yssarbsigconttimecontfreqnoncentnotone (\Yssconttimeval , \Ysscontfreqval)}

\newcommand{\yssarbsigconttimecontfreqnoncentnotoneindexone}
          {$\Yssarbsigconttimecontfreqnoncentnotoneindexone$}
\newcommand{\Yssarbsigconttimecontfreqnoncentnotoneindexone}
           {\Yssarbsigconttimecontfreqnoncentnotone _{1}}

\newcommand{\yssarbsigconttimecontfreqnoncentnotonevalindexone}
          {$\Yssarbsigconttimecontfreqnoncentnotonevalindexone$}
\newcommand{\Yssarbsigconttimecontfreqnoncentnotonevalindexone}
           {\Yssarbsigconttimecontfreqnoncentnotoneindexone (\Yssconttimeval , \Ysscontfreqval)}

\newcommand{\yssarbsigconttimecontfreqnoncentnotoneindextwo}
          {$\Yssarbsigconttimecontfreqnoncentnotoneindextwo$}
\newcommand{\Yssarbsigconttimecontfreqnoncentnotoneindextwo}
           {\Yssarbsigconttimecontfreqnoncentnotone _{2}}

\newcommand{\yssarbsigconttimecontfreqnoncentnotonevalindextwo}
          {$\Yssarbsigconttimecontfreqnoncentnotonevalindextwo$}
\newcommand{\Yssarbsigconttimecontfreqnoncentnotonevalindextwo}
           {\Yssarbsigconttimecontfreqnoncentnotoneindextwo (\Yssconttimeval , \Ysscontfreqval)}

\newcommand{\yssarbsigconttimecontfreqnoncentnottwo}
          {$\Yssarbsigconttimecontfreqnoncentnottwo$}
\newcommand{\Yssarbsigconttimecontfreqnoncentnottwo}
           {W}


\newcommand{\yssarbsigconttimecontscalenoncentnotone}
          {$\Yssarbsigconttimecontscalenoncentnotone$}
\newcommand{\Yssarbsigconttimecontscalenoncentnotone}
           {\Ysstimescalecontcoefnot _{\Yssarbsigconttimenotone}}

\newcommand{\yssarbsigconttimecontscalenoncentnotoneval}
          {$\Yssarbsigconttimecontscalenoncentnotoneval$}
\newcommand{\Yssarbsigconttimecontscalenoncentnotoneval}
           {\Yssarbsigconttimecontscalenoncentnotone
        ( \Ysstimescalecontshiftval , \Ysstimescalecontscaleval )
           }

\newcommand{\yssarbsigconttimecontscalenoncentnotoneindexone}
          {$\Yssarbsigconttimecontscalenoncentnotoneindexone$}
\newcommand{\Yssarbsigconttimecontscalenoncentnotoneindexone}
           {\Ysstimescalecontcoefnot _{\Yssarbsigconttimenotoneindexone} }

\newcommand{\yssarbsigconttimecontscalenoncentnotonevalindexone}
          {$\Yssarbsigconttimecontscalenoncentnotonevalindexone$}
\newcommand{\Yssarbsigconttimecontscalenoncentnotonevalindexone}
           {\Yssarbsigconttimecontscalenoncentnotoneindexone
        ( \Ysstimescalecontshiftval , \Ysstimescalecontscaleval )
       }

\newcommand{\yssarbsigconttimecontscalenoncentnotoneindextwo}
          {$\Yssarbsigconttimecontscalenoncentnotoneindextwo$}
\newcommand{\Yssarbsigconttimecontscalenoncentnotoneindextwo}
           {\Ysstimescalecontcoefnot _{\Yssarbsigconttimenotoneindextwo} }

\newcommand{\yssarbsigconttimecontscalenoncentnotonevalindextwo}
          {$\Yssarbsigconttimecontscalenoncentnotonevalindextwo$}
\newcommand{\Yssarbsigconttimecontscalenoncentnotonevalindextwo}
           {\Yssarbsigconttimecontscalenoncentnotoneindextwo
        ( \Ysstimescalecontshiftval , \Ysstimescalecontscaleval )
       }


\newcommand{\yssarbsigdisctimenotone}
          {$\Yssarbsigdisctimenotone$}
\newcommand{\Yssarbsigdisctimenotone}
           {v}

\newcommand{\yssarbsigdisctimenotoneval}
          {$\Yssarbsigdisctimenotoneval$}
\newcommand{\Yssarbsigdisctimenotoneval}
           {\Yssarbsigdisctimenotone (\Yssdisctimeval)}

\newcommand{\yssarbsigdisctimenottwo}
          {$\Yssarbsigdisctimenottwo$}
\newcommand{\Yssarbsigdisctimenottwo}
           {w}



\newcommand{\yssarbsigdisctimecentnotonezt}
          {$\Yssarbsigdisctimecentnotonezt$}
\newcommand{\Yssarbsigdisctimecentnotonezt}
           {V}

\newcommand{\yssarbsigdisctimecentnotoneztval}
          {$\Yssarbsigdisctimecentnotoneztval$}
\newcommand{\Yssarbsigdisctimecentnotoneztval}
           {\Yssarbsigdisctimecentnotonezt (z)}

\newcommand{\yssarbsigdisctimecentnotoneindexonezt}
          {$\Yssarbsigdisctimecentnotoneindexonezt$}
\newcommand{\Yssarbsigdisctimecentnotoneindexonezt}
           {\Yssarbsigdisctimecentnotonezt _{1}}

\newcommand{\yssarbsigdisctimecentnotoneindexoneztval}
          {$\Yssarbsigdisctimecentnotoneindexoneztval$}
\newcommand{\Yssarbsigdisctimecentnotoneindexoneztval}
           {\Yssarbsigdisctimecentnotoneindexonezt (z)}

\newcommand{\yssarbsigdisctimecentnotoneindextwozt}
          {$\Yssarbsigdisctimecentnotoneindextwozt$}
\newcommand{\Yssarbsigdisctimecentnotoneindextwozt}
           {\Yssarbsigdisctimecentnotonezt _{2}}

\newcommand{\yssarbsigdisctimecentnotoneindextwoztval}
          {$\Yssarbsigdisctimecentnotoneindextwoztval$}
\newcommand{\Yssarbsigdisctimecentnotoneindextwoztval}
           {\Yssarbsigdisctimecentnotoneindextwozt (z)}




\newcommand{\yssinnovtosrcfillengthnegnoindex}
           {$\Yssinnovtosrcfillengthnegnoindex$}
\newcommand{\Yssinnovtosrcfillengthnegnoindex}
           {L_{1}}

\newcommand{\yssinnovtosrcfillengthposnoindex}
           {$\Yssinnovtosrcfillengthposnoindex$}
\newcommand{\Yssinnovtosrcfillengthposnoindex}
           {L_{2}}


\newcommand{\yssinnovtosrcimprespnoindex}
           {$\Yssinnovtosrcimprespnoindex$}
\newcommand{\Yssinnovtosrcimprespnoindex}
           {d}

\newcommand{\yssinnovtosrcimprespone}
           {$\Yssinnovtosrcimprespone$}
\newcommand{\Yssinnovtosrcimprespone}
           {\Yssinnovtosrcimprespnoindex _{1}}

\newcommand{\yssinnovtosrcimpresptwo}
           {$\Yssinnovtosrcimpresptwo$}
\newcommand{\Yssinnovtosrcimpresptwo}
           {\Yssinnovtosrcimprespnoindex_{2}}

\newcommand{\yssinnovtosrcimprespindex}
           {$\Yssinnovtosrcimprespindex$}
\newcommand{\Yssinnovtosrcimprespindex}
           {\Yssinnovtosrcimprespnoindex_{\Ysswayindex}}


\newcommand{\yssinnovtosrctransfuncnoindex}
           {$\Yssinnovtosrctransfuncnoindex$}
\newcommand{\Yssinnovtosrctransfuncnoindex}
           {D}

\newcommand{\yssinnovtosrctransfuncindex}
           {$\Yssinnovtosrctransfuncindex$}
\newcommand{\Yssinnovtosrctransfuncindex}
           {\Yssinnovtosrctransfuncnoindex_{\Ysswayindex}}

\newcommand{\yssinnovtosrctransfuncindexval}
           {$\Yssinnovtosrctransfuncindexval$}
\newcommand{\Yssinnovtosrctransfuncindexval}
           {\Yssinnovtosrctransfuncindex (z)}



\newcommand{\yssmixmatrixscalar}
           {$\Yssmixmatrixscalar$}
\newcommand{\Yssmixmatrixscalar}
           {A}

%
%
\newcommand{\yssmixmatrixscalaradapt}
           {$\Yssmixmatrixscalaradapt$}
\newcommand{\Yssmixmatrixscalaradapt}
           {\check{A}_
\ell
}
%
%

\newcommand{\yssmixmatrixscalarestim}
           {$\Yssmixmatrixscalarestim$}
\newcommand{\Yssmixmatrixscalarestim}
           {\hat{\Yssmixmatrixscalar}}

\newcommand{\yssmixmatrixscalarelnoindex}
          {$\Yssmixmatrixscalarelnoindex$}
\newcommand{\Yssmixmatrixscalarelnoindex}
           {a}

%
\newcommand{\yssmixmatrixscalarelnoindexestim}
          {$\Yssmixmatrixscalarelnoindexestim$}
\newcommand{\Yssmixmatrixscalarelnoindexestim}
           {\hat{\Yssmixmatrixscalarelnoindex}}

\newcommand{\yssmixmatrixscalareloneone}
          {$\Yssmixmatrixscalareloneone$}
\newcommand{\Yssmixmatrixscalareloneone}
           {\Yssmixmatrixscalarelnoindex _{1 1}}

\newcommand{\yssmixmatrixscalarelonetwo}
          {$\Yssmixmatrixscalarelonetwo$}
\newcommand{\Yssmixmatrixscalarelonetwo}
           {\Yssmixmatrixscalarelnoindex _{1 2}}

\newcommand{\yssmixmatrixscalareloneindexother}
          {$\Yssmixmatrixscalareloneindexother$}
\newcommand{\Yssmixmatrixscalareloneindexother}
           {\Yssmixmatrixscalarelnoindex _{1 \Ysswayindexother}}

\newcommand{\yssmixmatrixscalareloneindexthird}
          {$\Yssmixmatrixscalareloneindexthird$}
\newcommand{\Yssmixmatrixscalareloneindexthird}
           {\Yssmixmatrixscalarelnoindex _{1 \Ysswayindexthird}}

\newcommand{\yssmixmatrixscalareloneindexfourth}
{$\Yssmixmatrixscalareloneindexfourth$}
\newcommand{\Yssmixmatrixscalareloneindexfourth}
{\Yssmixmatrixscalarelnoindex _{1 \Ysswayindexfourth}}

\newcommand{\yssmixmatrixscalareltwoone}
          {$\Yssmixmatrixscalareltwoone$}
\newcommand{\Yssmixmatrixscalareltwoone}
           {\Yssmixmatrixscalarelnoindex _{2 1}}

\newcommand{\yssmixmatrixscalareltwotwo}
          {$\Yssmixmatrixscalareltwotwo$}
\newcommand{\Yssmixmatrixscalareltwotwo}
           {\Yssmixmatrixscalarelnoindex _{2 2}}

\newcommand{\yssmixmatrixscalarelindexone}
          {$\Yssmixmatrixscalarelindexone$}
\newcommand{\Yssmixmatrixscalarelindexone}
           {\Yssmixmatrixscalarelnoindex _{\Ysswayindex 1}}

\newcommand{\yssmixmatrixscalarelindextwo}
          {$\Yssmixmatrixscalarelindextwo$}
\newcommand{\Yssmixmatrixscalarelindextwo}
           {\Yssmixmatrixscalarelnoindex _{\Ysswayindex 2}}

\newcommand{\yssmixmatrixscalarelindexthree}
          {$\Yssmixmatrixscalarelindexthree$}
\newcommand{\Yssmixmatrixscalarelindexthree}
           {\Yssmixmatrixscalarelnoindex _{\Ysswayindex 3}}

\newcommand{\yssmixmatrixscalarelindexindex}
          {$\Yssmixmatrixscalarelindexindex$}
\newcommand{\Yssmixmatrixscalarelindexindex}
           {\Yssmixmatrixscalarelnoindex _{\Ysswayindex \Ysswayindex}}

\newcommand{\yssmixmatrixscalarelindexindexother}
          {$\Yssmixmatrixscalarelindexindexother$}
\newcommand{\Yssmixmatrixscalarelindexindexother}
           {\Yssmixmatrixscalarelnoindex _{\Ysswayindex \Ysswayindexother}}

\newcommand{\yssmixmatrixscalarelindexindexotherestim}
          {$\Yssmixmatrixscalarelindexindexotherestim$}
\newcommand{\Yssmixmatrixscalarelindexindexotherestim}
           {\Yssmixmatrixscalarelnoindexestim _{\Ysswayindex \Ysswayindexother}}

\newcommand{\yssmixmatrixscalarelindexindexthird}
          {$\Yssmixmatrixscalarelindexindexthird$}
\newcommand{\Yssmixmatrixscalarelindexindexthird}
           {\Yssmixmatrixscalarelnoindex _{\Ysswayindex \Ysswayindexthird}}

\newcommand{\yssmixmatrixscalarelindexindexfourth}
{$\Yssmixmatrixscalarelindexindexfourth$}
\newcommand{\Yssmixmatrixscalarelindexindexfourth}
{\Yssmixmatrixscalarelnoindex _{\Ysswayindex \Ysswayindexfourth}}

\newcommand{\yssmixmatrixscalarelindexotherindexother}
{$\Yssmixmatrixscalarelindexotherindexother$}
\newcommand{\Yssmixmatrixscalarelindexotherindexother}
{\Yssmixmatrixscalarelnoindex _{\Ysswayindexother \Ysswayindexother}}

\newcommand{\yssmixmatrixscalarelindexthirdindexother}
          {$\Yssmixmatrixscalarelindexthirdindexother$}
\newcommand{\Yssmixmatrixscalarelindexthirdindexother}
           {\Yssmixmatrixscalarelnoindex _{\Ysswayindexthird \Ysswayindexother}}

\newcommand{\yssmixmatrixscalarelindexthirdindexfourth}
          {$\Yssmixmatrixscalarelindexthirdindexfourth$}
\newcommand{\Yssmixmatrixscalarelindexthirdindexfourth}
           {\Yssmixmatrixscalarelnoindex _{\Ysswayindexthird \Ysswayindexfourth}}

\newcommand{\yssinvmixmatrixscalar}
           {$\Yssinvmixmatrixscalar$}
\newcommand{\Yssinvmixmatrixscalar}
           {\Yssmixmatrixscalar ^{-1}}

\newcommand{\yssinvmixmatrixscalarestim}
           {$\Yssinvmixmatrixscalarestim$}
\newcommand{\Yssinvmixmatrixscalarestim}
           {\Yssmixmatrixscalarestim ^{-1}}

%
\newcommand{\yssinvmixmatrixscalarestiminv}
           {$\Yssinvmixmatrixscalarestiminv$}
\newcommand{\Yssinvmixmatrixscalarestiminv}
           {\widehat{\Yssmixmatrixscalar ^{-1}}}



\newcommand{\yssmixmatrixlagellagnoindex}
          {$\Yssmixmatrixlagellagnoindex$}
\newcommand{\Yssmixmatrixlagellagnoindex}
           {n}

\newcommand{\yssmixmatrixlagellagindexindexother}
          {$\Yssmixmatrixlagellagindexindexother$}
\newcommand{\Yssmixmatrixlagellagindexindexother}
           {\Yssmixmatrixlagellagnoindex _{\Ysswayindex \Ysswayindexother}}

\newcommand{\yssmixmatrixlagellagoneindexthird}
          {$\Yssmixmatrixlagellagoneindexthird$}
\newcommand{\Yssmixmatrixlagellagoneindexthird}
           {\Yssmixmatrixlagellagnoindex _{1 \Ysswayindexthird}}

\newcommand{\yssmixmatrixlagellagindexindexthird}
          {$\Yssmixmatrixlagellagindexindexthird$}
\newcommand{\Yssmixmatrixlagellagindexindexthird}
           {\Yssmixmatrixlagellagnoindex _{\Ysswayindex \Ysswayindexthird}}



\newcommand{\yssmixmatrixcontfreqnotonefunc}
           {$\Yssmixmatrixcontfreqnotonefunc$}
\newcommand{\Yssmixmatrixcontfreqnotonefunc}
           {A}


\newcommand{\yssmixmatrixcontfreqnotoneval}
           {$\Yssmixmatrixcontfreqnotoneval$}
\newcommand{\Yssmixmatrixcontfreqnotoneval}
           {\Yssmixmatrixcontfreqnotonefunc ( \Ysscontfreqval ) }




\newcommand{\yssmixtransfuncorder}
          {$\Yssmixtransfuncorder$}
\newcommand{\Yssmixtransfuncorder}
           {M}


\newcommand{\ymixmatrixzt}
           {$\Ymixmatrixzt$}
\newcommand{\Ymixmatrixzt}
           {A(z)}


\newcommand{\yssmixtransfuncnoindex}
           {$\Yssmixtransfuncnoindex$}
\newcommand{\Yssmixtransfuncnoindex}
           {A}

\newcommand{\yssmixtransfunconeone}
          {$\Yssmixtransfunconeone$}
\newcommand{\Yssmixtransfunconeone}
           {\Yssmixtransfuncnoindex _{1 1} (z)}

\newcommand{\yssmixtransfunconetwo}
          {$\Yssmixtransfunconetwo$}
\newcommand{\Yssmixtransfunconetwo}
           {\Yssmixtransfuncnoindex _{1 2} (z)}

\newcommand{\yssmixtransfunconethree}
          {$\Yssmixtransfunconethree$}
\newcommand{\Yssmixtransfunconethree}
           {\Yssmixtransfuncnoindex _{1 3} (z)}

\newcommand{\yssmixtransfunctwoone}
          {$\Yssmixtransfunctwoone$}
\newcommand{\Yssmixtransfunctwoone}
           {\Yssmixtransfuncnoindex _{2 1} (z)}

\newcommand{\yssmixtransfunctwotwo}
          {$\Yssmixtransfunctwotwo$}
\newcommand{\Yssmixtransfunctwotwo}
           {\Yssmixtransfuncnoindex _{2 2} (z)}

\newcommand{\yssmixtransfunctwothree}
          {$\Yssmixtransfunctwothree$}
\newcommand{\Yssmixtransfunctwothree}
           {\Yssmixtransfuncnoindex _{2 3} (z)}

\newcommand{\yssmixtransfunconeindexother}
          {$\Yssmixtransfunconeindexother$}
\newcommand{\Yssmixtransfunconeindexother}
           {\Yssmixtransfuncnoindex _{1 \Ysswayindexother} (z)}

\newcommand{\yssmixtransfunctwoindexother}
          {$\Yssmixtransfunctwoindexother$}
\newcommand{\Yssmixtransfunctwoindexother}
           {\Yssmixtransfuncnoindex _{2 \Ysswayindexother} (z)}

\newcommand{\yssmixtransfuncindexindexother}
          {$\Yssmixtransfuncindexindexother$}
\newcommand{\Yssmixtransfuncindexindexother}
           {\Yssmixtransfuncnoindex _{\Ysswayindex \Ysswayindexother} (z)}


\newcommand{\yssmixmatrixscalarquad}
{$\Yssmixmatrixscalarquad$}
\newcommand{\Yssmixmatrixscalarquad}
{B}

\newcommand{\yssmixmatrixscalarelquadnoindex}
          {$\Yssmixmatrixscalarelquadnoindex$}
\newcommand{\Yssmixmatrixscalarelquadnoindex}
          {b}

%
\newcommand{\yssmixmatrixscalarelquadnoindexestim}
          {$\Yssmixmatrixscalarelquadnoindexestim$}
\newcommand{\Yssmixmatrixscalarelquadnoindexestim}
          {\hat{\Yssmixmatrixscalarelquadnoindex}}

\newcommand{\yssmixmatrixscalarelquadone}
{$\Yssmixmatrixscalarelquadone$}
\newcommand{\Yssmixmatrixscalarelquadone}
{\Yssmixmatrixscalarelquadnoindex _{1}}

\newcommand{\yssmixmatrixscalarelquadtwo}
{$\Yssmixmatrixscalarelquadtwo$}
\newcommand{\Yssmixmatrixscalarelquadtwo}
{\Yssmixmatrixscalarelquadnoindex _{2}}

\newcommand{\yssmixmatrixscalarelquadindex}
{$\Yssmixmatrixscalarelquadindex$}
\newcommand{\Yssmixmatrixscalarelquadindex}
{\Yssmixmatrixscalarelquadnoindex _{\Ysswayindex}}

\newcommand{\yssmixmatrixscalarelquadoneindexotherindexthird}
{$\Yssmixmatrixscalarelquadoneindexotherindexthird$}
\newcommand{\Yssmixmatrixscalarelquadoneindexotherindexthird}
{\Yssmixmatrixscalarelquadnoindex _{1 \Ysswayindexother \Ysswayindexthird}}

\newcommand{\yssmixmatrixscalarelquadindexindexotherindexthird}
{$\Yssmixmatrixscalarelquadindexindexotherindexthird$}
\newcommand{\Yssmixmatrixscalarelquadindexindexotherindexthird}
{\Yssmixmatrixscalarelquadnoindex _{\Ysswayindex \Ysswayindexother \Ysswayindexthird}}

%
\newcommand{\yssmixmatrixscalarelquadindexindexotherindexthirdestim}
{$\Yssmixmatrixscalarelquadindexindexotherindexthirdestim$}
\newcommand{\Yssmixmatrixscalarelquadindexindexotherindexthirdestim}
{\Yssmixmatrixscalarelquadnoindexestim _{\Ysswayindex \Ysswayindexother \Ysswayindexthird}}

%
\newcommand{\yssmixmatrixscalarquadext}
{$\Yssmixmatrixscalarquadext$}
\newcommand{\Yssmixmatrixscalarquadext}
           {
\widetilde
{\Yssmixmatrixscalar}}
%
%
%
%
\newcommand{\yssmixmatrixscalarquadextadapt}
{$\Yssmixmatrixscalarquadextadapt$}
\newcommand{\Yssmixmatrixscalarquadextadapt}
           {
%
%
%
\check{\Yssmixmatrixscalar}
}
%
%
\newcommand{\yssmixmatrixscalarquadextestim}
{$\Yssmixmatrixscalarquadextestim$}
\newcommand{\Yssmixmatrixscalarquadextestim}
%
           {\widehat{\Yssmixmatrixscalarquadext}}


\newcommand{\yssmixmatrixscalarlinquadpartlinelnoindex}
{$\Yssmixmatrixscalarlinquadpartlinelnoindex$}
\newcommand{\Yssmixmatrixscalarlinquadpartlinelnoindex}
{L}

\newcommand{\yssmixmatrixscalarlinquadpartlinelonetwo}
{$\Yssmixmatrixscalarlinquadpartlinelonetwo$}
\newcommand{\Yssmixmatrixscalarlinquadpartlinelonetwo}
{\Yssmixmatrixscalarlinquadpartlinelnoindex_{12}}

\newcommand{\yssmixmatrixscalarlinquadpartlineltwoone}
{$\Yssmixmatrixscalarlinquadpartlineltwoone$}
\newcommand{\Yssmixmatrixscalarlinquadpartlineltwoone}
{\Yssmixmatrixscalarlinquadpartlinelnoindex_{21}}

\newcommand{\yssmixmatrixscalarlinquadpartlinelindexindexother}
{$\Yssmixmatrixscalarlinquadpartlinelindexindexother$}
\newcommand{\Yssmixmatrixscalarlinquadpartlinelindexindexother}
{\Yssmixmatrixscalarlinquadpartlinelnoindex_{\Ysswayindex \Ysswayindexother}}

\newcommand{\yssmixmatrixscalarlinquadpartquadelnoindex}
{$\Yssmixmatrixscalarlinquadpartquadelnoindex$}
\newcommand{\Yssmixmatrixscalarlinquadpartquadelnoindex}
{Q}

\newcommand{\yssmixmatrixscalarlinquadpartquadelone}
{$\Yssmixmatrixscalarlinquadpartquadelone$}
\newcommand{\Yssmixmatrixscalarlinquadpartquadelone}
{\Yssmixmatrixscalarlinquadpartquadelnoindex_{1}}

\newcommand{\yssmixmatrixscalarlinquadpartquadeltwo}
{$\Yssmixmatrixscalarlinquadpartquadeltwo$}
\newcommand{\Yssmixmatrixscalarlinquadpartquadeltwo}
{\Yssmixmatrixscalarlinquadpartquadelnoindex_{2}}

\newcommand{\yssmixmatrixscalarlinquadpartquadelindex}
{$\Yssmixmatrixscalarlinquadpartquadelindex$}
\newcommand{\Yssmixmatrixscalarlinquadpartquadelindex}
{\Yssmixmatrixscalarlinquadpartquadelnoindex_{\Ysswayindex}}



\newcommand{\ysssepsystmatrixscalar}
          {$\Ysssepsystmatrixscalar$}
\newcommand{\Ysssepsystmatrixscalar}
{C}

\newcommand{\ysssepsystmatrixscalarelnoindex}
          {$\Ysssepsystmatrixscalarelnoindex$}
\newcommand{\Ysssepsystmatrixscalarelnoindex}
           {c}

\newcommand{\ysssepsystmatrixscalareloneindexother}
          {$\Ysssepsystmatrixscalareloneindexother$}
\newcommand{\Ysssepsystmatrixscalareloneindexother}
           {\Ysssepsystmatrixscalarelnoindex _{1 \Ysswayindexother}}

\newcommand{\ysssepsystmatrixscalarelindexindexother}
          {$\Ysssepsystmatrixscalarelindexindexother$}
\newcommand{\Ysssepsystmatrixscalarelindexindexother}
           {\Ysssepsystmatrixscalarelnoindex _{\Ysswayindex \Ysswayindexother}}


\newcommand{\ysssepsystmatrixscalarelonetwo}
{$\Ysssepsystmatrixscalarelonetwo$}
\newcommand{\Ysssepsystmatrixscalarelonetwo}
{\Ysssepsystmatrixscalarelnoindex _{1 2}}

\newcommand{\ysssepsystmatrixscalareltwoone}
{$\Ysssepsystmatrixscalareltwoone$}
\newcommand{\Ysssepsystmatrixscalareltwoone}
{\Ysssepsystmatrixscalarelnoindex _{2 1}}



\newcommand{\ysssepsysmatrixscalarassocone}
           {$\Ysssepsysmatrixscalarassocone$}
\newcommand{\Ysssepsysmatrixscalarassocone}
           {C}

\newcommand{\ysssepsystmatrixscalarassoconeelnoindex}
          {$\Ysssepsystmatrixscalarassoconeelnoindex$}
\newcommand{\Ysssepsystmatrixscalarassoconeelnoindex}
           {\alpha}

\newcommand{\ysssepsystmatrixscalarassoconeeltwoone}
          {$\Ysssepsystmatrixscalarassoconeeltwoone$}
\newcommand{\Ysssepsystmatrixscalarassoconeeltwoone}
           {\Ysssepsystmatrixscalarassoconeelnoindex _{2 1}}

\newcommand{\ysssepsystmatrixscalarassoconeeltwotwo}
          {$\Ysssepsystmatrixscalarassoconeeltwotwo$}
\newcommand{\Ysssepsystmatrixscalarassoconeeltwotwo}
           {\Ysssepsystmatrixscalarassoconeelnoindex _{2 2}}

\newcommand{\ysssepsystmatrixscalarassoconeelindexindexother}
          {$\Ysssepsystmatrixscalarassoconeelindexindexother$}
\newcommand{\Ysssepsystmatrixscalarassoconeelindexindexother}
           {\Ysssepsystmatrixscalarassoconeelnoindex _{\Ysswayindex \Ysswayindexother}}

\newcommand{\ysssepsystmatrixscalarassoconeelindexindexthird}
          {$\Ysssepsystmatrixscalarassoconeelindexindexthird$}
\newcommand{\Ysssepsystmatrixscalarassoconeelindexindexthird}
           {\Ysssepsystmatrixscalarassoconeelnoindex _{\Ysswayindex \Ysswayindexthird}}


\newcommand{\ysssepsysmatrixscalarassoctwo}
           {$\Ysssepsysmatrixscalarassoctwo$}
\newcommand{\Ysssepsysmatrixscalarassoctwo}
           {D}

\newcommand{\ysssepsystmatrixscalarassoctwoelnoindex}
          {$\Ysssepsystmatrixscalarassoctwoelnoindex$}
\newcommand{\Ysssepsystmatrixscalarassoctwoelnoindex}
           {\beta}

\newcommand{\ysssepsystmatrixscalarassoctwoeltwoone}
          {$\Ysssepsystmatrixscalarassoctwoeltwoone$}
\newcommand{\Ysssepsystmatrixscalarassoctwoeltwoone}
           {\Ysssepsystmatrixscalarassoctwoelnoindex _{2 1}}

\newcommand{\ysssepsystmatrixscalarassoctwoeltwotwo}
          {$\Ysssepsystmatrixscalarassoctwoeltwotwo$}
\newcommand{\Ysssepsystmatrixscalarassoctwoeltwotwo}
           {\Ysssepsystmatrixscalarassoctwoelnoindex _{2 2}}

\newcommand{\ysssepsystmatrixscalarassoctwoelindexindexother}
          {$\Ysssepsystmatrixscalarassoctwoelindexindexother$}
\newcommand{\Ysssepsystmatrixscalarassoctwoelindexindexother}
           {\Ysssepsystmatrixscalarassoctwoelnoindex _{\Ysswayindex \Ysswayindexother}}

\newcommand{\ysssepsystmatrixscalarassoctwoelindexindexthird}
          {$\Ysssepsystmatrixscalarassoctwoelindexindexthird$}
\newcommand{\Ysssepsystmatrixscalarassoctwoelindexindexthird}
           {\Ysssepsystmatrixscalarassoctwoelnoindex _{\Ysswayindex \Ysswayindexthird}}


\newcommand{\ysssepsysmatrixscalarassocthree}
           {$\Ysssepsysmatrixscalarassocthree$}
\newcommand{\Ysssepsysmatrixscalarassocthree}
           {\Ysssepsysmatrixscalarassocone^{\prime}}

\newcommand{\ysssepsystmatrixscalarassocthreeelnoindex}
          {$\Ysssepsystmatrixscalarassocthreeelnoindex$}
\newcommand{\Ysssepsystmatrixscalarassocthreeelnoindex}
           {\Ysssepsystmatrixscalarassoconeelnoindex^{\prime}}

\newcommand{\ysssepsystmatrixscalarassocthreeelindexindexother}
          {$\Ysssepsystmatrixscalarassocthreeelindexindexother$}
\newcommand{\Ysssepsystmatrixscalarassocthreeelindexindexother}
           {\Ysssepsystmatrixscalarassocthreeelnoindex _{\Ysswayindex \Ysswayindexother}}

\newcommand{\ysssepsystmatrixscalarassocthreeelindexindexthird}
          {$\Ysssepsystmatrixscalarassocthreeelindexindexthird$}
\newcommand{\Ysssepsystmatrixscalarassocthreeelindexindexthird}
           {\Ysssepsystmatrixscalarassocthreeelnoindex _{\Ysswayindex \Ysswayindexthird}}


\newcommand{\ysssepsystmatrixcontfreqnotonefunc}
           {$\Ysssepsystmatrixcontfreqnotonefunc$}
\newcommand{\Ysssepsystmatrixcontfreqnotonefunc}
           {B}

\newcommand{\ysssepsystmatrixcontfreqnotoneval}
           {$\Ysssepsystmatrixcontfreqnotoneval$}
\newcommand{\Ysssepsystmatrixcontfreqnotoneval}
           {\Ysssepsystmatrixcontfreqnotonefunc ( \Ysscontfreqval ) }

\newcommand{\ysssepsystindexindexothercontfreqnotoneval}
          {$\Ysssepsystindexindexothercontfreqnotoneval$}
\newcommand{\Ysssepsystindexindexothercontfreqnotoneval}
           {\Ysssepsystmatrixcontfreqnotonefunc
        _{\Ysswayindex \Ysswayindexother} ( \Ysscontfreqval )}




\newcommand{\ysssepsysttransfuncnoindex}
           {$\Ysssepsysttransfuncnoindex$}
\newcommand{\Ysssepsysttransfuncnoindex}
           {B}

\newcommand{\ysssepsysttransfuncnoindexval}
           {$\Ysssepsysttransfuncnoindexval$}
\newcommand{\Ysssepsysttransfuncnoindexval}
           {\Ysssepsysttransfuncnoindex (z)}

\newcommand{\ysssepsysttransfunconetwo}
          {$\Ysssepsysttransfunconetwo$}
\newcommand{\Ysssepsysttransfunconetwo}
           {\Ysssepsysttransfuncnoindex _{1 2} (z)}

\newcommand{\ysssepsysttransfunctwoone}
          {$\Ysssepsysttransfunctwoone$}
\newcommand{\Ysssepsysttransfunctwoone}
           {\Ysssepsysttransfuncnoindex _{2 1} (z)}

\newcommand{\ysssepsysttransfuncindexindexother}
          {$\Ysssepsysttransfuncindexindexother$}
\newcommand{\Ysssepsysttransfuncindexindexother}
           {\Ysssepsysttransfuncnoindex _{\Ysswayindex \Ysswayindexother} (z)}


\newcommand{\ysssepsystcoefnoindex}
           {$\Ysssepsystcoefnoindex$}
\newcommand{\Ysssepsystcoefnoindex}
           {b}

\newcommand{\ysssepsystcoefindexindexother}
           {$\Ysssepsystcoefindexindexother$}
\newcommand{\Ysssepsystcoefindexindexother}
           {\Ysssepsystcoefnoindex _{\Ysswayindex \Ysswayindexother}}

\newcommand{\ysssepsystcoefonetwo}
           {$\Ysssepsystcoefonetwo$}
\newcommand{\Ysssepsystcoefonetwo}
           {\Ysssepsystcoefnoindex _{1 2}}

\newcommand{\ysssepsystcoeftwoone}
           {$\Ysssepsystcoeftwoone$}
\newcommand{\Ysssepsystcoeftwoone}
           {\Ysssepsystcoefnoindex _{2 1}}


\newcommand{\ysssepsystscalarlinquadpartlinelnoindex}
{$\Ysssepsystscalarlinquadpartlinelnoindex$}
\newcommand{\Ysssepsystscalarlinquadpartlinelnoindex}
{l}

\newcommand{\ysssepsystscalarlinquadpartlineloneone}
{$\Ysssepsystscalarlinquadpartlineloneone$}
\newcommand{\Ysssepsystscalarlinquadpartlineloneone}
{\Ysssepsystscalarlinquadpartlinelnoindex_{11}}

\newcommand{\ysssepsystscalarlinquadpartlineloneoneassoc}
{$\Ysssepsystscalarlinquadpartlineloneoneassoc$}
\newcommand{\Ysssepsystscalarlinquadpartlineloneoneassoc}
{\Ysssepsystscalarlinquadpartlineloneone ^{\prime}}

\newcommand{\ysssepsystscalarlinquadpartlinelonetwo}
{$\Ysssepsystscalarlinquadpartlinelonetwo$}
\newcommand{\Ysssepsystscalarlinquadpartlinelonetwo}
{\Ysssepsystscalarlinquadpartlinelnoindex_{12}}

\newcommand{\ysssepsystscalarlinquadpartlineltwoone}
{$\Ysssepsystscalarlinquadpartlineltwoone$}
\newcommand{\Ysssepsystscalarlinquadpartlineltwoone}
{\Ysssepsystscalarlinquadpartlinelnoindex_{21}}

\newcommand{\ysssepsystscalarlinquadpartlineltwotwo}
{$\Ysssepsystscalarlinquadpartlineltwotwo$}
\newcommand{\Ysssepsystscalarlinquadpartlineltwotwo}
{\Ysssepsystscalarlinquadpartlinelnoindex_{22}}

\newcommand{\ysssepsystscalarlinquadpartlineltwotwoassoc}
{$\Ysssepsystscalarlinquadpartlineltwotwoassoc$}
\newcommand{\Ysssepsystscalarlinquadpartlineltwotwoassoc}
{\Ysssepsystscalarlinquadpartlineltwotwo ^{\prime}}

\newcommand{\ysssepsystscalarlinquadpartlinelindexindex}
{$\Ysssepsystscalarlinquadpartlinelindexindex$}
\newcommand{\Ysssepsystscalarlinquadpartlinelindexindex}
{\Ysssepsystscalarlinquadpartlinelnoindex_{\Ysswayindex \Ysswayindex}}

\newcommand{\ysssepsystscalarlinquadpartlinelindexindexassoc}
{$\Ysssepsystscalarlinquadpartlinelindexindexassoc$}
\newcommand{\Ysssepsystscalarlinquadpartlinelindexindexassoc}
{\Ysssepsystscalarlinquadpartlinelindexindex ^{\prime}}

\newcommand{\ysssepsystscalarlinquadpartlinelindexindexother}
{$\Ysssepsystscalarlinquadpartlinelindexindexother$}
\newcommand{\Ysssepsystscalarlinquadpartlinelindexindexother}
{\Ysssepsystscalarlinquadpartlinelnoindex_{\Ysswayindex \Ysswayindexother}}

\newcommand{\ysssepsystscalarlinquadpartquadelnoindex}
{$\Ysssepsystscalarlinquadpartquadelnoindex$}
\newcommand{\Ysssepsystscalarlinquadpartquadelnoindex}
{q}

\newcommand{\ysssepsystscalarlinquadpartquadelone}
{$\Ysssepsystscalarlinquadpartquadelone$}
\newcommand{\Ysssepsystscalarlinquadpartquadelone}
{\Ysssepsystscalarlinquadpartquadelnoindex_{1}}

\newcommand{\ysssepsystscalarlinquadpartquadeltwo}
{$\Ysssepsystscalarlinquadpartquadeltwo$}
\newcommand{\Ysssepsystscalarlinquadpartquadeltwo}
{\Ysssepsystscalarlinquadpartquadelnoindex_{2}}

\newcommand{\ysssepsystscalarlinquadpartquadelindex}
{$\Ysssepsystscalarlinquadpartquadelindex$}
\newcommand{\Ysssepsystscalarlinquadpartquadelindex}
{\Ysssepsystscalarlinquadpartquadelnoindex_{\Ysswayindex}}



\newcommand{\ysscomplfiltcoefnoindex}
           {$\Ysscomplfiltcoefnoindex$}
\newcommand{\Ysscomplfiltcoefnoindex}
           {h}

\newcommand{\ysscomplfiltcoefindexone}
           {$\Ysscomplfiltcoefindexone$}
\newcommand{\Ysscomplfiltcoefindexone}
           {\Ysscomplfiltcoefnoindex _{1}}

\newcommand{\ysscomplfiltcoefindextwo}
           {$\Ysscomplfiltcoefindextwo$}
\newcommand{\Ysscomplfiltcoefindextwo}
           {\Ysscomplfiltcoefnoindex _{2}}

\newcommand{\ysscomplfiltcoefindexindex}
           {$\Ysscomplfiltcoefindexindex$}
\newcommand{\Ysscomplfiltcoefindexindex}
           {\Ysscomplfiltcoefnoindex _{\Ysswayindex}}

\newcommand{\ysscomplfiltcoefindexindexthird}
           {$\Ysscomplfiltcoefindexindexthird$}
\newcommand{\Ysscomplfiltcoefindexindexthird}
           {\Ysscomplfiltcoefnoindex _{\Ysswayindex \Ysswayindexthird}}

\newcommand{\ysscomplfiltcoefindexotherindexthird}
           {$\Ysscomplfiltcoefindexotherindexthird$}
\newcommand{\Ysscomplfiltcoefindexotherindexthird}
           {\Ysscomplfiltcoefnoindex _{\Ysswayindexother \Ysswayindexthird}}



\newcommand{\ysssepsystadaptgainnoindex}
           {$\Ysssepsystadaptgainnoindex$}
\newcommand{\Ysssepsystadaptgainnoindex}
           {\mu}

\newcommand{\ysssepsystadaptgainone}
           {$\Ysssepsystadaptgainone$}
\newcommand{\Ysssepsystadaptgainone}
           {\Ysssepsystadaptgainnoindex _{1}}

\newcommand{\ysssepsystadaptgaintwo}
           {$\Ysssepsystadaptgaintwo$}
\newcommand{\Ysssepsystadaptgaintwo}
           {\Ysssepsystadaptgainnoindex _{2}}

\newcommand{\ysssepsystadaptgainindex}
           {$\Ysssepsystadaptgainindex$}
\newcommand{\Ysssepsystadaptgainindex}
           {\Ysssepsystadaptgainnoindex _{\Ysswayindex}}



\newcommand{\yodedisctimeval}
          {$\Yodedisctimeval$}
\newcommand{\Yodedisctimeval}
           {\Yssdisctimeval}


\newcommand{\yodesystemstatenoindex}
          {$\Yodesystemstatenoindex$}
\newcommand{\Yodesystemstatenoindex}
           {\theta}

\newcommand{\yodesystemstateindex}
          {$\Yodesystemstateindex$}
\newcommand{\Yodesystemstateindex}
           {\Yodesystemstatenoindex _{\Yodedisctimeval}}

\newcommand{\yodesystemstateeq}
          {$\Yodesystemstateeq$}
\newcommand{\Yodesystemstateeq}
           {\Yodesystemstatenoindex ^{*}}

\newcommand{\yodesystemstatesep}
          {$\Yodesystemstatesep$}
\newcommand{\Yodesystemstatesep}
           {\Yodesystemstatenoindex ^{s}}


\newcommand{\yodesystemsignoindex}
          {$\Yodesystemsignoindex$}
\newcommand{\Yodesystemsignoindex}
           {\xi}

\newcommand{\yodesystemsigtimeindexplusone}
          {$\Yodesystemsigtimeindexplusone$}
\newcommand{\Yodesystemsigtimeindexplusone}
           {\Yodesystemsignoindex _{\Yodedisctimeval + 1}}


\newcommand{\yodesystemfuncnoindex}
          {$\Yodesystemfuncnoindex$}
\newcommand{\Yodesystemfuncnoindex}
           {H}

\newcommand{\yodesystemfuncindex}
          {$\Yodesystemfuncindex$}
\newcommand{\Yodesystemfuncindex}
           {\Yodesystemfuncnoindex ( \Yodesystemstateindex , \Yodesystemsigtimeindexplusone) }


\newcommand{\yodesystemjacobmatrix}
          {$\Yodesystemjacobmatrix$}
\newcommand{\Yodesystemjacobmatrix}
           {J}

\newcommand{\yodesystemjacobmatrixstatesep}
          {$\Yodesystemjacobmatrixstatesep$}
\newcommand{\Yodesystemjacobmatrixstatesep}
           {\Yodesystemjacobmatrix ( \Yodesystemstatesep )}

\newcommand{\yodesystemjacobmatrixlineindex}
          {$\Yodesystemjacobmatrixlineindex$}
\newcommand{\Yodesystemjacobmatrixlineindex}
           {i}

\newcommand{\yodesystemjacobmatrixcolumnindex}
          {$\Yodesystemjacobmatrixcolumnindex$}
\newcommand{\Yodesystemjacobmatrixcolumnindex}
           {j}

\newcommand{\yodesystemjacobmatrixellineindexcolumnindex}
          {$\Yodesystemjacobmatrixellineindexcolumnindex$}
\newcommand{\Yodesystemjacobmatrixellineindexcolumnindex}
           {\Yodesystemjacobmatrix _{\Yodesystemjacobmatrixlineindex \Yodesystemjacobmatrixcolumnindex}}

%
%
%
%
%
\newcommand{\yqubitonesevspace}
{$\Yqubitonesevspace$}
\newcommand{\Yqubitonesevspace}
{{\cal E}}
%
%
%
%
\newcommand{\yqubitoneindexstd}
{$\Yqubitoneindexstd$}
\newcommand{\Yqubitoneindexstd}
{j}
%
%
%
%
%
\newcommand{\yqubitonespaceindexone}
{$\Yqubitonespaceindexone$}
\newcommand{\Yqubitonespaceindexone}
{\Yqubitonesevspace_{1}}
%
\newcommand{\yqubitonespaceindextwo}
{$\Yqubitonespaceindextwo$}
\newcommand{\Yqubitonespaceindextwo}
{\Yqubitonesevspace_{2}}
%
\newcommand{\yqubitonespaceindexstd}
{$\Yqubitonespaceindexstd$}
\newcommand{\Yqubitonespaceindexstd}
{\Yqubitonesevspace_{\Yqubitoneindexstd}}
%
%
%
\newcommand{\yqubitonetimeinit}
{$\Yqubitonetimeinit$}
\newcommand{\Yqubitonetimeinit}
{t_0}
%
%
%
%
\newcommand{\yqubitonetimefinal}
{$\Yqubitonetimefinal$}
\newcommand{\Yqubitonetimefinal}
{t}
%
%
%
%
%
%
\newcommand{\yqubitonetimeinitstateindexone}
{$\Yqubitonetimeinitstateindexone$}
\newcommand{\Yqubitonetimeinitstateindexone}
{| \psi_1 ( \Yqubitonetimeinit ) \rangle}
%
\newcommand{\yqubitonetimeinitstateindextwo}
{$\Yqubitonetimeinitstateindextwo$}
\newcommand{\Yqubitonetimeinitstateindextwo}
{| \psi_2 ( \Yqubitonetimeinit ) \rangle}
%
%
%
\newcommand{\yqubitonetimeinitveccomp}
{$\Yqubitonetimeinitveccomp$}
\newcommand{\Yqubitonetimeinitveccomp}
{C_{\Yqubitoneindexstd}(\Yqubitonetimeinit)}
%
\newcommand{\yqubitonetimeanyveccomp}
{$\Yqubitonetimeanyveccomp$}
\newcommand{\Yqubitonetimeanyveccomp}
{C_{\Yqubitoneindexstd}(t)}
%
%
%
\newcommand{\yqubitonetimeinitanyhamilton}
{$\Yqubitonetimeinitanyhamilton$}
\newcommand{\Yqubitonetimeinitanyhamilton}
{
\Yopmix
_{\Yqubitoneindexstd}}
%
%
%
%
\newcommand{\yqubitonetimenonestateindexqubitoneindexstd}
{$\Yqubitonetimenonestateindexqubitoneindexstd$}
\newcommand{\Yqubitonetimenonestateindexqubitoneindexstd}
{| \psi_{\Yqubitoneindexstd} \rangle}
%
%
%
%
%
%
%
%
\newcommand{\yqubitbothspace}
{$\Yqubitbothspace$}
\newcommand{\Yqubitbothspace}
{\Yqubitonesevspace}
%
%
%
%
%
%
%
%
%
%
\newcommand{\ymixsyststateinitial}
{$\Ymixsyststateinitial$}
\newcommand{\Ymixsyststateinitial}
{| \psi 
( \Yqubitonetimeinit ) \rangle
}
%
%
%
%
%
%
%
%
%
%
%
\newcommand{\ytwoqubitwritereadtimeinterval}
{$\Ytwoqubitwritereadtimeinterval$}
\newcommand{\Ytwoqubitwritereadtimeinterval}
{\tau}
%
%
%
%
%
\newcommand{\ytwoqubitwritereadtimeintervalindexone}
{$\Ytwoqubitwritereadtimeintervalindexone$}
\newcommand{\Ytwoqubitwritereadtimeintervalindexone}
{\Ytwoqubitwritereadtimeinterval_{1}}
%
\newcommand{\ytwoqubitwritereadtimeintervalindextwo}
{$\Ytwoqubitwritereadtimeintervalindextwo$}
\newcommand{\Ytwoqubitwritereadtimeintervalindextwo}
{\Ytwoqubitwritereadtimeinterval_{2}}
%
%
\newcommand{\ytwoqubitwritereadtimeintervalindexthree}
{$\Ytwoqubitwritereadtimeintervalindexthree$}
\newcommand{\Ytwoqubitwritereadtimeintervalindexthree}
{\Ytwoqubitwritereadtimeinterval_{3}}
%
%
%
%
%
%
%
%
%
%
%
%
\newcommand{\yqubitbothtimeinitstatecoefnot}
{$\Yqubitbothtimeinitstatecoefnot$}
\newcommand{\Yqubitbothtimeinitstatecoefnot}
{c}
%
\newcommand{\yqubitbothtimeinitstatecoefplusplus}
{$\Yqubitbothtimeinitstatecoefplusplus$}
\newcommand{\Yqubitbothtimeinitstatecoefplusplus}
{\Yqubitbothtimeinitstatecoefnot_{
1
}( \Yqubitonetimeinit )}
%
\newcommand{\yqubitbothtimeinitstatecoefplusminus}
{$\Yqubitbothtimeinitstatecoefplusminus$}
\newcommand{\Yqubitbothtimeinitstatecoefplusminus}
{\Yqubitbothtimeinitstatecoefnot_{
2
}( \Yqubitonetimeinit )}
%
\newcommand{\yqubitbothtimeinitstatecoefminusplus}
{$\Yqubitbothtimeinitstatecoefminusplus$}
\newcommand{\Yqubitbothtimeinitstatecoefminusplus}
{\Yqubitbothtimeinitstatecoefnot_{
3
}( \Yqubitonetimeinit )}
%
\newcommand{\yqubitbothtimeinitstatecoefminusminus}
{$\Yqubitbothtimeinitstatecoefminusminus$}
\newcommand{\Yqubitbothtimeinitstatecoefminusminus}
{\Yqubitbothtimeinitstatecoefnot_{
4
}( \Yqubitonetimeinit )}
%
%
%
%
\newcommand{\yqubitbothtimeinitstatecoefindexequalstd}
{$\Yqubitbothtimeinitstatecoefindexequalstd$}
\newcommand{\Yqubitbothtimeinitstatecoefindexequalstd}
{k}
%
\newcommand{\yqubitbothtimeinitstatecoefwithindexstd}
{$\Yqubitbothtimeinitstatecoefwithindexstd$}
\newcommand{\Yqubitbothtimeinitstatecoefwithindexstd}
{\Yqubitbothtimeinitstatecoefnot
_{\Yqubitbothtimeinitstatecoefindexequalstd}( \Yqubitonetimeinit )}
%
%
%
%
%
%
%
%
%
%
%
\newcommand{\yqubitbothtimefinalstatecoefplusplus}
{$\Yqubitbothtimefinalstatecoefplusplus$}
\newcommand{\Yqubitbothtimefinalstatecoefplusplus}
{\Yqubitbothtimeinitstatecoefnot_{1}( \Yqubitonetimefinal )}
%
\newcommand{\yqubitbothtimefinalstatecoefplusminus}
{$\Yqubitbothtimefinalstatecoefplusminus$}
\newcommand{\Yqubitbothtimefinalstatecoefplusminus}
{\Yqubitbothtimeinitstatecoefnot_{2}( \Yqubitonetimefinal )}
%
\newcommand{\yqubitbothtimefinalstatecoefminusplus}
{$\Yqubitbothtimefinalstatecoefminusplus$}
\newcommand{\Yqubitbothtimefinalstatecoefminusplus}
{\Yqubitbothtimeinitstatecoefnot_{3}( \Yqubitonetimefinal )}
%
\newcommand{\yqubitbothtimefinalstatecoefminusminus}
{$\Yqubitbothtimefinalstatecoefminusminus$}
\newcommand{\Yqubitbothtimefinalstatecoefminusminus}
{\Yqubitbothtimeinitstatecoefnot_{4}( \Yqubitonetimefinal )}
%
%
%
%
%

%
\newcommand{\yqubitbothtimefinalstatecoefindexindexstd}
{$\Yqubitbothtimefinalstatecoefindexindexstd$}
\newcommand{\Yqubitbothtimefinalstatecoefindexindexstd}
{k}
%
\newcommand{\yqubitbothtimefinalstatecoefindexstd}
{$\Yqubitbothtimefinalstatecoefindexstd$}
\newcommand{\Yqubitbothtimefinalstatecoefindexstd}
{\Yqubitbothtimeinitstatecoefnot
_{\Yqubitbothtimefinalstatecoefindexindexstd}( \Yqubitonetimefinal )}
%
\newcommand{\yqubitbothtimefinalstatecoefindexstdpolarmod}
{$\Yqubitbothtimefinalstatecoefindexstdpolarmod$}
\newcommand{\Yqubitbothtimefinalstatecoefindexstdpolarmod}
{\rho
_{\Yqubitbothtimefinalstatecoefindexindexstd}
}
%
\newcommand{\yqubitbothtimefinalstatecoefpolarphasenot}
{$\Yqubitbothtimefinalstatecoefpolarphasenot$}
\newcommand{\Yqubitbothtimefinalstatecoefpolarphasenot}
{\xi}
%
\newcommand{\yqubitbothtimefinalstatecoefindexstdpolarphase}
{$\Yqubitbothtimefinalstatecoefindexstdpolarphase$}
\newcommand{\Yqubitbothtimefinalstatecoefindexstdpolarphase}
{\Yqubitbothtimefinalstatecoefpolarphasenot
_{\Yqubitbothtimefinalstatecoefindexindexstd}
}
%
%
%
%
\newcommand{\yqubitbothtimefinalstatecoefpluspluspolarphase}
{$\Yqubitbothtimefinalstatecoefpluspluspolarphase$}
\newcommand{\Yqubitbothtimefinalstatecoefpluspluspolarphase}
{\Yqubitbothtimefinalstatecoefpolarphasenot
_{1}
}
%
\newcommand{\yqubitbothtimefinalstatecoefplusminuspolarphase}
{$\Yqubitbothtimefinalstatecoefplusminuspolarphase$}
\newcommand{\Yqubitbothtimefinalstatecoefplusminuspolarphase}
{\Yqubitbothtimefinalstatecoefpolarphasenot
_{2}
}
%
\newcommand{\yqubitbothtimefinalstatecoefminuspluspolarphase}
{$\Yqubitbothtimefinalstatecoefminuspluspolarphase$}
\newcommand{\Yqubitbothtimefinalstatecoefminuspluspolarphase}
{\Yqubitbothtimefinalstatecoefpolarphasenot
_{3}
}
%
\newcommand{\yqubitbothtimefinalstatecoefminusminuspolarphase}
{$\Yqubitbothtimefinalstatecoefminusminuspolarphase$}
\newcommand{\Yqubitbothtimefinalstatecoefminusminuspolarphase}
{\Yqubitbothtimefinalstatecoefpolarphasenot
_{4}
}
%
%
%
%
%
%
%
%
%
\newcommand{\yveccompsyststatetinit}
{$\Yveccompsyststatetinit$}
\newcommand{\Yveccompsyststatetinit}
{C_{+
} 
(
\Yqubitonetimeinit
)
}
%
%
%
%
\newcommand{\yveccompsyststatetfinal}
{$\Yveccompsyststatetfinal$}
\newcommand{\Yveccompsyststatetfinal}
{C_{+
} (t)}
%
%
%
%
%
%
%
\newcommand{\ymixsyststatefinal}
{$\Ymixsyststatefinal$}
\newcommand{\Ymixsyststatefinal}
{| \psi (t) 
\rangle
}
%
%
%
%
%
%
%
%
\newcommand{\yopmix}
{$\Yopmix$}
\newcommand{\Yopmix}
{M}
%
%
%
%
\newcommand{\yopmixestim}
{$\Yopmixestim$}
\newcommand{\Yopmixestim}
{\widehat{\Yopmix}}
%
%
%
%
%
%
%
%
%
%
\newcommand{\yopmixbases}
{$\Yopmixbases$}
\newcommand{\Yopmixbases}
{Q}
%
\newcommand{\yopmixbasesimagin}
{$\Yopmixbasesimagin$}
\newcommand{\Yopmixbasesimagin}
{\Yopmixbases_{\Ysqrtminusone}}
%
\newcommand{\yopmixdiag}
{$\Yopmixdiag$}
\newcommand{\Yopmixdiag}
{D}
%
%
%
%
\newcommand{\yopmixdiagestim}
{$\Yopmixdiagestim$}
\newcommand{\Yopmixdiagestim}
{\widehat{\Yopmixdiag}}
%
%
%
%
%
%
%
%
%
%
\newcommand{\yopsep}
{$\Yopsep$}
\newcommand{\Yopsep}
{U}
%
%
%
%
%
%
%
%
%
%
\newcommand{\ysepsyststateoutnnot}
{$\Ysepsyststateoutnot$}
\newcommand{\Ysepsyststateoutnot}
{\Phi}
%
%
%
\newcommand{\ysepsyststateout}
{$\Ysepsyststateout$}
\newcommand{\Ysepsyststateout}
{|
\Ysepsyststateoutnot
\rangle
%
}
%
%
%
%
%
%
%
%
%
%
%
\newcommand{\yveccompsepsyststateout}
{$\Yveccompsepsyststateout$}
\newcommand{\Yveccompsepsyststateout}
{C
%
}
%
%
%
%
\newcommand{\ysepsyststateoutcoefnot}
{$\Ysepsyststateoutcoefnot$}
\newcommand{\Ysepsyststateoutcoefnot}
{c}
%
%
%
%
\newcommand{\ysepsystdedicstateoutcoefnot}
{$\Ysepsystdedicstateoutcoefnot$}
\newcommand{\Ysepsystdedicstateoutcoefnot}
{c}
%
\newcommand{\ysepsystdedicstateoutcoefplusplus}
{$\Ysepsystdedicstateoutcoefplusplus$}
\newcommand{\Ysepsystdedicstateoutcoefplusplus}
{\Ysepsystdedicstateoutcoefnot_{
1
}}
%
\newcommand{\ysepsystdedicstateoutcoefplusminus}
{$\Ysepsystdedicstateoutcoefplusminus$}
\newcommand{\Ysepsystdedicstateoutcoefplusminus}
{\Ysepsystdedicstateoutcoefnot_{
2
}}
%
\newcommand{\ysepsystdedicstateoutcoefminusplus}
{$\Ysepsystdedicstateoutcoefminusplus$}
\newcommand{\Ysepsystdedicstateoutcoefminusplus}
{\Ysepsystdedicstateoutcoefnot_{
3
}}
%
\newcommand{\ysepsystdedicstateoutcoefminusminus}
{$\Ysepsystdedicstateoutcoefminusminus$}
\newcommand{\Ysepsystdedicstateoutcoefminusminus}
{\Ysepsystdedicstateoutcoefnot_{
4
}}
%
%
%
%
%
\newcommand{\ysepsystdedicstateoutcoefpart}
{$\Ysepsystdedicstateoutcoefpart$}
\newcommand{\Ysepsystdedicstateoutcoefpart}
{\Ysepsystdedicstateoutcoefnot_{5}}
%
%
%
%
%
\newcommand{\ysepsystdedicstateoutcoefpartphase}
{$\Ysepsystdedicstateoutcoefpartphase$}
\newcommand{\Ysepsystdedicstateoutcoefpartphase}
{\phi_{5}}
%
%
%
%
%
\newcommand{\ysepsystdedicstateoutcoefwithindexstd}
{$\Ysepsystdedicstateoutcoefwithindexstd$}
\newcommand{\Ysepsystdedicstateoutcoefwithindexstd}
{\Ysepsystdedicstateoutcoefnot_{\Yqubitbothtimeinitstatecoefindexequalstd}}
%
%
%
%
%
%
%
%
%
%
%
%

%
\newcommand{\ysepsystdedicstateoutcoefplusplusseqindex}
{$\Ysepsystdedicstateoutcoefplusplusseqindex$}
\newcommand{\Ysepsystdedicstateoutcoefplusplusseqindex}
{
\Ysepsystdedicstateoutcoefplusplus
(
\Ytwoqubitseqindex
)
}
%
\newcommand{\ysepsystdedicstateoutcoefplusminusseqindex}
{$\Ysepsystdedicstateoutcoefplusminusseqindex$}
\newcommand{\Ysepsystdedicstateoutcoefplusminusseqindex}
{
\Ysepsystdedicstateoutcoefplusminus
(
\Ytwoqubitseqindex
)
}
%
\newcommand{\ysepsystdedicstateoutcoefminusplusseqindex}
{$\Ysepsystdedicstateoutcoefminusplusseqindex$}
\newcommand{\Ysepsystdedicstateoutcoefminusplusseqindex}
{
\Ysepsystdedicstateoutcoefminusplus
(
\Ytwoqubitseqindex
)
}
%
\newcommand{\ysepsystdedicstateoutcoefminusminusseqindex}
{$\Ysepsystdedicstateoutcoefminusminusseqindex$}
\newcommand{\Ysepsystdedicstateoutcoefminusminusseqindex}
{
\Ysepsystdedicstateoutcoefminusminus
(
\Ytwoqubitseqindex
)
}
%
%
%
%
%
%
%
%
%
%
%
%
\newcommand{\ysepsystdedicstateoutcoefprobnot}
{$\Ysepsystdedicstateoutcoefprobnot$}
\newcommand{\Ysepsystdedicstateoutcoefprobnot}
{P}
%
\newcommand{\ysepsystdedicstateoutcoefprobplusplus}
{$\Ysepsystdedicstateoutcoefprobplusplus$}
\newcommand{\Ysepsystdedicstateoutcoefprobplusplus}
{\Ysepsystdedicstateoutcoefprobnot_{
1
%
%
%
z
}}
%
\newcommand{\ysepsystdedicstateoutcoefprobplusminus}
{$\Ysepsystdedicstateoutcoefprobplusminus$}
\newcommand{\Ysepsystdedicstateoutcoefprobplusminus}
{\Ysepsystdedicstateoutcoefprobnot_{
2
%
%
%
z
}}
%
\newcommand{\ysepsystdedicstateoutcoefprobminusplus}
{$\Ysepsystdedicstateoutcoefprobminusplus$}
\newcommand{\Ysepsystdedicstateoutcoefprobminusplus}
{\Ysepsystdedicstateoutcoefprobnot_{
3
%
%
%
z
}}
%
\newcommand{\ysepsystdedicstateoutcoefprobminusminus}
{$\Ysepsystdedicstateoutcoefprobminusminus$}
\newcommand{\Ysepsystdedicstateoutcoefprobminusminus}
{\Ysepsystdedicstateoutcoefprobnot_{
4
%
%
%
z
}}
%
%
%
%
%
\newcommand{\ysepsystdedicstateoutcoefprobplusplusseqindex}
{$\Ysepsystdedicstateoutcoefprobplusplusseqindex$}
\newcommand{\Ysepsystdedicstateoutcoefprobplusplusseqindex}
{
\Ysepsystdedicstateoutcoefprobplusplus
(
\Ytwoqubitseqindex
)
}
%
\newcommand{\ysepsystdedicstateoutcoefprobplusminusseqindex}
{$\Ysepsystdedicstateoutcoefprobplusminusseqindex$}
\newcommand{\Ysepsystdedicstateoutcoefprobplusminusseqindex}
{
\Ysepsystdedicstateoutcoefprobplusminus
(
\Ytwoqubitseqindex
)
}
%
\newcommand{\ysepsystdedicstateoutcoefprobminusplusseqindex}
{$\Ysepsystdedicstateoutcoefprobminusplusseqindex$}
\newcommand{\Ysepsystdedicstateoutcoefprobminusplusseqindex}
{
\Ysepsystdedicstateoutcoefprobminusplus
(
\Ytwoqubitseqindex
)
}
%
\newcommand{\ysepsystdedicstateoutcoefprobminusminusseqindex}
{$\Ysepsystdedicstateoutcoefprobminusminusseqindex$}
\newcommand{\Ysepsystdedicstateoutcoefprobminusminusseqindex}
{
\Ysepsystdedicstateoutcoefprobminusminus
(
\Ytwoqubitseqindex
)
}
%
%
%
%
%
%
%
%
\newcommand{\ysepsystdedicstateoutcoefprobnotox}
{$\Ysepsystdedicstateoutcoefprobnotox$}
\newcommand{\Ysepsystdedicstateoutcoefprobnotox}
{P}
%
\newcommand{\ysepsystdedicstateoutcoefprobplusplusox}
{$\Ysepsystdedicstateoutcoefprobplusplusox$}
\newcommand{\Ysepsystdedicstateoutcoefprobplusplusox}
{\Ysepsystdedicstateoutcoefprobnot_{1x}}
%
\newcommand{\ysepsystdedicstateoutcoefprobplusminusox}
{$\Ysepsystdedicstateoutcoefprobplusminusox$}
\newcommand{\Ysepsystdedicstateoutcoefprobplusminusox}
{\Ysepsystdedicstateoutcoefprobnot_{2x}}
%
\newcommand{\ysepsystdedicstateoutcoefprobminusplusox}
{$\Ysepsystdedicstateoutcoefprobminusplusox$}
\newcommand{\Ysepsystdedicstateoutcoefprobminusplusox}
{\Ysepsystdedicstateoutcoefprobnot_{3x}}
%
\newcommand{\ysepsystdedicstateoutcoefprobminusminusox}
{$\Ysepsystdedicstateoutcoefprobminusminusox$}
\newcommand{\Ysepsystdedicstateoutcoefprobminusminusox}
{\Ysepsystdedicstateoutcoefprobnot_{4x}}
%
%
%
%
\newcommand{\ysepsystdedicstateoutcoefprobplusplusoxseqindex}
{$\Ysepsystdedicstateoutcoefprobplusplusoxseqindex$}
\newcommand{\Ysepsystdedicstateoutcoefprobplusplusoxseqindex}
{
\Ysepsystdedicstateoutcoefprobplusplusox
(
\Ytwoqubitseqindex
)
}
%
\newcommand{\ysepsystdedicstateoutcoefprobplusminusoxseqindex}
{$\Ysepsystdedicstateoutcoefprobplusminusoxseqindex$}
\newcommand{\Ysepsystdedicstateoutcoefprobplusminusoxseqindex}
{
\Ysepsystdedicstateoutcoefprobplusminusox
(
\Ytwoqubitseqindex
)
}
%
\newcommand{\ysepsystdedicstateoutcoefprobminusplusoxseqindex}
{$\Ysepsystdedicstateoutcoefprobminusplusoxseqindex$}
\newcommand{\Ysepsystdedicstateoutcoefprobminusplusoxseqindex}
{
\Ysepsystdedicstateoutcoefprobminusplusox
(
\Ytwoqubitseqindex
)
}
%
\newcommand{\ysepsystdedicstateoutcoefprobminusminusoxseqindex}
{$\Ysepsystdedicstateoutcoefprobminusminusoxseqindex$}
\newcommand{\Ysepsystdedicstateoutcoefprobminusminusoxseqindex}
{
\Ysepsystdedicstateoutcoefprobminusminusox
(
\Ytwoqubitseqindex
)
}
%
%
%
%
%
%
%
%
%
\newcommand{\ysepsystdedicopmixbasesonestateoutcoefprobzznot}
{$\Ysepsystdedicopmixbasesonestateoutcoefprobzznot$}
\newcommand{\Ysepsystdedicopmixbasesonestateoutcoefprobzznot}
{\Ysepsystdedicstateoutcoefprobnot}
%
\newcommand{\ysepsystdedicopmixbasesonestateoutcoefprobzzplusplus}
{$\Ysepsystdedicopmixbasesonestateoutcoefprobzzplusplus$}
\newcommand{\Ysepsystdedicopmixbasesonestateoutcoefprobzzplusplus}
{\Ysepsystdedicopmixbasesonestateoutcoefprobzznot
_{1z}
( \Yopmixbases )
}
%
\newcommand{\ysepsystdedicopmixbasesonestateoutcoefprobzzplusminus}
{$\Ysepsystdedicopmixbasesonestateoutcoefprobzzplusminus$}
\newcommand{\Ysepsystdedicopmixbasesonestateoutcoefprobzzplusminus}
{\Ysepsystdedicopmixbasesonestateoutcoefprobzznot
_{2z}
( \Yopmixbases )
}
%
\newcommand{\ysepsystdedicopmixbasesonestateoutcoefprobzzminusplus}
{$\Ysepsystdedicopmixbasesonestateoutcoefprobzzminusplus$}
\newcommand{\Ysepsystdedicopmixbasesonestateoutcoefprobzzminusplus}
{\Ysepsystdedicopmixbasesonestateoutcoefprobzznot
_{3z}
( \Yopmixbases )
}
%
\newcommand{\ysepsystdedicopmixbasesonestateoutcoefprobzzminusminus}
{$\Ysepsystdedicopmixbasesonestateoutcoefprobzzminusminus$}
\newcommand{\Ysepsystdedicopmixbasesonestateoutcoefprobzzminusminus}
{\Ysepsystdedicopmixbasesonestateoutcoefprobzznot
_{4z}
( \Yopmixbases )
}
%
%
%
%
\newcommand{\ysepsystdedicopmixbasesimaginonestateoutcoefprobzznot}
{$\Ysepsystdedicopmixbasesimaginonestateoutcoefprobzznot$}
\newcommand{\Ysepsystdedicopmixbasesimaginonestateoutcoefprobzznot}
{\Ysepsystdedicstateoutcoefprobnot}
%
\newcommand{\ysepsystdedicopmixbasesimaginonestateoutcoefprobzzplusplus}
{$\Ysepsystdedicopmixbasesimaginonestateoutcoefprobzzplusplus$}
\newcommand{\Ysepsystdedicopmixbasesimaginonestateoutcoefprobzzplusplus}
{\Ysepsystdedicopmixbasesimaginonestateoutcoefprobzznot
_{1z}
( \Yopmixbasesimagin )
}
%
\newcommand{\ysepsystdedicopmixbasesimaginonestateoutcoefprobzzplusminus}
{$\Ysepsystdedicopmixbasesimaginonestateoutcoefprobzzplusminus$}
\newcommand{\Ysepsystdedicopmixbasesimaginonestateoutcoefprobzzplusminus}
{\Ysepsystdedicopmixbasesimaginonestateoutcoefprobzznot
_{2z}
( \Yopmixbasesimagin )
}
%
\newcommand{\ysepsystdedicopmixbasesimaginonestateoutcoefprobzzminusplus}
{$\Ysepsystdedicopmixbasesimaginonestateoutcoefprobzzminusplus$}
\newcommand{\Ysepsystdedicopmixbasesimaginonestateoutcoefprobzzminusplus}
{\Ysepsystdedicopmixbasesimaginonestateoutcoefprobzznot
_{3z}
( \Yopmixbasesimagin )
}
%
\newcommand{\ysepsystdedicopmixbasesimaginonestateoutcoefprobzzminusminus}
{$\Ysepsystdedicopmixbasesimaginonestateoutcoefprobzzminusminus$}
\newcommand{\Ysepsystdedicopmixbasesimaginonestateoutcoefprobzzminusminus}
{\Ysepsystdedicopmixbasesimaginonestateoutcoefprobzznot
_{4z}
( \Yopmixbasesimagin )
}
%
%
%
%
%
%
%
%
%
%
%
\newcommand{\ysepmixdiag}
{$\Ysepmixdiag$}
\newcommand{\Ysepmixdiag}
{\tilde{\Yopmixdiag}}
%
%
\newcommand{\ysepmixdiagelnot}
{$\Ysepmixdiagelnot$}
\newcommand{\Ysepmixdiagelnot}
{\gamma}
%
%
\newcommand{\ysepmixdiagelomegaoneone}
{$\Ysepmixdiagelomegaoneone$}
\newcommand{\Ysepmixdiagelomegaoneone}
{
\Ysepmixdiagelnot
_1}
%
\newcommand{\ysepmixdiagelomegaonezero}
{$\Ysepmixdiagelomegaonezero$}
\newcommand{\Ysepmixdiagelomegaonezero}
{
\Ysepmixdiagelnot
_2}
%
\newcommand{\ysepmixdiagelomegazerozero}
{$\Ysepmixdiagelomegazerozero$}
\newcommand{\Ysepmixdiagelomegazerozero}
{
\Ysepmixdiagelnot
_3}
%
\newcommand{\ysepmixdiagelomegaoneminusone}
{$\Ysepmixdiagelomegaoneminusone$}
\newcommand{\Ysepmixdiagelomegaoneminusone}
{
\Ysepmixdiagelnot
_4}
%
%
%
%
%
\newcommand{\yindexstdforsepmixdiagel}
{$\Yindexstdforsepmixdiagel$}
\newcommand{\Yindexstdforsepmixdiagel}
{k}
%
\newcommand{\ysepmixdiagelindexstd}
{$\Ysepmixdiagelindexstd$}
\newcommand{\Ysepmixdiagelindexstd}
{\gamma_{\Yindexstdforsepmixdiagel}}
%
\newcommand{\ysepmixdiagelphysnot}
{$\Ysepmixdiagelphysnot$}
\newcommand{\Ysepmixdiagelphysnot}
{V}
%
\newcommand{\ysepmixdiagelomegaoneonephys}
{$\Ysepmixdiagelomegaoneonephys$}
\newcommand{\Ysepmixdiagelomegaoneonephys}
{\Ysepmixdiagelphysnot_1}
%
\newcommand{\ysepmixdiagelomegaonezerophys}
{$\Ysepmixdiagelomegaonezerophys$}
\newcommand{\Ysepmixdiagelomegaonezerophys}
{\Ysepmixdiagelphysnot_2}
%
\newcommand{\ysepmixdiagelomegazerozerophys}
{$\Ysepmixdiagelomegazerozerophys$}
\newcommand{\Ysepmixdiagelomegazerozerophys}
{\Ysepmixdiagelphysnot_3}
%
\newcommand{\ysepmixdiagelomegaoneminusonephys}
{$\Ysepmixdiagelomegaoneminusonephys$}
\newcommand{\Ysepmixdiagelomegaoneminusonephys}
{\Ysepmixdiagelphysnot_4}
%
\newcommand{\ysepmixdiagelindexstdphys}
{$\Ysepmixdiagelindexstdphys$}
\newcommand{\Ysepmixdiagelindexstdphys}
{\Ysepmixdiagelphysnot_{\Yindexstdforsepmixdiagel}}
%
\newcommand{\ysepmixdiagelindexstdfuncphystoel}
{$\Ysepmixdiagelindexstdfuncphystoel$}
\newcommand{\Ysepmixdiagelindexstdfuncphystoel}
{g_{\Yindexstdforsepmixdiagel}}
%
%
%
%
%
%
\newcommand{\ysepmixdiagelomegaoneoneideal}
{$\Ysepmixdiagelomegaoneoneideal$}
\newcommand{\Ysepmixdiagelomegaoneoneideal}
{
\Ysepmixdiagelnot
_{1d}
}
%
\newcommand{\ysepmixdiagelomegaonezeroideal}
{$\Ysepmixdiagelomegaonezeroideal$}
\newcommand{\Ysepmixdiagelomegaonezeroideal}
{
\Ysepmixdiagelnot
_{2d}}
%
\newcommand{\ysepmixdiagelomegazerozeroideal}
{$\Ysepmixdiagelomegazerozeroideal$}
\newcommand{\Ysepmixdiagelomegazerozeroideal}
{
\Ysepmixdiagelnot
_{3d}}
%
\newcommand{\ysepmixdiagelomegaoneminusoneideal}
{$\Ysepmixdiagelomegaoneminusoneideal$}
\newcommand{\Ysepmixdiagelomegaoneminusoneideal}
{
\Ysepmixdiagelnot
_{4d}}
%
%
%
%
%
%
%
%
%
\newcommand{\ysepmixdiagelindexstdideal}
{$\Ysepmixdiagelindexstdideal$}
\newcommand{\Ysepmixdiagelindexstdideal}
{
\Ysepmixdiagelnot
_{\Yindexstdforsepmixdiagel d}}
%
%
%
%
%
%
%
%
%
%
%

%
%
\newcommand{\ysepmixdiagelnotestim}
{$\Ysepmixdiagelnotestim$}
\newcommand{\Ysepmixdiagelnotestim}
{\widehat{\Ysepmixdiagelnot}}
%
\newcommand{\ysepmixdiagelomegaoneoneestim}
{$\Ysepmixdiagelomegaoneoneestim$}
\newcommand{\Ysepmixdiagelomegaoneoneestim}
{\Ysepmixdiagelnotestim _1}
%
\newcommand{\ysepmixdiagelomegaonezeroestim}
{$\Ysepmixdiagelomegaonezeroestim$}
\newcommand{\Ysepmixdiagelomegaonezeroestim}
{\Ysepmixdiagelnotestim _2}
%
\newcommand{\ysepmixdiagelomegazerozeroestim}
{$\Ysepmixdiagelomegazerozeroestim$}
\newcommand{\Ysepmixdiagelomegazerozeroestim}
{\Ysepmixdiagelnotestim _3}
%
\newcommand{\ysepmixdiagelomegaoneminusoneestim}
{$\Ysepmixdiagelomegaoneminusoneestim$}
\newcommand{\Ysepmixdiagelomegaoneminusoneestim}
{\Ysepmixdiagelnotestim _4}
%
%
%
%
%
%
%
%
%
%
%
%
%
%
%
%
\newcommand{\yopglob}
{$\Yopglob$}
\newcommand{\Yopglob}
{G}
%
%
%
%
\newcommand{\yopglobdiag}
{$\Yopglobdiag$}
\newcommand{\Yopglobdiag}
{\Delta}
%
%
\newcommand{\yopglobdiagelnot}
{$\Yopglobdiagelnot$}
\newcommand{\Yopglobdiagelnot}
{\delta}
%
%
\newcommand{\yopglobdiagelomegaoneone}
{$\Yopglobdiagelomegaoneone$}
\newcommand{\Yopglobdiagelomegaoneone}
{
\Yopglobdiagelnot
_1}
%
\newcommand{\yopglobdiagelomegaonezero}
{$\Yopglobdiagelomegaonezero$}
\newcommand{\Yopglobdiagelomegaonezero}
{
\Yopglobdiagelnot
_2}
%
\newcommand{\yopglobdiagelomegazerozero}
{$\Yopglobdiagelomegazerozero$}
\newcommand{\Yopglobdiagelomegazerozero}
{
\Yopglobdiagelnot
_3}
%
\newcommand{\yopglobdiagelomegaoneminusone}
{$\Yopglobdiagelomegaoneminusone$}
\newcommand{\Yopglobdiagelomegaoneminusone}
{
\Yopglobdiagelnot
_4}
%
%
%
%
\newcommand{\yopglobdiagelindexstdforsepmixdiagel}
{$\Yopglobdiagelindexstdforsepmixdiagel$}
\newcommand{\Yopglobdiagelindexstdforsepmixdiagel}
{
\Yopglobdiagelnot_{\Yindexstdforsepmixdiagel}}
%
%
%
%
\newcommand{\yopglobdiagelcombinone}
{$\Yopglobdiagelcombinone$}
\newcommand{\Yopglobdiagelcombinone}
{
\Yopglobdiagelnot
_5}
%
%
%
%
\newcommand{\yalphaonebetatwomod}
{$\Yalphaonebetatwomod$}
\newcommand{\Yalphaonebetatwomod}
{A_1}
%
\newcommand{\yalphaonebetatwophase}
{$\Yalphaonebetatwophase$}
\newcommand{\Yalphaonebetatwophase}
{\xi_1}
%
%
%
\newcommand{\yalphatwobetaonemod}
{$\Yalphatwobetaonemod$}
\newcommand{\Yalphatwobetaonemod}
{A_2}
%
\newcommand{\yalphatwobetaonephase}
{$\Yalphatwobetaonephase$}
\newcommand{\Yalphatwobetaonephase}
{\xi_2}
%
%
%
%
%
%
%
%
%
%
%
\newcommand{\ynonentangcondmodulusquantityone}
{$\Ynonentangcondmodulusquantityone$}
\newcommand{\Ynonentangcondmodulusquantityone}
{
B
_1}
%
\newcommand{\ynonentangcondmodulusquantitytwo}
{$\Ynonentangcondmodulusquantitytwo$}
\newcommand{\Ynonentangcondmodulusquantitytwo}
{
B
_2}
%
%
%
%
%
%
%
%
%
%
%
\newcommand{\ytwoqubitsbasisplusplus}
{$\Ytwoqubitsbasisplusplus$}
\newcommand{\Ytwoqubitsbasisplusplus}
{{\cal B} _+}
%
%
%
%
\newcommand{\ytwoqubitsbasisplusxplusx}
{$\Ytwoqubitsbasisplusxplusx$}
\newcommand{\Ytwoqubitsbasisplusxplusx}
{{\cal B} _{+x}}
%
%
%
%
%
%
%
%
%
%
%
%
%
%
%
%
\newcommand{\ysepsystoutnonentangcondmodulusmultiple}
{$\Ysepsystoutnonentangcondmodulusmultiple$}
\newcommand{\Ysepsystoutnonentangcondmodulusmultiple}
{m}
%
%
%
%
%
%
%
%
%
%
\newcommand{\ysepsystoutnonentangcondmodulusmultipleestim}
{$\Ysepsystoutnonentangcondmodulusmultipleestim$}
\newcommand{\Ysepsystoutnonentangcondmodulusmultipleestim}
{\widehat{\Ysepsystoutnonentangcondmodulusmultiple}}
%
\newcommand{\ysepsystoutnonentangcondmodulusmultipleestimminusactual}
{$\Ysepsystoutnonentangcondmodulusmultipleestimminusactual$}
\newcommand{\Ysepsystoutnonentangcondmodulusmultipleestimminusactual}
{\Delta_{\Ysepsystoutnonentangcondmodulusmultiple}}
%
%
%
%
%
%
%
%
%
%
%
\newcommand{\yqubitnbarb}
{$\Yqubitnbarb$}
\newcommand{\Yqubitnbarb}
{Q}
%
%
%
%
%
%
\newcommand{\yqubitindexstd}
{$\Yqubitindexstd$}
\newcommand{\Yqubitindexstd}
{j}
%
%
%
%
%
%
%
%
\newcommand{\yqubitindexstdtwo}
          {$\Yqubitindexstdtwo$}
\newcommand{\Yqubitindexstdtwo}
           {q}                 
%
%
%
\newcommand{\yqubitonespaceindexqubitnbarb}
{$\Yqubitonespaceindexqubitnbarb$}
\newcommand{\Yqubitonespaceindexqubitnbarb}
{\Yqubitonesevspace_{\Yqubitnbarb}}
%
%
%
\newcommand{\yqubitnbarbtimenonestatenot}
{$\Yqubitnbarbtimenonestatenot$}
\newcommand{\Yqubitnbarbtimenonestatenot}
{\psi}
%
%
%
\newcommand{\yqubitnbarbtimenonestate}
{$\Yqubitnbarbtimenonestate$}
\newcommand{\Yqubitnbarbtimenonestate}
{| \Yqubitnbarbtimenonestatenot \rangle}
%
%
%
\newcommand{\yqubitnbarbtimenonestateseqindex}
{$\Yqubitnbarbtimenonestateseqindex$}
\newcommand{\Yqubitnbarbtimenonestateseqindex}
{| \Yqubitnbarbtimenonestatenot
(
\Ytwoqubitseqindex
)
\rangle}
%
%
%
%
\newcommand{\yqubitnbarbstatecoefindexequalstd}
{$\Yqubitnbarbstatecoefindexequalstd$}
\newcommand{\Yqubitnbarbstatecoefindexequalstd}
{k}
%
%
%
\newcommand{\yqubitnbarbspaceoverallvecbasiscoefindexequalstd}
{$\Yqubitnbarbspaceoverallvecbasiscoefindexequalstd$}
\newcommand{\Yqubitnbarbspaceoverallvecbasiscoefindexequalstd}
{
|
%
{\Yqubitnbarbstatecoefindexequalstd}
\rangle
}
%
%
%
\newcommand{\yqubitnbarbspaceoverallcoefindexequalstdwithvecbasis}
{$\Yqubitnbarbspaceoverallcoefindexequalstdwithvecbasis$}
\newcommand{\Yqubitnbarbspaceoverallcoefindexequalstdwithvecbasis}
{
c
_
{\Yqubitnbarbstatecoefindexequalstd}
}
%
\newcommand{\yqubitnbarbspaceoverallcoefindexequalstdwithvecbasisseqindex}
{$\Yqubitnbarbspaceoverallcoefindexequalstdwithvecbasisseqindex$}
\newcommand{\Yqubitnbarbspaceoverallcoefindexequalstdwithvecbasisseqindex}
{
c
_
{\Yqubitnbarbstatecoefindexequalstd}
(
\Ytwoqubitseqindex
)
}
%
%
%
\newcommand{\yqubitnbarbspaceoveralleventindexequalstdwithvecbasis}
{$\Yqubitnbarbspaceoveralleventindexequalstdwithvecbasis$}
\newcommand{\Yqubitnbarbspaceoveralleventindexequalstdwithvecbasis}
{
A
_
{\Yqubitnbarbstatecoefindexequalstd}
}
%
%
%
\newcommand{\yqubitnbarbspaceoveralleventindexequalstdwithvecbasisprob}
{$\Yqubitnbarbspaceoveralleventindexequalstdwithvecbasisprob$}
\newcommand{\Yqubitnbarbspaceoveralleventindexequalstdwithvecbasisprob}
{
P
(
\Yqubitnbarbspaceoveralleventindexequalstdwithvecbasis
)
}
%
\newcommand{\yqubitnbarbspaceoveralleventindexequalstdwithvecbasisprobexpect}
{$\Yqubitnbarbspaceoveralleventindexequalstdwithvecbasisprobexpect$}
\newcommand{\Yqubitnbarbspaceoveralleventindexequalstdwithvecbasisprobexpect}
{
E
\{
\Yqubitnbarbspaceoveralleventindexequalstdwithvecbasisprob
\}
}
%
%
\newcommand{\yqubitnbarbspaceoveralleventindexequalstdwithvecbasisprobexpectapproxonetwoqubitseqnb}
{$\Yqubitnbarbspaceoveralleventindexequalstdwithvecbasisprobexpectapproxonetwoqubitseqnb$}
\newcommand{\Yqubitnbarbspaceoveralleventindexequalstdwithvecbasisprobexpectapproxonetwoqubitseqnb}
{
E
^{\prime}
\{
\Yqubitnbarbspaceoveralleventindexequalstdwithvecbasisprob
\}
}
%
%
\newcommand{\yqubitnbarbspaceoveralleventindexequalstdwithvecbasisprobexpectapproxtwotwoqubitseqnb}
{$\Yqubitnbarbspaceoveralleventindexequalstdwithvecbasisprobexpectapproxtwotwoqubitseqnb$}
\newcommand{\Yqubitnbarbspaceoveralleventindexequalstdwithvecbasisprobexpectapproxtwotwoqubitseqnb}
{
E
^{\prime \prime}
\{
\Yqubitnbarbspaceoveralleventindexequalstdwithvecbasisprob
\}
}
%
%
%
\newcommand{\yqubitnbarbspaceoveralleventindexequalstdwithvecbasisprobseqindex}
{$\Yqubitnbarbspaceoveralleventindexequalstdwithvecbasisprobseqindex$}
\newcommand{\Yqubitnbarbspaceoveralleventindexequalstdwithvecbasisprobseqindex}
{
P
(
\Yqubitnbarbspaceoveralleventindexequalstdwithvecbasis
,
\Ytwoqubitseqindex
)
}
%
%
%

\newcommand{\yqubitnbarbspaceoveralleventindexequalstdwithvecbasisprobstatewritereadseqindex}
{$\Yqubitnbarbspaceoveralleventindexequalstdwithvecbasisprobstatewritereadseqindex$}
\newcommand{\Yqubitnbarbspaceoveralleventindexequalstdwithvecbasisprobstatewritereadseqindex}
{
P
(
\Yqubitnbarbspaceoveralleventindexequalstdwithvecbasis
,
\Yqubitnbarbstatewritereadseqindex
)
}
%
%
%
\newcommand{\yqubitnbarbspaceoveralleventindexequalstdwithvecbasisprobseqindexapproxwritereadonestatenb}
{$\Yqubitnbarbspaceoveralleventindexequalstdwithvecbasisprobseqindexapproxwritereadonestatenb$}
\newcommand{\Yqubitnbarbspaceoveralleventindexequalstdwithvecbasisprobseqindexapproxwritereadonestatenb}
{
P
^{\prime}
(
\Yqubitnbarbspaceoveralleventindexequalstdwithvecbasis
,
\Ytwoqubitseqindex
,
\Ywritereadonestatenb
)
}
%
%
%
\newcommand{\yfuncnumberofofeventoccur}
{$\Yfuncnumberofofeventoccur$}
\newcommand{\Yfuncnumberofofeventoccur}
{{\cal N}}
%
\newcommand{\yqubitnbarbspaceoveralleventindexequalstdwithvecbasisnboccurseqindexwritereadonestatenb}
{$\Yqubitnbarbspaceoveralleventindexequalstdwithvecbasisnboccurseqindexwritereadonestatenb$}
\newcommand{\Yqubitnbarbspaceoveralleventindexequalstdwithvecbasisnboccurseqindexwritereadonestatenb}
{
\Yfuncnumberofofeventoccur
(
\Yqubitnbarbspaceoveralleventindexequalstdwithvecbasis
,
\Ytwoqubitseqindex
,
\Ywritereadonestatenb
)
}
%
\newcommand{\yqubitnbarbspaceoveralleventindexequalstdwithvecbasisnboccurqubitnbarbstatewritereadseqnb}
{$\Yqubitnbarbspaceoveralleventindexequalstdwithvecbasisnboccurqubitnbarbstatewritereadseqnb$}
\newcommand{\Yqubitnbarbspaceoveralleventindexequalstdwithvecbasisnboccurqubitnbarbstatewritereadseqnb}
{
\Yfuncnumberofofeventoccur
(
\Yqubitnbarbspaceoveralleventindexequalstdwithvecbasis
,
\Yqubitnbarbstatewritereadseqnb
)
}
%
\newcommand{\yqubitnbarbspaceoveralleventindexequalstdwithvecbasisindicfuncqubitnbarbstatewritereadseqindex}
{$\Yqubitnbarbspaceoveralleventindexequalstdwithvecbasisindicfuncqubitnbarbstatewritereadseqindex$}
\newcommand{\Yqubitnbarbspaceoveralleventindexequalstdwithvecbasisindicfuncqubitnbarbstatewritereadseqindex}
{
1\hspace{-.12cm}1
(
\Yqubitnbarbspaceoveralleventindexequalstdwithvecbasis
,
\Yqubitnbarbstatewritereadseqindex
)
}
%
%
%
%
%
%
%
%
%
%
%
\newcommand{\yregisterindexstdone}
          {$\Yregisterindexstdone$}
\newcommand{\Yregisterindexstdone}
           {r}           
%
%
%
\newcommand{\yqubitspaceregisterindexstdonequbitindexone}
          {$\Yqubitspaceregisterindexstdonequbitindexone$}
\newcommand{\Yqubitspaceregisterindexstdonequbitindexone}
           {\Yqubitonesevspace
           _{\Yregisterindexstdone 1}
           }                                 
%
%
%
\newcommand{\yqubitspaceregisterindexstdonequbitindexstdtwo}
          {$\Yqubitspaceregisterindexstdonequbitindexstdtwo$}
\newcommand{\Yqubitspaceregisterindexstdonequbitindexstdtwo}
           {\Yqubitonesevspace
           _{\Yregisterindexstdone \Yqubitindexstdtwo}
           }          
%
%
%
\newcommand{\yqubitspaceregisterindexstdonequbitindexqubitnbarb}
          {$\Yqubitspaceregisterindexstdonequbitindexqubitnbarb$}
\newcommand{\Yqubitspaceregisterindexstdonequbitindexqubitnbarb}
           {\Yqubitonesevspace
           _{\Yregisterindexstdone \Yqubitnbarb}
           } 																								
%
%
%
\newcommand{\yqubitspaceregisternoindexqubitnoindexbasisintstd}
          {$\Yqubitspaceregisternoindexqubitnoindexbasisintstd$}
\newcommand{\Yqubitspaceregisternoindexqubitnoindexbasisintstd}
           {k}
%
%
%
\newcommand{\yqubitspaceregisterindexstdonequbitonebasisintstd}
          {$\Yqubitspaceregisterindexstdonequbitonebasisintstd$}
\newcommand{\Yqubitspaceregisterindexstdonequbitonebasisintstd}
           {\Yqubitspaceregisternoindexqubitnoindexbasisintstd
           _{\Yregisterindexstdone 1}
           } 							
%
%
%
\newcommand{\yqubitspaceregisterindexstdonequbitindexstdtwobasisintstd}
          {$\Yqubitspaceregisterindexstdonequbitindexstdtwobasisintstd$}
\newcommand{\Yqubitspaceregisterindexstdonequbitindexstdtwobasisintstd}
           {\Yqubitspaceregisternoindexqubitnoindexbasisintstd
           _{\Yregisterindexstdone \Yqubitindexstdtwo}
           }                      
%
%
%
\newcommand{\yqubitspaceregisterindexstdonequbitlastbasisintstd}
          {$\Yqubitspaceregisterindexstdonequbitlastbasisintstd$}
\newcommand{\Yqubitspaceregisterindexstdonequbitlastbasisintstd}
           {\Yqubitspaceregisternoindexqubitnoindexbasisintstd
           _{\Yregisterindexstdone \Yqubitnbarb}
           }  
%
%
%
\newcommand{\yqubitspaceregisterindexstdonequbitanybasisintstd}
          {$\Yqubitspaceregisterindexstdonequbitanybasisintstd$}
\newcommand{\Yqubitspaceregisterindexstdonequbitanybasisintstd}
           {\Yqubitspaceregisternoindexqubitnoindexbasisintstd
           _{\Yregisterindexstdone \bullet}
           }						
%
%
%
\newcommand{\ysetnotnoarg}
          {$\Ysetnotnoarg$}
\newcommand{\Ysetnotnoarg}
           {{\cal S}}  																		
%
%
%
\newcommand{\ysetnotargqubitspaceregisterindexstdonequbitanybasisintstd}
          {$\Ysetnotargqubitspaceregisterindexstdonequbitanybasisintstd$}
\newcommand{\Ysetnotargqubitspaceregisterindexstdonequbitanybasisintstd}
           {\Ysetnotnoarg (
					 \Yqubitspaceregisterindexstdonequbitanybasisintstd )
					 } 																
%
%
%
\newcommand{\yqubitspaceregisterindexstdonequbitonebasisstateintstd}
          {$\Yqubitspaceregisterindexstdonequbitonebasisstateintstd$}
\newcommand{\Yqubitspaceregisterindexstdonequbitonebasisstateintstd}
           {| \Yqubitspaceregisterindexstdonequbitonebasisintstd
           \rangle
           _{\Yregisterindexstdone 1}
           }                	
%
%
%
\newcommand{\yqubitspaceregisterindexstdonequbitindexstdtwobasisstateintstd}
          {$\Yqubitspaceregisterindexstdonequbitindexstdtwobasisstateintstd$}
\newcommand{\Yqubitspaceregisterindexstdonequbitindexstdtwobasisstateintstd}
           {| \Yqubitspaceregisterindexstdonequbitindexstdtwobasisintstd
           \rangle
           _{\Yregisterindexstdone \Yqubitindexstdtwo}
           }     																										
%
%
%
\newcommand{\yqubitspaceregisterindexstdonequbitlastbasisstateintstd}
          {$\Yqubitspaceregisterindexstdonequbitlastbasisstateintstd$}
\newcommand{\Yqubitspaceregisterindexstdonequbitlastbasisstateintstd}
           {| \Yqubitspaceregisterindexstdonequbitlastbasisintstd
           \rangle
           _{\Yregisterindexstdone \Yqubitnbarb}
           } 																																			
%
%
%
\newcommand{\yqubitspaceregisterindexnoqubitsindexnobasisstatecoefnot}
          {$\Yqubitspaceregisterindexnoqubitsindexnobasisstatecoefnot$}
\newcommand{\Yqubitspaceregisterindexnoqubitsindexnobasisstatecoefnot}
           {c} 																																				
%
%
%
\newcommand{\yregisteronequbitspaceregisterindexstdonequbitsonetolastbasisstatecoefstd}
          {$\Yregisteronequbitspaceregisterindexstdonequbitsonetolastbasisstatecoefstd$}
\newcommand{\Yregisteronequbitspaceregisterindexstdonequbitsonetolastbasisstatecoefstd}
           {\Yqubitspaceregisterindexnoqubitsindexnobasisstatecoefnot
           _{1
					   \Yqubitspaceregisterindexstdonequbitonebasisintstd
					   \dots
						 \Yqubitspaceregisterindexstdonequbitlastbasisintstd}
           } 
%
%
%
\newcommand{\yregistertwoqubitspaceregisterindexstdonequbitsonetolastbasisstatecoefstd}
          {$\Yregistertwoqubitspaceregisterindexstdonequbitsonetolastbasisstatecoefstd$}
\newcommand{\Yregistertwoqubitspaceregisterindexstdonequbitsonetolastbasisstatecoefstd}
           {\Yqubitspaceregisterindexnoqubitsindexnobasisstatecoefnot
           _{2
					   \Yqubitspaceregisterindexstdonequbitonebasisintstd
					   \dots
						 \Yqubitspaceregisterindexstdonequbitlastbasisintstd}
           } 																		
%
%
%
\newcommand{\yregisterstdonequbitspaceregisterindexstdonequbitsonetolastbasisstatecoefstd}
          {$\Yregisterstdonequbitspaceregisterindexstdonequbitsonetolastbasisstatecoefstd$}
\newcommand{\Yregisterstdonequbitspaceregisterindexstdonequbitsonetolastbasisstatecoefstd}
           {\Yqubitspaceregisterindexnoqubitsindexnobasisstatecoefnot
           _{\Yregisterindexstdone
					   \Yqubitspaceregisterindexstdonequbitonebasisintstd
					   \dots
						 \Yqubitspaceregisterindexstdonequbitlastbasisintstd}
           } 
%
%
%
\newcommand{\ystateregisternot}
          {$\Ystateregisternot$}
\newcommand{\Ystateregisternot}
           {\psi}

%
%
%
\newcommand{\ystateregisterindexone}
          {$\Ystateregisterindexone$}
\newcommand{\Ystateregisterindexone}
           {| \Ystateregisternot
					_{1}
					  \rangle
           }
%
%
%
\newcommand{\ystateregisterindextwo}
          {$\Ystateregisterindextwo$}
\newcommand{\Ystateregisterindextwo}
           {| \Ystateregisternot
					_{2}
					  \rangle
           }
%
%
%
\newcommand{\ystateregisterindexstdone}
          {$\Ystateregisterindexstdone$}
\newcommand{\Ystateregisterindexstdone}
           {| \Ystateregisternot
					_{\Yregisterindexstdone}
					  \rangle
           }
%
%
%
%
%
%
%
%
%
%
%
%
%
%
\newcommand{\ytwoqubitseqindex}
{$\Ytwoqubitseqindex$}
\newcommand{\Ytwoqubitseqindex}
{n}
%
%
%
%
\newcommand{\ytwoqubitseqnb}
{$\Ytwoqubitseqnb$}
\newcommand{\Ytwoqubitseqnb}
{N}
%
%
%
\newcommand{\ytwoqubitseqnbox}
{$\Ytwoqubitseqnbox$}
\newcommand{\Ytwoqubitseqnbox}
{\Ytwoqubitseqnb _x}
%
%
%
\newcommand{\ytwoqubitseqnboz}
{$\Ytwoqubitseqnboz$}
\newcommand{\Ytwoqubitseqnboz}
{\Ytwoqubitseqnb _z}
%
%
%
%
%
\newcommand{\yqubitnbarbstatewritereadseqindex}
{$\Yqubitnbarbstatewritereadseqindex$}
\newcommand{\Yqubitnbarbstatewritereadseqindex}
{\ell}
%
%
%
%
\newcommand{\yqubitnbarbstatewritereadseqnb}
{$\Yqubitnbarbstatewritereadseqnb$}
\newcommand{\Yqubitnbarbstatewritereadseqnb}
{L}
%
%
%
%
%
\newcommand{\yalphaoneseqindex}
{$\Yalphaoneseqindex$}
\newcommand{\Yalphaoneseqindex}
{
\alpha _1
(
\Ytwoqubitseqindex
)
}
%
\newcommand{\ybetaoneseqindex}
{$\Ybetaoneseqindex$}
\newcommand{\Ybetaoneseqindex}
{
\beta _1
(
\Ytwoqubitseqindex
)
}
%
\newcommand{\yalphatwoseqindex}
{$\Yalphatwoseqindex$}
\newcommand{\Yalphatwoseqindex}
{
\alpha _2
(
\Ytwoqubitseqindex
)
}
%
\newcommand{\ybetatwoseqindex}
{$\Ybetatwoseqindex$}
\newcommand{\Ybetatwoseqindex}
{
\beta _2
(
\Ytwoqubitseqindex
)
}
%
%
%
%
%
%
%
%
%
\newcommand{\ycostfuncozone}
{$\Ycostfuncozone$}
\newcommand{\Ycostfuncozone}
{F_z}
%
\newcommand{\ycostfuncozoneeqindex}
{$\Ycostfuncozoneeqindex$}
\newcommand{\Ycostfuncozoneeqindex}
{f_z
(
\Ytwoqubitseqindex
)
}
%
\newcommand{\ycostfuncozoneeqindexcoefnot}
{$\Ycostfuncozoneeqindexcoefnot$}
\newcommand{\Ycostfuncozoneeqindexcoefnot}
{w}
%
\newcommand{\ycostfuncozoneeqindexcoefcos}
{$\Ycostfuncozoneeqindexcoefcos$}
\newcommand{\Ycostfuncozoneeqindexcoefcos}
{\Ycostfuncozoneeqindexcoefnot_1
(
\Ytwoqubitseqindex
)
}
%
\newcommand{\ycostfuncozoneeqindexcoefsin}
{$\Ycostfuncozoneeqindexcoefsin$}
\newcommand{\Ycostfuncozoneeqindexcoefsin}
{\Ycostfuncozoneeqindexcoefnot_2
(
\Ytwoqubitseqindex
)
}
%
\newcommand{\ycostfuncozoneeqindexcoefconst}
{$\Ycostfuncozoneeqindexcoefconst$}
\newcommand{\Ycostfuncozoneeqindexcoefconst}
{\Ycostfuncozoneeqindexcoefnot_3
(
\Ytwoqubitseqindex
)
}
%
%
%
\newcommand{\ycostfuncoxone}
{$\Ycostfuncoxone$}
\newcommand{\Ycostfuncoxone}
{F_x}
%
%
%
%
%
\newcommand{\ycostfuncoxoneeqindex}
{$\Ycostfuncoxoneeqindex$}
\newcommand{\Ycostfuncoxoneeqindex}
{f_x
(
\Ytwoqubitseqindex
)
}
%
%
%
%
%
%
%
%
\newcommand{\ypropertycomplexsingle}
{$\Ypropertycomplexsingle$}
\newcommand{\Ypropertycomplexsingle}
{d}
%
\newcommand{\ypropertycomplexseqindex}
{$\Ypropertycomplexseqindex$}
\newcommand{\Ypropertycomplexseqindex}
{\Ypropertycomplexsingle ( \Ytwoqubitseqindex )}
%
%
%
%
%
%
%
%
%
%
%
\newcommand{\ysepsystdensopoutqubitonesqrtracepartcondphasemultiple}
{$\Ysepsystdensopoutqubitonesqrtracepartcondphasemultiple$}
\newcommand{\Ysepsystdensopoutqubitonesqrtracepartcondphasemultiple}
{
k
}
%
%
%
%
%
%
%
%
%
%
%
%
%
%
\newcommand{\ysepsystdensopoutqubitonesqrtracepartcondphasemultipleestim}
{$\Ysepsystdensopoutqubitonesqrtracepartcondphasemultipleestim$}
\newcommand{\Ysepsystdensopoutqubitonesqrtracepartcondphasemultipleestim}
{\widehat{\Ysepsystdensopoutqubitonesqrtracepartcondphasemultiple}}
%
\newcommand{\ysepsystdensopoutqubitonesqrtracepartcondphasemultipleestimminusactual}
{$\Ysepsystdensopoutqubitonesqrtracepartcondphasemultipleestimminusactual$}
\newcommand{\Ysepsystdensopoutqubitonesqrtracepartcondphasemultipleestimminusactual}
{\Delta_{\Ysepsystdensopoutqubitonesqrtracepartcondphasemultiple}}
%
%
%
%
%
%
%
%
%
%
%


%
%
%
%
\newcommand{\ytwoqubitsprobaplusplus}
{$\Ytwoqubitsprobaplusplus$}
\newcommand{\Ytwoqubitsprobaplusplus}
{p_1}
%
\newcommand{\ytwoqubitsprobaplusminus}
{$\Ytwoqubitsprobaplusminus$}
\newcommand{\Ytwoqubitsprobaplusminus}
{p_2}
%
\newcommand{\ytwoqubitsprobaminusplus}
{$\Ytwoqubitsprobaminusplus$}
\newcommand{\Ytwoqubitsprobaminusplus}
{p_3}
%
\newcommand{\ytwoqubitsprobaminusminus}
{$\Ytwoqubitsprobaminusminus$}
\newcommand{\Ytwoqubitsprobaminusminus}
{p_4}
%
%
%
%
\newcommand{\ytwoqubitsprobaindexstd}
{$\Ytwoqubitsprobaindexstd$}
\newcommand{\Ytwoqubitsprobaindexstd}
{\Ytwoqubitsprobanot
_{\Yqubitbothtimefinalstatecoefindexindexstd}
}
%
%
%
%
\newcommand{\ytwoqubitsprobavec}
{$\Ytwoqubitsprobavec$}
\newcommand{\Ytwoqubitsprobavec}
{p}
%
%
%
%
%
%
\newcommand{\ytwoqubitsprobanot}
{$\Ytwoqubitsprobanot$}
\newcommand{\Ytwoqubitsprobanot}
{p}
%
\newcommand{\ytwoqubitsprobaplusplusdirzz}
{$\Ytwoqubitsprobaplusplusdirzz$}
\newcommand{\Ytwoqubitsprobaplusplusdirzz}
{\Ytwoqubitsprobanot
%
%
_{1zz}
}
%
\newcommand{\ytwoqubitsprobaplusminusdirzz}
{$\Ytwoqubitsprobaplusminusdirzz$}
\newcommand{\Ytwoqubitsprobaplusminusdirzz}
{\Ytwoqubitsprobanot
%
%
_{2zz}
}
%
\newcommand{\ytwoqubitsprobaminusplusdirzz}
{$\Ytwoqubitsprobaminusplusdirzz$}
\newcommand{\Ytwoqubitsprobaminusplusdirzz}
{\Ytwoqubitsprobanot
%
_{3zz}
}
%
\newcommand{\ytwoqubitsprobaminusminusdirzz}
{$\Ytwoqubitsprobaminusminusdirzz$}
\newcommand{\Ytwoqubitsprobaminusminusdirzz}
{\Ytwoqubitsprobanot
%
_{4zz}
}
%
%
%
%
\newcommand{\ytwoqubitsprobaindexstddirzz}
{$\Ytwoqubitsprobaindexstddirzz$}
\newcommand{\Ytwoqubitsprobaindexstddirzz}
{\Ytwoqubitsprobanot
_{\Yqubitbothtimefinalstatecoefindexindexstd zz}
}

%
%
\newcommand{\ytwoqubitsprobaplusplusdirxx}
{$\Ytwoqubitsprobaplusplusdirxx$}
\newcommand{\Ytwoqubitsprobaplusplusdirxx}
{\Ytwoqubitsprobanot 
_{1xx}
}
%
\newcommand{\ytwoqubitsprobaplusminusdirxx}
{$\Ytwoqubitsprobaplusminusdirxx$}
\newcommand{\Ytwoqubitsprobaplusminusdirxx}
{\Ytwoqubitsprobanot
_{2xx}
}
%
\newcommand{\ytwoqubitsprobaminusplusdirxx}
{$\Ytwoqubitsprobaminusplusdirxx$}
\newcommand{\Ytwoqubitsprobaminusplusdirxx}
{\Ytwoqubitsprobanot
_{3xx}
}
%
\newcommand{\ytwoqubitsprobaminusminusdirxx}
{$\Ytwoqubitsprobaminusminusdirxx$}
\newcommand{\Ytwoqubitsprobaminusminusdirxx}
{\Ytwoqubitsprobanot
_{4xx}
}
%
%
%
%
%
\newcommand{\ytwoqubitsprobaindexstddirxx}
{$\Ytwoqubitsprobaindexstddirxx$}
\newcommand{\Ytwoqubitsprobaindexstddirxx}
{\Ytwoqubitsprobanot
_{\Yqubitbothtimefinalstatecoefindexindexstd xx}
}
%
%
%
%
\newcommand{\ytwoqubitsprobaplusplusdirzx}
{$\Ytwoqubitsprobaplusplusdirzx$}
\newcommand{\Ytwoqubitsprobaplusplusdirzx}
{\Ytwoqubitsprobanot_{1zx}}
%
\newcommand{\ytwoqubitsprobaplusminusdirzx}
{$\Ytwoqubitsprobaplusminusdirzx$}
\newcommand{\Ytwoqubitsprobaplusminusdirzx}
{\Ytwoqubitsprobanot_{2zx}}
%
\newcommand{\ytwoqubitsprobaminusplusdirzx}
{$\Ytwoqubitsprobaminusplusdirzx$}
\newcommand{\Ytwoqubitsprobaminusplusdirzx}
{\Ytwoqubitsprobanot_{3zx}}
%
\newcommand{\ytwoqubitsprobaminusminusdirzx}
{$\Ytwoqubitsprobaminusminusdirzx$}
\newcommand{\Ytwoqubitsprobaminusminusdirzx}
{\Ytwoqubitsprobanot_{4zx}}
%
%
%
%
\newcommand{\ytwoqubitsprobaplusplusdirzy}
{$\Ytwoqubitsprobaplusplusdirzy$}
\newcommand{\Ytwoqubitsprobaplusplusdirzy}
{\Ytwoqubitsprobanot_{1zy}}
%
\newcommand{\ytwoqubitsprobaplusminusdirzy}
{$\Ytwoqubitsprobaplusminusdirzy$}
\newcommand{\Ytwoqubitsprobaplusminusdirzy}
{\Ytwoqubitsprobanot_{2zy}}
%
\newcommand{\ytwoqubitsprobaminusplusdirzy}
{$\Ytwoqubitsprobaminusplusdirzy$}
\newcommand{\Ytwoqubitsprobaminusplusdirzy}
{\Ytwoqubitsprobanot_{3zy}}
%
\newcommand{\ytwoqubitsprobaminusminusdirzy}
{$\Ytwoqubitsprobaminusminusdirzy$}
\newcommand{\Ytwoqubitsprobaminusminusdirzy}
{\Ytwoqubitsprobanot_{4zy}}
%
%
%
%
\newcommand{\ytwoqubitsprobaplusplusdirxz}
{$\Ytwoqubitsprobaplusplusdirxz$}
\newcommand{\Ytwoqubitsprobaplusplusdirxz}
{\Ytwoqubitsprobanot_{1xz}}
%
\newcommand{\ytwoqubitsprobaplusminusdirxz}
{$\Ytwoqubitsprobaplusminusdirxz$}
\newcommand{\Ytwoqubitsprobaplusminusdirxz}
{\Ytwoqubitsprobanot_{2xz}}
%
\newcommand{\ytwoqubitsprobaminusplusdirxz}
{$\Ytwoqubitsprobaminusplusdirxz$}
\newcommand{\Ytwoqubitsprobaminusplusdirxz}
{\Ytwoqubitsprobanot_{3xz}}
%
\newcommand{\ytwoqubitsprobaminusminusdirxz}
{$\Ytwoqubitsprobaminusminusdirxz$}
\newcommand{\Ytwoqubitsprobaminusminusdirxz}
{\Ytwoqubitsprobanot_{4xz}}
%
%
%
%
\newcommand{\ytwoqubitsprobaplusplusdiryz}
{$\Ytwoqubitsprobaplusplusdiryz$}
\newcommand{\Ytwoqubitsprobaplusplusdiryz}
{\Ytwoqubitsprobanot_{1yz}}
%
\newcommand{\ytwoqubitsprobaplusminusdiryz}
{$\Ytwoqubitsprobaplusminusdiryz$}
\newcommand{\Ytwoqubitsprobaplusminusdiryz}
{\Ytwoqubitsprobanot_{2yz}}
%
\newcommand{\ytwoqubitsprobaminusplusdiryz}
{$\Ytwoqubitsprobaminusplusdiryz$}
\newcommand{\Ytwoqubitsprobaminusplusdiryz}
{\Ytwoqubitsprobanot_{3yz}}
%
\newcommand{\ytwoqubitsprobaminusminusdiryz}
{$\Ytwoqubitsprobaminusminusdiryz$}
\newcommand{\Ytwoqubitsprobaminusminusdiryz}
{\Ytwoqubitsprobanot_{4yz}}
\newcommand{\ytwoqubitresultphaseinit}
{$\Ytwoqubitresultphaseinit$}
\newcommand{\Ytwoqubitresultphaseinit}
{\Delta _I}
%
%
%
\newcommand{\ytwoqubitresultphaseevol}
{$\Ytwoqubitresultphaseevol$}
\newcommand{\Ytwoqubitresultphaseevol}
{\Delta _E}
%
%
\newcommand{\ytwoqubitresultphaseevolindexd}
{$\Ytwoqubitresultphaseevolindexd$}
\newcommand{\Ytwoqubitresultphaseevolindexd}
{\Delta _{Ed}}
%
\newcommand{\ytwoqubitresultphaseevolindexdestim}
{$\Ytwoqubitresultphaseevolindexdestim$}
\newcommand{\Ytwoqubitresultphaseevolindexdestim}
{\widehat{\Delta} _{Ed}}
%
%
%
%
\newcommand{\ytwoqubitresultphaseevolsin}
{$\Ytwoqubitresultphaseevolsin$}
\newcommand{\Ytwoqubitresultphaseevolsin}
{v}
%
%
%
%
\newcommand{\ytwoqubitresultphaseevolsinindexone}
{$\Ytwoqubitresultphaseevolsinindexone$}
\newcommand{\Ytwoqubitresultphaseevolsinindexone}
{\Ytwoqubitresultphaseevolsin_{1}}
%
\newcommand{\ytwoqubitresultphaseevolsinindextwo}
{$\Ytwoqubitresultphaseevolsinindextwo$}
\newcommand{\Ytwoqubitresultphaseevolsinindextwo}
{\Ytwoqubitresultphaseevolsin_{2}}
\newcommand{\ytwoqubitresultphaseevolsinestim}
{$\Ytwoqubitresultphaseevolsinestim$}
\newcommand{\Ytwoqubitresultphaseevolsinestim}
{\overline{\Ytwoqubitresultphaseevolsin}}


%
%
\newcommand{\ytwoqubitresultphaseevolsinestimtwo}
{$\Ytwoqubitresultphaseevolsinestimtwo$}
\newcommand{\Ytwoqubitresultphaseevolsinestimtwo}
{\widehat{\Ytwoqubitresultphaseevolsin}}
%
%
\newcommand{\ytwoqubitresultphaseevolsinestimtwoindexone}
{$\Ytwoqubitresultphaseevolsinestimtwoindexone$}
\newcommand{\Ytwoqubitresultphaseevolsinestimtwoindexone}
{\Ytwoqubitresultphaseevolsinestimtwo_{1}}
%
%
\newcommand{\ytwoqubitresultphaseevolsinestimtwoindextwo}
{$\Ytwoqubitresultphaseevolsinestimtwoindextwo$}
\newcommand{\Ytwoqubitresultphaseevolsinestimtwoindextwo}
{\Ytwoqubitresultphaseevolsinestimtwo_{2}}
%


%
%
%
%
%
\newcommand{\yparamqubitbothstateplusmodulusnot}
{$\Yparamqubitbothstateplusmodulusnot$}
\newcommand{\Yparamqubitbothstateplusmodulusnot}
{r}
%
\newcommand{\yparamqubitonestateplusmodulus}
{$\Yparamqubitonestateplusmodulus$}
\newcommand{\Yparamqubitonestateplusmodulus}
{{\Yparamqubitbothstateplusmodulusnot}_1}
%
\newcommand{\yparamqubittwostateplusmodulus}
{$\Yparamqubittwostateplusmodulus$}
\newcommand{\Yparamqubittwostateplusmodulus}
{{\Yparamqubitbothstateplusmodulusnot}_2}
%
\newcommand{\yparamqubitindexstdstateplusmodulus}
{$\Yparamqubitindexstdstateplusmodulus$}
\newcommand{\Yparamqubitindexstdstateplusmodulus}
{{\Yparamqubitbothstateplusmodulusnot}_{\Yqubitindexstd}}
%
\newcommand{\yparamqubitbothstateminusmodulusnot}
{$\Yparamqubitbothstateminusmodulusnot$}
\newcommand{\Yparamqubitbothstateminusmodulusnot}
{q}
%
\newcommand{\yparamqubitonestateminusmodulus}
{$\Yparamqubitonestateminusmodulus$}
\newcommand{\Yparamqubitonestateminusmodulus}
{{\Yparamqubitbothstateminusmodulusnot}_1}
%
\newcommand{\yparamqubittwostateminusmodulus}
{$\Yparamqubittwostateminusmodulus$}
\newcommand{\Yparamqubittwostateminusmodulus}
{{\Yparamqubitbothstateminusmodulusnot}_2}
%
\newcommand{\yparamqubitindexstdstateminusmodulus}
{$\Yparamqubitindexstdstateminusmodulus$}
\newcommand{\Yparamqubitindexstdstateminusmodulus}
{{\Yparamqubitbothstateminusmodulusnot}_{\Yqubitindexstd}}
%
%
%
%
%
\newcommand{\yparamqubitbothstateplusphasenot}
{$\Yparamqubitbothstateplusphasenot$}
\newcommand{\Yparamqubitbothstateplusphasenot}
{\theta}
%
\newcommand{\yparamqubitonestateplusphase}
{$\Yparamqubitonestateplusphase$}
\newcommand{\Yparamqubitonestateplusphase}
{\Yparamqubitbothstateplusphasenot_1}
%
\newcommand{\yparamqubittwostateplusphase}
{$\Yparamqubittwostateplusphase$}
\newcommand{\Yparamqubittwostateplusphase}
{\Yparamqubitbothstateplusphasenot_2}
%
%
\newcommand{\yparamqubitindexstdstateplusphase}
{$\Yparamqubitindexstdstateplusphase$}
\newcommand{\Yparamqubitindexstdstateplusphase}
{{\Yparamqubitbothstateplusphasenot}_{\Yqubitindexstd}}
%
%
%
%
%
%
\newcommand{\yparamqubitbothstateminusphasenot}
{$\Yparamqubitbothstateminusphasenot$}
\newcommand{\Yparamqubitbothstateminusphasenot}
{\phi}
%
%
\newcommand{\yparamqubitonestateminusphase}
{$\Yparamqubitonestateminusphase$}
\newcommand{\Yparamqubitonestateminusphase}
{\Yparamqubitbothstateminusphasenot_1}
%
\newcommand{\yparamqubittwostateminusphase}
{$\Yparamqubittwostateminusphase$}
\newcommand{\Yparamqubittwostateminusphase}
{\Yparamqubitbothstateminusphasenot_2}
%
\newcommand{\yparamqubitindexstdstateminusphase}
{$\Yparamqubitindexstdstateminusphase$}
\newcommand{\Yparamqubitindexstdstateminusphase}
{{\Yparamqubitbothstateminusphasenot}_{\Yqubitindexstd}}
%
%
%
%
%
%
%
%
%
%
%
\newcommand{\ygennumberonerealpart}
{$\Ygennumberonerealpart$}
\newcommand{\Ygennumberonerealpart}
{a_1}
%
%
%
\newcommand{\ygennumbertworealpart}
{$\Ygennumbertworealpart$}
\newcommand{\Ygennumbertworealpart}
{a_2}
%
\newcommand{\ygennumbertwoimagpart}
{$\Ygennumbertwoimagpart$}
\newcommand{\Ygennumbertwoimagpart}
{b_2}
%
%
%
%
%
%
%
%
\newcommand{\ymagfieldnot}
{$\Ymagfieldnot$}
\newcommand{\Ymagfieldnot}
{B}
%
\newcommand{\ymagfieldvec}
{$\Ymagfieldvec$}
\newcommand{\Ymagfieldvec}
{\overrightarrow{\Ymagfieldnot}}
%
%
%
\newcommand{\yhamiltonfieldscale}
{$\Yhamiltonfieldscale$}
\newcommand{\Yhamiltonfieldscale}
{G}
%
%
%
%
%
%
%
%
%
%
%

\newcommand{\yexchangetensorppalvaluexy}
{$\Yexchangetensorppalvaluexy$}
\newcommand{\Yexchangetensorppalvaluexy}
{J_{xy}}

\newcommand{\yexchangetensorppalvaluexyestim}
{$\Yexchangetensorppalvaluexyestim$}
\newcommand{\Yexchangetensorppalvaluexyestim}
{\widehat{J}_{xy}}
%
%
%
%
\newcommand{\yexchangetensorppalvaluexyestimindexone}
{$\Yexchangetensorppalvaluexyestimindexone$}
\newcommand{\Yexchangetensorppalvaluexyestimindexone}
{\widehat{J}_{xy1}}
%
\newcommand{\yexchangetensorppalvaluexyestimindextwo}
{$\Yexchangetensorppalvaluexyestimindextwo$}
\newcommand{\Yexchangetensorppalvaluexyestimindextwo}
{\widehat{J}_{xy2}}
%

%
\newcommand{\yexchangetensorppalvaluexyestimargtwoqubitwritereadtimeintervalindexone}
{$\Yexchangetensorppalvaluexyestimargtwoqubitwritereadtimeintervalindexone$}
\newcommand{\Yexchangetensorppalvaluexyestimargtwoqubitwritereadtimeintervalindexone}
{\Yexchangetensorppalvaluexyestim
(
\Ytwoqubitwritereadtimeintervalindexone
)}
%

%
\newcommand{\yexchangetensorppalvaluexyshiftindetermint}
{$\Yexchangetensorppalvaluexyshiftindetermint$}
\newcommand{\Yexchangetensorppalvaluexyshiftindetermint}
{k_{xy}}
%
\newcommand{\yexchangetensorppalvaluexyshiftindetermintindexone}
{$\Yexchangetensorppalvaluexyshiftindetermintindexone$}
\newcommand{\Yexchangetensorppalvaluexyshiftindetermintindexone}
{k_{xy1}}
%
\newcommand{\yexchangetensorppalvaluexyshiftindetermintindextwo}
{$\Yexchangetensorppalvaluexyshiftindetermintindextwo$}
\newcommand{\Yexchangetensorppalvaluexyshiftindetermintindextwo}
{k_{xy2}}
%
%
%
\newcommand{\yexchangetensorppalvaluexyshiftindetermintestim}
{$\Yexchangetensorppalvaluexyshiftindetermintestim$}
\newcommand{\Yexchangetensorppalvaluexyshiftindetermintestim}
{\widehat{k}_{xy}}
%
\newcommand{\yexchangetensorppalvaluexyshiftindetermintestimindexone}
{$\Yexchangetensorppalvaluexyshiftindetermintestimindexone$}
\newcommand{\Yexchangetensorppalvaluexyshiftindetermintestimindexone}
{\widehat{k}_{xy1}}
%
\newcommand{\yexchangetensorppalvaluexyshiftindetermintestimindextwo}
{$\Yexchangetensorppalvaluexyshiftindetermintestimindextwo$}
\newcommand{\Yexchangetensorppalvaluexyshiftindetermintestimindextwo}
{\widehat{k}_{xy2}}
%
%
%
%
\newcommand{\yexchangetensorppalvaluexyshiftindetermintestimminusactual}
{$\Yexchangetensorppalvaluexyshiftindetermintestimminusactual$}
\newcommand{\Yexchangetensorppalvaluexyshiftindetermintestimminusactual}
{\Delta k _{xy}}
%
%
\newcommand{\yexchangetensorppalvaluexyshiftindetermintestimminusactualindexone}
{$\Yexchangetensorppalvaluexyshiftindetermintestimminusactualindexone$}
\newcommand{\Yexchangetensorppalvaluexyshiftindetermintestimminusactualindexone}
{\Delta k _{xy1}}
%
\newcommand{\yexchangetensorppalvaluexyshiftindetermintestimminusactualindextwo}
{$\Yexchangetensorppalvaluexyshiftindetermintestimminusactualindextwo$}
\newcommand{\Yexchangetensorppalvaluexyshiftindetermintestimminusactualindextwo}
{\Delta k _{xy2}}
%
%
%
\newcommand{\yexchangetensorppalvaluexyshiftindetermintestimmin}
{$\Yexchangetensorppalvaluexyshiftindetermintestimmin$}
\newcommand{\Yexchangetensorppalvaluexyshiftindetermintestimmin}
{\widehat{k}_{xy}^{min}}
%
\newcommand{\yexchangetensorppalvaluexyshiftindetermintestimminindexone}
{$\Yexchangetensorppalvaluexyshiftindetermintestimminindexone$}
\newcommand{\Yexchangetensorppalvaluexyshiftindetermintestimminindexone}
{\widehat{k}_{xy1}^{min}}
%
\newcommand{\yexchangetensorppalvaluexyshiftindetermintestimminindextwo}
{$\Yexchangetensorppalvaluexyshiftindetermintestimminindextwo$}
\newcommand{\Yexchangetensorppalvaluexyshiftindetermintestimminindextwo}
{\widehat{k}_{xy2}^{min}}
%
%
%
\newcommand{\yexchangetensorppalvaluexyshiftindetermintestimmax}
{$\Yexchangetensorppalvaluexyshiftindetermintestimmax$}
\newcommand{\Yexchangetensorppalvaluexyshiftindetermintestimmax}
{\widehat{k}_{xy}^{max}}
%
\newcommand{\yexchangetensorppalvaluexyshiftindetermintestimmaxindexone}
{$\Yexchangetensorppalvaluexyshiftindetermintestimmaxindexone$}
\newcommand{\Yexchangetensorppalvaluexyshiftindetermintestimmaxindexone}
{\widehat{k}_{xy1}^{max}}
%
\newcommand{\yexchangetensorppalvaluexyshiftindetermintestimmaxindextwo}
{$\Yexchangetensorppalvaluexyshiftindetermintestimmaxindextwo$}
\newcommand{\Yexchangetensorppalvaluexyshiftindetermintestimmaxindextwo}
{\widehat{k}_{xy2}^{max}}
%

\newcommand{\yexchangetensorppalvaluez}
{$\Yexchangetensorppalvaluez$}
\newcommand{\Yexchangetensorppalvaluez}
{J_z}

\newcommand{\yexchangetensorppalvaluezestim}
{$\Yexchangetensorppalvaluezestim$}
\newcommand{\Yexchangetensorppalvaluezestim}
{\widehat{J}_z}
%
%
%
%
%
\newcommand{\yexchangetensorppalvaluezestimindexone}
{$\Yexchangetensorppalvaluezestimindexone$}
\newcommand{\Yexchangetensorppalvaluezestimindexone}
{\widehat{J}_{z1}}
%
\newcommand{\yexchangetensorppalvaluezestimindextwo}
{$\Yexchangetensorppalvaluezestimindextwo$}
\newcommand{\Yexchangetensorppalvaluezestimindextwo}
{\widehat{J}_{z2}}
%

%
\newcommand{\yexchangetensorppalvaluezestimargtwoqubitwritereadtimeintervalindextwo}
{$\Yexchangetensorppalvaluezestimargtwoqubitwritereadtimeintervalindextwo$}
\newcommand{\Yexchangetensorppalvaluezestimargtwoqubitwritereadtimeintervalindextwo}
{\Yexchangetensorppalvaluezestim
(
\Ytwoqubitwritereadtimeintervalindextwo
)}
%
%
%
%
%
\newcommand{\yexchangetensorppalvaluezshiftindetermint}
{$\Yexchangetensorppalvaluezshiftindetermint$}
\newcommand{\Yexchangetensorppalvaluezshiftindetermint}
{k_{z}}
%
\newcommand{\yexchangetensorppalvaluezshiftindetermintindexone}
{$\Yexchangetensorppalvaluezshiftindetermintindexone$}
\newcommand{\Yexchangetensorppalvaluezshiftindetermintindexone}
{k_{z1}}
%
\newcommand{\yexchangetensorppalvaluezshiftindetermintindextwo}
{$\Yexchangetensorppalvaluezshiftindetermintindextwo$}
\newcommand{\Yexchangetensorppalvaluezshiftindetermintindextwo}
{k_{z2}}
%
%
%
\newcommand{\yexchangetensorppalvaluezshiftindetermintestim}
{$\Yexchangetensorppalvaluezshiftindetermintestim$}
\newcommand{\Yexchangetensorppalvaluezshiftindetermintestim}
{\widehat{k}_{z}}
%
\newcommand{\yexchangetensorppalvaluezshiftindetermintestimindexone}
{$\Yexchangetensorppalvaluezshiftindetermintestimindexone$}
\newcommand{\Yexchangetensorppalvaluezshiftindetermintestimindexone}
{\widehat{k}_{z1}}
%
\newcommand{\yexchangetensorppalvaluezshiftindetermintestimindextwo}
{$\Yexchangetensorppalvaluezshiftindetermintestimindextwo$}
\newcommand{\Yexchangetensorppalvaluezshiftindetermintestimindextwo}
{\widehat{k}_{z2}}
%
%
%
%
%
\newcommand{\yexchangetensorppalvaluezshiftindetermintestimminusactual}
{$\Yexchangetensorppalvaluezshiftindetermintestimminusactual$}
\newcommand{\Yexchangetensorppalvaluezshiftindetermintestimminusactual}
{\Delta k _{z}}
%
%
\newcommand{\yexchangetensorppalvaluezshiftindetermintestimminusactualindexone}
{$\Yexchangetensorppalvaluezshiftindetermintestimminusactualindexone$}
\newcommand{\Yexchangetensorppalvaluezshiftindetermintestimminusactualindexone}
{\Delta k _{z1}}
%
\newcommand{\yexchangetensorppalvaluezshiftindetermintestimminusactualindextwo}
{$\Yexchangetensorppalvaluezshiftindetermintestimminusactualindextwo$}
\newcommand{\Yexchangetensorppalvaluezshiftindetermintestimminusactualindextwo}
{\Delta k _{z2}}
%
%
%
\newcommand{\yexchangetensorppalvaluezshiftindetermintestimmin}
{$\Yexchangetensorppalvaluezshiftindetermintestimmin$}
\newcommand{\Yexchangetensorppalvaluezshiftindetermintestimmin}
{\widehat{k}_{z}^{min}}
%
\newcommand{\yexchangetensorppalvaluezshiftindetermintestimminindexone}
{$\Yexchangetensorppalvaluezshiftindetermintestimminindexone$}
\newcommand{\Yexchangetensorppalvaluezshiftindetermintestimminindexone}
{\widehat{k}_{z1}^{min}}
%
\newcommand{\yexchangetensorppalvaluezshiftindetermintestimminindextwo}
{$\Yexchangetensorppalvaluezshiftindetermintestimminindextwo$}
\newcommand{\Yexchangetensorppalvaluezshiftindetermintestimminindextwo}
{\widehat{k}_{z2}^{min}}
%
%
%
\newcommand{\yexchangetensorppalvaluezshiftindetermintestimmax}
{$\Yexchangetensorppalvaluezshiftindetermintestimmax$}
\newcommand{\Yexchangetensorppalvaluezshiftindetermintestimmax}
{\widehat{k}_{z}^{max}}
%
\newcommand{\yexchangetensorppalvaluezshiftindetermintestimmaxindexone}
{$\Yexchangetensorppalvaluezshiftindetermintestimmaxindexone$}
\newcommand{\Yexchangetensorppalvaluezshiftindetermintestimmaxindexone}
{\widehat{k}_{z1}^{max}}
%
\newcommand{\yexchangetensorppalvaluezshiftindetermintestimmaxindextwo}
{$\Yexchangetensorppalvaluezshiftindetermintestimmaxindextwo$}
\newcommand{\Yexchangetensorppalvaluezshiftindetermintestimmaxindextwo}
{\widehat{k}_{z2}^{max}}
%
%
%
%
%
%
%
%
%
%
%
%
%
%
%
%
%
%


%
%
%
\newcommand{\ytwoqubitsprobadirxxplusplussumminusminusphasediff}
{$\Ytwoqubitsprobadirxxplusplussumminusminusphasediff$}
\newcommand{\Ytwoqubitsprobadirxxplusplussumminusminusphasediff}
{\Delta \Phi_{1,-1}}
%
%
%
\newcommand{\ytwoqubitsprobadirxxplusplusdiffminusminusfactorreal}
{$\Ytwoqubitsprobadirxxplusplusdiffminusminusfactorreal$}
\newcommand{\Ytwoqubitsprobadirxxplusplusdiffminusminusfactorreal}
{R_{14}}
%
%
%
\newcommand{\ytwoqubitsprobadirxxplusplusdiffminusminusfactorimag}
{$\Ytwoqubitsprobadirxxplusplusdiffminusminusfactorimag$}
\newcommand{\Ytwoqubitsprobadirxxplusplusdiffminusminusfactorimag}
{I_{14}}
%
\newcommand{\ytwoqubitsprobadirxxplusplusdiffminusminusphasediff}
{$\Ytwoqubitsprobadirxxplusplusdiffminusminusphasediff$}
\newcommand{\Ytwoqubitsprobadirxxplusplusdiffminusminusphasediff}
{\Delta \Phi_{1,0}}
%
%
%
%
%
\newcommand{\ytwoqubitsprobadirxxplusplusdiffminusminusphasediffindexd}
{$\Ytwoqubitsprobadirxxplusplusdiffminusminusphasediffindexd$}
\newcommand{\Ytwoqubitsprobadirxxplusplusdiffminusminusphasediffindexd}
{\Delta \Phi_{1,0d}}
%
\newcommand{\ytwoqubitsprobadirxxplusplusdiffminusminusphasediffindexdestim}
{$\Ytwoqubitsprobadirxxplusplusdiffminusminusphasediffindexdestim$}
\newcommand{\Ytwoqubitsprobadirxxplusplusdiffminusminusphasediffindexdestim}
{\widehat{\Delta \Phi}_{1,0d}}
%
%
%
%
%
\newcommand{\ytwoqubitsprobadirxxplusplusdiffminusminusphasediffcosgen}
{$\Ytwoqubitsprobadirxxplusplusdiffminusminusphasediffcosgen$}
\newcommand{\Ytwoqubitsprobadirxxplusplusdiffminusminusphasediffcosgen}
{w_1}
%
%
\newcommand{\ytwoqubitsprobadirxxplusplusdiffminusminusphasediffcosgenestim}
{$\Ytwoqubitsprobadirxxplusplusdiffminusminusphasediffcosgenestim$}
\newcommand{\Ytwoqubitsprobadirxxplusplusdiffminusminusphasediffcosgenestim}
{\widehat{w}_1}
%
%
%
%
\newcommand{\ytwoqubitsprobadirxxplusplusdiffminusminusphasediffsingen}
{$\Ytwoqubitsprobadirxxplusplusdiffminusminusphasediffsingen$}
\newcommand{\Ytwoqubitsprobadirxxplusplusdiffminusminusphasediffsingen}
{w_2}
%
%
\newcommand{\ytwoqubitsprobadirxxplusplusdiffminusminusphasediffsingenestim}
{$\Ytwoqubitsprobadirxxplusplusdiffminusminusphasediffsingenestim$}
\newcommand{\Ytwoqubitsprobadirxxplusplusdiffminusminusphasediffsingenestim}
{\widehat{w}_2}
%
%
%
%

%
%
%
\newcommand{\ysqrtminusone}
{$\Ysqrtminusone$}
\newcommand{\Ysqrtminusone}
{i}
\newcommand{\yboltzmannconst}
{$\Yboltzmannconst$}
\newcommand{\Yboltzmannconst}
{k_B}
%
%
%
%
%
%
%
%
%
%
%
%
\newcommand{\ywritereadonestatenb}
{$\Ywritereadonestatenb$}
\newcommand{\Ywritereadonestatenb}
{K}
%
\newcommand{\ywritereadonestatenbmixident}
{$\Ywritereadonestatenbmixident$}
\newcommand{\Ywritereadonestatenbmixident}
{\Ywritereadonestatenb
}
%
%
%
%
%
%
%
%
%
%
%
\newcommand{\yphaseoptfitveccompsyststatetinitandveccompsepsyststateout}
{$\Yphaseoptfitveccompsyststatetinitandveccompsepsyststateout$}
\newcommand{\Yphaseoptfitveccompsyststatetinitandveccompsepsyststateout}
{\phi_{opt}}
%
\newcommand{\yphaseextfitveccompsyststatetinitandveccompsepsyststateout}
{$\Yphaseextfitveccompsyststatetinitandveccompsepsyststateout$}
\newcommand{\Yphaseextfitveccompsyststatetinitandveccompsepsyststateout}
{\phi_{ext}}
%
%
%
\newcommand{\yrmsemoduli}
{$\Yrmsemoduli$}
\newcommand{\Yrmsemoduli}
{RMSE_m}
%
\newcommand{\yrmsecoef}
{$\Yrmsecoef$}
\newcommand{\Yrmsecoef}
{RMSE_c}
%
%
%
%

\newcommand{\yeqdefhamiltonian}
{
\begin{eqnarray}
\nonumber
H
&
=
&
\Yhamiltonfieldscale s_{1z} \Ymagfieldnot
+ \Yhamiltonfieldscale s_{2z} \Ymagfieldnot
- 2 J_{xy} ( s_{1x} s_{2x} + s_{1y} s_{2y} )
\\
&
&
- 2 J_{z} s_{1z} s_{2z}
\label{eq-online-hamiltonian}
\end{eqnarray}
}
\newcommand{\yeqdeftwoqubitstateinitindexi}
{
\begin{equation}
\label{eq-twoqubit-state-init-index-i}
| 
\psi_{
\Yqubitindexstd
}
(
\Yqubitonetimeinit
)
\rangle
=
\alpha 
_{
\Yqubitindexstd
}
| + 
{\rangle}
+ \beta 
_{
\Yqubitindexstd
}
| - 
{\rangle}
\end{equation}
}
\newcommand{\yeqdefqubitpolarqubitindexstd}
{
\begin{equation}
\label{eq-def-qubit-polar-qubit-indexstd} 
\alpha
_{
\Yqubitindexstd
}
=
\Yparamqubitindexstdstateplusmodulus
e ^{ \Ysqrtminusone
\Yparamqubitindexstdstateplusphase }
\hspace{5mm} 
\beta 
_{
\Yqubitindexstd
}
=
\Yparamqubitindexstdstateminusmodulus e ^{ \Ysqrtminusone
\Yparamqubitindexstdstateminusphase }
\hspace{10mm} 
\Yqubitindexstd \in \{ 1 , 2 \}
\end{equation}
}
\newcommand{\yeqdefparamqubitindexstdstateminusmodulusvsparamqubitindexstdstateplusmodulus}
{
\begin{equation}
\Yparamqubitindexstdstateminusmodulus
=
\sqrt
{
1
-
\Yparamqubitindexstdstateplusmodulus
^2
}
\hspace{10mm} 
\Yqubitindexstd \in \{ 1 , 2 \}
\label{eq-paramqubitindexstdstateminusmodulus-vs-paramqubitindexstdstateplusmodulus}
\end{equation}
}
\newcommand{\yeqdefqubitbothtimeinitstatetensorprod}
{
\begin{eqnarray}
\label{eq-qubitbothtimeinitstate-tensor-prod}
\Ymixsyststateinitial
&
=
&
\Yqubitonetimeinitstateindexone
\otimes
\Yqubitonetimeinitstateindextwo
\\
\nonumber
&
=
&
\alpha _1
\alpha _2
| ++ 
\rangle
+
\alpha _1
\beta _2
| +- 
\rangle
\\
\label{eq-etat-deuxspin-plusplus-initial-decompos}
&
&
+
\beta _1
\alpha _2
| -+ 
\rangle
+
\beta _1
\beta _2
| -- 
\rangle
\end{eqnarray}
}
\newcommand{\yeqdefstatetfinalvsopmixstatetinitcomponents}
{
\begin{equation}
\label{eq-statetfinalvsopmixstatetinit-components}
\Yveccompsyststatetfinal
=
\Yopmix
\Yveccompsyststatetinit
\end{equation}
}
\newcommand{\yeqdefveccompsyststatetinit}
{
\begin{equation}
\Yveccompsyststatetinit
=
[
\alpha _1
\alpha _2
,
\alpha _1
\beta _2
,
\beta _1
\alpha _2
,
\beta _1
\beta _2
]
^T
\label{eq-def-veccompsyststatetinit}
\end{equation}
}
\newcommand{\yeqdefopmixmatrixdecompose}
{
\begin{equation}
\label{eq-opmixmatrixdecompose}
\Yopmix 
= \Yopmixbases \Yopmixdiag \Yopmixbases^{-1}
= \Yopmixbases \Yopmixdiag \Yopmixbases
\end{equation}
}
\newcommand{\yeqdefopmixbasesdef}
{
\begin{equation}
\label{eq-opmixbasesdef}
\Yopmixbases 
=
\Yopmixbases ^{-1}
=
\left[
\begin{tabular}{llll}
1
&
0
&
0
&
0
\\
0
&
$\frac{1}{\sqrt{2}}$
&
$\frac{1}{\sqrt{2}}$
&
0
\\
0
&
$\frac{1}{\sqrt{2}}$
&
$- \frac{1}{\sqrt{2}}$
&
0
\\
0
&
0
&
0
&
1
\end{tabular}
\right]
\end{equation}
}
\newcommand{\yeqdefopmixdiagdef}
{
{ 
\begin{eqnarray}
\left[
\begin{tabular}{llll}
$
e^{
-
\Ysqrtminusone
\omega _{1 , 1}
(t - t_0)
}
$
&
0
&
0
&
0
\\
0
&
$
e^{
- 
\Ysqrtminusone
\omega _{1 , 0}
(t - t_0)
}
$
&
0
&
0
\\
0
&
0&
$
e^{
- 
\Ysqrtminusone
\omega _{0 , 0}
(t - t_0)
}
$
&
0
\\
0
&
0
&
0
&
$
e^{
- 
\Ysqrtminusone
\omega _{1 , -1}
(t - t_0)
}
$
\end{tabular}
\right]
.
\nonumber
\\
\label{eq-opmixdiagdef}
\end{eqnarray}
}
}
\newcommand{\yeqdefomegaallexpress}
{
\begin{eqnarray}
\hspace{-3mm}
\omega _{1, 1}
=
\frac{1}{\hbar}
\left[
\Yhamiltonfieldscale \Ymagfieldnot - 
\frac{
\Yexchangetensorppalvaluez
}
{2}
\right]
,
&
\hspace{3mm}
&
\label{eq-omega-one-zero-express}
\omega _{1, 0}
=
\frac{1}{\hbar}
\left[
- 
\Yexchangetensorppalvaluexy
+ 
\frac{
\Yexchangetensorppalvaluez
}
{2}
\right]
,
\\
\hspace{-33mm}
\omega _{0, 0}
=
\frac{1}{\hbar}
\left[
\Yexchangetensorppalvaluexy
+ 
\frac{ 
\Yexchangetensorppalvaluez
}
{2}
\right]
,
&
\hspace{3mm}
&
\omega _{1, - 1}
=
\frac{1}{\hbar}
\left[
- \Yhamiltonfieldscale
\Ymagfieldnot - 
\frac{ 
\Yexchangetensorppalvaluez
}
{2}
\right]
.
\nonumber
\\
\label{eq-omega-zero-zero-express}
\end{eqnarray}
}
\newcommand{\yeqdefexchangetensorppalvaluexyvstwoqubitresultphaseevolindexdeqone}
{
\begin{equation}
\frac{ 
\Yexchangetensorppalvaluexy
\Ytwoqubitwritereadtimeintervalindexone
} 
{ \hbar }
=
-
\Ytwoqubitresultphaseevolindexd
+
\Yexchangetensorppalvaluexyshiftindetermint
\pi
\label{eq-exchangetensorppalvaluexy-vs-twoqubitresultphaseevolindexd}
\end{equation}
with
\begin{equation}
\Ytwoqubitresultphaseevolindexd
=
\mathrm{arcsin}
( \Ytwoqubitresultphaseevolsin )
\label{eq-twoqubitresultphaseevolindexd}
\end{equation}
}
\newcommand{\yeqdefexchangetensorppalvaluexyvstwoqubitresultphaseevolindexdestim}
{
\begin{equation}
\frac{ 
\Yexchangetensorppalvaluexyestim
\Ytwoqubitwritereadtimeintervalindexone
} 
{ \hbar }
=
-
\Ytwoqubitresultphaseevolindexdestim
+
\Yexchangetensorppalvaluexyshiftindetermintestim
\pi
\label{eq-exchangetensorppalvaluexy-vs-twoqubitresultphaseevolindexd-estim}
\end{equation}
}
\newcommand{\yeqdefexchangetensorppalvaluezvstwoqubitsprobadirxxplusplusdiffminusminusphasediffindexdestimintervalindextwo}
{
\begin{equation}
\frac{ 
\Yexchangetensorppalvaluezestim
\Ytwoqubitwritereadtimeintervalindextwo
} 
{ \hbar }
=
\Ytwoqubitsprobadirxxplusplusdiffminusminusphasediffindexdestim
+
2
\Yexchangetensorppalvaluezshiftindetermintestim
\pi
+
\frac{ 
\Yexchangetensorppalvaluexyestim
\Ytwoqubitwritereadtimeintervalindextwo
} 
{ \hbar }
+
\frac{ 
\Yhamiltonfieldscale
\Ymagfieldnot
\Ytwoqubitwritereadtimeintervalindextwo
} 
{ \hbar }
\label{eq-exchangetensorppalvaluez-vs-twoqubitsprobadirxxplusplusdiffminusminusphasediffindexd-estim-intervalindextwo}
\end{equation}
}
\newcommand{\yeqdefexchangetensorppalvaluezvstwoqubitsprobadirxxplusplusdiffminusminusphasediffindexdeqoneintervalindextwo}
{
\begin{equation}
\frac{ 
\Yexchangetensorppalvaluez
\Ytwoqubitwritereadtimeintervalindextwo
} 
{ \hbar }
=
\Ytwoqubitsprobadirxxplusplusdiffminusminusphasediffindexd
+
2
\Yexchangetensorppalvaluezshiftindetermint
\pi
+
\frac{ 
\Yexchangetensorppalvaluexy
\Ytwoqubitwritereadtimeintervalindextwo
} 
{ \hbar }
+
\frac{ 
\Yhamiltonfieldscale
\Ymagfieldnot
\Ytwoqubitwritereadtimeintervalindextwo
} 
{ \hbar }
\label{eq-exchangetensorppalvaluez-vs-twoqubitsprobadirxxplusplusdiffminusminusphasediffindexd-intervalindextwo}
\end{equation}
}
\newcommand{\yeqdefexchangetensorppalvaluexyestimminusactualvsexchangetensorppalvaluexyshiftindetermintestiminminusactualindexoneandnot}
{
\begin{equation}
\Yexchangetensorppalvaluexyestimindexone
=
\Yexchangetensorppalvaluexy
+
\frac{ \hbar }
{
\Ytwoqubitwritereadtimeintervalindexoneone
}
\left(
\Ytwoqubitresultphaseevolindexdone
-
\Ytwoqubitresultphaseevolindexdoneestim
+
\Yexchangetensorppalvaluexyshiftindetermintestimminusactualindexone
\pi
\right)
\label{eq-exchangetensorppalvaluexyestim-minus-actual-vs-exchangetensorppalvaluexyshiftindetermintestimin-minus-actual-indexone}
\end{equation}
with
\begin{equation}
\Yexchangetensorppalvaluexyshiftindetermintestimminusactualindexone
=
\Yexchangetensorppalvaluexyshiftindetermintestimindexone
-
\Yexchangetensorppalvaluexyshiftindetermintindexone
.
\end{equation}
}
\newcommand{\yeqdefexchangetensorppalvaluexyestimminusactualvsexchangetensorppalvaluexyshiftindetermintestiminminusactualindextwoandnot}
{
\begin{equation}
\Yexchangetensorppalvaluexyestimindextwo
=
\Yexchangetensorppalvaluexy
+
\frac{ \hbar }
{
\Ytwoqubitwritereadtimeintervalindexonetwo
}
\left(
\Ytwoqubitresultphaseevolindexdtwo
-
\Ytwoqubitresultphaseevolindexdtwoestim
+
\Yexchangetensorppalvaluexyshiftindetermintestimminusactualindextwo
\pi
\right)
\label{eq-exchangetensorppalvaluexyestim-minus-actual-vs-exchangetensorppalvaluexyshiftindetermintestimin-minus-actual-indextwo}
\end{equation}
with
\begin{equation}
\Yexchangetensorppalvaluexyshiftindetermintestimminusactualindextwo
=
\Yexchangetensorppalvaluexyshiftindetermintestimindextwo
-
\Yexchangetensorppalvaluexyshiftindetermintindextwo
.
\end{equation}
}
\newcommand{\yeqdeftwoqubitresultphaseevolindexdoneandtwoestimideal}
{
\begin{equation}
\Ytwoqubitresultphaseevolindexdoneestim
=
\Ytwoqubitresultphaseevolindexdone
\hspace{10mm}
\mathrm{and}
\hspace{10mm}
\Ytwoqubitresultphaseevolindexdtwoestim
=
\Ytwoqubitresultphaseevolindexdtwo
.
\label{eq-twoqubitresultphaseevolindexdoneandtwoestim-ideal}
\end{equation}
}
\newcommand{\yeqdefstatetfinalvsopmixstatetinitcomponentsversiontwo}
{
\begin{equation}
\label{eq-statetfinalvsopmixstatetinit-components-version-two}
\Yveccompsyststatetfinal
=
\Yopmix
\Yveccompsyststatetinit
\end{equation}
}
\newcommand{\yeqdefstatecoefvstwoqubitsprobathreepolar}
{
\begin{eqnarray}
\label{eq-statecoef-vs-twoqubitsprobaplusplus-polar}
\Ytwoqubitsprobaplusplusdirzz
&
=
& 
\Yparamqubitonestateplusmodulus ^2 \Yparamqubittwostateplusmodulus
^2 
\\
\nonumber
\Ytwoqubitsprobaplusminusdirzz
&
=
&
\Yparamqubitonestateplusmodulus ^2 ( 1
-
\Yparamqubittwostateplusmodulus ^2 ) ( 1 -
\Ytwoqubitresultphaseevolsin ^2 ) + ( 1
-
\Yparamqubitonestateplusmodulus ^2 )
\Yparamqubittwostateplusmodulus ^2 \Ytwoqubitresultphaseevolsin ^2
\\
&
&
{
-
2 \Yparamqubitonestateplusmodulus \Yparamqubittwostateplusmodulus
\sqrt{ 1
-
\Yparamqubitonestateplusmodulus ^2 } \sqrt{ 1
-
\Yparamqubittwostateplusmodulus ^2 } \sqrt{ 1 -
\Ytwoqubitresultphaseevolsin ^2 } \Ytwoqubitresultphaseevolsin
\sin \Ytwoqubitresultphaseinit
}
\label{eq-statecoef-vs-twoqubitsprobaplusminus-polar-versionthree}
\\
\label{eq-statecoef-vs-twoqubitsprobaminusminus-polar-vs-paramqubitindexstdstateplusmodulus}
\Ytwoqubitsprobaminusminusdirzz
&
=
&
( 1
-
\Yparamqubitonestateplusmodulus ^2 ) ( 1
-
\Yparamqubittwostateplusmodulus ^2 )
\end{eqnarray}
}
\newcommand{\yeqdefeqdeftwoqubitresultphaseinittwoqubitresultphaseevol}
{
\begin{eqnarray}
\label{eq-def-twoqubitresultphaseinit-twoqubitresultphaseevol}
\Ytwoqubitresultphaseinit
&
=
&
( \Yparamqubittwostateminusphase
-
\Yparamqubittwostateplusphase
)
-
(
\Yparamqubitonestateminusphase
-
\Yparamqubitonestateplusphase )
\\
\label{eq-def-Ytwoqubitresultphaseevol-versiontwo}
\Ytwoqubitresultphaseevol
&
=
&
- 
\frac{ 
\Yexchangetensorppalvaluexy
( t - t_0 )
} { \hbar } 
\\
\Ytwoqubitresultphaseevolsin
&
=
&
\mbox{sgn} ( \cos \Ytwoqubitresultphaseevol )
\sin \Ytwoqubitresultphaseevol
.
\label{eq-def-Ytwoqubitresultphaseevol-versionthree}
\end{eqnarray}
}
\newcommand{\yeqdefstatecoefvstwoqubitsprobaplusminuspolarversionthreeexpect}
{
\begin{eqnarray}
E
\{
\Ytwoqubitsprobaplusminusdirzz
\}
&
=
&
E
\{
\Yparamqubitonestateplusmodulus ^2 
\}
( 1
-
E
\{
\Yparamqubittwostateplusmodulus ^2 
\}
) ( 1 -
\Ytwoqubitresultphaseevolsin ^2 ) 
\nonumber
\\
&
&
+ ( 1
-
E
\{
\Yparamqubitonestateplusmodulus ^2 
\}
)
E
\{
\Yparamqubittwostateplusmodulus ^2 
\}
\Ytwoqubitresultphaseevolsin ^2
\nonumber
\\
&
&
-
2 
E
\{
\Yparamqubitonestateplusmodulus 
\sqrt{ 1
-
\Yparamqubitonestateplusmodulus ^2 } 
\}
E
\{
\Yparamqubittwostateplusmodulus
\sqrt{ 1
-
\Yparamqubittwostateplusmodulus ^2 } 
\}
\sqrt{ 1 -
\Ytwoqubitresultphaseevolsin ^2 } \Ytwoqubitresultphaseevolsin
\nonumber
\\
&
&
\hspace{2mm}
\times
E
\{
\sin \Ytwoqubitresultphaseinit
\}
.
\label{eq-statecoef-vs-twoqubitsprobaplusminus-polar-versionthree-expect}
\end{eqnarray}
}
\newcommand{\yeqdefopmixmatrixdecomposeversiontwo}
{
\begin{equation}
\label{eq-opmixmatrixdecompose-versiontwo}
\Yopmix 
= \Yopmixbases \Yopmixdiag \Yopmixbases
\end{equation}
}
\newcommand{\yeqdefopsepmatrixdecompose}
{
\begin{equation}
\label{eq-opsepmatrixdecompose}
\Yopsep
= \Yopmixbases \Ysepmixdiag \Yopmixbases
\end{equation}
}
\newcommand{\yeqdefsepmixdiagdef}
{
\begin{eqnarray}
\Ysepmixdiag
&
=
\left[
\begin{tabular}{llll}
$
e^{
\Ysqrtminusone
\Ysepmixdiagelomegaoneone
}
$
&
0
&
0
&
0
\\
0
&
$
e^{
\Ysqrtminusone
\Ysepmixdiagelomegaonezero
}
$
&
0
&
0
\\
0
&
0&
$
e^{
\Ysqrtminusone
\Ysepmixdiagelomegazerozero
}
$
&
0
\\
0
&
0
&
0
&
$
e^{
\Ysqrtminusone
\Ysepmixdiagelomegaoneminusone
}
$
\end{tabular}
\right]
\label{eq-sepmixdiag-def}
\end{eqnarray}
}
\newcommand{\yeqdefopmixdiagelementsexpress}
{
\begin{eqnarray}
\hspace{-3mm}
\Ysepmixdiagelomegaoneone
=
\frac{
\Yhamiltonfieldscale \Ymagfieldnot
\Ytwoqubitwritereadtimeintervalindexthree
}{\hbar}
-
\frac{
\Yexchangetensorppalvaluezestim
\Ytwoqubitwritereadtimeintervalindexthree
}{2\hbar}
,
&
\hspace{3mm}
&
\label{eq-sepmixdiagel-omega-one-zero}
\Ysepmixdiagelomegaonezero
=
-
\frac{
\Yexchangetensorppalvaluexyestim
\Ytwoqubitwritereadtimeintervalindexthree
}{\hbar}
+
\frac{
\Yexchangetensorppalvaluezestim
\Ytwoqubitwritereadtimeintervalindexthree
}{2\hbar}
,
\\
\label{eq-sepmixdiagel-omega-zero-zero}
\hspace{-33mm}
\Ysepmixdiagelomegazerozero
=
\frac{
\Yexchangetensorppalvaluexyestim
\Ytwoqubitwritereadtimeintervalindexthree
}{\hbar}
+
\frac{
\Yexchangetensorppalvaluezestim
\Ytwoqubitwritereadtimeintervalindexthree
}{2 \hbar}
,
&
\hspace{3mm}
&
\Ysepmixdiagelomegaoneminusone
=
-
\frac{
\Yhamiltonfieldscale
\Ymagfieldnot 
\Ytwoqubitwritereadtimeintervalindexthree
}{\hbar}
-
\frac{
\Yexchangetensorppalvaluezestim
\Ytwoqubitwritereadtimeintervalindexthree
}{2 \hbar}
\end{eqnarray}
}
\newcommand{\yeqdeftwoqubitwritereadtimeintervallinkthree}
{
\begin{equation}
\Ytwoqubitwritereadtimeintervalindextwo
=
2
\Ytwoqubitwritereadtimeintervalindexone
\hspace{10mm}
\mathrm{and}
\hspace{10mm}
\Ytwoqubitwritereadtimeintervalindexthree
=
2
\Ytwoqubitwritereadtimeintervalindextwo
.
\label{eq-twoqubitwritereadtimeinterval-link-three}
\end{equation}
}
\newcommand{\yeqdefexchangetensorppalvaluexyestimminusactualvsexchangetensorppalvaluexyshiftindetermintestiminminusactualindexno}
{
\begin{equation}
\Yexchangetensorppalvaluexyestim
=
\Yexchangetensorppalvaluexy
+
\frac{ \hbar }
{
\Ytwoqubitwritereadtimeintervalindexone
}
\Yexchangetensorppalvaluexyshiftindetermintestimminusactual
\pi
\label{eq-exchangetensorppalvaluexyestim-minus-actual-vs-exchangetensorppalvaluexyshiftindetermintestimin-minus-actual-indexno}
\end{equation}
}
\newcommand{\yeqdefexchangetensorppalvaluezestimminusactualvsexchangetensorppalvaluezshiftindetermintestiminminusactualindexno}
{
\begin{equation}
\Yexchangetensorppalvaluezestim
=
\Yexchangetensorppalvaluez
+
\frac{ \hbar }
{
\Ytwoqubitwritereadtimeintervalindextwo
}
2
\Yexchangetensorppalvaluezshiftindetermintestimminusactual
\pi
+
(
\Yexchangetensorppalvaluexyestim
-
\Yexchangetensorppalvaluexy
)
\label{eq-exchangetensorppalvaluezestim-minus-actual-vs-exchangetensorppalvaluezshiftindetermintestimin-minus-actual-indexno}
\end{equation}
}
\newcommand{\yeqdeftableidentifinvertterminology}
{
\begin{table}[htb]
\begin{center}
\begin{tabular}{|l|l|l|}
\hline
&
system identification
&
system inversion,
\\
&
&
signal restoration,
\\
&
&
source separation
\\
\hline
classical
&
$\bullet$
channel identification,
&
$\bullet$
channel equalization
\\
signals
&
\hspace{2mm}
channel estimation
&
\hspace{2mm}
(communication)
\\
&
\hspace{2mm}
(communication)
&
$\bullet$
deconvolution
\\
&
$\bullet$
impulse response
&
\hspace{2mm}
(image,
\ymodifartitionehundredthirtyninevonestepone{seismology}%
...)
\\
&
\hspace{2mm}
or transfer function
&
$\bullet$
dereverberation 
\\
&
\hspace{2mm}
estimation
&
\hspace{2mm}
(acoustics)
\\
&
\hspace{2mm}
(acoustics...)
&
$\bullet$
cocktail party 
\\
&
&
\hspace{2mm}
processing
\\
&
&
\hspace{2mm}
(audio)
\\
&
&
$\bullet$
deblurring 
\\
&
&
\hspace{2mm}
(image)
\\
\hline
quantum
&
$\bullet$
process tomography 
&
$\bullet$
source separation
\\
states
&
\hspace{2mm}
(Section \ref{sec-bqpt})
&
\hspace{2mm}
(Section \ref{sec-blind-quantum-source-sep})
\\
&
$\bullet$
Hamiltonian estimation
&
$\bullet$
state restoration
\\
&
\hspace{2mm}
(Section \ref{sec-blind-hamilton-estim})
&
\hspace{2mm}
(Section \ref{sec-bqrs})
\\
&
$\bullet$
channel estimation
&
$\bullet$
channel equalization
\\
&
\hspace{2mm}
(Section \ref{sec-quantum-channel-estim})
&
\hspace{2mm}
(Section \ref{sec-quantum-channel-invert})
\\
&
$\bullet$
phase estimation
&
\\
&
\hspace{2mm}
(Section \ref{sec-quantum-channel-estim})
&
\\
\hline
\end{tabular}
\end{center}
\caption{Application-dependent terminology of classical and quantum
machine learning
tasks 
(i.e. 
data-driven
learning/adaptation
methods)
related to system identification%
,
system inversion%
,
signal restoration and
source 
separation 
(apart from
classification/clustering and
regression). This yields non-blind and blind (i.e., supervised and
unsupervised) variants. The sections of this paper 
mainly dealing with the
blind 
(i.e., unsupervised) 
quantum single-preparation versions of these
methods are mentioned.}
\label{tab-identifinvertterminology}
\end{table}
}
\newcommand{\yeqdeflinkdotproductdistance}
{
\begin{equation}
||\Yvecnotstdoneindexstdone
-
\Yvecnotstdoneindexstdtwo||^2
=
||\Yvecnotstdoneindexstdone||^2
+
||\Yvecnotstdoneindexstdtwo||^2
-
2
\Yvecnotstdoneindexstdone\Ytranspose
\Yvecnotstdoneindexstdtwo
.
\label{eq-linkdotproductdistance}
\end{equation}
}
\newcommand{\yeqdeflinkdotproductdistanceunitnorm}
{
\begin{equation}
||\Yvecnotstdoneindexstdone
-
\Yvecnotstdoneindexstdtwo||^2
=
2
(
1
-
\Yvecnotstdoneindexstdone\Ytranspose
\Yvecnotstdoneindexstdtwo
)
.
\label{eq-linkdotproductdistanceunitnorm}
\end{equation}
}
\newcommand{\yeqdefcorrcoef}
{
\begin{equation}
\Ycorrcoefvecnotstdoneindicesstdonetwo
=
\frac{
\Yvecnotstdoneindexstdone\Ytranspose
\Yvecnotstdoneindexstdtwo
}
{
||\Yvecnotstdoneindexstdone||
.
||\Yvecnotstdoneindexstdtwo||
}
.
\label{eq-correlcoef}
\end{equation}
}
\newcommand{\yeqdefstateregisterindexstdone}
{
\begin{equation}
\Ystateregisterindexstdone
=
\sum
_{\Ysetnotargqubitspaceregisterindexstdonequbitanybasisintstd}
\Yregisterstdonequbitspaceregisterindexstdonequbitsonetolastbasisstatecoefstd
\Yqubitspaceregisterindexstdonequbitonebasisstateintstd
\otimes
\dots
\otimes
\Yqubitspaceregisterindexstdonequbitlastbasisstateintstd
\label{eq-stateregisterindexstdone}
\end{equation}
}
\newcommand{\yeqdefstateregisterindexonedottwo}
{
\begin{equation}
\langle
\Ystateregisternot
					_{1}
\Ystateregisterindextwo
=	
\sum
_{\Ysetnotargqubitspaceregisterindexstdonequbitanybasisintstd}
\Yregisteronequbitspaceregisterindexstdonequbitsonetolastbasisstatecoefstd
^*		
\Yregistertwoqubitspaceregisterindexstdonequbitsonetolastbasisstatecoefstd
\label{eq-stateregisterindexonedottwo}
\end{equation}
}

\newcommand{\ypcaprojecteddim}
          {$\Ypcaprojecteddim$}
\newcommand{\Ypcaprojecteddim}
           {D}
\newcommand{\ytwoqubitwritereadtimeintervalindexoneone}
{$\Ytwoqubitwritereadtimeintervalindexoneone$}
\newcommand{\Ytwoqubitwritereadtimeintervalindexoneone}
{\Ytwoqubitwritereadtimeinterval_{11}}
\newcommand{\ytwoqubitwritereadtimeintervalindexonetwo}
{$\Ytwoqubitwritereadtimeintervalindexonetwo$}
\newcommand{\Ytwoqubitwritereadtimeintervalindexonetwo}
{\Ytwoqubitwritereadtimeinterval_{12}}
\newcommand{\ytwoqubitwritereadtimeintervalindextwoone}
{$\Ytwoqubitwritereadtimeintervalindextwoone$}
\newcommand{\Ytwoqubitwritereadtimeintervalindextwoone}
{\Ytwoqubitwritereadtimeinterval_{21}}
\newcommand{\ytwoqubitwritereadtimeintervalindextwotwo}
{$\Ytwoqubitwritereadtimeintervalindextwotwo$}
\newcommand{\Ytwoqubitwritereadtimeintervalindextwotwo}
{\Ytwoqubitwritereadtimeinterval_{22}}
\newcommand{\ytwoqubitresultphaseevolindexdone}
{$\Ytwoqubitresultphaseevolindexdone$}
\newcommand{\Ytwoqubitresultphaseevolindexdone}
{\Delta _{Ed1}}
\newcommand{\ytwoqubitresultphaseevolindexdoneestim}
{$\Ytwoqubitresultphaseevolindexdoneestim$}
\newcommand{\Ytwoqubitresultphaseevolindexdoneestim}
{\widehat{\Delta} _{Ed1}}
\newcommand{\ytwoqubitresultphaseevolindexdtwo}
{$\Ytwoqubitresultphaseevolindexdtwo$}
\newcommand{\Ytwoqubitresultphaseevolindexdtwo}
{\Delta _{Ed2}}
\newcommand{\ytwoqubitresultphaseevolindexdtwoestim}
{$\Ytwoqubitresultphaseevolindexdtwoestim$}
\newcommand{\Ytwoqubitresultphaseevolindexdtwoestim}
{\widehat{\Delta} _{Ed2}}
\newcommand{\ytwoqubitsprobadirxxplusplusdiffminusminusphasediffindexdone}
{$\Ytwoqubitsprobadirxxplusplusdiffminusminusphasediffindexdone$}
\newcommand{\Ytwoqubitsprobadirxxplusplusdiffminusminusphasediffindexdone}
{\Delta \Phi_{1,0d1}}
\newcommand{\ytwoqubitsprobadirxxplusplusdiffminusminusphasediffindexdoneestim}
{$\Ytwoqubitsprobadirxxplusplusdiffminusminusphasediffindexdoneestim$}
\newcommand{\Ytwoqubitsprobadirxxplusplusdiffminusminusphasediffindexdoneestim}
{\widehat{\Delta \Phi}_{1,0d1}}
\newcommand{\ymapfunc}
          {$\Ymapfunc$}
\newcommand{\Ymapfunc}
           {{\cal M}}
\newcommand{\yclassnb}
          {$\Yclassnb$}
\newcommand{\Yclassnb}
           {C}  
\newcommand{\yclassindexstd}
          {$\Yclassindexstd$}
\newcommand{\Yclassindexstd}
           {c} 
\newcommand{\yvecindexstdone}
          {$\Yvecindexstdone$}
\newcommand{\Yvecindexstdone}
           {j}
\newcommand{\yvecindexstdtwo}
          {$\Yvecindexstdtwo$}
\newcommand{\Yvecindexstdtwo}
           {k}
\newcommand{\yvecnotstdone}
          {$\Yvecnotstdone$}
\newcommand{\Yvecnotstdone}
           {v}            
\newcommand{\yvecnotstdoneindexone}
          {$\Yvecnotstdoneindexone$}
\newcommand{\Yvecnotstdoneindexone}
           {\Yvecnotstdone_{1}}       
\newcommand{\yvecnotstdoneindextwo}
          {$\Yvecnotstdoneindextwo$}
\newcommand{\Yvecnotstdoneindextwo}
           {\Yvecnotstdone_{2}} 
\newcommand{\yvecnotstdoneindexthree}
          {$\Yvecnotstdoneindexthree$}
\newcommand{\Yvecnotstdoneindexthree}
           {\Yvecnotstdone_{3}}    
\newcommand{\yvecnotstdoneindexthreescalefact}
          {$\Yvecnotstdoneindexthreescalefact$}
\newcommand{\Yvecnotstdoneindexthreescalefact}
           {\mu_{3}}    
\newcommand{\yvecnotstdoneindexfour}
          {$\Yvecnotstdoneindexfour$}
\newcommand{\Yvecnotstdoneindexfour}
           {\Yvecnotstdone_{4}}    
\newcommand{\yvecnotstdoneindexfourscalefact}
          {$\Yvecnotstdoneindexfourscalefact$}
\newcommand{\Yvecnotstdoneindexfourscalefact}
           {\mu_{4}}          
\newcommand{\yvecnotstdoneindexstdone}
          {$\Yvecnotstdoneindexstdone$}
\newcommand{\Yvecnotstdoneindexstdone}
           {\Yvecnotstdone_{\Yvecindexstdone}}
\newcommand{\yvecnotstdoneindexstdtwo}
          {$\Yvecnotstdoneindexstdtwo$}
\newcommand{\Yvecnotstdoneindexstdtwo}
           {\Yvecnotstdone_{\Yvecindexstdtwo}} 
\newcommand{\yvecnotstdoneclassstdvecone}
          {$\Yvecnotstdoneclassstdvecone$}
\newcommand{\Yvecnotstdoneclassstdvecone}
           {\Yvecnotstdone_{\Yclassindexstd 1}}					
\newcommand{\yvecnotstdoneclassstdvecindexstdone}
          {$\Yvecnotstdoneclassstdvecindexstdone$}
\newcommand{\Yvecnotstdoneclassstdvecindexstdone}
           {\Yvecnotstdone_{\Yclassindexstd \Yvecindexstdone}}		
\newcommand{\ycorrcoefvecnotstdoneindicesstdonetwo}
          {$\Ycorrcoefvecnotstdoneindicesstdonetwo$}
\newcommand{\Ycorrcoefvecnotstdoneindicesstdonetwo}
           {\rho (
           \Yvecnotstdoneindexstdone
           ,
           \Yvecnotstdoneindexstdtwo)}                   
\newcommand{\ytranspose}
          {$\Ytranspose$}
\newcommand{\Ytranspose}
           {^{T}} 
\newcommand{\ystateregisterindexthree}
          {$\Ystateregisterindexthree$}
\newcommand{\Ystateregisterindexthree}
           {| \Ystateregisternot
					_{3}
					  \rangle
           }
\newcommand{\ystateregisterindexfour}
          {$\Ystateregisterindexfour$}
\newcommand{\Ystateregisterindexfour}
           {| \Ystateregisternot
					_{4}
					  \rangle
           }
\newcommand{\ystatevecnotstdoneclassstdvecone}
          {$\Ystatevecnotstdoneclassstdvecone$}
\newcommand{\Ystatevecnotstdoneclassstdvecone}
           {| \Ystateregisternot										
				  	_{\Yclassindexstd 1}
					  \rangle
           }

\newcommand{\ystatevecnotstdoneclassstdvecindexstdone}
          {$\Ystatevecnotstdoneclassstdvecindexstdone$}
\newcommand{\Ystatevecnotstdoneclassstdvecindexstdone}
           {| \Ystateregisternot										
				  	_{\Yclassindexstd \Yvecindexstdone}
					  \rangle
           }									
\newcommand{\ystatevectobeclassifiednot}
          {$\Ystatevectobeclassifiednot$}
\newcommand{\Ystatevectobeclassifiednot}
           {\phi}
\newcommand{\ystatevectobeclassified}
          {$\Ystatevectobeclassified$}
\newcommand{\Ystatevectobeclassified}
           {| \Ystatevectobeclassifiednot \rangle}
\newcommand{\yqubitnbarbspaceoveralleventindexequalstdwithvecbasisrvnot}
{$\Yqubitnbarbspaceoveralleventindexequalstdwithvecbasisrvnot$}
\newcommand{\Yqubitnbarbspaceoveralleventindexequalstdwithvecbasisrvnot}
{
{\rm X}
}
\newcommand{\yqubitnbarbspaceoveralleventindexequalstdwithvecbasisrvqubitnbarbstatewritereadseqindex}
{$\Yqubitnbarbspaceoveralleventindexequalstdwithvecbasisrvqubitnbarbstatewritereadseqindex$}
\newcommand{\Yqubitnbarbspaceoveralleventindexequalstdwithvecbasisrvqubitnbarbstatewritereadseqindex}
{
\Yqubitnbarbspaceoveralleventindexequalstdwithvecbasisrvnot
_{\Yqubitnbarbstatewritereadseqindex}
}
\newcommand{\yqubitnbarbspaceoveralleventindexequalstdwithvecbasisrvmeanqubitnbarbstatewritereadseqnb}
{$\Yqubitnbarbspaceoveralleventindexequalstdwithvecbasisrvmeanqubitnbarbstatewritereadseqnb$}
\newcommand{\Yqubitnbarbspaceoveralleventindexequalstdwithvecbasisrvmeanqubitnbarbstatewritereadseqnb}
{
\overline{
\Yqubitnbarbspaceoveralleventindexequalstdwithvecbasisrvnot
}
}

\title{%
Single-preparation unsupervised 
quantum machine learning%
: concepts and applications%
}

\author{Yannick Deville}
\email[]{yannick.deville@irap.omp.eu}
\homepage[]{\\
http://userpages.irap.omp.eu/$\sim$ydeville/}
\affiliation{%
Universit\'e de Toulouse%
,
UPS, CNRS, CNES,
OMP,\\
IRAP
(Institut de Recherche en Astrophysique et Plan\'etologie),
F-31400
Toulouse,
France%
}

\author{Alain Deville}
\email[]{alain.deville@univ-amu.fr}
\affiliation{%
Aix-Marseille Universit\'e,
CNRS%
,
IM2NP UMR 7334%
,
F-%
13397
Marseille, France%
}

\date{\today}

\begin{abstract}
The term ``machine learning'' especially refers to
algorithms (and associated systems) that derive mappings,
i.e. intput/output transforms, by using
numerical data 
that
provide information about
the transform 
of
interest in the
considered application.
The data processing tasks to be performed in these applications
not only include classification/clustering
and regression, but also
various 
problems
related to
system identification, system inversion and input signal
restoration or source separation
(i.e. signal separation) when considering several signals.
In this paper, we first analyze
the connections that exist between
all
these problems, in the classical and quantum frameworks.
We then focus 
on their most challenging versions,
where quantum data and/or quantum processing means are
considered and learning is performed in the unsupervised mode,
also called the blind mode,
i.e. without ``reference values'' (at the input or output
of the 
mapping,
depending on the considered task).
Moreover, we propose 
the quite general concept 
of
SIngle-Preparation Quantum Information Processing (SIPQIP).
The term ``preparation'' refers to the
initialization of qubit states. As explained in the introduction, 
it is here used
in a more general sense 
than usually in quantum mechanics: 
we 
mainly
consider kets with random-valued coefficients. The resulting
methods only require one to
estimate
expectations of probabilities
of measurement outcomes associated with 
considered
states.
This may be achieved
with
a \emph{single} instance of each 
state
in our SIPQIP framework.
This
avoids the burden 
of usual 
methods, that have to 
very accurately
create
many 
copies
of each fixed
state, 
to
estimate
statistical features
associated with that state.
We
detail or discuss the application of
this SIPQIP concept to various 
tasks 
and
systems
that 
fulfill 
quantum mechanical principles, 
related to
system identification
(blind quantum process tomography or BQPT,
blind Hamiltonian parameter estimation or BHPE,
blind quantum channel identification/estimation,
blind phase estimation),
system inversion and state estimation
(blind quantum source separation or BQSS,
blind quantum entangled state restoration or BQSR,
blind quantum channel equalization)
and 
classification.
Part of these methods are detailed for a specific class of quantum
processes, which then allows one to extend them to other processes.
These processes correspond
to
two 
qubits
implemented as electron spins 
1/2,
internally
coupled according
to the cylindrical-symmetry Heisenberg model,
with unknown principal values for the exchange tensor.
The resulting
numerical performance of 
these methods is reported,
thus showing that the proposed SIPQIP framework moreover yields
much more accurate estimation than the standard multiple-preparation
approach, for a given total number of state preparations.
Several types of proposed methods are especially of interest 
when used in
a 
\emph{qua}ntum co\emph{mputer}%
, that
we propose
to more briefly call a ``quamputer'':
BQPT and BHPE 
simplify 
the characterization 
of 
the gates of quamputers,
whereas 
BQSS and BQSR 
allow
one to design
quantum gates that may be used to compensate for
the non-idealities that may alter states stored in quantum
registers.
BQSS/BQSR moreover
opens the way to the much more general
concept of 
self-adaptive
quantum gates, that could automatically
adapt their behavior, according to predefined rules that would
allow them to compensate for
various non-idealities in 
quamputers.
\end{abstract}

\maketitle

\section{Introduction}\label{sec-intro}
Classical machine learning is currently a 
booming field
\cite{a616}
and various quantum machine learning
extensions are also being considered
\cite{amq94,
amq73,
amq72bis,
amq74}.
The processing tasks that involve data-driven learning not
only include widespread
classification/clustering 
\cite{elivrebishop,
elivrebishop2006,
livredudahart2000,
a616,
eelevenlivretheodoridistheory,
amq73,
amq84}
and regression
\cite{elivrebishop,
elivrebishop2006,
eelevenlivretheodoridistheory}, 
but also
especially:
(a)
classical
system identification
\cite{a593,
book-equalization-ding-li,
book-ljung}
and its quantum extension,
called
(non-blind
\cite{booknielsen,
amq-baldwin-physreva-2014,
amq75,
amq45official,
amq50-physical-review,
amq59,
amq48,
amq52-physical-review,
amq56,
amq41}
or
blind 
\cite{amoi6-46,
amoi6-79,
amoi6-118}%
)
quantum process tomography,
(b) 
system inversion and signal restoration
and
(c)
blind source separation (BSS)
e.g. based on independent component analysis (ICA)
\cite{icabook-cichocki,
book-cichocki-nmf,
book-comon-jutten-ap,
amoi6-7,
amoi6-48,
icabook-oja,
book-makino-bss-speech}
(with a close connection 
\ymodifartitionehundredthirtyninevonesteptwo{with}
principal
component analysis
\cite{paper-abdi-williams-pca,
book-jolliffe-pca,
amq72})
and quantum extensions of BSS/ICA
\cite{amoi5-31,
amoi6-18,
amoi6-42,
amoi6-34,
amoi6-37,
amoi6-64}.
Moreover, in various application fields, these tasks are 
given different names
(see the summary in Table
\ref{tab-identifinvertterminology}),
especially
channel identification
or channel estimation
\cite{book-equalization-ding-li},
(channel)
equalization
\cite{book-equalization-ding-li,
livreproakis},
dereverbation
\cite{book-de-mori},
deconvolution 
\cite{book-starck},
deblurring
\cite{book-starck}
or
cocktail party problem
\cite{acherry1953}.

\yeqdeftableidentifinvertterminology

\ymodifartitionehundredthirtyninevonestepone{Beyond 
their apparent diversity, the above data processing tasks}
share major features,
that are analyzed in Section
\ref{sec-machine-learning}:
they involve mappings 
from input data to output data, 
and these mappings are derived
from a set of known values of these input and/or output quantities,
depending whether that learning is performed in the so-called
supervised or unsupervised (i.e. non-blind or
blind) modes, whose definitions depend on the
considered task and are also analyzed in Section
\ref{sec-machine-learning}.
These approaches are developed in order
to characterize the mapping performed by a given natural or artificial
system, and/or 
in order
to build an artificial system that performs a mapping
(i.e. a data transformation)
suited to the considered application.

In this paper, we investigate a variety of the above-defined data
processing tasks and we focus on advanced configurations from the 
following points of view.
First, we only 
consider a quantum framework, in terms of the
nature of the data to be processed and/or of the means used to
process them.
Second, we
almost only
address unsupervised learning, which is more
challenging than supervised learning because it consists of
learning mappings 
without
known values (but with a few known
properties) for the input or output of that mapping.
The overview of classical and quantum machine learning provided in Section
\ref{sec-machine-learning}
includes references to the currently
quite limited 
set
of works from the literature
which is dedicated to
that quantum and 
unsupervised learning framework 
that we tackle in
this paper.
Moreover, we here proceed beyond that framework, by adding
another feature: we focus on 
\ymodifartitionehundredthirtyninevonestepone{what we call
``single-preparation operation''.}
This concept is detailed in Section \ref{sec-single-prep}.
To put it briefly, various quantum machine learning methods
from the literature%
,
e.g. intended for system identification (i.e., say,
quantum process tomography) or system inversion,
use
multiple-preparation approaches 
in the sense that, for each
quantum state value that they consider, they estimate the
probabilities of corresponding measurement
outcomes by using the sample
frequencies of these outcomes over a set of measurements, which requires
a set of copies of the considered quantum state, to perform one
measurement for each copy
(see details in Section \ref{sec-stochastic-QIP-standard-concepts}).
In contrast,
we very recently introduced
a statistical approach 
which
yields much higher flexiblity, since
it avoids the burden of 
very
acccurately preparing
many 
ideally
identical copies of the same known state, by
allowing one to replace these copies
by a set of states whose values are possibly different and unknown but
only requested to belong to a general known class
\cite{amoi6-104,
amoi6-118}.
This concept is quite general but, in
\cite{amoi6-104,
amoi6-118},
we only detailed 
its application to a single data processing task,
namely single-preparation blind (i.e. unsupervised) quantum process
tomography.
In the present paper, we aim at showing how this 
single-preparation processing 
concept may be applied to a variety of
other quantum information processing tasks of interest,
thus yielding a general ``SIngle-Preparation Quantum Information Processing''
(SIPQIP) framework.

\ymodifartitionehundredthirtyninevonestepone{The 
terminology used in this paper
deserves the following comments.}
\ymodifartitionehundredthirtyninevonestepone{Quantum mechanics 
(QM)
considers that an 
isolated quantum system may be either in a pure 
state - the result of some preparation -, described by a ket with deterministic 
coefficients (in the 
Schr\"odinger 
picture), or more generally in a 
state called a mixed state or a statistical mixture, usually described by a 
density operator. When developing our methods, first in the BQSS then in the 
BQPT fields, we were led to distinguish between a ``deterministic pure state''
(the usual pure state of QM), and a ``random pure state'', 
described by
a ket with 
\ymodifartitionehundredthirtyninevonestepone{random-valued}
coefficients when 
developed
over an orthonormal basis of fixed vectors. 
It has been shown that the system is then in a mixed state 
\cite{amoi6-67}.
\ymodifartitionehundredthirtyninevonestepone{Deterministic pure states may
also be considered as a specific subset of random pure states, corresponding
to the case when the random variables that define the ket coefficients
of random pure states
reduce to fixed values, i.e. with ``no uncertainty''.}
In the context of BQSS or BQPT, 
depending on the considered method,
the system of interest 
is
initialized 
either in a deterministic or in a random pure state. Most 
of the methods 
proposed in this paper are based on qubits initialized with random pure states.}
\ymodifartitionehundredthirtyninevonestepone{Such an
initialization is also called a 
state preparation hereafter.}

The remainder of this paper 
\ymodifartitionehundredthirtyninevonestepone{is}
organized as follows.
In Section
\ref{sec-machine-learning},
we first provide an overview of major
classical and quantum machine learning tasks, and we 
analyze
their connections,
that are of interest for our subsequent original contributions in this paper.
Then, in Section \ref{sec-single-prep},
we define the single-preparation quantum processing concept,
which is 
an
original general
feature then used in all
processing methods
proposed in this paper. 
These methods
deal with various problems 
related to
quantum system identification
(Section \ref{sec-system-ident}),
quantum system inversion and
state restoration
(Section \ref{sec-system-invert})
and quantum classification
(Section
\ref{sec-classif}).
Finally, Section
\ref{sec-concl}
contains conclusions about the processing tasks
addressed in this paper and
a discussion of potential extensions 
of the proposed methods to
other quantum information processing problems.

\section{Classical and quantum
machine learning approaches for data mapping}
\label{sec-machine-learning}
Many
classical and quantum information processing systems 
aim at applying transforms, 
defined by
mathematical functions, to their input data, in order to map them to output data.
These transforms are 
often called mappings or maps, both in the classical
\cite{elivrebishop}
and quantum
\cite{preskill-http-chap3}
information processing literature
(quantum maps are also called quantum channels
\cite{preskill-http-chap3},
with a reference to 
communications).
In basic systems, the considered transform is predefined by the
human
system designer, depending on the target application. 
In contrast,
more advanced systems, that are considered 
hereafter, are referred to as 
(self-)adaptive systems,
since they adapt their behavior 
(i.e. the mapping they perform)
to the data they receive
\cite{elivrejohnson, 
bibref-efive-ta-chap-adap-book-haykin},
by means of algorithms
which perform so-called adaptation, training or (machine) learning
\cite{elivrebishop,
elivrebishop2006,
livredudahart2000,
a616,
eelevenlivretheodoridistheory}.
In other applications, input/output mappings are also learnt
from data but with other goals, especially to 
characterize the behavior of a given natural medium or artificial
system, as detailed below.
\ymodifartitionehundredthirtyninevonestepone{The
classical and quantum versions of
machine learning thus involve}
various
types of applications and associated types of transforms,
that are analyzed in more detail hereafter.

Machine learning is first used in classical
classification and regression systems, 
whose
transforms map a set of input quantities
(each of which has its own nature)
to output quantities which often have a different nature
from input ones.
This is especially true for classification systems
\cite{elivrebishop,
elivrebishop2006,
livredudahart2000,
a616,
eelevenlivretheodoridistheory},
which receive a set of input quantities that are
most often continuous valued,
whereas their 
(possibly thresholded)
outputs are binary valued.
More precisely,
\ymodifartitionehundredthirtyninevonestepone{let us first consider a
classification system without the rejection capability that is
defined further. Such a system}
generally
outputs
\yclassnb\
values, where 
\yclassnb\
is the number of classes involved in the
considered application. 
Only one of these outputs is equal to 1, say output with index
\yclassindexstd,
whereas all other outputs are equal to 0.
The index 
\yclassindexstd\
of the active output defines the decision made by the
classifier:
it considers that its input values correspond to a
case when the input belongs to class 
\yclassindexstd.
During the final use of the classifier, called the
``resolution phase'',
``classification phase'' or ``test phase'',
the above output values are provided to
the target application.
A typical use
of this framework is Optical Character Recognition
(OCR)
\cite{elivrebishop,
livredudahart2000,
eelevenlivretheodoridistheory}.
The classifier then 
receives an image, i.e. a
set of pixel values (or features, i.e. parameters, extracted from them),
where a letter or symbol belonging to a given alphabet is written.
The classifier
sets its 
\yclassindexstd%
-th output to 1 
if it considers that this
particular input image
contains the 
\yclassindexstd%
-th symbol of that alphabet.
\ymodifartitionehundredthirtyninevonestepone{Moreover,
improved classification algorithms are able to
detect when they consider that the input that they receive during
the resolution phase does not belong to any of
the considered classes, e.g. when the received image
is not similar enough to any character of the
considered alphabet (indeed, an image may contain
a shape which is not a character of any written
language). 
Such a classifier then decides
that it is not able to classify the considered
input ``object''
and it rejects it. This may be expressed e.g. by
setting all
\yclassnb\
outputs to zero or by adding a
$
(
\Yclassnb
+
1
)
$-th
output to the classifier, which is equal to 1 when
this classifier succeeds in classifying the considered
input and to 0 otherwise.}

Before the above 
resolution
phase,
machine learning algorithms are typically used,
e.g. in OCR systems, to initially
build an adequate input-output mapping, during the so-called
``learning phase'',
``training phase''
or ``adaptation phase''
(this possibly includes a so-called ``validation''),
by using data
composed of training examples.
In supervised learning approaches,
each 
example consists of 
an input (e.g., an image containing a character
for OCR)
and 
its correct class, i.e. the associated desired values of 
the classifier
outputs (called labels), which are provided by a supervisor, i.e.
typically a human expert of the considered
application.
In contrast,
in unsupervised learning approaches
for classification
(called clustering
\cite{eelevenlivretheodoridistheory}), 
the system
self-organizes by using only inputs
(e.g., images in the OCR example), i.e. unlabeled data.
Various system architectures and 
supervised or unsupervised
learning algorithms have thus been developed,
especially including (artificial)
neural networks 
\cite{elivrebishop,
elivrebishop2006,
livredudahart2000,
eelevenlivretheodoridistheory}
and their recent deep extensions
\cite{a616},
as well as Support Vector Machines (SVM)
\cite{elivrebishop2006,
livredudahart2000,
eelevenlivretheodoridistheory}.

Regression systems 
\cite{elivrebishop,
elivrebishop2006,
eelevenlivretheodoridistheory}
are similar to the above classifiers, except that
their outputs are continuous-valued. 
They 
typically 
first use a supervised training phase in order to
learn mappings
from data samples, which are
examples of adequate pairs composed of
input values and correct corresponding output values in the considered
applications.
Once this mapping has been fixed,
such a regression system
may eventually
be used e.g. to control an industrial setup in
a 
factory:
the regression system then
receives, as its inputs, different types of measured quantities
provided by sensors available in the 
factory,
and this system maps its inputs to possibly different types of 
continuous-valued quantities,
used to drive the actuators that control the industrial setup.
More specifically, regression systems where the inputs and output(s)
have the same nature especially concern prediction tasks
for time series, where the
system aims at providing the expected future value(s) of a quantity
(e.g. a currency exchange rate or streamflow)
from its past values.

Whereas the above concepts were initially developed for classical data,
they are currently being extended to 
quantum data and/or quantum
processing means
\cite{amq73}, especially
because the Quantum Information Processing (QIP) 
\cite{booknielsen}
community
is investigating quantum extensions of
classifiers to handle the huge processing power and amount of data
involved in current real-world applications.
These extensions include the implementation of SVM classifiers
on quantum computers with 
very low
computational complexity
\cite{amq84}.
Besides, the
versatile quantum optical neural networks proposed in
\cite{amq85}
can perform different
related tasks, 
including reinforcement learning.

Although one may first have in mind the above 
general classification and
regression/prediction tasks
when thinking of classical and quantum machine learning,
data-driven algorithms are also widely used 
in a 
partly 
related
set
of processing tasks, called system identification 
and
system inversion, with an extension to
source separation.
First considering the classical framework,
this e.g. includes situations when an
electromagnetic or acoustic signal is emitted
from a first location, then
transferred through a medium, which may be seen as an
electromagnetic or acoustic ``channel'' that transforms
its input composed of the emitted signal.
The
output of that channel is then
the signal received by an
electromagnetic antenna or microphone in a second
location.
These and other practical situations yield 
two types of machine learning 
problems.
The first one is often called system identification
\cite{a593,
book-equalization-ding-li,
book-ljung},
and more specifically channel identification
or channel estimation
in the field
of electromagnetic communications
\cite{book-equalization-ding-li}.
Its simpler 
version \cite{book-ljung},
called the ``non-blind version'' by the signal 
and image processing community
and the ``supervised version'' by the machine learning and
data analysis community,
operates as follows.
As in the above regression task,
it uses 
a set of
known
continuous-valued
pairs composed of
input values 
and corresponding output values of the considered
``system'',
i.e. of the above-defined channel,
in order to estimate, i.e. learn, the 
unknown transform 
(i.e. mapping) performed
by this system. 
It should be noted that, in the 
above examples involving
electromagnetic or acoustic signals, the considered ```system''
is not an artificial system to be built by human beings in
order to perform a given type of data processing
(as in the above classification and regression tasks), but the
``natural system'' formed by the considered electromagnetic
or acoustic propagation medium.
The goal of
system identification is then
to 
characterize
this medium.

System identification methods have also been extended to 
the more challenging blind,
i.e. unsupervised,
configuration \cite{a593,
book-equalization-ding-li}. 
In that case,
during the learning or estimation phase,
only the values of the system output 
(i.e. the values of the received channel output
in the above examples)
are known,
whereas its input values are unknown.
However, the system input is most often
required to have some known properties,
e.g. some known statistical features,
so that this configuration is sometimes stated to be semi-blind
or semi-supervised.
This
problem may therefore be seen as a non-conventional
form of regression, with
only partial knowledge about the input data.
Besides, 
it should be noted that
the known values 
are only the \emph{output} values
in blind system identification,
whereas they 
are only the \emph{input} values
in
the above unsupervised classification problem.
Whereas non-blind or blind system identification is applied to
single-input single output (SISO) systems in its basic form,
it 
may then be extended to
multiple-input multiple-output (MIMO) systems.

The second considered machine learning
problem, 
in connection with system identification, 
deals with
system inversion and signal restoration.
One
then again considers an unknown ``system'', called the direct system,
but one here
aims at building an artificial system that 
essentially
performs a
transform equal to (an estimate of) the inverse of that of the direct
system (assuming that direct transform is invertible). 
This inverse transform is first learnt during the training/adaptation phase,
either
directly or by first learning the direct transform
(using
system identification methods)
and then deriving its inverse
(in low-noise scenarios%
).
Then, 
in the ``inversion phase'',
which corresponds to the final use of the inversion system,
the (estimated) inverse transform
is applied to known output values of the direct system, in order to 
recover
(estimates of)
corresponding unknown input values of that direct system.
This approach e.g. applies to the above two configurations
involving channels.
The direct system then corresponds
to the physical propagation medium, such as an
electromagnetic communication channel, which alters the emitted
signal in a initially unknown way. 
One then aims at restoring the emitted signal from its
altered received version,
by learning an adequate inverse transform
from data samples. 
In the field of radio-frequency 
communications,
this signal processing task is often referred to as
(channel)
equalization
\cite{book-equalization-ding-li,
livreproakis}.
Similarly, the restoration of an unknown emitted acoustic signal
from its received version altered by reverberation
during propagation is often called
dereverbation
\cite{book-de-mori}.
More generally speaking,
the problem
of 
restoring 
a source signal only from an observation which is a
transformed version 
of that source signal
is called deconvolution 
(for a linear invariant transform)
or deblurring in
various fields, such as
astronomical image analysis
\cite{book-starck}.
Whatever the considered application field, the initial learning
procedure for estimating the inverse transform may be applied
in the non-blind or blind mode, i.e.
respectively with known or unknown input values for the direct system,
whereas the output values of that system are known in both modes.
This ``(unknown) system inversion'' task may also be
extended to MIMO configurations, in order to restore a set of
unknown signals from a set of their transformed versions.

The blind MIMO system inversion
problem is also closely related to the field of
blind source separation (BSS) 
\cite{icabook-cichocki,
book-cichocki-nmf,
book-comon-jutten-ap,
amoi6-7,
amoi6-48,
icabook-oja,
book-makino-bss-speech},
whose quantum extension is one of the major topics
tackled
further in this paper,
together with quantum extensions of system identification.
In BSS, the goal is also to
restore a set of unknown source signals from a set of available
\ymodifartitionehundredthirtyninevonesteptwo{combinations
of these signals (called mixtures in BSS),}
that result from an unknown transform
which combines (i.e. mixes) these source signals.
However, in BSS, one most often
allows each restored signal to be equal to a
source signal only up to an acceptable
residual transform (called an
indeterminacy), because such transforms cannot be
avoided, due to the 
limited constraints that are
set on the considered classes of signals and mixing
transforms.
When applied to the separation of acoustic/audio signals from
their mixtures 
recorded by a set of microphones,
BSS is often referred to as the ``cocktail party problem''
\cite{acherry1953}.

The first class of BSS methods that was developed and that is still of
major importance is Independent Component Analysis, or ICA
\cite{book-comon-jutten-ap,
amoi6-7,
amoi6-48,
icabook-oja}.
ICA is a statistical approach, which essentially
requires statistically independent random source signals.
Thus,
ICA is guaranteed to restore
the source signals up to limited indeterminacies
for the simplest class of mixtures,
that is when
the available signals are linear instantaneous
(i.e. memoryless) combinations of the unknown source signals
\cite{book-comon-jutten-ap,
amoi6-7,
amoi6-48,
icabook-oja}.
For such mixtures, 
ICA
may be seen as an extension of more conventional Principal
Component Analysis, or PCA
\cite{paper-abdi-williams-pca,
book-jolliffe-pca}.

PCA and ICA may both be used to
perform mappings from the available 
\ysssensnb\
variables to
\ysssensnb\
output variables 
that
are linear instantaneous mixtures of these
available variables. In other words, they yield a representation 
of the same data in a new basis.
The selected bases are different in PCA and ICA. 
PCA uses one of the bases 
that are such
that the output variables are uncorrelated.
ICA uses one of the bases
that are such
that these output variables are statistically independent,
which includes uncorrelatedness but is more constraining
(for non-Gaussian signals).
This is the reason why
PCA alone cannot achieve
BSS
\cite{icabook-oja},
but is often 
used
as a first stage in ICA algorithms.
Outside the framework of ICA,
PCA is most often used
as a mapping that projects
the available data 
onto 
a lower-dimensional
space,
i.e. with dimension 
\ypcaprojecteddim\
lower than
\ysssensnb , by keeping only the first 
\ypcaprojecteddim\
coordinates in the output basis,
for visualization 
(with
$
\Ypcaprojecteddim
=
2
$,
i.e. projection onto 
a plane,
or
$
\Ypcaprojecteddim
=
3
$,
i.e. 3-dimensional visualization)
or compression tasks.
Such a projection
may also be used
as a preprocessing stage 
of
ICA,
in order
to reduce the influence of noise,
when the available mixed signals contain noise and their
number
\ysssensnb\
is higher than the number
\ysssrcnb\
of source signals:
one then keeps the first
$
\Ypcaprojecteddim
=
\Ysssrcnb
$
output components of PCA.

ICA also has connections with the above fields of classification and regression
in the sense that a significant part of the algorithms developed in all these
fields are based on the same class of tools, namely
neural networks.
More precisely,
when initially developing ICA methods for linear instantaneous mixtures,
one of the very first proposed approaches was the well-known
H{\'e}rault-Jutten neural network
(see e.g. 
\cite{hj-cras,
a140,
a141,
a142,
amoi3-22}
for its definition and analysis)
and extended versions of that network
were then introduced and analyzed
(see e.g.
\cite{a286,
amoi4-10}).
Neural approaches were 
then
proposed for specific classes of
\emph{nonlinear} mixtures or without considering 
any restrictions
on the type of mixture
(see e.g.
\cite{aalmeidajmlr,
amoi5-49,
a563,
amoi6-19}).
Finally, the interest in neural methods recently raised again
also
in the field of BSS/ICA.
For instance,
generative adversarial networks (GANs) were used to perform
linear and non-linear ICA
\cite{brakelbengioarxiv2017}.

The above connected fields of classical system identification, system inversion, (B)SS
and PCA
have been partly extended as follows to the quantum framework.
Among these problems,
the one 
which was
first
studied is the quantum version 
of non-blind 
system identification,
especially
\footnote{%
See also
\cite{booknielsen}
p. 398 for the other earliest references.%
}
introduced in 1997 in 
\cite{amq30official}
and
called
``quantum process tomography'' or QPT
by the QIP 
community: see
e.g.
\cite{booknielsen,
amq-baldwin-physreva-2014,
amq75,
amq45official,
amq50-physical-review,
amq59,
amq48,
amq52-physical-review,
amq56,
amq41}.
The connection between non-blind system identification and regression
(and hence, to a lower extent, classification),
that we highlighted above, 
was by the way mentioned for
the quantum framework in 
\cite{amq106},
which states 
``Quantum process tomography is able to
learn an unknown function within well-defined symmetry and physical constraints
-
this is useful for regression analysis''
and further 
considers
``Regression based on quantum process tomography''.

The quantum version of the above-mentioned classical
source separation,
called Quantum Source Separation, or QSS,
and especially its blind version, or BQSS, were then introduced in
2007 in
\cite{amoi5-31}.
Two main classes of BQSS methods were developed since then.
The first one may be seen as a quantum extension of the above-mentioned
classical ICA methods, since it
takes advantage of the statistical independence of
the parameters that define random source quantum states
(qubit states).
It is called Quantum Independent Component Analysis
(or QICA, see e.g.
\cite{amoi5-31},\cite{amoi6-18})
or, more precisely,
Quantum-Source Independent Component Analysis
(or QSICA, see e.g.
\cite{amoi6-42})
to insist on the quantum nature of the considered source data,
whereas it uses classical processing means
(after quantum/classical data conversion).
The second main class of BQSS methods was introduced in 2013-2014 in
\cite{amoi6-34},\cite{amoi6-37}
and then especially detailed in
\cite{amoi6-64}.
It is based on the unentanglement of the considered source quantum states
and it typically
uses quantum processing means to restore these unknown states
from their coupled version.
Independently from the above quantum extensions of 
BSS/%
ICA, a quantum version
of PCA was introduced in 2014 in
\cite{amq72}.
Finally, the blind extension of QPT was introduced in 2015 in
\cite{amoi6-46}
and then especially extended in
\cite{amoi6-79}
for its multiple-preparation version.

Indeed, the above blind or non-blind QPT and 
(B)QSS
methods
are restricted to ``multiple-preparation'' operation, as defined in
Section \ref{sec-intro}.
Beyond these approaches,
we hereafter proceed to
the general ``single-preparation'' QIP
(or SIPQIP)
framework that may be built
to obtain a more efficient
operation and 
we 
then
present
its 
application
to various QIP tasks.

\section{Single-preparation quantum information processing (QIP)}
\label{sec-single-prep}
\subsection{%
Multiple-preparation
{QIP}%
}
\label{sec-stochastic-QIP-standard-concepts}
Let us 
consider an arbitrary number
\yqubitnbarb\ of 
distinguishable
\cite{amoi6-64}
qubits,
physically
implemented as 
spins
1/2.
If the 
quantum
state
\yqubitnbarbtimenonestate\ of this 
{set}
of qubits 
at a given time
is pure
and 
\ymodifartitionehundredthirtyninevonesteptwo{deterministic}%
, 
it belongs to
the 
$
2^{
\Yqubitnbarb
}
$-dimensional
space
\yqubitbothspace\
defined as the tensor product
of
the 
2-dimensional
spaces
respectively associated with each of the 
considered qubits.
Moreover,
this state
reads
\begin{equation}
\Yqubitnbarbtimenonestate
=
\sum
_{
\Yqubitnbarbstatecoefindexequalstd
=
1
}
^{
2^{\Yqubitnbarb}
}
\Yqubitnbarbspaceoverallcoefindexequalstdwithvecbasis
\Yqubitnbarbspaceoverallvecbasiscoefindexequalstd
\label{eq-def-qubitnbarbtimenonestate}
\end{equation}
where the vectors
\yqubitnbarbspaceoverallvecbasiscoefindexequalstd
,
with
$
\Yqubitnbarbstatecoefindexequalstd
\in
\{
1,
\dots
,
2
^
{
\Yqubitnbarb
}
\}
$,
form the standard basis of
\yqubitbothspace ,
i.e.
they are respectively equal to
\mbox{$
| + 
\rangle
_{1}
\otimes
| + 
\rangle
_{2}
\otimes
\dots
\otimes
| + 
\rangle
_{\Yqubitnbarb - 1}
\otimes
| + 
\rangle
_{\Yqubitnbarb}
$}
to
\mbox{$
| - 
\rangle
_{1}
\otimes
| - 
\rangle
_{2}
\otimes
\dots
\otimes
| - 
\rangle
_{\Yqubitnbarb - 1}
\otimes
| - 
\rangle
_{\Yqubitnbarb}
$}%
,
where
$
| + 
\rangle
_{\Yqubitoneindexstd}
$
and
$
| -
\rangle
_{\Yqubitoneindexstd}
$
form the standard basis of the state space
associated with the qubit with
index
\yqubitoneindexstd\
and
$
\otimes
$
is the tensor product.
The 
complex-valued coefficients
\yqubitnbarbspaceoverallcoefindexequalstdwithvecbasis\
{are 
fixed and arbitrary, 
except that
they}
meet the normalization condition
\begin{equation}
\sum
_{
\Yqubitnbarbstatecoefindexequalstd
=
1
}
^{
2^{\Yqubitnbarb}
}
|
\Yqubitnbarbspaceoverallcoefindexequalstdwithvecbasis
|
^2
=
1
\label{eq-qubitnbarbspaceoverallcoefindexequalstdwithvecbasis-normalization-cond}
.
\end{equation}

When simultaneously 
measuring the spin components of all
\yqubitnbarb\
qubits along the quantization axis,
the obtained result is a vector of \yqubitnbarb\ values
respectively associated with each of the qubits.
That vector has a random nature and its
$
2
^
{\Yqubitnbarb}
$
possible values 
(in normalized units)
are
$
[
+
\frac{1}{2}
,
+
\frac{1}{2}
,
\dots
,
+
\frac{1}{2}
,
+
\frac{1}{2}
]
$,
$
[
+
\frac{1}{2}
,
+
\frac{1}{2}
,
\dots
,
+
\frac{1}{2}
,
-
\frac{1}{2}
]
$,
and so on,
these values being respectively associated with the basis
vectors
\yqubitnbarbspaceoverallvecbasiscoefindexequalstd\
and hereafter indexed by \yqubitnbarbstatecoefindexequalstd .
Thus, the experiment consisting of this \yqubitnbarb -qubit
measurement yields a random result, and 
each 
elementary
event
\cite{book-papoulis}
\yqubitnbarbspaceoveralleventindexequalstdwithvecbasis\
is defined as:
the result of the experiment is equal to the
\yqubitnbarbstatecoefindexequalstd -th
\yqubitnbarb -entry vector in the above series
of possible values 
$
[
+
\frac{1}{2}
,
+
\frac{1}{2}
,
\dots
,
+
\frac{1}{2}
,
+
\frac{1}{2}
]
$
and so on.
Moreover, the probabilities of these 
events
are 
equal to
\begin{equation}
\Yqubitnbarbspaceoveralleventindexequalstdwithvecbasisprob
=
|
\Yqubitnbarbspaceoverallcoefindexequalstdwithvecbasis
|
^2
\hspace{10mm}
\forall
\
\Yqubitnbarbstatecoefindexequalstd
\in
\{
1,
\dots
,
2
^
{
\Yqubitnbarb
}
\}
.
\label{eq-def-qubitnbarbspaceoveralleventindexequalstdwithvecbasisprob}
\end{equation}

The
standard
procedure,
applied
in practice to estimate the above probabilities
for a given \yqubitnbarb -qubit state,
requires one to prepare
a large number
(typically from a few 
{thousand}
up to 
a few
hundred 
{thousand}
\cite{amoi6-18,
amoi6-64})
of copies of that state,
so that we hereafter call this
standard
approach 
``multiple-preparation QIP''
(this terminology and the connection between these methods and
classical adaptive processing are discussed in Appendix
\ref{sec-stochastic-QIP-link-with-classical}).
These copies may be
obtained in parallel from an ensemble 
of systems
or successively for the 
{same}
system
(``repeated write/read'', or RWR, procedure
\cite{amoi5-31,
amoi6-18,
amoi6-42}).
The above type of measurement is performed for each of these copies
and one counts the number of occurrences of each of the possible
results
$
[
+
\frac{1}{2}
,
+
\frac{1}{2}
,
\dots
,
+
\frac{1}{2}
,
+
\frac{1}{2}
]
$
and so on.
The associated sample relative
frequencies are then used as estimates
of the 
probabilities
\yqubitnbarbspaceoveralleventindexequalstdwithvecbasisprob .

As stated above, this approach requires many copies of the \emph{same}
quantum state.
This is therefore constraining, especially 
in the framework of
\emph{blind} QIP,
where the processing methods should operate with unknown values
of some quantum states (e.g. unknown inputs of the process to
be identified with QPT): being able to operate without
requiring known values of quantum states in blind methods
is attractive,
but then
requesting many 
copies of
each such state to be available is still a limitation,
because it still requires some form of control of these states,
that we would like to avoid, 
in order to 
{simplify}
the practical operation of 
the considered
methods and
to make them ``%
{blinder}''.
We hereafter provide a solution to this problem.
\subsection{%
Single-preparation 
{QIP}
based on 
probability expectations%
}
\label{sec-stochastic-QIP-introduce}
The above description was provided for an arbitrarily
selected 
{deterministic pure}
quantum state
\yqubitnbarbtimenonestate .
When developing our first class of BQSS
methods
(see e.g.
\cite{amoi5-31,
amoi6-18,
amoi6-42,
amoi6-67})
and associated 
BQPT methods
(see e.g.
\cite{amoi6-46,
amoi6-79,
amoi6-67})%
,
we had to extend that framework 
to
\emph{random}
{pure quantum}
states.
We especially detailed that concept in
\cite{amoi6-67}.
Briefly,
{the}
coefficients
\yqubitnbarbspaceoverallcoefindexequalstdwithvecbasis\
in
(\ref{eq-def-qubitnbarbtimenonestate})
{then}
become complex-valued random variables%
, instead of fixed parameters%
.
Hence, 
the
probabilities in
(\ref{eq-def-qubitnbarbspaceoveralleventindexequalstdwithvecbasisprob})
also become 
random variables!

The problem tackled
in this section is the estimation of some statistical parameters
of these 
random variables
defined by
(\ref{eq-def-qubitnbarbspaceoveralleventindexequalstdwithvecbasisprob}),
namely 
their expectations.
The natural 
(global)
procedure
that may be used to this end,
and that we used in  
our above-mentioned first BQSS and BQPT investigations,
consists of the following two levels.
The lower level only concerns 
{one deterministic}
state
(\ref{eq-def-qubitnbarbtimenonestate})
and the associated
probabilities
(\ref{eq-def-qubitnbarbspaceoveralleventindexequalstdwithvecbasisprob})
{which}
are estimated from a large number of copies
of the considered state, using the 
multiple-preparation
QIP framework 
of
Section
\ref{sec-stochastic-QIP-standard-concepts}.
This is repeated for different states 
(\ref{eq-def-qubitnbarbtimenonestate})
and then, at the higher level,
the sample mean over all these states is separately computed for
each probability
\yqubitnbarbspaceoveralleventindexequalstdwithvecbasisprob\
{%
(with samples
supposedly drawn from the same 
{statistical}
distribution).}

Beyond the above natural procedure, we hereafter focus on a more advanced
approach, that we recently introduced in the short conference paper
\cite{amoi6-104}.
We then
only partly described it and applied it to 
a single
QIP task (namely BQPT) 
in the journal paper
\cite{amoi6-118}
whereas, in the present paper,
we define 
it and analyze 
it
in more detail
and we then show that it also applies to 
a wide range
of
other QIP tasks.
That modified
procedure is based on the following principle:
At
the above%
{-defined}
lower level, we aim at using \emph{a small number}
of copies of the considered state, or ultimately
\emph{a single} 
instance
of that state,
thus developing 
{what}
we 
call 
``SIngle-Preparation 
QIP''%
or 
more 
briefly SIPQIP
(this terminology and the connection between these methods and
classical adaptive processing are also discussed in Appendix
\ref{sec-stochastic-QIP-link-with-classical}).
At first sight, it might seem that this is not possible, because the
lower level would thus not provide accurate estimates, 
that one could
then confidently gather at the higher level.
However, 
we claim 
and 
show below 
that this 
approach can
be used
{if}
one only aims at estimating some 
{statistical parameters}
of 
the considered quantum states.

We now first 
build 
the proposed approach by starting from
the frequentist 
view 
of probabilities
(see e.g.
\cite{book-papoulis})
at the above-defined two levels of the considered 
procedure,
that is:
\begin{itemize}
\item At the higher level, where one combines the contributions
associated with
\ytwoqubitseqnb\
states
of the set of
\yqubitnbarb\
qubits.
These states
are
indexed by
$
\Ytwoqubitseqindex 
\in
\{
1
,
\dots
,
\Ytwoqubitseqnb
\}
$
and denoted as
\yqubitnbarbtimenonestateseqindex .
\item At the lower level, 
{which 
concerns
one
deterministic
}
state
\yqubitnbarbtimenonestateseqindex\
and the associated probabilities
\yqubitnbarbspaceoveralleventindexequalstdwithvecbasisprobseqindex\
defined 
{by
(\ref{eq-def-qubitnbarbspaceoveralleventindexequalstdwithvecbasisprob}) but}
with coefficients
\yqubitnbarbspaceoverallcoefindexequalstdwithvecbasisseqindex\
{which depend}
on 
{state}
\yqubitnbarbtimenonestateseqindex .
\end{itemize}
At the 
lower level, each probability 
\yqubitnbarbspaceoveralleventindexequalstdwithvecbasisprobseqindex\
is defined as
\begin{equation}
\Yqubitnbarbspaceoveralleventindexequalstdwithvecbasisprobseqindex
=
\lim
_{
\Ywritereadonestatenb
\rightarrow
+
\infty
}
\frac
{
\Yqubitnbarbspaceoveralleventindexequalstdwithvecbasisnboccurseqindexwritereadonestatenb
}
{
\Ywritereadonestatenb
}
\label{eq-def-qubitnbarbspaceoveralleventindexequalstdwithvecbasisprobseqindex}
\end{equation}
{provided this limit exists.}
\yqubitnbarbspaceoveralleventindexequalstdwithvecbasisnboccurseqindexwritereadonestatenb\
is the number of occurrences 
of event
\yqubitnbarbspaceoveralleventindexequalstdwithvecbasis\
for the state
\yqubitnbarbtimenonestateseqindex\
when performing measurements for a set of
\ywritereadonestatenb\
copies
of that state
\yqubitnbarbtimenonestateseqindex .
In practice, one uses only a \emph{finite} number
\ywritereadonestatenb\
of copies
of state
\yqubitnbarbtimenonestateseqindex\
and therefore only accesses the following approximation
of the above probability:
\begin{equation}
\Yqubitnbarbspaceoveralleventindexequalstdwithvecbasisprobseqindexapproxwritereadonestatenb
=
\frac
{
\Yqubitnbarbspaceoveralleventindexequalstdwithvecbasisnboccurseqindexwritereadonestatenb
}
{
\Ywritereadonestatenb
}
.
\label{eq-def-qubitnbarbspaceoveralleventindexequalstdwithvecbasisprobseqindexapproxwritereadonestatenb}
\end{equation}

The higher level of the considered procedure
then addresses
the statistical mean
associated with
samples,
indexed by
\ytwoqubitseqindex ,
of a given quantity,
which is here theoretically
\yqubitnbarbspaceoveralleventindexequalstdwithvecbasisprobseqindex .
In the frequentist approach, this statistical mean is
defined 
{(if the limit exists)}
as
\begin{equation}
\Yqubitnbarbspaceoveralleventindexequalstdwithvecbasisprobexpect
=
\lim
_{
\Ytwoqubitseqnb
\rightarrow
+
\infty
}
\frac
{
\sum
_{
\Ytwoqubitseqindex
=
1
}
^{
\Ytwoqubitseqnb
}
\Yqubitnbarbspaceoveralleventindexequalstdwithvecbasisprobseqindex
}
{
\Ytwoqubitseqnb
}
.
\end{equation}
At the higher level too,
in practice one uses only a \emph{finite} 
number
\ytwoqubitseqnb\
of states
\yqubitnbarbtimenonestateseqindex ,
which 
first
yields
the following approximation
if only 
{performing an}
approximation at the higher level
of the procedure:
\begin{equation}
\Yqubitnbarbspaceoveralleventindexequalstdwithvecbasisprobexpectapproxonetwoqubitseqnb
=
\frac
{
\sum
_{
\Ytwoqubitseqindex
=
1
}
^{
\Ytwoqubitseqnb
}
\Yqubitnbarbspaceoveralleventindexequalstdwithvecbasisprobseqindex
}
{
\Ytwoqubitseqnb
}
.
\label{eq-def-qubitnbarbspaceoveralleventindexequalstdwithvecbasisprobexpectapproxonetwoqubitseqnb}
\end{equation}
The latter expression may then be modified by replacing
its term
\yqubitnbarbspaceoveralleventindexequalstdwithvecbasisprobseqindex\
by its approximation
(\ref{eq-def-qubitnbarbspaceoveralleventindexequalstdwithvecbasisprobseqindexapproxwritereadonestatenb}).
This yields
\begin{equation}
\Yqubitnbarbspaceoveralleventindexequalstdwithvecbasisprobexpectapproxtwotwoqubitseqnb
=
\frac
{
\sum
_{
\Ytwoqubitseqindex
=
1
}
^{
\Ytwoqubitseqnb
}
\Yqubitnbarbspaceoveralleventindexequalstdwithvecbasisnboccurseqindexwritereadonestatenb
}
{
\Ytwoqubitseqnb
\Ywritereadonestatenb
}
.
\label{eq-def-qubitnbarbspaceoveralleventindexequalstdwithvecbasisprobexpectapproxtwotwoqubitseqnb}
\end{equation}
$
\sum
_{
\Ytwoqubitseqindex
=
1
}
^{
\Ytwoqubitseqnb
}
\Yqubitnbarbspaceoveralleventindexequalstdwithvecbasisnboccurseqindexwritereadonestatenb
$
is nothing but the 
number,
hereafter denoted as
\yqubitnbarbspaceoveralleventindexequalstdwithvecbasisnboccurqubitnbarbstatewritereadseqnb
,
of occurrences of event
\yqubitnbarbspaceoveralleventindexequalstdwithvecbasis\
for the
complete
considered set of
$
\Yqubitnbarbstatewritereadseqnb
=
\Ytwoqubitseqnb
\Ywritereadonestatenb
$
measurements.
Therefore,
\yqubitnbarbspaceoveralleventindexequalstdwithvecbasisprobexpectapproxtwotwoqubitseqnb\
is the 
relative frequency of occurrence of that
event over these
$
\Yqubitnbarbstatewritereadseqnb
$
measurements,
or ```trials'', 
using standard probabilistic terms
{\cite{book-papoulis}.}
This quantity (\ref{eq-def-qubitnbarbspaceoveralleventindexequalstdwithvecbasisprobexpectapproxtwotwoqubitseqnb}) may therefore also be expressed
as
\begin{eqnarray}
\Yqubitnbarbspaceoveralleventindexequalstdwithvecbasisprobexpectapproxtwotwoqubitseqnb
&
=
&
\frac
{
\Yqubitnbarbspaceoveralleventindexequalstdwithvecbasisnboccurqubitnbarbstatewritereadseqnb
}
{
\Yqubitnbarbstatewritereadseqnb
}
\label{eq-qubitnbarbspaceoveralleventindexequalstdwithvecbasisprobexpectapproxtwotwoqubitseqnb-number-in-qubitnbarbstatewritereadseqnb}
\\
&
=
&
\frac
{
\sum
_{
\Yqubitnbarbstatewritereadseqindex
=
1
}
^{
\Yqubitnbarbstatewritereadseqnb
}
\Yqubitnbarbspaceoveralleventindexequalstdwithvecbasisindicfuncqubitnbarbstatewritereadseqindex
}
{
\Yqubitnbarbstatewritereadseqnb
}
\label{eq-qubitnbarbspaceoveralleventindexequalstdwithvecbasisprobexpectapproxtwotwoqubitseqnb-indicfunc}
\end{eqnarray}
where
\yqubitnbarbspaceoveralleventindexequalstdwithvecbasisindicfuncqubitnbarbstatewritereadseqindex\
is the value of the indicator function 
of
event
\yqubitnbarbspaceoveralleventindexequalstdwithvecbasis\
for trial
\yqubitnbarbstatewritereadseqindex ,
which takes the value 1 if
\yqubitnbarbspaceoveralleventindexequalstdwithvecbasis\
occurs during that trial, and 0 otherwise.
When using
(\ref{eq-qubitnbarbspaceoveralleventindexequalstdwithvecbasisprobexpectapproxtwotwoqubitseqnb-indicfunc}),
one now considers the
$
\Yqubitnbarbstatewritereadseqnb
=
\Ytwoqubitseqnb
\Ywritereadonestatenb
$
trials as organized as a single series, with trials
indexed by 
\yqubitnbarbstatewritereadseqindex .
One thus
fuses
the above%
{-defined}
two levels of the
procedure into a single one, thus disregarding the fact
that, in this series, each block of 
\ywritereadonestatenb\
consecutive
trials uses the same state
\yqubitnbarbtimenonestateseqindex .
{One may therefore wonder whether
the number
\ywritereadonestatenb\
of used copies of
each state
\yqubitnbarbtimenonestateseqindex\
may be freely decreased, and even set to one,
while possibly keeping the same
total
number 
\yqubitnbarbstatewritereadseqnb\
of trials.}
A 
formal proof
of the relevance 
{of that approach%
}%
,
using Kolmogorov's view of probabilities%
, is provided in 
Appendix
\ref{sec-stochastic-QIP-validate}.
Moreover,
Appendix
\ref{sec-stochastic-QIP-validate}
thus proves that the proposed estimator 
(\ref{eq-qubitnbarbspaceoveralleventindexequalstdwithvecbasisprobexpectapproxtwotwoqubitseqnb-indicfunc})
of
\yqubitnbarbspaceoveralleventindexequalstdwithvecbasisprobexpect\
is attractive because,
for states independently randomly drawn with the same
distribution and with one 
instance of each
state,
this estimator is asymptotically efficient.

It should be clear that
this procedure for estimating
\yqubitnbarbspaceoveralleventindexequalstdwithvecbasisprobexpect ,
and hence the resulting 
SIPQIP
methods,
can be freely used with
either one instance or several (e.g. many) copies per state, i.e. 
this 
SIPQIP
terminology means that
these methods
\emph{allow} one to use a single instance of each state.
In contrast,
so-called 
{multiple-preparation}
QIP
methods \emph{force} one to use many state copies to achieve
good performance.
\subsection{Single-preparation 
QIP
based on 
sample means of
probabilities}
\label{sec-stochastic-QIP-introduce-sample-mean}
As explained in Section
\ref{sec-stochastic-QIP-introduce},
the framework introduced in that section is intended for
a 
formalism 
based on \emph{random} pure states, that we
use in most of this paper.
In addition, we 
employ
the more standard formalism of
\emph{deterministic} pure states in Section
\ref{sec-classif}, where the
considered QIP task
eventually 
boils down 
to estimating
the mean, over a finite
number (i.e. the sample mean) of deterministic pure states,
of the (hence deterministic) probabilities respectively
associated with each of these states.
Although this framework is conceptually different from
the one of Section
\ref{sec-stochastic-QIP-introduce},
it eventually yields the same implementation
as will now be shown.
Here again, at the lower level, each of the considered
probabilities is theoretically defined by
(\ref{eq-def-qubitnbarbspaceoveralleventindexequalstdwithvecbasisprobseqindex})
but then replaced by
(\ref{eq-def-qubitnbarbspaceoveralleventindexequalstdwithvecbasisprobseqindexapproxwritereadonestatenb})
in practice.
The difference with respect to
Section
\ref{sec-stochastic-QIP-introduce}
then appears at the higher level of the approach,
since we here directly aim at handling a \emph{finite} number of
quantum states, so that we directly consider the quantity in
(\ref{eq-def-qubitnbarbspaceoveralleventindexequalstdwithvecbasisprobexpectapproxonetwoqubitseqnb}).
The remainder of the analysis of 
Section
\ref{sec-stochastic-QIP-introduce}
then also applies to the framework considered here,
so that our SIPQIP concept
also applies to this framework
and thus here again yields
the above-defined advantages.

\section{QIP tasks related to system identification}\label{sec-system-ident}

\subsection{Blind quantum process tomography}\label{sec-bqpt}
As explained in Sections \ref{sec-intro} and \ref{sec-machine-learning},
the quantum version of system identification is often referred to as
Quantum Process Tomography (QPT).
It is 
of major importance, 
especially
for characterizing the
actual behavior of quantum gates
(see e.g.
\cite{booknielsen,
amq45official,
amq50-physical-review,
amq48,
amq52-physical-review,
amq41}%
)%
, which are the building blocks of a
\emph{qua}ntum co\emph{mputer},
that, by the way, we propose
to more briefly call a ``quamputer''.
In this section, we consider the
blind and single-preparation
extension of QPT. 
This corresponds to the only SIPQIP task that we detailed
in our previous papers
(see
\cite{amoi6-104} for a partial version
and 
\cite{amoi6-118}
for complete extensions).
We hereafter summarize the major features of the main method 
that we proposed in
\cite{amoi6-118},
because several 
original contributions introduced further in
this paper build upon that single-preparation blind QPT
method.

Various papers from the literature dealing with
conventional (i.e. non-blind and multiple-preparation)
QPT are focused on
specific processes
or classes of
processes: 
see e.g.
\cite{amq59,
amq52-physical-review,
amq87}.
Similarly, the method that we introduced in
\cite{amoi6-118}
is dedicated to the class of configurations involving
two 
distinguishable
\cite{amoi6-64}
qubits
implemented as electron spins 1/2, that are 
internally
coupled according to the
cylindrical-symmetry Heisenberg model,
with unknown 
principal values 
\yexchangetensorppalvaluexy\
and
\yexchangetensorppalvaluez\
of the exchange tensor.
We stress that this type of
coupling is only used as a concrete example
\footnote{%
We do not focus on whether
Heisenberg coupling
could be used as a desired phenomenon,
to build suitable gates for 
quamputers,
as already mentioned in \cite{amoi6-118}.}%
, 
to show
how to fully implement the proposed 
{general}
concepts in a relevant case, but that
these concepts and resulting practical 
algorithms (for performing BQPT and other QIP tasks detailed
further in this paper)
may then
be 
{extended%
}
to other classes of quantum processes
and associated applications.

The above Heisenberg model 
is detailed in 
Appendix
\ref{sec-heisenberg process}.
This shows that the associated quantum process,
from its input (i.e. initial) quantum state
\ymixsyststateinitial\
to its output (i.e. final) quantum state
\ymixsyststatefinal ,
is represented by a matrix
\yopmix ,
and that
the only quantities
that must be estimated in order to obtain an estimate of
\yopmix\
are
$
\exp
\left[
\Ysqrtminusone
\frac{ 
\Yexchangetensorppalvaluexy
(t - t_0)
} 
{ \hbar }
\right]
$
and
$
\exp
\left[
\Ysqrtminusone
\frac{ 
\Yexchangetensorppalvaluez
(t - t_0)
} 
{ 2 \hbar }
\right]
$.
The main method proposed in
\cite{amoi6-118}
to estimate
\yopmix\
uses
three
values of the time interval
$(t - t_0)$,
denoted as
\ytwoqubitwritereadtimeintervalindexone,
\ytwoqubitwritereadtimeintervalindextwo\
and
\ytwoqubitwritereadtimeintervalindexthree,
with
\yeqdeftwoqubitwritereadtimeintervallinkthree
These values are respectively used to first estimate
$
\exp
\left[
\Ysqrtminusone
\frac{ 
\Yexchangetensorppalvaluexy
\Ytwoqubitwritereadtimeintervalindexone
} 
{ \hbar }
\right]
$,
then estimate
$
\exp
\left[
\Ysqrtminusone
\frac{ 
\Yexchangetensorppalvaluez
\Ytwoqubitwritereadtimeintervalindextwo
} 
{ 2 \hbar }
\right]
$
and finally obtain an estimate of
\yopmix\
which is
non-ambiguous only from the point of view
of the final use of this process with
$(t - t_0)
=
\Ytwoqubitwritereadtimeintervalindexthree
$
(%
\cite{amoi6-118}
discusses
the relevance of finally using a quantum process
in conditions,
i.e. here with a
value of
$
(t - t_0)
$,
different from those
initially
used to identify that process,
e.g. when that process corresponds to a gate of
a 
quamputer).

More precisely, when applying the considered BQPT method
in the purely single-preparation mode, the
first part 
of this method uses one instance of each
output quantum state
\ymixsyststatefinal .
For each such state,
it measures the components of the considered two spins
along the $Oz$ axis.
As discussed in
\cite{amoi6-118}
and in Section
\ref{sec-stochastic-QIP-standard-concepts}
of the present paper,
the result of
each such measurement has
four possible values,
that is
$(+\frac{1}{2},+\frac{1}{2})$,
$(+\frac{1}{2},-\frac{1}{2})$,
$(-\frac{1}{2},+\frac{1}{2} )$
or
$(-\frac{1}{2},-\frac{1}{2})$
in normalized units.
Their probabilities are respectively denoted as
\ytwoqubitsprobaplusplusdirzz\
to
\ytwoqubitsprobaminusminusdirzz\
hereafter.
Using the 
moduli
\yparamqubitindexstdstateplusmodulus\
and the phases
\yparamqubitindexstdstateplusphase\
and
\yparamqubitindexstdstateminusphase\
of the
polar representation
(\ref{eq-def-qubit-polar-qubit-indexstd})
of the qubit parameters that define the input
state \ymixsyststateinitial,
these
probabilities read
\cite{amoi6-18,amoi6-42}
\yeqdefstatecoefvstwoqubitsprobathreepolar
with
\yeqdefeqdeftwoqubitresultphaseinittwoqubitresultphaseevol
Probability
\ytwoqubitsprobaminusplusdirzz\
is not considered hereafter because
the sum of
\ytwoqubitsprobaplusplusdirzz\
to
\ytwoqubitsprobaminusminusdirzz\
is equal to 1.

Using
$(t - t_0)
=
\Ytwoqubitwritereadtimeintervalindexone
$
in the first part of this method,
(\ref{eq-def-Ytwoqubitresultphaseevol-versiontwo})-%
(\ref{eq-def-Ytwoqubitresultphaseevol-versionthree})
may then be inverted as
\yeqdefexchangetensorppalvaluexyvstwoqubitresultphaseevolindexdeqone
where
\ytwoqubitresultphaseevolindexd\ is 
a
determination associated
with the actual value
\ytwoqubitresultphaseevol,
i.e. 
\ytwoqubitresultphaseevolindexd\
is equal to
\ytwoqubitresultphaseevol\
up to the additive constant
$- \Yexchangetensorppalvaluexyshiftindetermint
\pi$,
where
\yexchangetensorppalvaluexyshiftindetermint\
is an integer.

The SIPQIP framework defined in Section
\ref{sec-stochastic-QIP-introduce}
then makes it possible to derive an estimate
\ytwoqubitresultphaseevolindexdestim\
of
\ytwoqubitresultphaseevolindexd\
as follows.
We consider the case when
\yparamqubitonestateplusmodulus,
\yparamqubittwostateplusmodulus\
and
\ytwoqubitresultphaseinit\
are random valued
and when these 
random variables
are
statistically independent.
Eq.
(\ref{eq-statecoef-vs-twoqubitsprobaplusminus-polar-versionthree})
then
yields
\yeqdefstatecoefvstwoqubitsprobaplusminuspolarversionthreeexpect
In this equation,
$E \{ 
\Ytwoqubitsprobaplusminusdirzz \}$
is known:
in practice, it is
estimated by using the
SIPQIP approach 
of Section
\ref{sec-stochastic-QIP-introduce},
i.e.
by using the sample mean of the estimates of
all values of \ytwoqubitsprobaplusminusdirzz, themselves typically
estimated
with sample frequencies
(possibly each reduced to one measurement outcome).
Similarly, 
$E \{ \Yparamqubitonestateplusmodulus ^2 \}$
and
$E \{ \Yparamqubittwostateplusmodulus ^2 \}$
are known:
as detailed in
\cite{amoi6-118},
they may be derived by solving the two equations
obtained by taking the expectation of
(\ref{eq-statecoef-vs-twoqubitsprobaplusplus-polar})
and
(\ref{eq-statecoef-vs-twoqubitsprobaminusminus-polar-vs-paramqubitindexstdstateplusmodulus}),
which involve
$E\{
\Ytwoqubitsprobaplusplusdirzz 
\}$
and
$E\{
\Ytwoqubitsprobaminusminusdirzz
\}$,
that are also estimated with the SIPQIP approach.
Finally, the blind version of QPT concerns the case when the 
individual values of the input 
quantum states of the considered process are unknown, but 
it
allows
one to request some of the statistical parameters of these
inputs to be known.
Therefore, we here request 
the states 
\ymixsyststateinitial\
to be prepared with a procedure which is such that the value of
$
E
\{
\sin \Ytwoqubitresultphaseinit
\}
$,
or at least its sign,
is known.
Thus, 
(\ref{eq-statecoef-vs-twoqubitsprobaplusminus-polar-versionthree-expect})
can be exploited
so that
the 
only unknown
is
\ytwoqubitresultphaseevolsin .
Ref.
\cite{amoi6-118}
shows how to
solve this equation.
More precisely,
two instances of this equation,
with different values of
$
E
\{
\sin \Ytwoqubitresultphaseinit
\}
$, are used:
the first one yields an estimate of the
absolute value of
\ytwoqubitresultphaseevolsin\
and the second equation provides an
estimate of the sign of
\ytwoqubitresultphaseevolsin.
Combining these two results
yields an estimate
\ytwoqubitresultphaseevolsinestimtwo\
of
\ytwoqubitresultphaseevolsin\
and hence
an estimate
\ytwoqubitresultphaseevolindexdestim\
of
\ytwoqubitresultphaseevolindexd\
by using
\ytwoqubitresultphaseevolsinestimtwo\
in
(\ref{eq-twoqubitresultphaseevolindexd}).

Based on
(\ref{eq-exchangetensorppalvaluexy-vs-twoqubitresultphaseevolindexd}),
once the 
above
estimate
\ytwoqubitresultphaseevolindexdestim\
has been obtained,
corresponding shifted estimates of
$
\frac{ 
\Yexchangetensorppalvaluexy
\Ytwoqubitwritereadtimeintervalindexone
} 
{ \hbar }
$
are 
derived as
\yeqdefexchangetensorppalvaluexyvstwoqubitresultphaseevolindexdestim
where
\yexchangetensorppalvaluexyshiftindetermintestim\
is an integer,
that corresponds to
\yexchangetensorppalvaluexyshiftindetermint\
in
(\ref{eq-exchangetensorppalvaluexy-vs-twoqubitresultphaseevolindexd}).
The value of 
\yexchangetensorppalvaluexyshiftindetermintestim\
has to be selected without knowing
the actual value of
\yexchangetensorppalvaluexyshiftindetermint\
in the fully blind case considered here,
i.e. when no prior information is available about the
value of
\yexchangetensorppalvaluexy.
But this is not an issue from the point of view of
the considered BQPT method,
because that method is designed so that the
obtained estimate of the process matrix \yopmix ,
for
$(t - t_0)
=
\Ytwoqubitwritereadtimeintervalindexthree
$,
does not depend on the integer value of
\yexchangetensorppalvaluexyshiftindetermintestim\
\cite{amoi6-118}.
The simplest approach therefore consists of setting
$
\Yexchangetensorppalvaluexyshiftindetermintestim
=
0
$
in
(\ref{eq-exchangetensorppalvaluexy-vs-twoqubitresultphaseevolindexd-estim}).

Similarly,
\cite{amoi6-118}
shows that
\yeqdefexchangetensorppalvaluezvstwoqubitsprobadirxxplusplusdiffminusminusphasediffindexdeqoneintervalindextwo
where
\yexchangetensorppalvaluezshiftindetermint\
is an integer.
This
is used to derive the estimate
\yeqdefexchangetensorppalvaluezvstwoqubitsprobadirxxplusplusdiffminusminusphasediffindexdestimintervalindextwo
where
the
SIPQIP framework of Section
\ref{sec-stochastic-QIP-introduce}
is again used to obtain an estimate
\ytwoqubitsprobadirxxplusplusdiffminusminusphasediffindexdestim\
of the quantity
\ytwoqubitsprobadirxxplusplusdiffminusminusphasediffindexd\
that may be derived from the same type of probability expectations
$
E\{
\Ytwoqubitsprobaindexstddirzz
\}
$
as above
and from the probability expectations
$
E\{
\Ytwoqubitsprobaindexstddirxx
\}
$
of results of additional measurements of spin components
along the $Ox$ axis
(see details in 
\cite{amoi6-118}%
).
Besides, 
\yexchangetensorppalvaluezshiftindetermintestim\
is an integer whose value has no influence on the final estimate
of the process matrix \yopmix ,
and that may therefore be set to zero.
Moreover,
$
\frac{ 
\Yexchangetensorppalvaluexyestim
\Ytwoqubitwritereadtimeintervalindextwo
} 
{ \hbar }
$
is equal to twice the value previously computed with
(\ref{eq-exchangetensorppalvaluexy-vs-twoqubitresultphaseevolindexd-estim})
and the other parameters have known values.

It should be noted that this QPT method only uses the known outputs of the
considered process 
and
general known properties of its inputs, not its input \emph{values},
which are unknown. This is therefore indeed a \emph{blind}
QPT method (moreover operating in the single-preparation mode).
In contrast,
non-blind methods are supposed to operate with
predefined values of their input states and
are in practice very sensitive to errors in the preparation
of these value,
as detailed in
\cite{amoi6-118}.
Their performance for actual preparations
is therefore significantly degraded, 
whereas the above blind operation yields
much better accuracy in the tests reported in 
\cite{amoi6-118}.

Beyond BQPT itself, we 
hereafter
move to one of the new methods that we propose
in this paper, for other QIP tasks, 
that
further exploit the results of the above algorithm.

\subsection{Blind Hamiltonian parameter estimation}\label{sec-blind-hamilton-estim}
\subsubsection{Proposed method}\label{sec-blind-hamilton-estim-method}
As shown in Appendix \ref{sec-heisenberg process}, the behavior of the
device composed of two Heisenberg-coupled qubits
that we considered above is primarily defined by its Hamiltonian,
whereas the above process matrix \yopmix\ follows when considering the evolution
of the state of that system from a fixed time
\yqubitonetimeinit\ to a fixed time
\yqubitonetimefinal .
Therefore, beyond the estimation of the process matrix \yopmix , a related
QIP task consists of estimating
the primary unknown parameters of the Hamiltonian of the studied device,
namely the
principal values 
\yexchangetensorppalvaluexy\
and
\yexchangetensorppalvaluez\
of the exchange tensor
(similar considerations are also provided in \cite{amq102}).
This type of task (for the parameters of this or other
Hamiltonians)
is called
Hamiltonian parameter estimation
hereafter and also 
especially in
\cite{amq99,
amq103,
amq102}.
Such parameter estimation problems
are also addressed but often referred to as Hamiltonian identification
e.g. in
\cite{amq97,
amq56,
amq101}
and partly
\cite{amq100}.
To our knowledge, in the literature
this task has been studied only in the
non-blind or ``controlled'' mode and/or
using
multiple preparations 
in approaches that are closely connected 
\ymodifartitionehundredthirtyninevonesteptwo{with}
conventional QPT
\cite{amq100}
or that are based on specific
protocols,
such as
periodical sampling
(hence with a potentially quite high total number of required state
preparations)
\cite{amq56,
amq101,
amq102},
use of a
closed-loop
\cite{amq97}
or
optimal feedback
\cite{amq103}
structure,
or curve fitting with respect to the experimental results
obtained for various
angles of the magnetic field \cite{amq99}.
In contrast,
we hereafter investigate a single-preparation and blind 
(without control) version
of this Hamiltonian parameter estimation task,
based on measurements along the $Oz$ and $Ox$ axes,
that
we did not address in our previous papers,
that has direct connections with the above-defined 
single-preparation and blind
version
of QPT and
that yields the same
type of attractive features as for QPT:
it avoids the burden of very accurately and repeatedly preparing
predefined states to estimate the unknown parameters of the
considered Hamiltonian.
Here again, we show how to develop such an extension for the
specific class of Hamiltonians defined in
Appendix \ref{sec-heisenberg process}, namely
Heisenberg coupling with unknown
\yexchangetensorppalvaluexy\
and
\yexchangetensorppalvaluez ,
but this should be considered only as an example, that the reader may
then extend to other types of Hamiltonians.
Similarly, various investigations 
in the literature considered
specific 
parametrized Hamiltonians, 
with a limited number of
unknown parameters, as the core of the proposed approaches or to illustrate them:
see e.g.
\cite{amq99,
amq97,
amq56,
amq101,
amq100,
amq102,
amq103}.

The blind Hamiltonian parameter estimation 
(BHPE) 
method that we propose 
builds upon the BQPT algorithm summarized in Section \ref{sec-bqpt}
but it requires subsequent developments for the following reason.
As explained in Section \ref{sec-bqpt}, that BQPT method
is strongly connected 
\ymodifartitionehundredthirtyninevonesteptwo{with}
estimating the quantities
$
\exp
\left[
\Ysqrtminusone
\frac{ 
\Yexchangetensorppalvaluexy
(t - t_0)
} 
{ \hbar }
\right]
$
and
$
\exp
\left[
\Ysqrtminusone
\frac{ 
\Yexchangetensorppalvaluez
(t - t_0)
} 
{ 2 \hbar }
\right]
$,
using some types of measurements.
The estimation of these very quantities would define their phase
arguments
$
\frac{ 
\Yexchangetensorppalvaluexy
(t - t_0)
} 
{ \hbar }
$
and
$
\frac{ 
\Yexchangetensorppalvaluez
(t - t_0)
} 
{ 2 \hbar }
$,
only up to 
\ymodifartitionehundredthirtyninevonestepone{additive integer multiples}
of
$2 \pi$,
that would 
\ymodifartitionehundredthirtyninevonestepone{yield
the indeterminacies}
of this estimation
procedure
\ymodifartitionehundredthirtyninevonestepone{from the point of view of
BHPE.}
More precisely, using the data provided by the
considered measurements, 
\ymodifartitionehundredthirtyninevonestepone{the above}
BQPT method yields
the
indeterminacies that  
consist of the additive constants
$\Yexchangetensorppalvaluexyshiftindetermintestim
\pi$
and
$2
\Yexchangetensorppalvaluezshiftindetermintestim
\pi$
of
(\ref{eq-exchangetensorppalvaluexy-vs-twoqubitresultphaseevolindexd-estim})
and
(\ref{eq-exchangetensorppalvaluez-vs-twoqubitsprobadirxxplusplusdiffminusminusphasediffindexd-estim-intervalindextwo}).
It thus does not provide a unique solution with respect to
$\frac{ 
\Yexchangetensorppalvaluexyestim
\Ytwoqubitwritereadtimeintervalindexone
} 
{ \hbar }$
and
$\frac{ 
\Yexchangetensorppalvaluezestim
\Ytwoqubitwritereadtimeintervalindextwo
} 
{ \hbar }
$,
and hence
\yexchangetensorppalvaluexyestim\
and
\yexchangetensorppalvaluezestim,
so that it does not solve the Hamiltonian parameter estimation problem
considered here
(related comments may be found in \cite{amq102}).
For instance, let us consider the test conditions defined in
Appendix
\ref{sec-appendix-test-conditions},
including the available prior knowledge about the range of values
to which
\yexchangetensorppalvaluexy\
is guaranteed to belong.
Then, 
a single
run of our BQPT method yields
32 acceptable determinations 
of
\yexchangetensorppalvaluexy\
in that range and no means to know which of 
these numerous potential solutions
corresponds
to the actual value
\yexchangetensorppalvaluexy .

We here aim at
\ymodifartitionehundredthirtyninevonestepone{developing a
BHPE method that takes advantage of the above
BQPT algorithm so as estimate
\yexchangetensorppalvaluexyestim\
and
\yexchangetensorppalvaluezestim\
without indeterminacies.}
The trick that we propose to 
this end
is based on estimating
each of the parameters 
\yexchangetensorppalvaluexy\
and
\yexchangetensorppalvaluez\
by using \emph{two} values of the above-defined time interval
$(\Yqubitonetimefinal
-
\Yqubitonetimeinit)$,
instead of one in the fundamental principle
of the above BQPT method.
This trick
also has relationships with the 
practical
approach
that
we used 
\ymodifartitionehundredthirtyninevonestepone{in
\cite{amoi6-118}, for BQPT only:
starting from a basic BQPT method that uses a single value
of
$(\Yqubitonetimefinal
-
\Yqubitonetimeinit)$
and that thus yields some indeterminacies with respect
to
\yopmix,
we then moved to a more advanced BQPT method,
that 
uses several values
of
$(\Yqubitonetimefinal
-
\Yqubitonetimeinit)$
and thus avoids all indeterminacies with respect
to
\yopmix\
(this is the method summarized in Section 
\ref{sec-bqpt}).}
However, for BQPT, we 
\ymodifartitionehundredthirtyninevonestepone{thus}
eventually used several values of
$(\Yqubitonetimefinal
-
\Yqubitonetimeinit)$
for the complete practical procedure but only one value for each
part
of 
\ymodifartitionehundredthirtyninevonestepone{that BQPT method,}
e.g. associated with 
\ymodifartitionehundredthirtyninevonestepone{the phase factor
involving}
one of the
parameters
\yexchangetensorppalvaluexy\
and
\yexchangetensorppalvaluez\
\ymodifartitionehundredthirtyninevonestepone{(see
(\ref{eq-exchangetensorppalvaluexy-vs-twoqubitresultphaseevolindexd-estim})
and
(\ref{eq-exchangetensorppalvaluez-vs-twoqubitsprobadirxxplusplusdiffminusminusphasediffindexd-estim-intervalindextwo}),
respectively),}
whereas%
\ymodifartitionehundredthirtyninevonestepone{, for BHPE,}
we here move to two values of
$(\Yqubitonetimefinal
-
\Yqubitonetimeinit)$
per parameter
\yexchangetensorppalvaluexy\
and
\yexchangetensorppalvaluez, these values therefore being exploited
in a new way, that we describe hereafter.

Let us first consider the estimation of
\yexchangetensorppalvaluexy .
To this end, we use the procedure of the first
part
of the
BQPT method of Section \ref{sec-bqpt}.
We apply it twice, 
with
\ytwoqubitwritereadtimeintervalindexone\
 of Section \ref{sec-bqpt}
successively replaced by two values denoted as
\ytwoqubitwritereadtimeintervalindexoneone\
and
\ytwoqubitwritereadtimeintervalindexonetwo .
Combining
(\ref{eq-exchangetensorppalvaluexy-vs-twoqubitresultphaseevolindexd})
and
(\ref{eq-exchangetensorppalvaluexy-vs-twoqubitresultphaseevolindexd-estim}),
with
\ytwoqubitwritereadtimeintervalindexone\
replaced by
\ytwoqubitwritereadtimeintervalindexoneone\
and similarly with an additional index ``1'' for the other variables 
whose values are specific to that first application of the procedure,
yields
\yeqdefexchangetensorppalvaluexyestimminusactualvsexchangetensorppalvaluexyshiftindetermintestiminminusactualindexoneandnot
This
shows
that
the procedure applied
with the time interval
\ytwoqubitwritereadtimeintervalindexoneone\
yields a regular one-dimensional grid of possible estimates
\yexchangetensorppalvaluexyestimindexone\
of
\yexchangetensorppalvaluexy\
(associated with the values of
\yexchangetensorppalvaluexyshiftindetermintestimindexone),
with 
a step equal
to
$\frac{\hbar \pi}
{\Ytwoqubitwritereadtimeintervalindexoneone}$.
Similarly, the second application of that procedure, with a
time interval
\ytwoqubitwritereadtimeintervalindexonetwo ,
yields
\yeqdefexchangetensorppalvaluexyestimminusactualvsexchangetensorppalvaluexyshiftindetermintestiminminusactualindextwoandnot
The corresponding estimates
\yexchangetensorppalvaluexyestimindextwo\
of
\yexchangetensorppalvaluexy\
therefore form a regular grid
with 
a step equal
to
$\frac{\hbar \pi}
{\Ytwoqubitwritereadtimeintervalindexonetwo}$.

We here aim at exploiting the differences 
between
the above two
\footnote{This approach might be further extended to more
than two grids, 
to make it more robust.}
grids of values.
As a preliminary stage, let us consider the ideal case, i.e. when
\yeqdeftwoqubitresultphaseevolindexdoneandtwoestimideal
Then, the above two grids share at least one value, equal to
\yexchangetensorppalvaluexy\
and obtained when
\yexchangetensorppalvaluexyshiftindetermintestimindexone\
and
\yexchangetensorppalvaluexyshiftindetermintestimindextwo\
are respectively set to
\yexchangetensorppalvaluexyshiftindetermintindexone\
and
\yexchangetensorppalvaluexyshiftindetermintindextwo ,
which results in
$\Yexchangetensorppalvaluexyshiftindetermintestimminusactualindexone
=
0$
and
$\Yexchangetensorppalvaluexyshiftindetermintestimminusactualindextwo
=
0$.
Moreover, let us consider the case
when 
$\Ytwoqubitwritereadtimeintervalindexonetwo
/
\Ytwoqubitwritereadtimeintervalindexoneone$
is set
to an irrational value.
Then, the above grids only share the value
\yexchangetensorppalvaluexy,
because 
(\ref{eq-exchangetensorppalvaluexyestim-minus-actual-vs-exchangetensorppalvaluexyshiftindetermintestimin-minus-actual-indexone})
and
(\ref{eq-exchangetensorppalvaluexyestim-minus-actual-vs-exchangetensorppalvaluexyshiftindetermintestimin-minus-actual-indextwo})
show that,
when
(\ref{eq-twoqubitresultphaseevolindexdoneandtwoestim-ideal})
is met,
the values of
\yexchangetensorppalvaluexyshiftindetermintestimindexone\
and
\yexchangetensorppalvaluexyshiftindetermintestimindextwo\
that are such that the corresponding estimates
\yexchangetensorppalvaluexyestimindexone\
and
\yexchangetensorppalvaluexyestimindextwo\
are 
equal
are those that meet
$\Yexchangetensorppalvaluexyshiftindetermintestimminusactualindexone
/
\Ytwoqubitwritereadtimeintervalindexoneone
=
\Yexchangetensorppalvaluexyshiftindetermintestimminusactualindextwo
/
\Ytwoqubitwritereadtimeintervalindexonetwo$
so that,
when
\yexchangetensorppalvaluexyshiftindetermintestimminusactualindexone\
and
\yexchangetensorppalvaluexyshiftindetermintestimminusactualindextwo\
are
nonzero,
this requires
$
\Ytwoqubitwritereadtimeintervalindexonetwo
/
\Ytwoqubitwritereadtimeintervalindexoneone
$
to be equal to the rational value
$\Yexchangetensorppalvaluexyshiftindetermintestimminusactualindextwo
/
\Yexchangetensorppalvaluexyshiftindetermintestimminusactualindexone$.
So, when
(\ref{eq-twoqubitresultphaseevolindexdoneandtwoestim-ideal})
is met and
$\Ytwoqubitwritereadtimeintervalindexonetwo
/
\Ytwoqubitwritereadtimeintervalindexoneone$
is set
to an irrational value,
a simple criterion for determining
\yexchangetensorppalvaluexy\
is: it is the only value shared by the above two grids.
This behavior has a relationship with the influence
of the sampling period when sampling a sine wave,
as e.g. detailed in
\cite{elivreoppenheim},
which thus also indirectly
shows that the above attractive behavior
of our grids
is obtained \emph{only}
if
$\Ytwoqubitwritereadtimeintervalindexonetwo
/
\Ytwoqubitwritereadtimeintervalindexoneone$
is irrational.

The above criterion must then be modified when moving to
practical situations, because the available estimates
\ytwoqubitresultphaseevolindexdoneestim\
and
\ytwoqubitresultphaseevolindexdtwoestim\
are somewhat and independently shifted with respect
to the corresponding actual values.
Therefore, to
estimate
\yexchangetensorppalvaluexy ,
instead of looking for values of both
grids which are identical in the ideal case, this
here suggests to compare each value
\yexchangetensorppalvaluexyestimindexone\
of the first grid to each value
\yexchangetensorppalvaluexyestimindextwo\
of the second grid
in order to derive the couple of closest values.
Moreover, in practical configurations,
one usually has prior knowledge about a range
of values to which
\yexchangetensorppalvaluexy\
and hence its relevant estimates
are guaranteed to belong.
This known range may be exploited in such a way that the values of
\yexchangetensorppalvaluexyestimindexone\
and
\yexchangetensorppalvaluexyestimindextwo\
which are the closest to one another in this range are also those which
are the closest to
\yexchangetensorppalvaluexy ,
by using the method detailed in Appendix
\ref{sec-estim-exchangetensorppalvaluexyz}.
From these two specific values,
an estimate of
\yexchangetensorppalvaluexy\
is eventually derived as
$\displaystyle\frac{
\Yexchangetensorppalvaluexyestimindexone
+
\Yexchangetensorppalvaluexyestimindextwo
}
{2}$.

Similarly, the parameter
\yexchangetensorppalvaluez\
is estimated by using
the procedure of the second 
part
of the
BQPT method of Section \ref{sec-bqpt},
based on
(\ref{eq-exchangetensorppalvaluez-vs-twoqubitsprobadirxxplusplusdiffminusminusphasediffindexd-intervalindextwo})
and
(\ref{eq-exchangetensorppalvaluez-vs-twoqubitsprobadirxxplusplusdiffminusminusphasediffindexd-estim-intervalindextwo}).
This procedure is here
applied
twice, with different values of the parameter \ytwoqubitwritereadtimeintervalindextwo\
 of Section \ref{sec-bqpt}.
The resulting method is described in Appendix
\ref{sec-estim-exchangetensorppalvaluexyz}.

\subsubsection{Test results}\label{sec-blind-hamilton-estim-tests}
The
physical implementation of qubits is an emerging
topic which is 
beyond the scope of this 
paper.
We therefore
assessed
the performance of the 
proposed BHPE method
by means of numerical tests performed with
data derived from
a software simulation of the 
considered configuration.
Each elementary
test
consists of 
the following
stages.
We first
create
a set of
\ytwoqubitseqnb\
input states
\ymixsyststateinitial.
Each such state is obtained
by randomly drawing its 
six parameters
\yparamqubitindexstdstateplusmodulus,
\yparamqubitindexstdstateplusphase\
and
\yparamqubitindexstdstateminusphase,
with
$\Yqubitindexstd \in \{ 1 , 2 \}$,
and then using
(\ref{eq-def-qubit-polar-qubit-indexstd}),
(\ref{eq-paramqubitindexstdstateminusmodulus-vs-paramqubitindexstdstateplusmodulus}),
(\ref{eq-etat-deuxspin-plusplus-initial-decompos})
(the state
(\ref{eq-etat-deuxspin-plusplus-initial-decompos})
is defined by the above six parameters, but only
the four parameters 
\yparamqubitindexstdstateplusmodulus\
and
$
\Yparamqubitindexstdstateminusphase
-
\Yparamqubitindexstdstateplusphase
$
have a physical meaning).
We then
process
the
states
\ymixsyststateinitial\
according to
(\ref{eq-statetfinalvsopmixstatetinit-components}),
with given values of the parameters of the
Hamiltonian
(\ref{eq-online-hamiltonian})
and hence of
the matrix
\yopmix\ 
involved in
(\ref{eq-statetfinalvsopmixstatetinit-components}).
This
yields the states 
\ymixsyststatefinal.
More precisely, we eventually
use simulated measurements of spin components
associated with these states 
\ymixsyststatefinal.
For measurements along the $Oz$ axis, this means that
we use the model
(\ref{eq-statecoef-vs-twoqubitsprobaplusplus-polar})-(\ref{eq-statecoef-vs-twoqubitsprobaminusminus-polar-vs-paramqubitindexstdstateplusmodulus})
with a given 
value of the
parameter
\ytwoqubitresultphaseevolsin,
corresponding to the above values of the parameters of the 
Hamiltonian
(\ref{eq-online-hamiltonian}).
For each of the
\ytwoqubitseqnb\
states
\ymixsyststateinitial,
corresponding to parameter values
$(
\Yparamqubitonestateplusmodulus ,
\Yparamqubittwostateplusmodulus ,
\Ytwoqubitresultphaseinit
)$,
Eq.
(\ref{eq-statecoef-vs-twoqubitsprobaplusplus-polar})-%
(\ref{eq-statecoef-vs-twoqubitsprobaminusminus-polar-vs-paramqubitindexstdstateplusmodulus})
thus yield the corresponding set of probability values
$(
\Ytwoqubitsprobaplusplusdirzz
,
\Ytwoqubitsprobaplusminusdirzz
,
\Ytwoqubitsprobaminusminusdirzz
)$,
which are used as follows.
We use \ywritereadonestatenb\
prepared copies of
the considered state
\ymixsyststateinitial\
to
simulate \ywritereadonestatenb\
random-valued
two-qubit
spin component measurements along the
$Oz$ axis,
drawn with the above probabilities
$(
\Ytwoqubitsprobaplusplusdirzz
,
\Ytwoqubitsprobaplusminusdirzz
,
\Ytwoqubitsprobaminusminusdirzz
)$.
We then derive the 
sample 
frequencies 
of the results of 
these
\ywritereadonestatenb\
measurements,
which are
estimates
of
\ytwoqubitsprobaplusplusdirzz,
\ytwoqubitsprobaplusminusdirzz\
and
\ytwoqubitsprobaminusminusdirzz\
for the considered 
state
\ymixsyststateinitial\
(see
(\ref{eq-def-qubitnbarbspaceoveralleventindexequalstdwithvecbasisprobseqindexapproxwritereadonestatenb})).
Then
computing the averages 
of these 
\ywritereadonestatenb-preparation
estimates over all
\ytwoqubitseqnb\
source vectors 
\ymixsyststateinitial\
yields
$(
\Ytwoqubitseqnb
\Ywritereadonestatenb
)$-preparation
estimates
of probability expectations
$E \{ 
\Ytwoqubitsprobaindexstddirzz \}$
(see  (\ref{eq-def-qubitnbarbspaceoveralleventindexequalstdwithvecbasisprobexpectapproxtwotwoqubitseqnb})).
Spin component measurements
for the $Ox$ axis are handled similarly (with other state preparations),
thus yielding 
estimates
of probability expectations
$E \{ 
\Ytwoqubitsprobaindexstddirxx \}$.
Both types of estimates
of probability expectations
are then used by our BHPE method defined in Section
\ref{sec-blind-hamilton-estim-method},
to derive
the estimates
\yexchangetensorppalvaluexyestim\
and
\yexchangetensorppalvaluezestim.

In these tests,
the above
parameters 
\ytwoqubitseqnb\
and
\ywritereadonestatenb\
were varied as described further in this section,
whereas
the numerical values of the other parameters
were fixed
as explained in Appendix \ref{sec-appendix-test-conditions},
so that we used
the same values for the parameters
of the
Hamiltonian
(\ref{eq-online-hamiltonian})
in all tests.
For each considered set of conditions defined by the values of
\ytwoqubitseqnb\
and
\ywritereadonestatenb,
we performed
100 above-defined elementary
tests, with different sets of states
\ymixsyststateinitial, in order to assess
the statistical performance of the considered BHPE method
over up to 100 estimations of the same set
$\{
\Yexchangetensorppalvaluexy
,
\Yexchangetensorppalvaluez
\}$
of parameter values.
More precisely, all 100 estimates
of
\yexchangetensorppalvaluexy\
were real-valued and were kept.
In constrast, for some test conditions, some estimates
of
\yexchangetensorppalvaluez\
were complex-valued (because they were derived from
trigonometric equations, where some \emph{estimates}
of sines or cosines may be situated
out of 
the interval
$[-1,1]$).
Since these false values can actually be detected and rejected in
practice,
the estimation performance for
\yexchangetensorppalvaluez\
was computed only over its real-valued estimates.

\begin{figure}[t]
\centerline{\epsfig{file=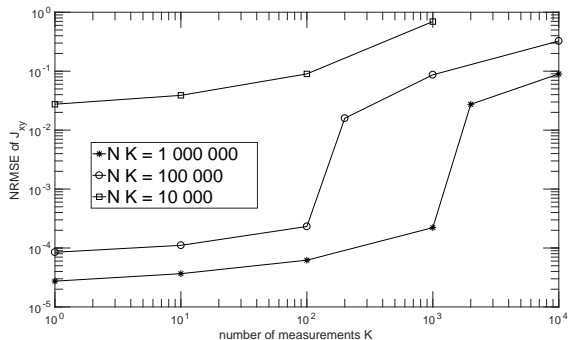,width=\linewidth}}
\caption{Normalized Root Mean Square
Error (NRMSE) of
estimation of parameter
\yexchangetensorppalvaluexy\
vs. number
\ywritereadonestatenb\
of preparations of each of the
\ytwoqubitseqnb\
used
states.}\label{fig-NRMSE-Jxy}\end{figure}

\begin{figure}[t]
\centerline{\epsfig{file=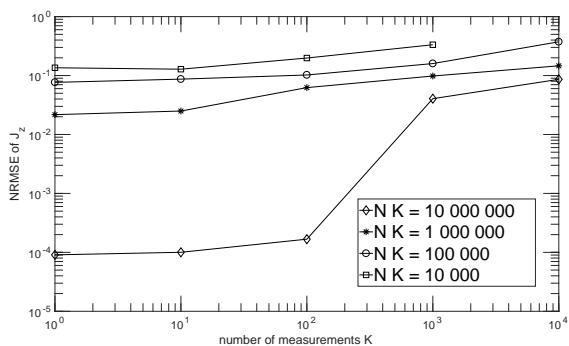,width=\linewidth}}
\caption{NRMSE of
estimation of parameter
\yexchangetensorppalvaluez\
vs. number
\ywritereadonestatenb\
of preparations of each of the
\ytwoqubitseqnb\
used
states.}\label{fig-NRMSE-Jz}\end{figure}

The considered performance criteria are defined as follows.
Separately for each of the parameters
\yexchangetensorppalvaluexy\
and
\yexchangetensorppalvaluez,
we computed the Normalized Root Mean Square Error (NRMSE) of that
parameter over all considered 
estimates,
defined as the ratio of its
RMSE to its actual (positive) value.
The values of these two performance criteria are shown in
Fig. \ref{fig-NRMSE-Jxy}
and
\ref{fig-NRMSE-Jz},
where
each plot corresponds to a fixed value of the product
$
\Ytwoqubitseqnb
\Ywritereadonestatenb
$,
i.e. of the complexity
of the BHPE method in terms of the total number of state preparations.
Each plot shows the variations of the 
considered performance criterion
vs. 
\ywritereadonestatenb,
hence with 
\ytwoqubitseqnb\
varied accordingly,
to keep the considered fixed value of
$
\Ytwoqubitseqnb
\Ywritereadonestatenb
$.

\ymodifartitionehundredthirtyninevonestepone{Fig. \ref{fig-NRMSE-Jxy}
and
\ref{fig-NRMSE-Jz}}
first show
that the proposed BHPE method is able to operate
with a number
\ywritereadonestatenb\
of preparations 
per state
\ymixsyststateinitial\
decreased down to one, as expected.
Moreover, 
for a fixed value of
$
\Ytwoqubitseqnb
\Ywritereadonestatenb
$,
the errors
decrease
when
\ywritereadonestatenb\
decreases, 
which is expected to be due to the fact that the
number
\ytwoqubitseqnb\
of \emph{different} used states thus increases,
allowing the estimation method to better
explore the statistics of the considered random process.
The magnitude of the error reduction from the highest value of
\ywritereadonestatenb\
down
to
$\Ywritereadonestatenb
=
1$
is often quite large, especially for
\yexchangetensorppalvaluexy,
that is,
between one and two orders of magnitude even when disregarding the
``discontinuity'' in some plots discussed hereafter.
This means that the proposed SIPQIP framework is then of high interest
not only in terms of simplicity of operation of QIP methods, but
also with respect to their accuracy.

Moreover, some of the plots contain the above-mentioned type
of discontinuity. For example, 
in Fig. \ref{fig-NRMSE-Jxy},
the NRMSE of
\yexchangetensorppalvaluexy\
for the fixed value
$\Ytwoqubitseqnb
\Ywritereadonestatenb
=
100,000$
abruptly decreases from around
$2
\times
10^{-2}$
when
$\Ywritereadonestatenb
=
200$
to
around
$2
\times
10^{-4}$
when
$\Ywritereadonestatenb
=
100$.
This behavior is normal: it is due to the intrinsically
discontinuous
nature of the specific type of estimation algorithm used here
for
\yexchangetensorppalvaluexy\
(the same considerations apply to 
\yexchangetensorppalvaluez, as confirmed by
Fig. \ref{fig-NRMSE-Jz}).
More precisely, in conditions when
\yexchangetensorppalvaluexy\
is estimated with a low accuracy, 
the following phenomenon may occur
for one or several runs of the
estimation procedure:
that procedure may select a false 
determination
of the estimate of
\yexchangetensorppalvaluexy,
that is,
a value corresponding to false
(i.e. nonzero)
values of
\yexchangetensorppalvaluexyshiftindetermintestimminusactualindexone\
in
(\ref{eq-exchangetensorppalvaluexyestim-minus-actual-vs-exchangetensorppalvaluexyshiftindetermintestimin-minus-actual-indexone})
and
\yexchangetensorppalvaluexyshiftindetermintestimminusactualindextwo\
in
(\ref{eq-exchangetensorppalvaluexyestim-minus-actual-vs-exchangetensorppalvaluexyshiftindetermintestimin-minus-actual-indextwo}).
The estimated value of
\yexchangetensorppalvaluexy\
is thus strongly shifted, because e.g. the corresponding values on
the first grid 
(\ref{eq-exchangetensorppalvaluexyestim-minus-actual-vs-exchangetensorppalvaluexyshiftindetermintestimin-minus-actual-indexone})
are shifted by multiples of the step
$\frac{\hbar \pi}
{\Ytwoqubitwritereadtimeintervalindexoneone}$
as explained above.
For the numerical values considered here, 
the corresponding step for the determinations of
$\Yexchangetensorppalvaluexyestim
/
k_{B}$
is
$\frac{\hbar \pi}
{\Ytwoqubitwritereadtimeintervalindexoneone
k_{B}}
\simeq
0.048$~K,
as compared to the actual value of
$\Yexchangetensorppalvaluexy
/
k_{B}
$
equal to 0.3~K in these tests.
Therefore, a shift
equal to one step,
i.e. obtained with
$\Yexchangetensorppalvaluexyshiftindetermintestimminusactualindexone
= 1$,
corresponds to a relative
error for
$\Yexchangetensorppalvaluexyestim
/
k_{B}$,
and hence for
\yexchangetensorppalvaluexyestim,
around 16 \% for the considered estimate of
\yexchangetensorppalvaluexy .
The overall error for 100 
estimates
then depends on the number of
runs where such false determinations are selected, but as long as
at least one of them is selected, the NRMSE of
\yexchangetensorppalvaluexy\
is lower bounded to a significant value.
In constrast,
in conditions when
\yexchangetensorppalvaluexy\
is estimated with a better accuracy, 
the correct determination of
\yexchangetensorppalvaluexy\
is selected for all 100 runs of the procedure and the NRMSE of
\yexchangetensorppalvaluexy\
is not lower bounded anymore: 
it regularly decreases when
$\Ytwoqubitseqnb
\Ywritereadonestatenb
$
increases or when
\ywritereadonestatenb\
decreases.
This is precisely what occurs in the above-mentioned example
of
Fig. \ref{fig-NRMSE-Jxy}
with
$\Ytwoqubitseqnb
\Ywritereadonestatenb
=
100,000$:
we manually checked all 100 estimates of
\yexchangetensorppalvaluexy\
(not shown here),
which proved that one of them
corresponds to a false determination
(with a shift equal to
a single step in the above-mentioned grid)
for
$\Ywritereadonestatenb
=
200$
and no false determination
for
$\Ywritereadonestatenb
=
100$.
The main conclusion of this analysis is that, when using
enough state preparations, the proposed procedure avoids
false determinations and thus has the usual behavior, with
performance regularly increasing when the conditions
(values of
$\Ytwoqubitseqnb
\Ywritereadonestatenb
$
and/or
\ywritereadonestatenb)
are improved.

By considering a wide range of test conditions,
Fig. 
\ref{fig-NRMSE-Jxy}
and
\ref{fig-NRMSE-Jz}
show that a wide range of estimation accuracies may be obtained for
\yexchangetensorppalvaluexy\
and
\yexchangetensorppalvaluez.
Focusing on the most interesting cases, namely when
$
\Ywritereadonestatenb
=
1
$,
the NRMSE of
\yexchangetensorppalvaluexy\
can e.g. here be made equal to 
$
2.75
\times
10^{-2}
=
2.75$~\%
for only
$\Ytwoqubitseqnb
=
10^4$
state preparations
or
$
8.46
\times
10^{-5}
$
for
$\Ytwoqubitseqnb
=
10^5$
or
$
2.74
\times
10^{-5}
$
for
$\Ytwoqubitseqnb
=
10^6$.
Similarly,
when
$
\Ywritereadonestatenb
=
1
$,
the NRMSE of
\yexchangetensorppalvaluez\
can e.g. 
be made equal to 
7.66~\%
for
$\Ytwoqubitseqnb
=
10^5$
or
2.17~\%
for
$\Ytwoqubitseqnb
=
10^6$
or
$
9.07
\times
10^{-5}
$
for
$\Ytwoqubitseqnb
=
10^7$.
The ``very low'' NRMSE values, corresponding to the
absence of false determinations
and to the parts ``below possible discontinuities''
in the plots of
Fig. 
\ref{fig-NRMSE-Jxy}
and
\ref{fig-NRMSE-Jz},
are thus achieved for
\ytwoqubitseqnb\
higher than
$
10^4$
for
\yexchangetensorppalvaluexy\
and
$
10^6$
for
\yexchangetensorppalvaluez.

All above 
results show that,
for 
given values of
\ywritereadonestatenb\
and
\ytwoqubitseqnb,
the parameter
\yexchangetensorppalvaluexy\
is often estimated much more accurately than
\yexchangetensorppalvaluez.
This is  
\ymodifartitionehundredthirtyninevonestepone{reasonable because,
on the one hand, \yexchangetensorppalvaluexy\ is estimated by using
only measurements along the $Oz$ axis, that lead to a 
relatively simple data
model and hence a simple estimation
procedure, which is likely to yield good estimation accuracy, whereas,
on the other hand, \yexchangetensorppalvaluez\ is estimated by combining
measurements along the $Ox$ and $Oz$ axes, and those
along the $Ox$ axis involve a 
more complex data
model, which yields an estimation procedure with possibly degraded estimation accuracy.}
This also means that, 
whereas
we here used a simple protocol
by considering the same values of
the set of parameters
$
\{
\Ywritereadonestatenb
,
\Ytwoqubitseqnb
\}
$
in the series of state preparations used for estimating
\yexchangetensorppalvaluexy\
and
\yexchangetensorppalvaluez,
one might instead use lower values of
the number 
\ytwoqubitseqnb\
of state preparations
(preferably 
with
$\Ywritereadonestatenb =1$)
in the series of preparations
performed for estimating
\yexchangetensorppalvaluexy\
than in those used for
\yexchangetensorppalvaluez,
in order to balance the estimation accuracies achieved for
\yexchangetensorppalvaluexy\
and
\yexchangetensorppalvaluez\
while reducing the total number of state preparations
(the BQPT method used here yields related considerations,
that were detailed in
\cite{amoi6-118}%
).

\subsection{%
\ymodifartitionehundredthirtyninevonestepone{Channel}
estimation and phase estimation}\label{sec-quantum-channel-estim}
In Sections
\ref{sec-intro}
and
\ref{sec-machine-learning},
we explained that,
in the classical framework,
the same information processing task is given different
names, depending on the considered application field.
In particular, the system identification task in the
field of automatic control corresponds to the
channel estimation task in the field of communications.
The same phenomenon occurs in the quantum framework.
In particular, QPT, and hence our blind
(and possibly single-preparation)
extension addressed in Section \ref{sec-bqpt},
is often stated to be the
quantum counterpart of classical system identification
(see e.g. \cite{booknielsen}
p. 389).
\ymodifartitionehundredthirtyninevonestepone{QPT}
applies to general quantum systems, not necessarily
defined by a small set of parameters, and could therefore
be called nonparametric system identification.
But the Hamiltonian parameter estimation task, and hence
our blind 
(and single-preparation)
extension introduced in Section
\ref{sec-blind-hamilton-estim}, is also closely
connected 
\ymodifartitionehundredthirtyninevonesteptwo{with}
system identification, and more precisely to
parametric system identification, since it estimates a
small set of parameters (e.g., the principal values of the
exchange tensor in the case of Heisenberg coupling
that was considered above as an example), and these
parameters then completely define the behavior of that
system, including the resulting process matrix in
the associated QPT task.

Moreover, although a different terminology is used for
other quantum information processing tasks, some of
these tasks actually address 
the same type of
problems 
as above. 
This first concerns the
quantum channel estimation task:
as explained e.g. in \cite{preskill-http-chap3},
a map
from the density operator 
associated with a
quantum state to 
another density operator is often called
a quantum channel, as a reference to classical
communication scenarios. The identification of such a map
may therefore be called quantum channel estimation and
is closely linked to the 
QPT
problem that we
considered above,
possibly in its blind and single-preparation form.
Similarly, a standard quantum information processing
procedure is phase estimation. In \cite{booknielsen} p. 221,
it is defined
as the estimation of the phase
$\Phi$ of an eigenvalue
$e^{2 \pi i \Phi}$ of a unitary operator.
This task is therefore related as follows to both investigations
reported in Sections
\ref{sec-bqpt}
and
\ref{sec-blind-hamilton-estim}.
First, as explained in Section
\ref{sec-bqpt},
the 
considered
(B)QPT problem essentially consists of estimating
the parameters
$
\exp
\left[
\Ysqrtminusone
\frac{ 
\Yexchangetensorppalvaluexy
(t - t_0)
} 
{ \hbar }
\right]
$
and
$
\exp
\left[
\Ysqrtminusone
\frac{ 
\Yexchangetensorppalvaluez
(t - t_0)
} 
{ 2 \hbar }
\right]
$
and hence the exponential terms of the
diagonal representation
\yopmixdiag\
of the considered operator
(see
(\ref{eq-opmixmatrixdecompose})-%
(\ref{eq-omega-zero-zero-express})).
This is therefore equivalent to estimating the phases of these
exponentials, up to a multiple of 
$2 \pi$.
Moreover, the method introduced in
Section
\ref{sec-blind-hamilton-estim}
is directly connected 
\ymodifartitionehundredthirtyninevonesteptwo{with}
removing the 
additive indeterminacy due to
this multiple of 
$2 \pi$.

This discussion shows that the blind and single-preparation
extensions that we proposed above in this paper for
quantum information processing tasks related to system identification
are expected to be of importance not only for the
scientific communities focused on QPT and Hamiltonian
parameter estimation
but also for quantum scientists
who investigate a variety of related problems, such as
quantum channel estimation and phase estimation.
Moreover, in this Section
\ref{sec-system-ident},
we restricted ourselves to problems related to the
characterization (i.e. identification) of the 
considered
quantum
process itself.
As explained in Sections
\ref{sec-intro} and
\ref{sec-machine-learning},
related QIP problems consist of building processing systems,
with quantum and/or classical means, that essentially implement
the \emph{inverse} of an initially unknown quantum process.
This corresponds to the quantum source separation
and related tasks, that we investigate 
in the next section,
still aiming at
extending the considered configurations to blind and single-preparation
ones.

\section{QIP tasks related to system inversion
and state restoration%
}\label{sec-system-invert}

\subsection{Blind quantum source separation}\label{sec-blind-quantum-source-sep}
A 
\ymodifartitionehundredthirtyninevonestepone{rather}
general version of the 
blind quantum source separation (BQSS) problem addressed here may be defined
as follows. A set of qubits 
with indices
\yqubitindexstd\
are independently prepared
with states 
$|\Yqubitnbarbtimenonestatenot_{\Yqubitindexstd}\rangle$.
The state 
\yqubitnbarbtimenonestate\
of the system composed of these
qubits,
which is equal to the tensor product of the above 
single-qubit states
$|\Yqubitnbarbtimenonestatenot_{\Yqubitindexstd}\rangle$,
is then transformed, i.e. mapped to another state
$|\Yqubitnbarbtimenonestatenot^{\prime}\rangle
=
\Ymapfunc(\Yqubitnbarbtimenonestate)$,
where the mapping function
\ymapfunc\
e.g. corresponds to temporal evolution
\ymodifartitionehundredthirtyninevonesteptwo{with coupling between qubits}%
, as detailed below.
In the blind configuration, the user is given a set of
transformed states
$|\Yqubitnbarbtimenonestatenot^{\prime}\rangle$
but does not know the corresponding set of original 
states
\yqubitnbarbtimenonestate,
and hence the source states
$|\Yqubitnbarbtimenonestatenot_{\Yqubitindexstd}\rangle$,
nor the mapping function
\ymapfunc.
The user then eventually wants to restore the 
information contained in (at least part of) the source
states, 
either in quantum form, by deriving estimates of these states
$|\Yqubitnbarbtimenonestatenot_{\Yqubitindexstd}\rangle$,
or in classical form,
typically by eventually using a classical computer to derive 
estimates of the coefficients 
of the states
$|\Yqubitnbarbtimenonestatenot_{\Yqubitindexstd}\rangle$
in a given basis.

This generic problem is connected 
\ymodifartitionehundredthirtyninevonesteptwo{with}
various application fields.
The first one, on which we focus hereafter, 
is related to
the operation of
quamputers.
In such a future quamputer,
data will
be stored
in registers of qubits, for subsequent use.
Due to non-idealities of the physical implementation of 
such a register,
the qubits which form it may 
have undesired 
coupling with
one another,
such as Heisenberg coupling, e.g.
if considering quamputer implementations
related to spintronics 
\cite{amq88,
Delgado_2017,
amq111}.
As time goes on, 
the register state will therefore
evolve in a complicated way
due to this undesired qubit coupling,
thus making 
the final value of that register state
not directly usable in the target 
quantum 
algorithm executed on that quamputer.
BQSS may
then be used as a preprocessing 
stage,
to restore the 
initial
register state,
before providing it
to the target
application of that quamputer.

To analyze this BQSS problem in more detail, we 
hereafter focus on a
basic case, from which the reader may then extend this analysis
to other configurations.
In the considered case, the device
(e.g., the qubit register) is restricted to two qubits,
implemented as electron spins 1/2,
and the 
undesired 
coupling which exists between them is again based on
the cylindrical-symmetry Heisenberg model defined in Appendix
\ref{sec-heisenberg process}.
Using the notations of that appendix, the initial state
\ymixsyststateinitial\ of the device
(e.g., the state stored at time
\yqubitonetimeinit\ in the register),
which corresponds to state
\yqubitnbarbtimenonestate\
in the above general definition of BQSS,
may be represented by the column vector
\yveccompsyststatetinit\
of the components 
of
\ymixsyststateinitial\
in the standard basis,
defined by
(\ref{eq-def-veccompsyststatetinit}).
Similarly,
the final state
\ymixsyststatefinal\ of the device
(e.g., the only state available to the user, at a later time
\yqubitonetimefinal, in the register),
which corresponds to state
$|\Yqubitnbarbtimenonestatenot^{\prime}\rangle$
in the above general definition of BQSS,
may be represented by the column vector
\yveccompsyststatetfinal\
of the components of
\ymixsyststatefinal\
in the standard basis.
The effect of coupling is then represented by the
relationship
\yeqdefstatetfinalvsopmixstatetinitcomponentsversiontwo
where
\yopmix\
is the matrix defined in Appendix
\ref{sec-heisenberg process}.

\begin{figure}[t]
\centerline{\epsfig{file=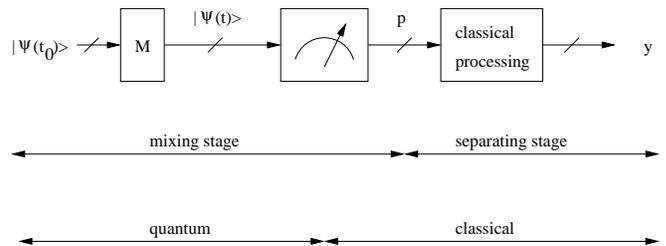,width=\linewidth}}
\caption{Global
(i.e. mixing + separating) 
blind quantum source separation (BQSS)
configuration
when only applying
measurements and classical processing
to the available coupled quantum state.}
\label{fig-quantmix_conv_classsep}
\end{figure}

The first class of BQSS methods that we previously developed for
handling this configuration (see especially
\cite{amoi5-31,
amoi6-18,
amoi6-42})
is the ``least quantum'' one, in the sense that,
starting from the available quantum states
\ymixsyststatefinal ,
it first converts them into classical-form data 
(probability estimates) by means
of measurements
and then processes the latter data with only classical
means, as shown in Fig.
\ref{fig-quantmix_conv_classsep}.
More precisely, in the reported investigations, only measurements
of the components of the two spins along the $Oz$ axis
were considered.
The probabilities of the outcomes of these measurements are therefore
again defined by
(\ref{eq-statecoef-vs-twoqubitsprobaplusplus-polar})-%
(\ref{eq-statecoef-vs-twoqubitsprobaminusminus-polar-vs-paramqubitindexstdstateplusmodulus}).
Unlike the BQPT method of Section
\ref{sec-bqpt},
the BQSS methods 
summarized here do not use 
the SIPQIP framework.
Instead, they separately derive
an
estimate 
of each set of probabilities
\ytwoqubitsprobaindexstddirzz,
with
$\Yqubitbothtimefinalstatecoefindexindexstd =1$, 2 and 4,
associated with one final state
\ymixsyststatefinal ,
so that they require each initial state
\ymixsyststateinitial\
to be prepared many times.

This class of BQSS methods therefore uses the mapping
(\ref{eq-statetfinalvsopmixstatetinit-components-version-two}),
from a state
\ymixsyststateinitial\
to a state
\ymixsyststatefinal,
\emph{indirectly}:
it only involves the mapping
(\ref{eq-statecoef-vs-twoqubitsprobaplusplus-polar})-%
(\ref{eq-statecoef-vs-twoqubitsprobaminusminus-polar-vs-paramqubitindexstdstateplusmodulus}),
which goes from the set of initial qubit parameters
$\{
\Yparamqubitonestateplusmodulus , \Yparamqubittwostateplusmodulus ,
\Ytwoqubitresultphaseinit
\}$,
to the set of probabilities
$\{
\Ytwoqubitsprobaplusplusdirzz
,
\Ytwoqubitsprobaplusminusdirzz
,
\Ytwoqubitsprobaminusminusdirzz
\}$.
The transform
(\ref{eq-statecoef-vs-twoqubitsprobaplusplus-polar})-%
(\ref{eq-statecoef-vs-twoqubitsprobaminusminus-polar-vs-paramqubitindexstdstateplusmodulus})
is then the ``mixing model'', using the classical BSS terminology
and, indeed, even if they are derived from an intrinsically
quantum phenomenon, the inputs and outputs of this transform
may be stored in classical form, on a classical computer:
the 
moduli
\yparamqubitindexstdstateplusmodulus\
and the phases
\yparamqubitindexstdstateplusphase\
and
\yparamqubitindexstdstateminusphase\
(in fact, only their differences
$
\Yparamqubitindexstdstateminusphase
-
\Yparamqubitindexstdstateplusphase
$
have a physical meaning)
may be stored on a classical computer before they are used 
by the procedure that
prepares the corresponding state
\ymixsyststateinitial.
The output
of this ``mixing stage'',
equal to an estimate of
$\{
\Ytwoqubitsprobaplusplusdirzz
,
\Ytwoqubitsprobaplusminusdirzz
,
\Ytwoqubitsprobaminusminusdirzz
\}$,
is then connected to the input of the ``separating
stage''
(see Fig.
\ref{fig-quantmix_conv_classsep} again),
which is composed of the separating system that we proposed
for restoring an estimate of
$\{
\Yparamqubitonestateplusmodulus , \Yparamqubittwostateplusmodulus ,
\Ytwoqubitresultphaseinit
\}$
from its input.
In other words, this 
separating
system ideally
aims at implementing the inverse of
the mapping
(\ref{eq-statecoef-vs-twoqubitsprobaplusplus-polar})-%
(\ref{eq-statecoef-vs-twoqubitsprobaminusminus-polar-vs-paramqubitindexstdstateplusmodulus}).
This can be done (up to 
an
approximation due to estimating the probabilities
\ytwoqubitsprobaindexstddirzz)
in the ideal case when the exact value of the parameter
\ytwoqubitresultphaseevolsin\ 
of
(\ref{eq-statecoef-vs-twoqubitsprobaplusminus-polar-versionthree})
is known, because the inverse
of
(\ref{eq-statecoef-vs-twoqubitsprobaplusplus-polar})-%
(\ref{eq-statecoef-vs-twoqubitsprobaminusminus-polar-vs-paramqubitindexstdstateplusmodulus})
can be analytically determined: see details in
\cite{amoi5-31,
amoi6-18,
amoi6-42}.
In contrast, the blind version of the problem, i.e.
when the value of
\ytwoqubitresultphaseevolsin\ is unknown,
is handled as follows.
One considers the class of direct mappings obtained by replacing 
\ytwoqubitresultphaseevolsin\ 
by a free parameter
\ytwoqubitresultphaseevolsinestim\
in
(\ref{eq-statecoef-vs-twoqubitsprobaplusplus-polar})-%
(\ref{eq-statecoef-vs-twoqubitsprobaminusminus-polar-vs-paramqubitindexstdstateplusmodulus}).
One then determines the analytical expression of the corresponding class
of inverse mappings, which is derived by replacing
\ytwoqubitresultphaseevolsin\ by
\ytwoqubitresultphaseevolsinestim\
in the above ideal inverse mapping.
The idea is then to derive an estimate
\ytwoqubitresultphaseevolsinestimtwo\
of 
\ytwoqubitresultphaseevolsin,
in order to use it as the value of
\ytwoqubitresultphaseevolsinestim\ in the inverse mapping.
Various methods have been proposed to this end
(see e.g.
\cite{amoi5-31,
amoi6-18,
amoi6-42}),
by extending differents concepts 
used in
classical Independent
Component Analysis (ICA) to the considered quantum problem.

The complete operation of the above
class
of BQSS methods consists of two phases, which correspond to
the general features that we provided in Section
\ref{sec-machine-learning}
for classical and quantum machine learning methods:
\begin{enumerate}
\item First, in the adaptation (or training) phase, a set of
states
\ymixsyststatefinal\ is used to derive the 
above
estimate
\ytwoqubitresultphaseevolsinestimtwo ,
i.e. to learn the (direct and) inverse mapping.
\item Then, in the inversion phase
(which corresponds to the final, useful, operation of the
separating system)%
, the probabilities
estimated for each new state
\ymixsyststatefinal\
are transferred through the above estimated inverse mapping,
to restore the considered parameters
of the corresponding state
\ymixsyststateinitial .
\end{enumerate}

As mentioned above, 
a major 
constraint in
that first class
of BQSS methods is that it requires the same state
\ymixsyststateinitial\ to be prepared many times,
both in the adaptation and inversion phases.
This
makes
these methods ``less blind''
because, although these states \ymixsyststateinitial\
are allowed to be unknown from the point of view of
the adaptation procedure, some control is required so that
the \emph{same} value is repeatedly prepared for each of these states.

\begin{figure}[t]
\centerline{\epsfig{file=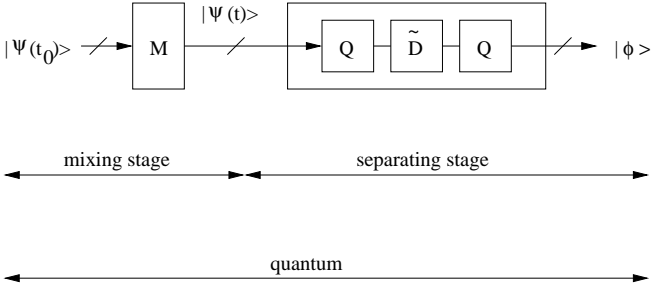,width=\linewidth}}
\caption{Mixing stage + quantum-processing
inverting block of separating system.}
\label{fig-quantmix_Q_Dtilde_Q}
\end{figure}

A solution to the above
problem was introduced, 
but
only for the
inversion phase, in our second class of BQSS methods,
especially described in
\cite{amoi6-34,
amoi6-37,
amoi6-64}.
We now detail it, since we take advantage of it in
the fully SIPQIP methods 
that we introduce further
in this paper for BQSS.
In that second class of BQSS methods, during the inversion
phase, each state
\ymixsyststatefinal\
available as the input of the separating system is directly
used in quantum form, i.e. without performing measurements,
so that this separating system outputs a quantum state
\ysepsyststateout\
that should ideally be equal to
the multi-qubit source state
\ymixsyststateinitial\
that one aims at restoring.
That part of the separating system,
called the inverting block, is thus a 
global
quantum gate
(see Fig. \ref{fig-quantmix_Q_Dtilde_Q}),
which only requires a \emph{single} instance of its
input state
\ymixsyststatefinal\
to derive its corresponding output state
\ysepsyststateout.

\begin{figure}[t]
\centerline{\epsfig{file=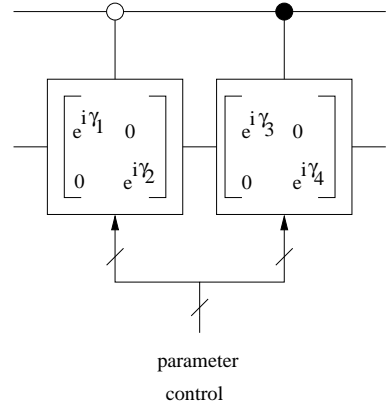,width=5cm}}
\caption{Implementation of quantum operator defined by matrix
\ysepmixdiag, used in inverting block.}
\label{fig-cq_sepmixdiag_adapt}
\end{figure}

The above gate
is designed as follows.
Although we here do not keep all the features of the above
first class of BQSS methods, we build upon some of its principles.
In particular, we exploit the fact that, although the actual
value of the mixing
matrix \yopmix\ 
of
(\ref{eq-statetfinalvsopmixstatetinit-components-version-two})
is not known in the blind configuration,
from
Appendix
\ref{sec-heisenberg process}
one knows
that it belongs to the class of matrices defined as
\yeqdefopmixmatrixdecomposeversiontwo
where
$\Yopmixbases = \Yopmixbases^{-1}$
is a known, fixed, matrix and
\yopmixdiag\ is a diagonal matrix, 
whose diagonal entries have unit modulus (and
a structure that is disregarded in this approach).
\ymodifartitionehundredthirtyninevonestepone{We therefore use
an inverting block of the separating
system which is adaptive (or tunable),
i.e. such that some
of the values of the parameters that define its behavior may
be modified. More precisely, this block is}
designed so that it is able to
implement the inverse of any transform in the above-defined
class,
\ymodifartitionehundredthirtyninevonestepone{depending on its
parameter values.}
Its operation is therefore
represented by a matrix
defined as
\yeqdefopsepmatrixdecompose
with
\yeqdefsepmixdiagdef
where
\ysepmixdiagelomegaoneone\ to
\ysepmixdiagelomegaoneminusone\
are free real-valued parameters.
This inverting block is thus the cascade of three 
simpler
quantum gates,
as shown in Fig.
\ref{fig-quantmix_Q_Dtilde_Q}.
The implementation of each gate corresponding to the matrix
\yopmixbases, as a combination of 
even
simpler gates,
was detailed in
\cite{amoi6-18}.
Moreover,
the 
\ymodifartitionehundredthirtyninevonestepone{adaptive}
gate
corresponding to
(\ref{eq-sepmixdiag-def}),
introduced in 
\cite{amoi6-64},
may be decomposed as shown in
Fig.
\ref{fig-cq_sepmixdiag_adapt},
where
the closed (i.e. black) and
open circle notations respectively indicate conditioning on the qubit
being set to one or zero, as in
\cite{booknielsen}
p. 184.
In 
\cite{amoi6-64}
and in the new use of that gate introduced further in this paper,
the
values of the
parameters
\ysepmixdiagelomegaoneone\ to
\ysepmixdiagelomegaoneminusone\
are
controlled
by classical-form
signals.
These parameters 
may e.g. be
independent,
known but
arbitrary,
increasing,
functions of control voltages.
Such control voltages are e.g. used 
in
the real device
described in
\cite{amq111}.

\begin{figure*}[t]
\centerline{\epsfig{file=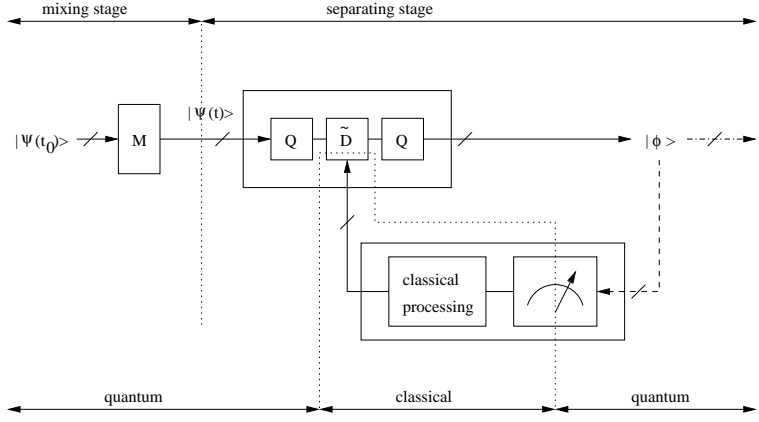,width=10cm}}
\caption{Global
(i.e. mixing + separating)
configuration, which a feedback separating system that includes
a quantum-processing
inverting block
and a classical-processing
adapting block.
Each quantum state \ysepsyststateout\ is 
used only once
(no cloning):
see 
\cite{amoi6-64}.}
\label{fig-quantmix_Q_Dtilde_Q_adapt_out_conv}
\end{figure*}

The complete operation of this second class
of BQSS methods therefore
consists of the same phases
as for
the above first class of 
methods:
\begin{enumerate}
\item First, in the adaptation 
phase, a set of
states
\ymixsyststatefinal\ is used to
\ymodifartitionehundredthirtyninevonestepone{adapt}
the matrix \ysepmixdiag,
i.e. to learn the 
inverse mapping.
\item Then, in the inversion phase, each new state
\ymixsyststatefinal\
is transferred through the gates of
Fig.
\ref{fig-quantmix_Q_Dtilde_Q},
which perform the
above estimated inverse mapping,
to restore
the corresponding state
\ymixsyststateinitial .
\end{enumerate}
The method used in
\cite{amoi6-64}
to
\ymodifartitionehundredthirtyninevonestepone{adapt}
the matrix \ysepmixdiag\
is based on the probabilities of measurements associated with a
set of
output states 
\ysepsyststateout\
of the inverting block of Fig. \ref{fig-quantmix_Q_Dtilde_Q}.
These probabilities are essentially
\ymodifartitionehundredthirtyninevonestepone{used}
to measure the degree of
entanglement of these states
\ysepsyststateout.
The matrix
\ysepmixdiag\
is 
\ymodifartitionehundredthirtyninevonestepone{adapted}
so as to essentially make these states
\ysepsyststateout\
unentangled, so that this type of methods performs an
``Unentangled Component Analysis''
\cite{amoi6-64},
as opposed to the above-mentioned classical and quantum
Principal Component Analysis and
Independent Component Analysis.
The complete structure of the resulting separating system
is shown in Fig.
\ref{fig-quantmix_Q_Dtilde_Q_adapt_out_conv}.
Unlike in the inversion phase, during the adaptation phase
this structure requires many copies of each of its
input states
\ymixsyststatefinal,
in order to derive the corresponding copies of the
output states
\ysepsyststateout\
and hence the corresponding probability estimates based on
sample frequencies of measurement outcomes.
This second class of BQSS methods is thus ``more quantum'' than
the first one, 
first
because it uses quantum processing means
in the inverting block, 
and second
because it is based on the 
quantum concept
of entanglement, which has no classical counterpart.

\begin{figure}[t]
\centerline{\epsfig{file=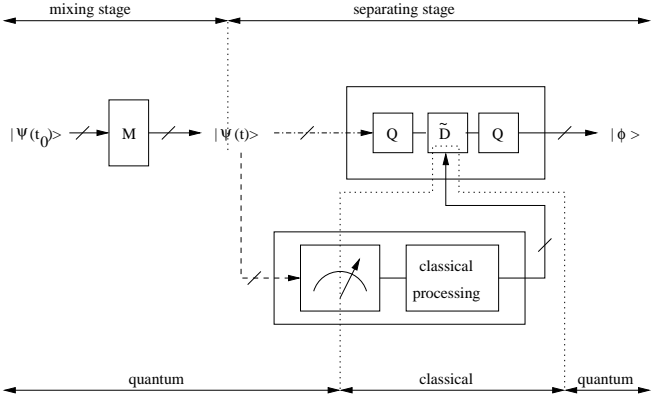,width=\linewidth}}
\caption{Global
(i.e. mixing + separating)
configuration, with a feedforward separating system that includes
a quantum-processing
inverting block
and a classical-processing
adapting block.
Each quantum state \ymixsyststatefinal\ is 
used only once
(no cloning):
see text.}
\label{fig-quantmix_Q_Dtilde_Q_adapt_in_conv_v2}
\end{figure}

In the present paper, we 
introduce a third class of BQSS methods,
which proceeds further than the above two classes, by using the SIPQIP
framework in all the operation of the separating system,
i.e. by using a single preparation of each state also during
the adaptation phase.
To this end, we exploit the structure of the matrix
\yopmixdiag , defined in
(\ref{eq-opmixdiagdef})-(\ref{eq-omega-zero-zero-express}).
We take into account the fact that the matrix
\ysepmixdiag\
of the separating system should ideally be set to the
inverse of
\yopmixdiag.
Therefore,
by replacing
\yexchangetensorppalvaluexy\
and
\yexchangetensorppalvaluez\
by their estimates
\yexchangetensorppalvaluexyestim\
and
\yexchangetensorppalvaluezestim\
in
(\ref{eq-opmixdiagdef})-(\ref{eq-omega-zero-zero-express}),
we set 
\ysepmixdiag\
as in
(\ref{eq-sepmixdiag-def}),
but here with
the following structure for the phases of its diagonal
elements:
\yeqdefopmixdiagelementsexpress
where
\ytwoqubitwritereadtimeintervalindexthree\
is the value of the time interval
$(\Yqubitonetimefinal
-
\Yqubitonetimeinit
)$
used in the inversion phase of the proposed BQSS method.
In this new BQSS method, we take advantage of
the 
BQPT method that we described
in Section
\ref{sec-bqpt}:
when applying the latter method
with
time intervals
$(\Yqubitonetimefinal
-
\Yqubitonetimeinit
)$
set according
to (\ref{eq-twoqubitwritereadtimeinterval-link-three})
\ymodifartitionehundredthirtyninevonestepone{with a freely selected value of
\ytwoqubitwritereadtimeintervalindexone}%
,
we get
(\ref{eq-exchangetensorppalvaluexy-vs-twoqubitresultphaseevolindexd-estim})
and
(\ref{eq-exchangetensorppalvaluez-vs-twoqubitsprobadirxxplusplusdiffminusminusphasediffindexd-estim-intervalindextwo}),
which
shows that all the quantities in
(\ref{eq-sepmixdiagel-omega-one-zero})-(\ref{eq-sepmixdiagel-omega-zero-zero})
required to assign
\ysepmixdiagelomegaoneone\
to
\ysepmixdiagelomegaoneminusone\
are known or can be estimated.
The adaptation phase of the proposed BQSS method therefore consists of
applying the above BQPT method, then
using
(\ref{eq-sepmixdiagel-omega-one-zero})-(\ref{eq-sepmixdiagel-omega-zero-zero})
to derive the selected values of
\ysepmixdiagelomegaoneone\
to
\ysepmixdiagelomegaoneminusone\
and finally using the supposedly known 
correspondence function which
makes it possible to convert these values of
\ysepmixdiagelomegaoneone\
to
\ysepmixdiagelomegaoneminusone\
into the practical control signals (e.g., voltages)
of the gates of Fig.
\ref{fig-cq_sepmixdiag_adapt}
which make these gates operate with these desired
values of
\ysepmixdiagelomegaoneone\
to
\ysepmixdiagelomegaoneminusone.
The resulting global configuration is shown in Fig. 
\ref{fig-quantmix_Q_Dtilde_Q_adapt_in_conv_v2}.
Each state \ymixsyststatefinal\ is thus used only once
(see also
\cite{amoi6-64}
about the no-cloning theorem):
during the adaptation phase, these states are sent to the part
of the system which performs measurements and BQPT
(dashed line and lower part of
Fig. 
\ref{fig-quantmix_Q_Dtilde_Q_adapt_in_conv_v2});
then, during the inversion phase,
they are sent to the inverting block
(dash-dotted line and upper right part of
Fig. 
\ref{fig-quantmix_Q_Dtilde_Q_adapt_in_conv_v2}).

Thanks to the properties of the BQPT method reused here
as a building block of the proposed BQSS method, the
output of the
latter method does not depend on the values used in
(\ref{eq-exchangetensorppalvaluexy-vs-twoqubitresultphaseevolindexd-estim})
and
(\ref{eq-exchangetensorppalvaluez-vs-twoqubitsprobadirxxplusplusdiffminusminusphasediffindexd-estim-intervalindextwo})
for the integers
\yexchangetensorppalvaluexyshiftindetermintestim\
and
\yexchangetensorppalvaluezshiftindetermintestim.
This may be seen by inserting
(\ref{eq-exchangetensorppalvaluexy-vs-twoqubitresultphaseevolindexd-estim})-%
(\ref{eq-exchangetensorppalvaluez-vs-twoqubitsprobadirxxplusplusdiffminusminusphasediffindexd-estim-intervalindextwo})
in
(\ref{eq-sepmixdiagel-omega-one-zero})-(\ref{eq-sepmixdiagel-omega-zero-zero})
and then in
(\ref{eq-sepmixdiag-def}),
with
(\ref{eq-twoqubitwritereadtimeinterval-link-three}),
which
shows that the terms of
(\ref{eq-exchangetensorppalvaluexy-vs-twoqubitresultphaseevolindexd-estim})-%
(\ref{eq-exchangetensorppalvaluez-vs-twoqubitsprobadirxxplusplusdiffminusminusphasediffindexd-estim-intervalindextwo})
that include
\yexchangetensorppalvaluexyshiftindetermintestim\
and
\yexchangetensorppalvaluezshiftindetermintestim\
yield 
terms in
(\ref{eq-sepmixdiagel-omega-one-zero})-(\ref{eq-sepmixdiagel-omega-zero-zero})
that are integer multiples of
$2 \pi$,
and that therefore have no influence on the value of 
\ysepmixdiag .

\subsection{Blind 
quantum 
(entangled)
state restoration}\label{sec-bqrs}
In the classical framework, the concept of blind source separation (BSS)
intrinsically refers to situations involving several (unknown)
source signals, created by several ``sources'' that may be
some kinds of ``objects''.
Such a situation may be mathematically described by gathering
all the values of these source signals, e.g. at a given time,
as the elements of an overall source vector.
In our quantum extensions of this classical BSS, we started from
a similar situation, involving several objects, such as qubits
implemented as spins 1/2, and we first independently 
considered the ``signal value'',
i.e. the initial quantum state
$|\psi_{\Yqubitindexstd}
(\Yqubitonetimeinit)\rangle$,
of each of them (see
(\ref{eq-twoqubit-state-init-index-i})).
We then ``gathered'' these individual states by defining
the state of the complete 
system
(see 
(\ref{eq-qubitbothtimeinitstate-tensor-prod}))
as the tensor product of
the states
$|\psi_{\Yqubitindexstd}
(\Yqubitonetimeinit)\rangle$,
somehow as the quantum counterpart of the above vector
of classical signal values.
However, the quantum framework opens the way to much
richer situations, because the possible states of a complete
system are not restricted to the tensor products of the
individual states of independent parts of this system: they also
include entangled states.
An 
extension of the above BQSS problem is therefore
blind quantum state restoration (BQSR),
\ymodifartitionehundredthirtyninevonestepone{aiming at}
restoring a \emph{possibly entangled} 
\ymodifartitionehundredthirtyninevonestepone{deterministic pure}
state of a multi-qubit system,
starting from an altered version of it.
One thus conceptually considers
a 
{single} arbitrary
multi-qubit source state, instead of 
{several} single-qubit source
states.
In particular, this includes restoring the possibly
entangled initial state
\ymixsyststateinitial\
of a multi-qubit system, from the state
\ymixsyststatefinal\
of that system at a later time.

We here investigate this extension of BQSS to possibly entangled 
source states,
by 
\ymodifartitionehundredthirtyninevonestepone{considering}
the use of such states in the second phase
of the
operation of this system, i.e. in the above-defined inversion phase,
after the transform performed by this system has been fixed by means of
the adaptation phase.
We claim that the 
BQSS system defined in 
Fig. \ref{fig-quantmix_Q_Dtilde_Q_adapt_in_conv_v2}
is 
directly able to 
\ymodifartitionehundredthirtyninevonesteptwo{perform}
the considered BQSR task,
since it operates as follows.
An unknown state
\ymixsyststateinitial\
is created,
then
modified by an operator represented by the
matrix
\yopmix, thus yielding the state
\ymixsyststatefinal .
The latter state is the state processed by the new separating
system that we designed in Section 
\ref{sec-blind-quantum-source-sep}.
The transform performed by this system is represented by
the matrix
$\Yopsep
= \Yopmixbases \Ysepmixdiag \Yopmixbases$
(see
(\ref{eq-opsepmatrixdecompose})).
But, during the adaptation phase of that separating system
that occurred before this inversion phase, 
\yopsep\
was made equal to the inverse of
\yopmix\ (up to estimation errors)
by the proposed adaptation method.
Therefore, when applying that separating transform
\yopsep\
to
\ymixsyststatefinal,
the output state of the separating system becomes equal to
\ymixsyststateinitial\
(up to estimation errors)
and this analysis does not depend at all whether
\ymixsyststateinitial\
is entangled or not.
In particular, if the proposed BQSS (and hence BQSR) method
is used to restore the initial state of a qubit register
by compensating for the undesired coupling between its
qubits, as discussed above, this means that this method
also applies when an entangled state is stored in this
qubit 
\ymodifartitionehundredthirtyninevonestepone{register.}

\subsection{Blind quantum channel equalization}\label{sec-quantum-channel-invert}
As discussed
in Sections
\ref{sec-intro},
\ref{sec-machine-learning}
and
\ref{sec-quantum-channel-estim},
in the classical and quantum frameworks,
the same information processing task is given different
names depending on the considered application field.
This 
is also true for
the generic QIP
problem, related to system inversion, that we initially defined for
non-entangled states at the beginning of Section
\ref{sec-blind-quantum-source-sep} ,
that we then extended to possibly entangled states in
Section
\ref{sec-bqrs} and that may be summarized as follows:
a user is given a set of
transformed states
$|\Yqubitnbarbtimenonestatenot^{\prime}\rangle
=
\Ymapfunc(\Yqubitnbarbtimenonestate)$,
but does not know the
original 
states
\yqubitnbarbtimenonestate,
nor
the mapping function
\ymapfunc;
the user wants to restore the 
original
states
\yqubitnbarbtimenonestate.
Whereas we illustrated that QIP task with one type of
application in Sections
\ref{sec-blind-quantum-source-sep}
and
\ref{sec-bqrs},
we anticipate that it will have various other applications
as the quite general field of QIP keeps on growing.
In particular, this problem too
may be rephrased as a 
quantum communication scenario.
The above user is then the receiver, who only knows the received
states 
$|\Yqubitnbarbtimenonestatenot^{\prime}\rangle$, that have been
altered by the channel
\ymapfunc .
Without knowing that channel, the receiver aims at restoring the
emitted states
\yqubitnbarbtimenonestate\
that he does not know either
(or the information, in classical form, contained in these
states \yqubitnbarbtimenonestate).
This therefore corresponds to the
blind quantum channel equalization problem
and the generic methods that we proposed in 
Sections
\ref{sec-blind-quantum-source-sep}
and
\ref{sec-bqrs}
also allow one to solve this problem, moreover with
the advantages of the proposed SIPQIP framework.

\section{%
Contributing
to 
quantum classification}
\label{sec-classif}
We now come back to the other main aspect
of machine learning
discussed in Section
\ref{sec-machine-learning},
namely classification.
In the classical framework, 
many classification algorithms
receive data that consist of vectors which contain
features that characterize the ``objects'' to be classified
\cite{amq106}.
These
algorithms
heavily rely on
computing the 
dot (i.e., scalar or inner) product
$\Yvecnotstdoneindexstdone\Ytranspose
\Yvecnotstdoneindexstdtwo$
of two
column vectors
\yvecnotstdoneindexstdone\
and
\yvecnotstdoneindexstdtwo,
where
\ytranspose\
stands for transpose,
or on computing the distance
$||\Yvecnotstdoneindexstdone
-
\Yvecnotstdoneindexstdtwo||$
between the associated two data points
\cite{amq106}.
These two quantities are moreover directly connected, since
\yeqdeflinkdotproductdistance
\ymodifartitionehundredthirtyninevonestepone{In particular, for any unit-norm vectors, this}
yields
\yeqdeflinkdotproductdistanceunitnorm
Using the signal%
/data
processing terminology,
$\Yvecnotstdoneindexstdone\Ytranspose
\Yvecnotstdoneindexstdtwo$
is also the basic, i.e. non-centered and non-normalized, correlation
parameter of the data vectors
\yvecnotstdoneindexstdone\
and
\yvecnotstdoneindexstdtwo,
whereas their non-centered correlation coefficient 
(also called the cosine similarity
\cite{amq106})
is
\yeqdefcorrcoef
These two correlation parameters 
coincide
for
unit-norm vectors.
The correlation coefficient
\ycorrcoefvecnotstdoneindicesstdonetwo\
is e.g. widely used for data characterization and classification by the
Earth observation (i.e. remote sensing) community.
Each of the vectors
\yvecnotstdoneindexstdone\
and
\yvecnotstdoneindexstdtwo\
then typically defines
a spectrum, which consists of
the light reflectance 
values of a
material at a set of 
\ymodifartitionehundredthirtyninevonestepone{``frequencies''}
(in fact, narrow spectral bands)
\cite{amoi6-48}.
More precisely,
one then computes
the arccosine of 
\ycorrcoefvecnotstdoneindicesstdonetwo,
which is equal to the angle between
\yvecnotstdoneindexstdone\
and
\yvecnotstdoneindexstdtwo\
\cite{amq106}
and is therefore
called the ``Spectral Angle Mapper'' (SAM)
between these vectors
\cite{a648}.
A 
low
value of that SAM corresponds
to a 
high
value of the ``spectral similarity''
of the considered materials,
i.e. of their similarity
in terms of the shape 
of the variations of their reflectance functions with respect to
\ymodifartitionehundredthirtyninevonestepone{frequency}
(where ``shape'' means regardless of
their global scale:
that scale
has no influence
on
(\ref{eq-correlcoef}) and hence on SAM).
Similar approaches are used in the field of Astrophysics,
with
spectral data vectors which consist of luminance values
(i.e. direct light flux from the observed object),
instead of reflectance.

Let us now consider the situation when data that are initially in
classical form are to be classified by using a quantum classifier,
in order to achieve higher classification speed
\cite{amq96,
amq84,
amq106}.
This first requires one to transform the initial classical data into
quantum states.
To this end,
each 
classical-form vector
is stored in a qubit register with index \yregisterindexstdone,
that consists of
\yqubitnbarb\ qubits.
Each individual qubit is thus indexed by
\yregisterindexstdone\
and
$
\Yqubitindexstdtwo
\in
\{1,\dots,\Yqubitnbarb\}$.
Its state space is denoted as
\yqubitspaceregisterindexstdonequbitindexstdtwo\
and a basis of this space is composed of the two kets
\yqubitspaceregisterindexstdonequbitindexstdtwobasisstateintstd\
with
$\Yqubitspaceregisterindexstdonequbitindexstdtwobasisintstd
\in
\{0,1\}$.
All 
\ymodifartitionehundredthirtyninevonestepone{deterministic}
pure states of the 
\ymodifartitionehundredthirtyninevonestepone{qubit}
register 
\yregisterindexstdone\
then belong to the
space
$\Yqubitspaceregisterindexstdonequbitindexone
\otimes
\dots
\Yqubitspaceregisterindexstdonequbitindexqubitnbarb$
and read
\yeqdefstateregisterindexstdone
where the compact notation
\ysetnotargqubitspaceregisterindexstdonequbitanybasisintstd\
means:
the set of
all values
of the ordered set of
\yqubitnbarb\
integers
\yqubitspaceregisterindexstdonequbitindexstdtwobasisintstd\
corresponding to the fixed value
\yregisterindexstdone\
and to all values
$
\Yqubitindexstdtwo
\in
\{1,\dots,\Yqubitnbarb\}$,
again
with
$\Yqubitspaceregisterindexstdonequbitindexstdtwobasisintstd
\in
\{0,1\}$. Besides,
the 
$2^{\Yqubitnbarb}$
complex-valued coefficients
\yregisterstdonequbitspaceregisterindexstdonequbitsonetolastbasisstatecoefstd\
are indexed by the index
\yregisterindexstdone\
of the considered register and by
all integers
\yqubitspaceregisterindexstdonequbitindexstdtwobasisintstd\
which define to which basis state
each coefficient
\yregisterstdonequbitspaceregisterindexstdonequbitsonetolastbasisstatecoefstd\
corresponds.
These coefficients are
such that
\ystateregisterindexstdone\
has unit norm.
Let us then consider
a classical-form complex-valued 
unit-norm vector
\yvecnotstdoneindexstdone\
with dimension
$2^{\Yqubitnbarb}$
(or lower: zero-valued components are then added to
\yvecnotstdoneindexstdone\
to reach
$2^{\Yqubitnbarb}$
components).
This vector may be 
\ymodifartitionehundredthirtyninevonesteptwo{stored in}
a ket
\ystateregisterindexstdone\
defined by
(\ref{eq-stateregisterindexstdone}),
by setting the coefficients
\yregisterstdonequbitspaceregisterindexstdonequbitsonetolastbasisstatecoefstd\
of
\ystateregisterindexstdone\
respectively to the values of the components
of
\yvecnotstdoneindexstdone\
\ymodifartitionehundredthirtyninevonesteptwo{(a common phase
reference may be used for all considered kets).}
If the norm of \yvecnotstdoneindexstdone\
is not equal to one, it may be
handled
separately, 
while
$\Yvecnotstdoneindexstdone / 
||\Yvecnotstdoneindexstdone||$
is stored in
\ystateregisterindexstdone ,
as stated in
\cite{amq96}.

The above-defined kets
(\ref{eq-stateregisterindexstdone})
may then be 
employed
in quantum classifiers, which often use
(i)
the dot product
$
\langle
\Ystateregisternot
					_{1}
\Ystateregisterindextwo
$
of 
such kets
or
(ii)
the squared modulus
of this dot product, 
which is called the overlap
of 
these kets
(see
\cite{amq108arxiv,
livreperes1995}
or
\cite{amq106} p. 120),
or (iii) the distance between
the points associated with such kets,
as e.g. discussed in
\cite{amq96,
amq84,
amq106}.
The
dot product formally associated with
two states
(\ref{eq-stateregisterindexstdone})
respectively stored in registers
with indices
$\Yregisterindexstdone = 1$
and
$\Yregisterindexstdone = 2$
is 
defined as if these kets belonged to the same
state space. This dot product therefore reads
\yeqdefstateregisterindexonedottwo
where
$^*$ stands for complex conjugate.
Some quantum circuits were proposed in the literature
for computing the corresponding state overlap.
A 
widely used approach,
called the swap test, 
was 
proposed 
in
\cite{amq107}
to essentially test
the equality of two states. 
The quantity used to this end is 
the probability of an
outcome of a measurement 
performed at the output of the
considered circuit. 
This quantity is equal to 0 if the
considered states are equal, and 
essentially equal to 1/2 otherwise
(more precisely, it is higher than a bound close to 1/2
if the states are far enough from one another).
Beyond this binary decision, this probability is a
continuous-valued quantity, which may be shown to
be 
linearly
related
to the overlap
of the considered
quantum
states
(part of the corresponding calculations are provided in
\cite{amq107,
amq106}).
Another approach for 
computing the overlap of quantum states
is based on
the circuit of Fig. 6(B) of 
\cite{amq108arxiv}.
The behavior of that circuit is only briefly defined in
\cite{amq108arxiv}, 
which outlines how to express the overlap 
associated with
the density operators 
of the two multi-qubit inputs
of the considered
circuit
as the result of classical
post-processing
applied to the results of measurements performed at the
output of that
circuit.
Our own calculations
(to be 
detailed
elsewhere),
performed for 
\ymodifartitionehundredthirtyninevonestepone{deterministic}
pure input states
\ystateregisterindexone\
and
\ystateregisterindextwo,
confirm that the 
squared modulus of
(\ref{eq-stateregisterindexonedottwo})
may be expressed as a 
(multistage)
linear combination
of probabilities of outcomes of measurements
performed at the output of that
quantum circuit.
\ymodifartitionehundredthirtyninevonesteptwo{In appendix
\ref{sec-compute-dot-product},
we show how the above quantum circuits may be further exploited
in order to compute
dot products
$
\langle
\Ystateregisternot
					_{1}
\Ystateregisterindextwo
$,
not only their (squared) moduli.}

The above 
\ymodifartitionehundredthirtyninevonesteptwo{dot products or
overlaps}
may 
be 
used in various ways in the general
framework of quantum classification.
More specifically,
we hereafter show how
enhanced approaches may be 
developed
by combining 
quantum classification principles
that use 
\ymodifartitionehundredthirtyninevonesteptwo{dot products or
overlaps}
with
our SIPQIP, i.e.
single-preparation, concept.
We 
illustrate this approach 
with a first original contribution
to single-preparation
quantum classification, that will be extended
in future papers.
In this contribution, we focus on the second phase
of the operation of a 
classifier, that is
on the ``resolution phase''
defined in Section
\ref{sec-machine-learning},
which takes place after the (unsupervised or supervised) learning phase.
This is, by the way, similar to what we did for BQSS, 
by first
applying our SIPQIP concept to the second phase of operation,
i.e. the inversion phase, before we extended it
to the first, i.e. adaptation, phase as explained in Section
\ref{sec-blind-quantum-source-sep}.

In the proposed approach,
we consider the situation when
the classical-form vectors to be classified are
characterized by their shapes, not their magnitudes,
e.g. as in the Earth observation and Astrophysics applications
outlined
at the beginning of the present section.
Therefore, these vectors may initially
be rescaled 
to
have unit norm,
so that this norm is not an issue when transforming these
classical vectors into
quantum states.
We hereafter
address
the general situation when the considered
classification
problem involves
\yclassnb\
classes,
indexed by
\yclassindexstd,
with
$\Yclassindexstd
\in
\{1,\dots,\Yclassnb\}$.
\ymodifartitionehundredthirtyninevonesteptwo{Moreover, we
consider the usual case when the classical-form data vectors,
and hence the associated dot products of kets, are real-valued.}

To describe
how classification is here performed, let us first
consider the non-realistic situation when each class
with index 
\yclassindexstd\
is initially defined by a single known classical-form vector
\yvecnotstdoneclassstdvecone\
and hence a single associated quantum state
\ystatevecnotstdoneclassstdvecone.
When analyzing
a new ``object'' of the considered application
(e.g. the spectrum of an unknown material in the
above Earth observation or Astrophysics applications),
represented by a quantum state
\ystatevectobeclassified,
a basic method for classifying that object consists
of separately estimating its 
\ymodifartitionehundredthirtyninevonesteptwo{dot product
$\langle \Ystatevectobeclassifiednot \Ystatevecnotstdoneclassstdvecone$}
with each of the states
\ystatevecnotstdoneclassstdvecone\
and in deciding that this object belongs to
the
class 
which
yields the highest estimated value of 
\ymodifartitionehundredthirtyninevonesteptwo{the dot product
$\langle \Ystatevectobeclassifiednot \Ystatevecnotstdoneclassstdvecone$,}
i.e. the best
\ymodifartitionehundredthirtyninevonesteptwo{similarity}
with
\ystatevectobeclassified.
\ymodifartitionehundredthirtyninevonesteptwo{
This approach may be simplified as follows in
the case when
the components of the
considered classical-form data vectors are
nonnegative,
which e.g. applies to
the reflectance
or luminance values that compose the above-mentioned
spectra.
In that case,
the 
square root of the overlap 
$|\langle \Ystatevectobeclassifiednot \Ystatevecnotstdoneclassstdvecone|^2$
coincides with the
corresponding dot product
$\langle \Ystatevectobeclassifiednot \Ystatevecnotstdoneclassstdvecone$.
This overlap is 
therefore sufficient for measuring
similarity
in that case (as opposed to the sign indeterminacy
that it yields with respect to the dot product for possibly negative data).
The above classifier then operates equivalently by
deciding that the considered object belongs to
the
class 
which
yields the highest estimated value of 
overlap
$|\langle \Ystatevectobeclassifiednot \Ystatevecnotstdoneclassstdvecone|^2$.
This is attractive, because an overlap is computed more easily that
the corresponding dot product, as shown in Appendix 
\ref{sec-compute-dot-product}.}

An improved variant of the above
classification method
employs
a 
user-defined threshold in addition%
\ymodifartitionehundredthirtyninevonestepone{, in order to
achieve the rejection capability defined in Section
\ref{sec-machine-learning}:}
if the highest of the above
\ymodifartitionehundredthirtyninevonesteptwo{dot products
(or overlaps, in the simplified version)}
remains lower than this threshold,
the considered object is ``rejected'',
i.e. the classifier decides that it is not able
to classify that object, because it is not similar
enough to any of the classes of objects that are
known in the considered problem.

\ymodifartitionehundredthirtyninevonesteptwo{%
All these classifiers
are based on computing overlaps, 
because their
decisions are either directly based on such overlaps or 
based on dot products, that may be derived
from overlaps, as explained
in Appendix 
\ref{sec-compute-dot-product}.
Each of these overlaps, such as
$|\langle \Ystatevectobeclassifiednot \Ystatevecnotstdoneclassstdvecone|^2$,}
is typically estimated by using the sample
frequency estimate(s) of one or several types of probabilities associated
with overlap in the quantum circuits that were
defined above for estimating overlaps.
This
\ymodifartitionehundredthirtyninevonesteptwo{then}
means that, for each class
\yclassindexstd,
many copies 
(typically $10^5$, as explained in Section
\ref{sec-stochastic-QIP-standard-concepts})
of the state
\ystatevecnotstdoneclassstdvecone\
must be prepared to estimate these probabilities.

Now consider the realistic version of the above problem,
when
each class
with index 
\yclassindexstd\
is initially defined by a full set 
of classical-form vectors
\yvecnotstdoneclassstdvecindexstdone,
with a vector index
\yvecindexstdone\ ranging from 1 to a maximum value that may
depend on the class.
These vectors are then transformed into
quantum states
\ystatevecnotstdoneclassstdvecindexstdone.
\ymodifartitionehundredthirtyninevonesteptwo{The above classifiers
may be extended as follows for this situation, focusing on
their version that directly bases its decisions on overlaps, for the
sake of clarity. 
For}
each class, one may first
compute a full set of (estimates of) overlaps
$|\langle \Ystatevectobeclassifiednot \Ystatevecnotstdoneclassstdvecindexstdone|^2$
and these quantities should then be reduced to a single parameter
that characterizes the overall similarity of the considered class
with
\ystatevectobeclassified.
A natural parameter that may be used to this end is the
mean
of all 
overlaps
$|\langle \Ystatevectobeclassifiednot \Ystatevecnotstdoneclassstdvecindexstdone|^2$
associated with the considered class.
Here again, 
in practice, 
only an estimate of this mean overlap is obtained,
by using
various
quantum state preparations and measurements.
However, unlike in the above non-realistic scenario, this may
here be achieved with two quite different approaches.
The first approach, which might
be considered as the most
natural one if disregarding our previous contributions in this paper,
consists of separately estimating each of the 
overlaps
$|\langle \Ystatevectobeclassifiednot \Ystatevecnotstdoneclassstdvecindexstdone|^2$
as above,
therefore
typically preparing
$10^5$ 
copies of each state
\ystatevecnotstdoneclassstdvecindexstdone\
(and of
\ystatevectobeclassified),
and then computing 
(on a classical computer)
the mean of these estimated overlaps.
This is an application of
the standard, multiple-preparation, approach
defined in Section
\ref{sec-stochastic-QIP-standard-concepts}.
However, we stress that we here
only aim at
computing
the \emph{mean} of this finite set of overlaps, 
so that we only need to estimate
the mean(s) of the corresponding set(s) of probabilities
(as explained above,
this involves one or several types of probabilities, depending on
the considered quantum circuit).
In Section
\ref{sec-stochastic-QIP-introduce-sample-mean},
we showed that this may
be performed much more efficiently 
by using our SIPQIP framework, 
which here means decreasing 
the number
of 
preparations per state
\ystatevecnotstdoneclassstdvecindexstdone\
and taking advantage of the averaging of measurement results
that is then performed over all these states
(thus still requesting one copy of
\ystatevectobeclassified\
per measurement).
This number of preparations per state
may even be decreased down to one 
if enough different states 
\ystatevecnotstdoneclassstdvecindexstdone\
are available
to reach 
a high enough estimation accuracy:
in 
\cite{amoi6-118}
and Section
\ref{sec-blind-hamilton-estim-tests}
of the present paper,
we 
analyzed
the 
numerical performance achieved by this
SIPQIP approach 
for the BQPT and BHPE tasks,
and we plan to 
investigate it
for classification in future papers.
In the literature, quantum classifiers have especially been
considered for big-data, 
i.e. 
large-scale, 
applications
\cite{amq84,
amq96}.
In such applications,
the above-mentioned large number of states
\ystatevecnotstdoneclassstdvecindexstdone\
will actually be available
and
our SIPQIP framework
will take full advantage of it
(besides, it can also attractively 
operate with a somewhat lower
total
number of states 
\ystatevecnotstdoneclassstdvecindexstdone\
and a number of preparations per state
somewhat higher than one).

\section{Conclusion}
\label{sec-concl}
The term ``machine learning'' especially refers to
algorithms (and associated systems) that derive mappings,
i.e. intput/output transforms, by using
numerical data 
that
provide information about
the transform which is of interest in the
considered application.
The data processing tasks to be performed in these applications
not only include classification and regression, but also
system identification, system inversion and input signal
restoration (or source separation when considering several signals).
Whereas these problems have been and are still widely
investigated in a purely classical framework,
part
of them are currently being extended
to configurations which involve quantum-form
data and/or quantum processing means.
Within this general quantum framework, we here
tackled
the most challenging
configurations from two points of view.
First, almost all this paper
is devoted to
\emph{unsupervised}, i.e. blind, configurations,
which have not been addressed 
in the literature
for 
most of the tasks
considered 
here.
Unsupervised learning is very attractive because,
as detailed in Sections
\ref{sec-intro}
and
\ref{sec-machine-learning},
it avoids
the need for known ``reference values'' (e.g.,
input values
for system identification)
to learn the required mappings.
Second, we here
mainly aim at extending a variety of aspects of quantum
machine learning 
by introducing new
algorithms
which can operate with only one instance of each
prepared 
state
\ymodifartitionehundredthirtyninevonestepone{(where the term
``preparation'' is used for both deterministic and random
pure states, as explained in Section \ref{sec-intro}).}
This 
approach first
avoids the burden of having to 
prepare many ideally identical
copies of each used state
in
order to 
compute statistical parameters separately for each such state.
Moreover, this approach
yields 
much better 
performance 
than the multiple-preparation approach
for a given total number
of 
state
preparations, 
as shown for blind quantum process tomography
in our very recent
paper
\cite{amoi6-118}
and confirmed here
by our new results for 
blind
Hamiltonian parameter estimation.
Besides,
this original single-preparation
approach
is especially of interest when combined with unsupervised
learning, because 
using 
the multiple-preparation approach 
instead,
in the unsupervised framework,
would mean allowing 
the ``reference values'' 
to be unknown
but 
still requesting 
that the \emph{same} (unknown) reference value be prepared many times,
which would still
require significant control 
in
the considered 
quantum learning procedure,
so that this procedure 
would 
be ``less unsupervised''.

The above 
concepts thus result in
a general 
SIngle-Preparation
Quantum Information Processing (SIPQIP) framework.
We illustrated it for various processing
tasks, 
including with a quantitative
evaluation of the numerical performance that it yields.
For the tasks related to
the blind, i.e. unsupervised, version of
system identification
(including quantum process tomography
and Hamiltonian parameter estimation), 
system inversion and
signal restoration (including source separation),
we 
showed how to apply the proposed approach to a concrete example,
related to spintronics,
which involves Heisenberg coupling
between two qubits.
Starting from the explicit algorithms and system architectures
that we detailed for this configuration, the reader may then adapt
them to other types of processes, and we 
will also 
extend this approach to other
processes in the future.
Similarly, we 
provided a first 
illustration
of the application of this SIPQIP framework to quantum
classification and we plan to report extensions 
of this approach in future papers.

Moreover,
when aiming at
compensating for undesired
Heisenberg coupling between qubits,
we proposed two
quantum system architectures:
see 
Fig.
\ref{fig-quantmix_Q_Dtilde_Q_adapt_out_conv}
for a 
feedback structure
and
Fig.
\ref{fig-quantmix_Q_Dtilde_Q_adapt_in_conv_v2}
for a 
feedforward structure.
These architectures
open the way to the much more general concept of 
\ymodifartitionehundredthirtyninevonestepone{``self-adaptive}
quantum gates'', 
i.e. gates which include
the following two features:
\begin{enumerate}
\item
some means for
controlling the values of parameters that define
the quantum state transform that such a gate performs
within a predefined class of transforms,
\item
an autonomous (i.e. blind
or unsupervised) algorithm which controls the adaptation
of these parameter values, 
so as to achieve a predefined condition,
that 
could consist of 
ensuring
output disentanglement, as in the above example,
or that could be a
counterpart of that condition, depending on
the available data and on
the type of undesired behavior
that one wants to
compensate for.
\end{enumerate}
Such gates would especially be of interest for a 
\emph{qua}ntum co\emph{mputer} 
that, by the way, we proposed
to more briefly call a ``quamputer'':
by 
adequately 
selecting the above-mentioned adaptation condition
and designing an associated adaptation
algorithm, 
one could create a 
\ymodifartitionehundredthirtyninevonestepone{self-adaptive}
quantum gate that automatically compensates for 
a given type of
non-ideality 
(instead of undesired Heisenberg coupling in the above example)
that occurs
e.g. in a gate that precedes the considered 
\ymodifartitionehundredthirtyninevonestepone{self-adaptive}
gate,
thus allowing practical future quamputers 
to operate correctly despite these
non-idealities,
thanks to their internal compensation means.

\appendix
\section
{%
Connection 
of single-preparation 
and multiple-preparation
QIP
with
classical adaptive
processing}
\label{sec-stochastic-QIP-link-with-classical}
Various types of
classical 
data processing methods are based on adapting the parameter
values
of a system, e.g. of 
a filter,
an artificial neural network
(including 
the above-mentioned
deep learning approaches)
or a blind source separation system
(see e.g.
\cite{book-comon-jutten-ap,
amoi6-48,
bibref-efive-ta-chap-adap-book-haykin,
elivrehertz,
icabook-oja,
a616,
eelevenlivretheodoridistheory,
a235,
book-widrow-stearns}
and
references in Sections
\ref{sec-intro}
and
\ref{sec-machine-learning}).
This adaptation is based on
a set of (often multidimensional)
data samples,
which are typically 
used
to minimize a cost function that
depends on the
considered problem.
This yields the following two approaches.

The first approach
corresponds to
so-called 
batch 
algorithms
\cite{icabook-oja,
eelevenlivretheodoridistheory}.
These algorithms are often iterative, 
and 
each of their 
steps 
uses the complete set of available data
samples.
This especially includes
gradient
descent algorithms
\cite{elivrehertz}
(i.e., steepest descent algorithms
\cite{bibref-efive-ta-chap-adap-book-haykin}),
where each
step 
uses the gradient of the
cost function defined by all data samples to perform one update for 
all parameters.

The cost functions of
the above 
algorithms,
and hence their gradients,  are often defined as the
mean of an ``elementary term'' 
associated with a single
data sample, 
where this mean is the expectation
if considering a probabilistic representation, 
or
the mean 
(or sum)
over
all data samples if using an empirical framework.
The second type of adaptation algorithms is 
then 
derived from
the first one
essentially
by removing
the above mean 
from the update rule used in each adaptation step
\cite{bibref-efive-ta-chap-adap-book-haykin,
elivrehertz,
book-widrow-stearns}
(see e.g.
\cite{eelevenlivretheodoridistheory}
for comments about Robbins-Monro algorithms
and 
stochastic approximation theory).
Each such step thus 
uses only the ``elementary term'' associated with
a \emph{single} data sample 
(or a few samples \cite{a616}), 
where this term
corresponds to a random
variable 
in a probabilistic framework 
(as opposed to the
expectation of these terms used in 
deterministic,
batch, 
algorithms).
This yields so-called 
online
\cite{icabook-oja,
eelevenlivretheodoridistheory}
or stochastic 
algorithms, 
and especially
stochastic gradient descent algorithms
\cite{bibref-efive-ta-chap-adap-book-haykin,
a616},
including 
the famous
least-mean-square (LMS) 
or Widrow-Hoff algorithm
\cite{bibref-efive-ta-chap-adap-book-haykin,
elivrehertz,
eelevenlivretheodoridistheory,
a235,
book-widrow-stearns}.

If only considering a \emph{single} step of the above second type
of algorithms, these algorithms may at first glance appear not to
be sound, because such a step only uses the elementary term
associated with
a single data sample
to estimate the relevant expectation derived for the above
first type of algorithms.
However, one should instead consider the \emph{complete} set of
adaptation steps, which uses 
(once or repeatedly
\cite{eelevenlivretheodoridistheory})
\emph{all} data points: 
this overall set of steps
of the
algorithm eventually yields relevant behavior.

Unlike most
above classical methods, the procedure that we 
described
in
Section
\ref{sec-stochastic-QIP-introduce} 
does not concern iterative algorithms for
optimizing a cost function.
However, these two 
types of methods share a two-level structure,
where the lower level uses only one 
\ymodifartitionehundredthirtyninevonesteptwo{(or}
a few) data sample(s),
and
the higher level gathers all
the contributions 
derived from
the lower level, thus extracting
overall
information
that is relevant for the considered applications.

The lower level of these
quantum methods might therefore be stated to be ``stochastic''
(in the sense ``using a single data sample''),
but that might be misleading for some QIP tasks, 
so that we instead
call them ``single-preparation (QIP) methods'', for the sake of clarity.
Therefore, the QIP methods based on the approach of Section
\ref{sec-stochastic-QIP-standard-concepts}
are called ``multiple-preparation (QIP) methods'', whereas they might have
been called ``batch (QIP) methods'', as a reference to the
above batch classical methods.
\section
{Validating
the 
single-preparation
QIP 
framework
with Kolmogorov's approach%
}
\label{sec-stochastic-QIP-validate}
One may decide to 
use the 
approach defined in
Section
\ref{sec-stochastic-QIP-introduce}
only as a means for \emph{proposing} to use
the expression
obtained in
{(\ref{eq-qubitnbarbspaceoveralleventindexequalstdwithvecbasisprobexpectapproxtwotwoqubitseqnb-indicfunc})}
as an estimator 
of the quantity of interest,
that is of
\yqubitnbarbspaceoveralleventindexequalstdwithvecbasisprobexpect\
for any
$
\Yqubitnbarbstatecoefindexequalstd
\in
\{
1,
\dots
,
2
^
{
\Yqubitnbarb
}
\}
$,
and to then \emph{validate} the relevance
of this estimator from scratch,
by using Kolmogorov's view of probabilities
(see e.g.
\cite{book-papoulis}).
This validation consists of deriving the mean 
(and hence bias) 
and
variance of this estimator.
It is provided hereafter, first for a finite number
\yqubitnbarbstatewritereadseqnb\ of trials,
and then for the asymptotic case
$
\Yqubitnbarbstatewritereadseqnb
\rightarrow
+
\infty
$.

The result of the
trial with index
\yqubitnbarbstatewritereadseqindex ,
which corresponds to the indicator function in
(\ref{eq-qubitnbarbspaceoveralleventindexequalstdwithvecbasisprobexpectapproxtwotwoqubitseqnb-indicfunc}),
is here represented by
a binary-valued 
random variable 
denoted as
\yqubitnbarbspaceoveralleventindexequalstdwithvecbasisrvqubitnbarbstatewritereadseqindex ,
which takes the values 1 and 0 respectively with
probabilities
\yqubitnbarbspaceoveralleventindexequalstdwithvecbasisprobstatewritereadseqindex\
(defined as 
in
Section
\ref{sec-stochastic-QIP-introduce})
and
$
[
1
-
\Yqubitnbarbspaceoveralleventindexequalstdwithvecbasisprobstatewritereadseqindex
]
$.
Then, the overall result
(\ref{eq-qubitnbarbspaceoveralleventindexequalstdwithvecbasisprobexpectapproxtwotwoqubitseqnb-indicfunc})
for all
\yqubitnbarbstatewritereadseqnb\
trials
is represented by
the 
random variable
\begin{equation}
\Yqubitnbarbspaceoveralleventindexequalstdwithvecbasisrvmeanqubitnbarbstatewritereadseqnb
=
\frac
{
\sum
_{
\Yqubitnbarbstatewritereadseqindex
=
1
}
^{
\Yqubitnbarbstatewritereadseqnb
}
\Yqubitnbarbspaceoveralleventindexequalstdwithvecbasisrvqubitnbarbstatewritereadseqindex
}
{
\Yqubitnbarbstatewritereadseqnb
}
\label{eq-def-qubitnbarbspaceoveralleventindexequalstdwithvecbasisrvmeanqubitnbarbstatewritereadseqnb}
\end{equation}
{where 
we omit in notations that
all
\yqubitnbarbspaceoveralleventindexequalstdwithvecbasisrvqubitnbarbstatewritereadseqindex\
and hence
\yqubitnbarbspaceoveralleventindexequalstdwithvecbasisrvmeanqubitnbarbstatewritereadseqnb\
depend on \yqubitnbarbstatecoefindexequalstd .}
~\\
~\\
\subsection{Mean
of
\yqubitnbarbspaceoveralleventindexequalstdwithvecbasisrvmeanqubitnbarbstatewritereadseqnb\
for a finite
\yqubitnbarbstatewritereadseqnb
}
As a first
scenario,
let us consider the case when
{%
the pure states
\yqubitnbarbtimenonestateseqindex\
and hence%
}
the quantities
\yqubitnbarbspaceoveralleventindexequalstdwithvecbasisprobstatewritereadseqindex\
are deterministic
{and the number
\ywritereadonestatenb\
of
copies
of each state
\yqubitnbarbtimenonestateseqindex\ is arbitrary
(including
$
\Ywritereadonestatenb
=
1
$)%
.}
It is then easily shown that
\begin{equation}
E_{1}
\{
\Yqubitnbarbspaceoveralleventindexequalstdwithvecbasisrvqubitnbarbstatewritereadseqindex
\}
=
\Yqubitnbarbspaceoveralleventindexequalstdwithvecbasisprobstatewritereadseqindex
\end{equation}
where
$
E_{1}
\{
.
\}
$
stands for expectation, 
i.e. statistical averaging with respect to the possible values
{(1 and 0)}
of
\yqubitnbarbspaceoveralleventindexequalstdwithvecbasisrvqubitnbarbstatewritereadseqindex ,
with their above fixed probabilities,
in this first scenario.
Eq.
(\ref{eq-def-qubitnbarbspaceoveralleventindexequalstdwithvecbasisrvmeanqubitnbarbstatewritereadseqnb})
then yields
\begin{equation}
E_{1}
\{
\Yqubitnbarbspaceoveralleventindexequalstdwithvecbasisrvmeanqubitnbarbstatewritereadseqnb
\}
=
\frac
{
\sum
_{
\Yqubitnbarbstatewritereadseqindex
=
1
}
^{
\Yqubitnbarbstatewritereadseqnb
}
\Yqubitnbarbspaceoveralleventindexequalstdwithvecbasisprobstatewritereadseqindex
}
{
\Yqubitnbarbstatewritereadseqnb
}
.
\label{eq-qubitnbarbspaceoveralleventindexequalstdwithvecbasisrvmeanqubitnbarbstatewritereadseqnb-mean-scenario-one}
\end{equation}
Then, the second scenario, which is the one of interest here, is
when the 
quantities
\yqubitnbarbspaceoveralleventindexequalstdwithvecbasisprobstatewritereadseqindex\
are stochastic and 
their samples are drawn with 
the same 
{statistical}
distribution
for all values of
\yqubitnbarbstatewritereadseqindex .
The 
{expectation}
of this 
distribution 
{is%
}
denoted as
\yqubitnbarbspaceoveralleventindexequalstdwithvecbasisprobexpect
{, since it does not depend on
\yqubitnbarbstatewritereadseqindex .}
The 
{expectation}
$
E_{2}
\{
\Yqubitnbarbspaceoveralleventindexequalstdwithvecbasisrvmeanqubitnbarbstatewritereadseqnb
\}
$
of
\yqubitnbarbspaceoveralleventindexequalstdwithvecbasisrvmeanqubitnbarbstatewritereadseqnb\
is then obtained
by performing
statistical averaging
not only
with respect to the possible values
of
\yqubitnbarbspaceoveralleventindexequalstdwithvecbasisrvqubitnbarbstatewritereadseqindex ,
as above,
but also
with respect to the possible values of the
quantities
\yqubitnbarbspaceoveralleventindexequalstdwithvecbasisprobstatewritereadseqindex .
Applying the latter averaging to
(\ref{eq-qubitnbarbspaceoveralleventindexequalstdwithvecbasisrvmeanqubitnbarbstatewritereadseqnb-mean-scenario-one})
yields
\begin{eqnarray}
E_{2}
\{
\Yqubitnbarbspaceoveralleventindexequalstdwithvecbasisrvmeanqubitnbarbstatewritereadseqnb
\}
&
=
&
\frac
{
\sum
_{
\Yqubitnbarbstatewritereadseqindex
=
1
}
^{
\Yqubitnbarbstatewritereadseqnb
}
E
\{
\Yqubitnbarbspaceoveralleventindexequalstdwithvecbasisprobstatewritereadseqindex
\}
}
{
\Yqubitnbarbstatewritereadseqnb
}
\\
&
=
&
\Yqubitnbarbspaceoveralleventindexequalstdwithvecbasisprobexpect
.
\label{eq-qubitnbarbspaceoveralleventindexequalstdwithvecbasisrvmeanqubitnbarbstatewritereadseqnb-mean-scenario-two-express-two}
\end{eqnarray}
So, for a finite 
\yqubitnbarbstatewritereadseqnb ,
the 
{expectation}
of the proposed estimator
\yqubitnbarbspaceoveralleventindexequalstdwithvecbasisrvmeanqubitnbarbstatewritereadseqnb\
of
\yqubitnbarbspaceoveralleventindexequalstdwithvecbasisprobexpect\
is equal to
\yqubitnbarbspaceoveralleventindexequalstdwithvecbasisprobexpect ,
i.e. this estimator is unbiased.

\subsection{Asymptotic
mean 
of
\yqubitnbarbspaceoveralleventindexequalstdwithvecbasisrvmeanqubitnbarbstatewritereadseqnb 
}
Starting from the above second scenario, 
when
\yqubitnbarbstatewritereadseqnb\
tends to infinity,
the above mean
(\ref{eq-qubitnbarbspaceoveralleventindexequalstdwithvecbasisrvmeanqubitnbarbstatewritereadseqnb-mean-scenario-two-express-two})
of course remains equal to
\yqubitnbarbspaceoveralleventindexequalstdwithvecbasisprobexpect .
The proposed estimator
{%
\yqubitnbarbspaceoveralleventindexequalstdwithvecbasisrvmeanqubitnbarbstatewritereadseqnb
}
of
\yqubitnbarbspaceoveralleventindexequalstdwithvecbasisprobexpect\
is therefore asymptotically unbiased.

It should be noted that, in the 
above first scenario,
(\ref{eq-qubitnbarbspaceoveralleventindexequalstdwithvecbasisrvmeanqubitnbarbstatewritereadseqnb-mean-scenario-one})
similarly
yields
\begin{equation}
\lim_{
\Yqubitnbarbstatewritereadseqnb
\rightarrow
+
\infty
}
E_{1}
\{
\Yqubitnbarbspaceoveralleventindexequalstdwithvecbasisrvmeanqubitnbarbstatewritereadseqnb
\}
=
\lim_{
\Yqubitnbarbstatewritereadseqnb
\rightarrow
+
\infty
}
\frac
{
\sum
_{
\Yqubitnbarbstatewritereadseqindex
=
1
}
^{
\Yqubitnbarbstatewritereadseqnb
}
\Yqubitnbarbspaceoveralleventindexequalstdwithvecbasisprobstatewritereadseqindex
}
{
\Yqubitnbarbstatewritereadseqnb
}
\label{eq-qubitnbarbspaceoveralleventindexequalstdwithvecbasisrvmeanqubitnbarbstatewritereadseqnb-mean-scenario-one-asymptot}
\end{equation}
provided this limit exists.
This result is consistent with the frequentist view of probabilities
{and 
statistical
mean}
because, in the latter view,
the right-hand term of
(\ref{eq-qubitnbarbspaceoveralleventindexequalstdwithvecbasisrvmeanqubitnbarbstatewritereadseqnb-mean-scenario-one-asymptot})
is the statistical mean
of
the samples
\yqubitnbarbspaceoveralleventindexequalstdwithvecbasisprobstatewritereadseqindex .
\subsection{Variance
of
\yqubitnbarbspaceoveralleventindexequalstdwithvecbasisrvmeanqubitnbarbstatewritereadseqnb\
for a finite
\yqubitnbarbstatewritereadseqnb
}
We here consider the 
case when
the number
\ywritereadonestatenb\
of
copies
of each state
\yqubitnbarbtimenonestateseqindex\ is equal to one,
assuming these states 
are independently drawn
{(%
with 
the
same 
distribution
again%
)}
in the second scenario.
For both scenarios,
it may be shown
that
the 
random variables
\yqubitnbarbspaceoveralleventindexequalstdwithvecbasisrvqubitnbarbstatewritereadseqindex\
are 
{uncorrelated}
and that
the variance 
of
\yqubitnbarbspaceoveralleventindexequalstdwithvecbasisrvmeanqubitnbarbstatewritereadseqnb
,
defined with
$
E_{1}
\{
.
\}
$
or
$
E_{2}
\{
.
\}
$ depending on the 
considered 
scenario,
meets
\begin{equation}
\mathrm{var}
\{
\Yqubitnbarbspaceoveralleventindexequalstdwithvecbasisrvmeanqubitnbarbstatewritereadseqnb
\}
=
\frac
{
\sum
_{
\Yqubitnbarbstatewritereadseqindex
=
1
}
^{
\Yqubitnbarbstatewritereadseqnb
}
\mathrm{var}
\{
\Yqubitnbarbspaceoveralleventindexequalstdwithvecbasisrvqubitnbarbstatewritereadseqindex
\}
}
{
\Yqubitnbarbstatewritereadseqnb
^2
}
.
\label{eq-qubitnbarbspaceoveralleventindexequalstdwithvecbasisrvmeanqubitnbarbstatewritereadseqnb-var-vs-each-RV}
\end{equation}
For deterministic values of
the 
quantities
\yqubitnbarbspaceoveralleventindexequalstdwithvecbasisprobstatewritereadseqindex ,
the variance of each 
random variable
\yqubitnbarbspaceoveralleventindexequalstdwithvecbasisrvqubitnbarbstatewritereadseqindex\
is easily calculated,
and
(\ref{eq-qubitnbarbspaceoveralleventindexequalstdwithvecbasisrvmeanqubitnbarbstatewritereadseqnb-var-vs-each-RV})
then yields
\begin{equation}
\mathrm{var}_{1}
\{
\Yqubitnbarbspaceoveralleventindexequalstdwithvecbasisrvmeanqubitnbarbstatewritereadseqnb
\}
=
\frac
{
\sum
_{
\Yqubitnbarbstatewritereadseqindex
=
1
}
^{
\Yqubitnbarbstatewritereadseqnb
}
[
\Yqubitnbarbspaceoveralleventindexequalstdwithvecbasisprobstatewritereadseqindex
-
\Yqubitnbarbspaceoveralleventindexequalstdwithvecbasisprobstatewritereadseqindex
^2
]
}
{
\Yqubitnbarbstatewritereadseqnb
^2
}
.
\label{eq-qubitnbarbspaceoveralleventindexequalstdwithvecbasisrvmeanqubitnbarbstatewritereadseqnb-var-deterministic}
\end{equation}
Since
$
\Yqubitnbarbspaceoveralleventindexequalstdwithvecbasisprobstatewritereadseqindex
\in
[
0
,
1
]
$,
Eq.
(\ref{eq-qubitnbarbspaceoveralleventindexequalstdwithvecbasisrvmeanqubitnbarbstatewritereadseqnb-var-deterministic})
yields
\begin{equation}
0
\leq
\mathrm{var}_{1}
\{
\Yqubitnbarbspaceoveralleventindexequalstdwithvecbasisrvmeanqubitnbarbstatewritereadseqnb
\}
\leq
\frac
{
1
}
{
4
\Yqubitnbarbstatewritereadseqnb
}
.
\label{eq-qubitnbarbspaceoveralleventindexequalstdwithvecbasisrvmeanqubitnbarbstatewritereadseqnb-var-scenario-one-bound}
\end{equation}
For 
stochastic values of
the 
quantities
\yqubitnbarbspaceoveralleventindexequalstdwithvecbasisprobstatewritereadseqindex ,
one similarly obtains
\begin{equation}
\mathrm{var}_{2}
\{
\Yqubitnbarbspaceoveralleventindexequalstdwithvecbasisrvmeanqubitnbarbstatewritereadseqnb
\}
=
\frac
{
\Yqubitnbarbspaceoveralleventindexequalstdwithvecbasisprobexpect
-
{%
(
\Yqubitnbarbspaceoveralleventindexequalstdwithvecbasisprobexpect
)^2
}
}
{
\Yqubitnbarbstatewritereadseqnb
}
\end{equation}
and
\begin{equation}
0
\leq
\mathrm{var}_{2}
\{
\Yqubitnbarbspaceoveralleventindexequalstdwithvecbasisrvmeanqubitnbarbstatewritereadseqnb
\}
\leq
\frac
{
1
}
{
4
\Yqubitnbarbstatewritereadseqnb
}
.
\label{eq-qubitnbarbspaceoveralleventindexequalstdwithvecbasisrvmeanqubitnbarbstatewritereadseqnb-var-scenario-two-bound}
\end{equation}
\subsection{Asymptotic
variance 
of
\yqubitnbarbspaceoveralleventindexequalstdwithvecbasisrvmeanqubitnbarbstatewritereadseqnb 
}
Eq.
(\ref{eq-qubitnbarbspaceoveralleventindexequalstdwithvecbasisrvmeanqubitnbarbstatewritereadseqnb-var-scenario-one-bound})
and
(\ref{eq-qubitnbarbspaceoveralleventindexequalstdwithvecbasisrvmeanqubitnbarbstatewritereadseqnb-var-scenario-two-bound})
directly yield
\begin{eqnarray}
\lim_{
\Yqubitnbarbstatewritereadseqnb
\rightarrow
+
\infty
}
\mathrm{var}_{1}
\{
\Yqubitnbarbspaceoveralleventindexequalstdwithvecbasisrvmeanqubitnbarbstatewritereadseqnb
\}
&
=
&
0
\\
\lim_{
\Yqubitnbarbstatewritereadseqnb
\rightarrow
+
\infty
}
\mathrm{var}_{2}
\{
\Yqubitnbarbspaceoveralleventindexequalstdwithvecbasisrvmeanqubitnbarbstatewritereadseqnb
\}
&
=
&
0
.
\end{eqnarray}
This, together with the above result concerning the 
asymptotic
mean of the
proposed estimator
\yqubitnbarbspaceoveralleventindexequalstdwithvecbasisrvmeanqubitnbarbstatewritereadseqnb
, shows that this estimator is asymptotically
efficient.
{%
This is the counterpart, for our framework, of the
convergence/stability analysis of 
the non-quantum adaptive systems considered in Appendix
\ref{sec-stochastic-QIP-link-with-classical}.}
\section{Considered quantum process and
state properties}\label{sec-heisenberg process}
In this
paper, we 
consider
a device composed of two 
distinguishable
\cite{amoi6-64}
qubits
implemented as electron spins 1/2, that are 
{internally}
coupled according to the
cylindrical-symmetry Heisenberg model,
which is e.g. relevant for spintronics 
applications
\cite{Delgado_2017,
amq88,
amq111}.
The symmetry axis of 
this
model is here denoted as $Oz$. 
The considered
spins are supposed to be placed in a magnetic
field (also oriented
along $Oz$ and with a magnitude \ymagfieldnot) and thus coupled to it.
Moreover, we assume an isotropic
$\overline{\overline{g}}$
tensor,
with principal value $g$.
The time interval when these spins are
considered
is 
supposed to be short
enough for 
their coupling with
their environment 
to be
negligible.
In these conditions, 
the temporal evolution of the state of
the device composed of these two spins
is governed by the following
Hamiltonian:
\yeqdefhamiltonian
where:
\begin{list}{}{\setlength{\leftmargin}{10mm}
	       \setlength{\labelwidth}{8mm}
	       \setlength{\labelsep}{2mm}}
\item[$\bullet$]
$\Yhamiltonfieldscale = g \mu_e$, where
$\mu_e$ is the Bohr magneton, i.e. $\mu_e = e \hbar / 2 m_e =
0.927 \times 10^{-23} J T^{-1}$
and
$
\hbar
$
is the reduced Planck constant,
\item[$\bullet$]
$s_{
\Yqubitindexstd
x}, \ s_{
\Yqubitindexstd
y}, \ s_{
\Yqubitindexstd
z}$, with 
$
\Yqubitindexstd
\in
\{
1,
2
\},
$
are the three components of the vector operator 
\overrightarrow{s_{
\Yqubitindexstd
}}
associated 
with
spin $
\Yqubitindexstd
$ 
in a cartesian frame,
\item[$\bullet$]
$J_{xy}$ and $J_{z}$ are the principal values of the exchange tensor.
\end{list}
Among the above parameters,
the value of $g$
may be experimentally determined,
and
\ymagfieldnot\
can be 
measured.
The values of
$J_{xy}$ and $J_{z}$ 
are 
here assumed to be
unknown.

We 
here suppose that 
each spin
\yqubitoneindexstd ,
with
$
\Yqubitoneindexstd
\in
\{
1,
2
\},
$
is prepared, i.e. initialized,
at a given time
\yqubitonetimeinit,
{in}
the pure
state
\yeqdeftwoqubitstateinitindexi
where
$
|+ \rangle
$
and
$
|- \rangle
$
are
eigenkets of
$s_{
\Yqubitindexstd
z}$,
for the eigenvalues
$1/2$ and $- 1/2$ respectively.
We will further use the polar representation of the
qubit parameters
$\alpha
_{
\Yqubitindexstd
}
$
and
$\beta
_{
\Yqubitindexstd
}
$,
which reads
\yeqdefqubitpolarqubitindexstd
where \ysqrtminusone\ is the imaginary unit, and
with
$
0
\leq
\Yparamqubitindexstdstateplusmodulus
\leq
1
$
and
\yeqdefparamqubitindexstdstateminusmodulusvsparamqubitindexstdstateplusmodulus
because
each spin state
$| \psi
_{
\Yqubitindexstd
}
(
\Yqubitonetimeinit
)
\rangle$
has unit norm.
Moreover, for each couple of phase parameters
\yparamqubitindexstdstateplusphase\
and
\yparamqubitindexstdstateminusphase ,
only their difference has a physical meaning.
After they have been prepared,
these
spins are coupled 
according to
the above-defined
model for
$
t
\geq
\Yqubitonetimeinit
$.

Hereafter,
we consider the state
of the overall system composed of these two
spins.
At time
\yqubitonetimeinit, this state is equal to the tensor product of the
states
of both spins
defined in
(\ref{eq-twoqubit-state-init-index-i}).
It therefore reads
\yeqdefqubitbothtimeinitstatetensorprod
in the four-dimensional basis
$
\Ytwoqubitsbasisplusplus
=
\{ | ++ 
\rangle
, | +- 
\rangle
, | -+ 
\rangle
, | -- 
\rangle
\}
$.

The state of this two-spin system then evolves with time.
Its value 
\ymixsyststatefinal\
at any subsequent time $t$
may be derived from its above-defined Hamiltonian. 
{It is defined
\cite{amoi6-18}} by
\yeqdefstatetfinalvsopmixstatetinitcomponents
where
\yveccompsyststatetinit\
and
\yveccompsyststatetfinal\
are the column vectors of components of
\ymixsyststateinitial\ and
\ymixsyststatefinal , respectively,
in basis
\ytwoqubitsbasisplusplus 
. For instance,
as shown by
(\ref{eq-etat-deuxspin-plusplus-initial-decompos}),
\yeqdefveccompsyststatetinit
where 
$
^T
$ 
stands for transpose.
Moreover, the matrix
\yopmix\ of 
(\ref{eq-statetfinalvsopmixstatetinit-components}),
which defines
the transform
applied to
\ymixsyststateinitial,
reads
\yeqdefopmixmatrixdecompose
with
\yeqdefopmixbasesdef
and
{\yopmixdiag\
equal to} 
\yeqdefopmixdiagdef
The four real (angular) frequencies
$
\omega _{1 , 1}
$
to
$
\omega _{1 , -1}
$
in
(\ref{eq-opmixdiagdef})
depend on 
the
physical setup.
In
\cite{amoi6-18},
it was
shown that they read
\yeqdefomegaallexpress
Since
the values of the parameters
$J_{xy}$ and $J_{z}$
of the Hamiltonian of 
(\ref{eq-online-hamiltonian})
are 
presently unknown,
the values of the parameters
$
\omega _{1 , 1}
$
to
$
\omega _{1 , -1}
$
of the quantum process involved in
(\ref{eq-statetfinalvsopmixstatetinit-components})
are also unknown.
Combining
(\ref{eq-opmixdiagdef})
and
(\ref{eq-omega-one-zero-express})-(\ref{eq-omega-zero-zero-express})
shows that the only quantities
that must be estimated in order to obtain an estimate of
\yopmixdiag\
and hence of
\yopmix\
are
$
\exp
\left[
\Ysqrtminusone
\frac{ 
\Yexchangetensorppalvaluexy
(t - t_0)
} 
{ \hbar }
\right]
$
and
$
\exp
\left[
\Ysqrtminusone
\frac{ 
\Yexchangetensorppalvaluez
(t - t_0)
} 
{ 2 \hbar }
\right]
$.

The (B)QPT problem then consists of
estimating
the matrix
\yopmix\
involved in
(\ref{eq-statetfinalvsopmixstatetinit-components}),
which defines the considered quantum process.
More precisely, its blind, i.e. unsupervised, version
proposed
in
\cite{amoi6-118}
operates as follows:
\begin{itemize}
\item
It uses
values of the output state
\ymixsyststatefinal\
of this process.
\item
It does not use nor know
values of its input state
\ymixsyststateinitial .
\item
But it knows and exploits
some properties of
these states
\ymixsyststateinitial.
In 
\cite{amoi6-118},
these requested properties are as follows.
The states
\ymixsyststateinitial\
are required to be unentangled (as shown by
(\ref{eq-qubitbothtimeinitstate-tensor-prod})).
Besides, the proposed BQPT methods
are statistical approaches and the six parameters
\yparamqubitindexstdstateplusmodulus,
\yparamqubitindexstdstateplusphase\
and
\yparamqubitindexstdstateminusphase,
with
$\Yqubitindexstd \in \{ 1 , 2 \}$,
defined in
(\ref{eq-def-qubit-polar-qubit-indexstd})
are constrained to have
properties that are similar
to those requested in the above-mentioned QSICA methods:
(i) these parameters are random valued, so that we 
consider
random pure quantum states
$| \psi_i 
(
\Yqubitonetimeinit
)
\rangle$
(see 
\cite{amoi6-67}
for more details)
and
(ii) some combinations of
the random variables 
\yparamqubitindexstdstateplusmodulus,
\yparamqubitindexstdstateplusphase\
and
\yparamqubitindexstdstateminusphase\
are statistically independent
and have a few known statistical features,
as detailed 
in
\cite{amoi6-118}.
\end{itemize}
\section{A method for estimating
\yexchangetensorppalvaluexy\
and
\yexchangetensorppalvaluez }
\label{sec-estim-exchangetensorppalvaluexyz}
\subsection{Estimating
\yexchangetensorppalvaluexy }
We here consider the problem of estimating the
Hamiltonian parameter
\yexchangetensorppalvaluexy ,
defined in Section
\ref{sec-blind-hamilton-estim}.
We focus on the practical situation with estimation errors 
for
\ytwoqubitresultphaseevolindexdone\
and
\ytwoqubitresultphaseevolindexdtwo ,
and with a known range for
\yexchangetensorppalvaluexy .
We hereafter show how to exploit
this range in such a way that the values of
\yexchangetensorppalvaluexyestimindexone\
and
\yexchangetensorppalvaluexyestimindextwo\
which are the closest to one another are also those which
are the closest to
\yexchangetensorppalvaluexy .
To this end, one takes into account that
\ytwoqubitresultphaseevolindexdoneestim\
and
\ytwoqubitresultphaseevolindexdtwoestim\
are always in the interval
$\displaystyle\left[
-
\frac{\pi}{2},
\frac{\pi}{2}
\right]$
(because they are values of the arcsin function%
: see
(\ref{eq-twoqubitresultphaseevolindexd})).
This, together with the known range of possible values of
\yexchangetensorppalvaluexyestimindexone ,
the known value of
\ytwoqubitwritereadtimeintervalindexoneone\
and the corresponding version of
(\ref{eq-exchangetensorppalvaluexy-vs-twoqubitresultphaseevolindexd-estim})
defines the range
$\{\Yexchangetensorppalvaluexyshiftindetermintestimminindexone,
\dots,
\Yexchangetensorppalvaluexyshiftindetermintestimmaxindexone\}$
of integers in which it is guaranteed that
\yexchangetensorppalvaluexyshiftindetermintestimindexone\
should be selected.
Similarly, 
the value of
\ytwoqubitwritereadtimeintervalindexonetwo\
is to be selected as explained hereafter, 
and
for the application of the procedure with
any given value
\ytwoqubitwritereadtimeintervalindexonetwo ,
the integer
\yexchangetensorppalvaluexyshiftindetermintestimindextwo\
should be selected in a known interval
$\{\Yexchangetensorppalvaluexyshiftindetermintestimminindextwo,
\dots,
\Yexchangetensorppalvaluexyshiftindetermintestimmaxindextwo\}$.
When the estimation errors 
for
\ytwoqubitresultphaseevolindexdone\
and
\ytwoqubitresultphaseevolindexdtwo\
remain low enough,
the values 
\yexchangetensorppalvaluexyestimindexone\
and
\yexchangetensorppalvaluexyestimindextwo\
of the grids respectively corresponding to
$\Yexchangetensorppalvaluexyshiftindetermintestimminusactualindexone
=
0$
and
$\Yexchangetensorppalvaluexyshiftindetermintestimminusactualindextwo
=
0$
both remain close to their theoretical value
\yexchangetensorppalvaluexy .
Around these values, the two grids almost coincide.
Then, 
for larger values of
$|\Yexchangetensorppalvaluexyshiftindetermintestimminusactualindexone|$
and
$|\Yexchangetensorppalvaluexyshiftindetermintestimminusactualindextwo|$
corresponding to the above-defined intervals,
we here want the associated parts of the two grids to become more
``desynchronized'', i.e. we want the
gaps between the values of the two grids to become
larger.
This is obtained by
adequately
selecting
\ytwoqubitwritereadtimeintervalindexonetwo\
for an arbitrarily 
chosen
value
\ytwoqubitwritereadtimeintervalindexoneone ,
but 
this
should be performed
without knowing where
\yexchangetensorppalvaluexy\ is in the considered interval.
We therefore use a worst-case approach in terms of desynchronization,
for the ideal estimation 
(\ref{eq-twoqubitresultphaseevolindexdoneandtwoestim-ideal})%
, as follows.
The reference point, shared by both grids,
is equal to
\yexchangetensorppalvaluexy\
and is obtained when
$\Yexchangetensorppalvaluexyshiftindetermintestimminusactualindexone
=
0$
and
$\Yexchangetensorppalvaluexyshiftindetermintestimminusactualindextwo
=
0$.
We consider the case when this reference point
is the lowest value in both bounded grids,
i.e. 
$
\Yexchangetensorppalvaluexyshiftindetermintestimminindexone
=
\Yexchangetensorppalvaluexyshiftindetermintindexone
$
and
$
\Yexchangetensorppalvaluexyshiftindetermintestimminindextwo
=
\Yexchangetensorppalvaluexyshiftindetermintindextwo
$.
For any given
\ytwoqubitwritereadtimeintervalindexoneone,
we select
a value
\ytwoqubitwritereadtimeintervalindexonetwo\
which is only
somewhat
larger than
\ytwoqubitwritereadtimeintervalindexoneone,
thus 
considering 
that
$\Yexchangetensorppalvaluexyshiftindetermintestimminindextwo
=
\Yexchangetensorppalvaluexyshiftindetermintestimminindexone$
and
$
\Yexchangetensorppalvaluexyshiftindetermintestimmaxindextwo
=
\Yexchangetensorppalvaluexyshiftindetermintestimmaxindexone$.
The values are then
somewhat closer to one another
in 
the second grid than in the first one.
Moreover,
we select
\ytwoqubitwritereadtimeintervalindexonetwo\
so that, when moving towards the higher values in both
bounded grids,
the gaps between the corresponding points
of the two grids increase,
until they reach the maximum possible gap for the
highest values.
This means that we set
\ytwoqubitwritereadtimeintervalindexonetwo\
so that
the highest value in the
first bounded grid 
(i.e. the value of
\yexchangetensorppalvaluexyestimindexone\
in
(\ref{eq-exchangetensorppalvaluexyestim-minus-actual-vs-exchangetensorppalvaluexyshiftindetermintestimin-minus-actual-indexone})
corresponding to
$
\Yexchangetensorppalvaluexyshiftindetermintestimindexone
=
\Yexchangetensorppalvaluexyshiftindetermintestimmaxindexone
$,
moreover taking into account
(\ref{eq-twoqubitresultphaseevolindexdoneandtwoestim-ideal}))
is equal to 
\ymodifartitionehundredthirtyninevonesteptwo{that of}
the middle of
the interval of the second grid defined as follows:
the lower bound of that interval is the highest
value in the bounded part of that grid
considered here
(i.e. the value of
\yexchangetensorppalvaluexyestimindextwo\
in
(\ref{eq-exchangetensorppalvaluexyestim-minus-actual-vs-exchangetensorppalvaluexyshiftindetermintestimin-minus-actual-indextwo})
corresponding to
$
\Yexchangetensorppalvaluexyshiftindetermintestimindextwo
=
\Yexchangetensorppalvaluexyshiftindetermintestimmaxindextwo
=
\Yexchangetensorppalvaluexyshiftindetermintestimmaxindexone
$,
moreover taking into account
(\ref{eq-twoqubitresultphaseevolindexdoneandtwoestim-ideal}))
and the higher bound of that interval is the
next value that would be found in that grid,
when moving towards higher values,
if that grid 
\ymodifartitionehundredthirtyninevonesteptwo{were}
complete,
i.e. this upper bound is equal to the lower bound plus
$\frac{\hbar \pi}
{\Ytwoqubitwritereadtimeintervalindexonetwo}$.
Using
(\ref{eq-exchangetensorppalvaluexyestim-minus-actual-vs-exchangetensorppalvaluexyshiftindetermintestimin-minus-actual-indexone})
to
(\ref{eq-twoqubitresultphaseevolindexdoneandtwoestim-ideal}),
it may easily be shown that the above desynchronization condition
for the highest values of the two grids yields
\begin{equation}
\frac{ \hbar }
{
\Ytwoqubitwritereadtimeintervalindexoneone
}
(
\Yexchangetensorppalvaluexyshiftindetermintestimmaxindexone
-
\Yexchangetensorppalvaluexyshiftindetermintestimminindexone
)
\pi
=
\frac{ \hbar }
{
\Ytwoqubitwritereadtimeintervalindexonetwo
}
(
\Yexchangetensorppalvaluexyshiftindetermintestimmaxindexone
-
\Yexchangetensorppalvaluexyshiftindetermintestimminindexone
+
\frac{1}{2}
)
\pi
.
\end{equation}
Therefore, for a given value
\ytwoqubitwritereadtimeintervalindexoneone ,
one should set
\ytwoqubitwritereadtimeintervalindexonetwo\
so that
\begin{eqnarray}
\frac{ 
\Ytwoqubitwritereadtimeintervalindexonetwo
 }
{
\Ytwoqubitwritereadtimeintervalindexoneone
}
&
=
&
\frac{  
\Yexchangetensorppalvaluexyshiftindetermintestimmaxindexone
-
\Yexchangetensorppalvaluexyshiftindetermintestimminindexone
+
\frac{1}{2}
}
{
\Yexchangetensorppalvaluexyshiftindetermintestimmaxindexone
-
\Yexchangetensorppalvaluexyshiftindetermintestimminindexone
}
\label{eq-twoqubitwritereadtimeintervalindextwodivindexone-rational-method-one}
\\
&
=
&
\frac{  
2
(
\Yexchangetensorppalvaluexyshiftindetermintestimmaxindexone
-
\Yexchangetensorppalvaluexyshiftindetermintestimminindexone
)
+
1
}
{
2
(
\Yexchangetensorppalvaluexyshiftindetermintestimmaxindexone
-
\Yexchangetensorppalvaluexyshiftindetermintestimminindexone
)
}
.
\end{eqnarray}
The latter expression shows that the value thus obtained in this practical procedure
for a bounded interval on
\yexchangetensorppalvaluexy\
yields a rational value of
$
\frac{ 
\Ytwoqubitwritereadtimeintervalindexonetwo
 }
{
\Ytwoqubitwritereadtimeintervalindexoneone
}
$
(unlike the above preliminary procedure for the ideal case and without
restrictions on the domain of
\yexchangetensorppalvaluexy ).
\subsection{Estimating
\yexchangetensorppalvaluez }
\label{sec-estim-exchangetensorppalvaluez}
The method used for estimating \yexchangetensorppalvaluez\ is very similar to
the approach described above for
\yexchangetensorppalvaluexy .
It is therefore 
more briefly outlined
hereafter.
It uses
the procedure of the second 
part
of the
BQPT method of Section \ref{sec-bqpt},
based on
(\ref{eq-exchangetensorppalvaluez-vs-twoqubitsprobadirxxplusplusdiffminusminusphasediffindexd-intervalindextwo})
and
(\ref{eq-exchangetensorppalvaluez-vs-twoqubitsprobadirxxplusplusdiffminusminusphasediffindexd-estim-intervalindextwo}).
This procedure is here
applied
twice, i.e. with
\ytwoqubitwritereadtimeintervalindextwo\
 of Section \ref{sec-bqpt}
successively replaced by two values denoted as
\ytwoqubitwritereadtimeintervalindextwoone\
and
\ytwoqubitwritereadtimeintervalindextwotwo .
For 
\ytwoqubitwritereadtimeintervalindextwoone ,
combining 
(\ref{eq-exchangetensorppalvaluez-vs-twoqubitsprobadirxxplusplusdiffminusminusphasediffindexd-intervalindextwo})
and
(\ref{eq-exchangetensorppalvaluez-vs-twoqubitsprobadirxxplusplusdiffminusminusphasediffindexd-estim-intervalindextwo})
and using the same type of notations as for
\yexchangetensorppalvaluexy\
yields
\begin{equation}
\Yexchangetensorppalvaluezestimindexone
=
\Yexchangetensorppalvaluez
+
\frac{ \hbar }
{
\Ytwoqubitwritereadtimeintervalindextwoone
}
\left[
\Ytwoqubitsprobadirxxplusplusdiffminusminusphasediffindexdoneestim
-
\Ytwoqubitsprobadirxxplusplusdiffminusminusphasediffindexdone
+
2
\Yexchangetensorppalvaluezshiftindetermintestimminusactualindexone
\pi
\right]
+
(
\Yexchangetensorppalvaluexyestim
-
\Yexchangetensorppalvaluexy
)
\label{eq-exchangetensorppalvaluezestim-minus-actual-vs-exchangetensorppalvaluezshiftindetermintestimin-minus-actual-indexone}
\end{equation}
with
\begin{equation}
\Yexchangetensorppalvaluezshiftindetermintestimminusactualindexone
=
\Yexchangetensorppalvaluezshiftindetermintestimindexone
-
\Yexchangetensorppalvaluezshiftindetermintindexone
\end{equation}
and where
\yexchangetensorppalvaluexyestim\
is the estimate
of
\yexchangetensorppalvaluexy\
without any indeterminacy that was obtained in the first 
part
of
this Hamiltonian parameter estimation method.
This
shows
that
the procedure applied
with the time interval
\ytwoqubitwritereadtimeintervalindextwoone\
yields a regular one-dimensional grid of possible estimates
\yexchangetensorppalvaluezestimindexone\
of
\yexchangetensorppalvaluez ,
with 
a step equal
to
$\frac{2 \hbar \pi}
{\Ytwoqubitwritereadtimeintervalindextwoone}
$.
Its application
with the time interval
\ytwoqubitwritereadtimeintervalindextwotwo\
is analyzed in the same way.
We here exploit the differences between these two grids, by transposing
the approach that we described above for
\yexchangetensorppalvaluexy .
Thus, first
considering the case with no estimation errors 
and with
\yexchangetensorppalvaluez\ equal to the lowest value of both
bounded grids
leads one to select
\ytwoqubitwritereadtimeintervalindextwoone\
and
\ytwoqubitwritereadtimeintervalindextwotwo\
so that
\begin{equation}
\frac{ \hbar }
{
\Ytwoqubitwritereadtimeintervalindextwoone
}
2
(
\Yexchangetensorppalvaluezshiftindetermintestimmaxindexone
-
\Yexchangetensorppalvaluezshiftindetermintestimminindexone
)
\pi
=
\frac{ \hbar }
{
\Ytwoqubitwritereadtimeintervalindextwotwo
}
2
(
\Yexchangetensorppalvaluezshiftindetermintestimmaxindexone
-
\Yexchangetensorppalvaluezshiftindetermintestimminindexone
+
\frac{1}{2}
)
\pi
\end{equation}
and hence
\begin{eqnarray}
\frac{ 
\Ytwoqubitwritereadtimeintervalindextwotwo
 }
{
\Ytwoqubitwritereadtimeintervalindextwoone
}
&
=
&
\frac{  
\Yexchangetensorppalvaluezshiftindetermintestimmaxindexone
-
\Yexchangetensorppalvaluezshiftindetermintestimminindexone
+
\frac{1}{2}
}
{
\Yexchangetensorppalvaluezshiftindetermintestimmaxindexone
-
\Yexchangetensorppalvaluezshiftindetermintestimminindexone
}
\label{eq-twoqubitwritereadtimeintervalindextwodivindexone-rational-method-one-oz}
\end{eqnarray}
where 
the integers
\yexchangetensorppalvaluezshiftindetermintestimminindexone\
and
\yexchangetensorppalvaluezshiftindetermintestimmaxindexone\
are defined by using the same approach as for
\yexchangetensorppalvaluexy ,
here taking into account that
\ytwoqubitsprobadirxxplusplusdiffminusminusphasediffindexdone\
and hence its relevant estimates
are guaranteed to be in the interval
$[ -\pi, \pi ]$
(see 
the expression 
of
\ytwoqubitsprobadirxxplusplusdiffminusminusphasediffindexdone\
in
\cite{amoi6-118})
and that
\yexchangetensorppalvaluexyestim\
and
$
\Yhamiltonfieldscale
\Ymagfieldnot
$
are known.

Then, for the practical situation with estimation errors,
and still with prior knowledge about an interval which contains the
actual value
\yexchangetensorppalvaluez ,
the method proposed for determining \yexchangetensorppalvaluez\
consists of 
comparing each value
\yexchangetensorppalvaluezestimindexone\
of the first bounded grid to each value
\yexchangetensorppalvaluezestimindextwo\
of the second bounded grid
in order to derive the couple of closest values
and then the corresponding estimate
$\displaystyle\frac{
\Yexchangetensorppalvaluezestimindexone
+
\Yexchangetensorppalvaluezestimindextwo
}
{2}$.

\section{Test conditions}
\label{sec-appendix-test-conditions}
We here define the conditions used for
all the tests
reported in Section 
\ref{sec-blind-hamilton-estim-tests}.
The actual values of the parameters of the Hamiltonian
(\ref{eq-online-hamiltonian}) were first selected by using
the following properties.
{Conventional 
Electron Spin Resonance
generally operates at 
$X$ or $Q$ 
bands
(around $10$ and $%
35$
GHz
respectively). For electron spins with 
$g=2$, at $35$
GHz,
the resonance field is near $1.25$~%
T.
In the simulations, we used the
values 
$g=2$ 
and
$B=0.99$~T.
Concerning the exchange coupling, 
we chose 
$J_{z}/k_{B}
\simeq
1$~%
K
and 
$%
J_{xy}/k_{B}=0.3$~%
K.
These values were motivated by
\cite{amoi6-118},
Appendix E of 
\cite{amoi6-18}
and 
\cite{ferretti-phys-rev-2005}%
.
}
As in 
\cite{amoi6-118},
we selected part of the parameters defined above and below
so as to avoid 
specific cases
(see footnote [50] 
of
\cite{amoi6-118}%
),
but this here led us to slightly shift some of these values
as compared with those of
\cite{amoi6-118},
because we here have to take 
\ymodifartitionehundredthirtyninevonestepone{these specific cases into account}
for \emph{four}
time intervals
(\ytwoqubitwritereadtimeintervalindexoneone,
\ytwoqubitwritereadtimeintervalindexonetwo,
\ytwoqubitwritereadtimeintervalindextwoone\
and
\ytwoqubitwritereadtimeintervalindextwotwo)
instead of only three
(\ytwoqubitwritereadtimeintervalindexone\
\ytwoqubitwritereadtimeintervalindextwo,
\ytwoqubitwritereadtimeintervalindexthree)
in
\cite{amoi6-118},
so that the 
blind Hamiltonian parameter estimation (BHPE)
method proposed here is somewhat more constraining
that the BQPT method of
\cite{amoi6-118}.

The parameters of the BHPE method were then set as follows.
The 
six parameters
\yparamqubitindexstdstateplusmodulus,
\yparamqubitindexstdstateplusphase\
and
\yparamqubitindexstdstateminusphase,
with
$\Yqubitindexstd \in \{ 1 , 2 \}$,
of each initial state
\ymixsyststateinitial\
were randomly drawn with a uniform distribution,
over an interval which depends on the
part
of the considered 
BHPE
method, in order to meet
the constraints on the statistics of these parameters
that are imposed by that 
BHPE
method.
The parameters
\yparamqubitonestateminusmodulus\
and
\yparamqubittwostateminusmodulus\
were
then
derived from
(\ref{eq-paramqubitindexstdstateminusmodulus-vs-paramqubitindexstdstateplusmodulus}).
More precisely,
the parameter
\yexchangetensorppalvaluexy\
was first estimated by applying
the procedure of the first 
part
of the
BQPT method of Section \ref{sec-bqpt}
successively to each of the two values
\ytwoqubitwritereadtimeintervalindexoneone\
and
\ytwoqubitwritereadtimeintervalindexonetwo.
For each of these values,
as a first 
step,
to estimate
the absolute value of
\ytwoqubitresultphaseevolsin\
as 
detailed in
\cite{amoi6-118},
the
qubit 
parameter values 
\yparamqubitonestateplusmodulus\
and
\yparamqubittwostateplusmodulus\
were
selected within the 
20\%-80\%
sub-range of their
0\%-100\%
allowed range 
defined 
in
\cite{amoi6-118},
that is,
$
[
0.1
,
0.4
[
$
for
\yparamqubitonestateplusmodulus\
and
$
[
0.6
,
0.9
[
$
for
\yparamqubittwostateplusmodulus ,
as in
\cite{amoi6-42}.
Besides,
\yparamqubitonestateminusphase\
and
\yparamqubittwostateminusphase\
were
drawn
over
$
[
0
,
{2 \pi [}
$
whereas
$
\Yparamqubitonestateplusphase
$
and
\yparamqubittwostateplusphase\
were fixed to 0
(%
{as stated above,}
the parameters which have a physical meaning are
$
\Yparamqubitindexstdstateminusphase
-
\Yparamqubitindexstdstateplusphase
$).
These data are thus such
that
$
E
\{
\sin \Ytwoqubitresultphaseinit
\}
=
0
$%
,
as required by 
this 
step 
of
the considered BQPT method.
Then,
as a second 
step,
to estimate
the sign of
\ytwoqubitresultphaseevolsin\
as 
detailed in
\cite{amoi6-118},
the same conditions as in the above first 
step
were used for
\yparamqubitindexstdstateplusmodulus,
\yparamqubitindexstdstateplusphase\
and
\yparamqubitindexstdstateminusphase,
with
$\Yqubitindexstd \in \{ 1 , 2 \}$,
except that
\yparamqubitonestateminusphase\
was fixed to 0
and
\yparamqubittwostateminusphase\
was drawn
over
$
[
0
,
\pi [
$.
These data are thus such
that
$
E
\{
\sin \Ytwoqubitresultphaseinit
\}
$
is non-zero 
and has a known sign (here, it is positive),
as required by 
this 
step 
of
the considered BQPT method.
The 
above two 
steps 
were performed with
$
\Ytwoqubitwritereadtimeintervalindexoneone
=
0.5
$
ns
and then
\ytwoqubitwritereadtimeintervalindexonetwo\
defined by
(\ref{eq-twoqubitwritereadtimeintervalindextwodivindexone-rational-method-one}),
with
$
\Yexchangetensorppalvaluexyshiftindetermintestimminindexone
=
0
$
and
$
\Yexchangetensorppalvaluexyshiftindetermintestimmaxindexone
=
31
$
because the only prior knowledge 
about
\yexchangetensorppalvaluexy\
which is provided to this
BHPE method is that
$
\Yexchangetensorppalvaluexy
/
k_{B}
$
is in the range
$
[
0,
1.5
\mathrm{K}
]
$
(the upper bound
1.5~K
was selected as 5 times the value
0.3~K,
which was
actually used to create the data
processed in these tests
as explained above).
For
\ytwoqubitwritereadtimeintervalindexonetwo,
the above interval of values of
$
\Yexchangetensorppalvaluexy
/
k_{B}
$
results in
$
\Yexchangetensorppalvaluexyshiftindetermintestimminindextwo
=
0
$
and
$
\Yexchangetensorppalvaluexyshiftindetermintestimmaxindextwo
=
32.
$

The parameter
\yexchangetensorppalvaluez\
was then estimated by applying
the procedure of the second 
part
of the
BQPT method of Section \ref{sec-bqpt}
successively to each of the two values
\ytwoqubitwritereadtimeintervalindextwoone\
and
\ytwoqubitwritereadtimeintervalindextwotwo,
with
$
\Ytwoqubitwritereadtimeintervalindextwoone
=
0.53
$
ns
and then
\ytwoqubitwritereadtimeintervalindextwotwo\
defined by
(\ref{eq-twoqubitwritereadtimeintervalindextwodivindexone-rational-method-one-oz}),
with
$
\Yexchangetensorppalvaluezshiftindetermintestimminindexone
=
-13
$
and
$
\Yexchangetensorppalvaluezshiftindetermintestimmaxindexone
=
7
$
because the only prior knowledge 
about
\yexchangetensorppalvaluez\
which is provided to this
BHPE method is that
$
\Yexchangetensorppalvaluez
/
k_{B}
$
is in the range
$
[
\simeq
0.45
\mathrm{K}
,
\simeq
2.24
\mathrm{K}
]
$
(these two bounds were selected as
the actual value
$
\simeq
1
$~K
respectively divided and multiplied
by
$
\sqrt{5}
$).
For
\ytwoqubitwritereadtimeintervalindextwotwo,
the above interval of values of
$
\Yexchangetensorppalvaluez
/
k_{B}
$
results in
$
\Yexchangetensorppalvaluezshiftindetermintestimminindextwo
=
-13
$
and
$
\Yexchangetensorppalvaluezshiftindetermintestimmaxindextwo
=
7
$.
The proposed method
uses measurements along the
$Oz$ and $Ox$ axes.
For each of the parameters
\yparamqubitindexstdstateplusmodulus,
\yparamqubitindexstdstateplusphase\
and
\yparamqubitindexstdstateminusphase,
with
$\Yqubitindexstd \in \{ 1 , 2 \}$,
we used the same statistics for
measurements along the
$Oz$ and $Ox$ axes.
These statistics were also the same when using
\ytwoqubitwritereadtimeintervalindextwoone\
and
\ytwoqubitwritereadtimeintervalindextwotwo,
and they are
defined as follows.
The proposed method is based on two instances
of Eq. (40) of
\cite{amoi6-118}.
For the first 
instance of this equation,
\yparamqubitonestateplusmodulus\
and
\yparamqubittwostateplusmodulus\
were drawn over
$
[
0.1
,
0.4
[
$
and
\yparamqubitonestateminusphase\
and
\yparamqubittwostateminusphase\
were drawn over
$
[
-
\pi
/
2
,
\pi
/
2
[
$,
whereas
\yparamqubitonestateplusphase\
and
\yparamqubittwostateplusphase\
were fixed to 0.
For the second 
instance of the above equation,
\yparamqubitonestateplusmodulus\
and
\yparamqubittwostateplusmodulus\
were drawn over
$
[
0.6
,
0.9
[
$,
whereas
\yparamqubitonestateminusphase ,
\yparamqubittwostateminusphase,
\yparamqubitonestateplusphase,
and
\yparamqubittwostateplusphase\
were selected in the same way as for the first 
instance of that equation.

\section{Computing the dot product of two kets}
\label{sec-compute-dot-product}
In Section
\ref{sec-classif},
we considered the situation when two known unit-norm
classical-form vectors
\yvecnotstdoneindexone\
and
\yvecnotstdoneindextwo\
are stored in two unit-norm kets
\ystateregisterindexone\
and
\ystateregisterindextwo,
and one then uses quantum circuits from the literature to
compute
the corresponding overlap
$
|
\langle
\Ystateregisternot
					_{1}
\Ystateregisterindextwo
|
^2
$.
We here propose an extension of this approach,
that has not been
reported in the literature to our knowledge, and that allows one to
compute the complex-valued dot product
$
\langle
\Ystateregisternot
					_{1}
\Ystateregisterindextwo
$ itself,
not only its (squared) modulus.
To this end, we first consider 
the classical-form vector
\begin{equation}
\Yvecnotstdoneindexthree
=
\Yvecnotstdoneindexthreescalefact
(
\Yvecnotstdoneindexone
+
\Yvecnotstdoneindextwo
)
\end{equation}
where
\yvecnotstdoneindexthreescalefact\
is real-valued and selected so that
\yvecnotstdoneindexthree\
has unit norm.
\yvecnotstdoneindexthree\
is stored in the unit-norm ket
\begin{equation}
\Ystateregisterindexthree
=
\Yvecnotstdoneindexthreescalefact
(
\Ystateregisterindexone
+
\Ystateregisterindextwo
)
.
\end{equation}
Simple calculations then yield
\begin{equation}
|
\langle
\Ystateregisternot
					_{1}
\Ystateregisterindexthree
|
^2
=
\Yvecnotstdoneindexthreescalefact
^2
(
1
+
|
\langle
\Ystateregisternot
					_{1}
\Ystateregisterindextwo
|
^2
+
2
\Re
(
\langle
\Ystateregisternot
					_{1}
\Ystateregisterindextwo
)
.
\label{eq-dot-product}
\end{equation}
The above-mentioned quantum circuits allow one to compute
the
overlaps
$
|
\langle
\Ystateregisternot
					_{1}
\Ystateregisterindexthree
|
^2
$
and
$
|
\langle
\Ystateregisternot
					_{1}
\Ystateregisterindextwo
|
^2
$,
and
(\ref{eq-dot-product})
then yields
$
\Re
(
\langle
\Ystateregisternot
					_{1}
\Ystateregisterindextwo
)
$.
Similarly,
$
\Im
(
\langle
\Ystateregisternot
					_{1}
\Ystateregisterindextwo
)
$
is obtained by using
the ket
\ystateregisterindexfour\
corresponding to the 
vector
\begin{equation}
\Yvecnotstdoneindexfour
=
\Yvecnotstdoneindexfourscalefact
(
\Yvecnotstdoneindexone
+
\Ysqrtminusone
\Yvecnotstdoneindextwo
)
\end{equation}
where
\yvecnotstdoneindexfourscalefact\
is real-valued and selected so that
\yvecnotstdoneindexfour\
has unit norm.
Similar
calculations then yield
\begin{equation}
|
\langle
\Ystateregisternot
					_{1}
\Ystateregisterindexfour
|
^2
=
\Yvecnotstdoneindexfourscalefact
^2
(
1
+
|
\langle
\Ystateregisternot
					_{1}
\Ystateregisterindextwo
|
^2
-
2
\Im
(
\langle
\Ystateregisternot
					_{1}
\Ystateregisterindextwo
)
.
\end{equation}
The dot
product
$
\langle
\Ystateregisternot
					_{1}
\Ystateregisterindextwo
$
is eventually derived from its above real and imaginary parts.


%

\end{document}